\documentclass[12pt]{article}
\usepackage{graphicx}
\usepackage{float}
\date{}

\newcommand{\ba}{\begin{array}}
\newcommand{\ea}{\end{array}}
\newcommand{\bd}{\begin{displaymath}}
\newcommand{\ed}{\end{displaymath}}
\newcommand{\bi}{\begin{itemize}}
\newcommand{\ei}{\end{itemize}}
\newcommand{\benu}{\begin{enumerate}}
\newcommand{\eenu}{\end{enumerate}}
\newcommand{\be}{\begin{equation}}
\newcommand{\ee}{\end{equation}}
\newcommand{\bea}{\begin{eqnarray}}
\newcommand{\eea}{\end{eqnarray}}
\usepackage{amsfonts}
\usepackage{amssymb}
\usepackage[centertags]{amsmath}
\usepackage{graphicx}
\usepackage{hyperref}
\usepackage{booktabs}
\usepackage{theorem}
\usepackage{array}
\usepackage{epsfig}
\usepackage{colordvi}
\usepackage{color}
\usepackage{ulem}
\usepackage{cite}
\usepackage{color}
\usepackage{subfigure}
\usepackage{pdflscape}
\def\1{\mathbf{1}}
\def\3{\mathbf{3}}
\def\2{\mathbf{2}}

\def\ltap{\ \raisebox{-.4ex}{\rlap{$\sim$}} \raisebox{.4ex}{$<$}\ }
\def\gtap{\ \raisebox{-.4ex}{\rlap{$\sim$}} \raisebox{.4ex}{$>$}\ }

\allowdisplaybreaks[1]

\newcommand{\bec}{\begin{cases}}
\newcommand{\eec}{\end{cases}}
\newcommand{\beq}{\begin{equation*}}
\newcommand{\eeq}{\end{equation*}}
\makeatletter

\newcommand{\Rmnum}[1]{\expandafter\@slowromancap\romannumeral #1@}
\makeatother

\begin{document}

\begin{titlepage}
\vspace*{-15mm}
\begin{flushright}
SISSA 20/2021/FISI\\
IFIC/21-47\\
IPMU21-0082
%arXiv:
\end{flushright}
\vspace*{0.7cm}

\begin{center}
{\bf{\large Neutrino Tomography of the Earth with ORCA Detector}}\\

\vspace{0.4cm} 
F. Capozzi$\mbox{}^{a,b)}$ and S. T. Petcov$\mbox{}^{c,d)}$\footnote{Also at:
Institute of Nuclear Research and Nuclear Energy,
Bulgarian Academy of Sciences, 1784 Sofia, Bulgaria.} 
\\[1mm]
\end{center}
\vspace*{0.50cm}
\centerline{$^{a}$ \it Center for Neutrino Physics, 
Department of Physics, Virginia Tech, Blacksburg, VA 24061, USA}
\vspace*{0.20cm}
\centerline{$^{b}$ \it Instituto de Fisica Corpuscular, Universidad de 
Valencia and} 
\vspace*{0.20cm}
 \centerline{ \it CSIC, Edificio Institutos de Investigacion, 
Calle Catedratico Jose Beltran 2, 46980 Paterna, Spain}
\vspace*{0.2cm}
\centerline{$^{c}$ \it SISSA/INFN, Via Bonomea 265, 34136 Trieste, Italy}
\vspace*{0.2cm}
\centerline{$^{d}$ \it Kavli IPMU (WPI), UTIAS,
% The University of Tokyo Institutes for Advanced Study,
The University of Tokyo, 
Kashiwa, Chiba 277-8583, Japan}
% \vspace*{1.20cm}
\vspace*{0.8cm}

\begin{abstract}
\noindent   Using PREM as a reference model for the 
Earth density distribution we investigate   
the sensitivity of ORCA detector to deviations of the Earth 
i) outer core (OC) density, ii) inner core (IC) density, 
iii) total core density, and iv) mantle density,  
from their respective PREM densities.
The analysis is performed by studying the effects of the Earth matter 
on the oscillations of atmospheric $\nu_{\mu}$, $\nu_e$, $\bar{\nu}_\mu$ 
and $\bar{\nu}_e$. We present results which
illustrate the dependence of the ORCA sensitivity 
to the OC, IC, core and mantle densities 
on the type of systematic uncertainties 
used in the analysis, on the value of the atmospheric neutrino mixing 
angle $\theta_{23}$, on whether the Earth mass constraint is implemented 
or not, and on the way it is implemented, 
and on the type - with normal ordering (NO) or 
inverted ordering (IO) -  of the light neutrino mass spectrum. 
We show, in particular, that in the ``most favorable'' NO case of 
implemented Earth mass constraint, 
``minimal'' systematic errors and $\sin^2\theta_{23}=0.58$,
ORCA can determine, e.g., the OC (mantle) 
density at $3\sigma$ C.L. after 10 years of operation 
with an uncertainty of (-18\%)/+15\%  (of (-6\%)/+8\%).  
In the  ``most disfavorable'' NO 
case of ``conservative'' systematic errors
and  $\sin^2\theta_{23}=0.42$,  
the uncertainty on OC (mantle) density 
reads (-43\%)/+39\% ((-17\%/+20\%), while for  
for $\sin^2\theta_{23} = 0.50$ and 0.58 it is noticeably smaller:
  (-37)\%/+30\% and (-30\%)/+24\%  ((-13\%)/+16\%  and  (-11\%/+14\%)). 
We find also that the sensitivity 
of ORCA to the OC, core and mantle densities 
is significantly worse for IO neutrino mass spectrum.
\end{abstract}

% \vspace{0.2cm}
% Keywords: neutrino physics, neutrino oscillations, 
% matter effects, tomography of the Earth core density.

\end{titlepage}
\setcounter{footnote}{0}

\vspace{-0.4cm}

\newpage

\section{Introduction}

A precise knowledge of the Earth's density distribution 
and of the average densities of the Earth's three different major structures - 
the mantle, outer core and inner core -
is essential for understanding the physical conditions and fundamental 
aspects of the structure and properties of the Earth's interior
(including the dynamics of mantle and core, the bulk composition of 
the Earth's three structures, the generation, properties and evolution 
of the Earth's magnetic field and the gravity field of the Earth) 
\cite{Bolt:1991,Yoder:1995,McDonough:2003,McDonough:2008zz}.
The thermal evolution of the Earth's core, in particular,
depends critically on the density change across the inner core - 
outer core boundary (see, e.g., \cite{Baffet:1991}).

At present our knowledge about the interior composition of the Earth 
and its density structure is based primarily on seismological and 
geophysical data (see, e.g., 
\cite{McDonough:2003,Kennett:1998,Masters:2003}). 
These data were used to construct the 
Preliminary Reference Earth Model (PREM)  \cite{PREM} of the density 
distribution of the Earth. In the PREM model, 
the Earth density distribution $\rho_{\rm E}$ is assumed to be 
spherically symmetric, $\rho_{\rm E} = \rho_{\rm E}(r)$, 
$r$ being the distance from the Earth center,
and there are two major density structures -
the core and the mantle, and
a certain number of substructures (shells or layers).
The mantle has seven  shells in the model, while the core 
is divided into an Inner Core (IC) and Outer Core (OC).  
The mean Earth radius is $R_{\oplus} = 6371$ km; the Earth core
has a radius of $R_c= 3480$ km, 
with the IC and OC extending respectively 
from $r = 0$ to $r = 1221.5$ km, 
and from $r = 1221.5$ km to $r = 3480$ km.
The mean densities of the mantle and the core are 
respectively $\bar{\rho}_{man} = 4.45$ g/cm$^3$ and 
$\bar{\rho}_{c} = 10.99$ g/cm$^3$.

 The determination of the radial density distributions in the mantle and core,
$\rho_{man}(r)$ and $\rho_{c}(r)$, from 
seismological and geophysical data is not direct and suffers from 
uncertainties 
\cite{Bolt:1991,Kennett:1998,Masters:2003}.  
An approximate and perhaps rather conservative estimate of this uncertainty 
for $\rho_{man}(r)$ is $\sim 5\%$; for the core density  
$\rho_{c}(r)$ it is larger and can 
be significantly larger 
\cite{Bolt:1991,Kennett:1998,Masters:2003}
It was concluded in   
\cite{Masters:2003}, in particular, that 
the density increase across the inner core - outer core boundary is 
known with an uncertainty of about 20\%.

 A unique alternative method of determination of the density profile 
of the Earth is the neutrino tomography of the Earth 
\cite{PlaZavatt1973,VolkovaZatsepin1974,Nedyalkov:1981,Nedyalkov:1981yy,Nedyalkov:1982,Nedyalkov:1983,DeRujula:1983ya,Wilson:1983an,Askar:1984,Borisov:1986sm,Borisov:1989kh,Winter:2006vg,Kuo95,Jain:1999kp,Reynoso:2004dt,Gonzalez-Garcia:2007wfs}.
The propagation of the active flavour neutrinos 
and antineutrinos 
$\nu_\alpha$ and $\bar{\nu}_\alpha$, $\alpha=e,\mu,\tau$,
in the Earth is affected by the Earth matter.
The original idea of neutrino Earth tomography is based on 
the observation that the cross section of the neutrino-nucleon 
interaction rises with energy. For neutrinos with energies 
$E_\nu \gtrsim$ a few TeV, the inelastic scattering off protons and neutrons 
leads to absorption of neutrinos and thus to 
attenuation of the initial neutrino flux. The magnitude of the 
attenuation depends on the Earth matter density profile along the neutrino path.
Attenuation data for neutrinos with different path-lengths 
in the Earth carry information about the matter density distribution 
in the Earth interior. 
The absorption method of Earth tomography with accelerator 
neutrino beams, which is difficult (if not impossible) to realise 
in practice was discussed first in \cite{PlaZavatt1973,VolkovaZatsepin1974}
and later in grater detail in  
\cite{Nedyalkov:1981,Nedyalkov:1981yy,Nedyalkov:1982,Nedyalkov:1983,DeRujula:1983ya,Wilson:1983an,Askar:1984,Borisov:1986sm,Borisov:1989kh,Winter:2006vg,Kuo95,Jain:1999kp,Reynoso:2004dt}.

  The oscillations between the active flavour neutrinos and antineutrinos, 
$\nu_{\alpha} \leftrightarrow \nu_{\beta}$ 
and $\bar{\nu}_{\alpha} \leftrightarrow \bar{\nu}_{\beta}$, 
$\alpha,\beta=e,\mu$ 
having energies in the range $E \sim (0.1 - 15.0)$ GeV and 
traversing the Earth can be strongly modified by the Earth matter effects 
(see, e.g., \cite{ParticleDataGroup:2018ovx}).
These modifications depend on the Earth matter density (more precisely,
the electron number density $N_e(r)$, see further)
along the path of the neutrinos. Thus, by studying the effects of Earth matter 
on the oscillations of, e.g., $\nu_\mu$ and $\nu_e$ ($\bar{\nu}_\mu$ and 
$\bar{\nu}_e$) neutrinos traversing the Earth along different trajectories 
it is possible to obtain information about the Earth (electron number) 
density distribution. 

Atmospheric neutrinos (see, e.g., \cite{Gaisser:2002jj}) 
are a perfect tool  for performing Earth tomography.
Consisting of significant fluxes of 
muon and electron neutrinos and antineutrinos,
$\nu_\mu$, $\nu_e$, $\bar{\nu}_\mu$ and $\bar{\nu}_e$, 
produced in the interactions of cosmic rays with the Earth 
atmosphere, they have a wide range of energies 
spanning the interval from a few MeV to multi-GeV to multi-TeV. 
Being produced isotropically in the upper part of the Earth atmosphere 
at a height of $\sim 15$ km, they travel 
distances from $\sim 15$ km to 12742 km before reaching 
detectors located on the Earth surface, 
crossing the Earth along all possible directions 
and thus ``scanning'' the Earth interior.  
The interaction rates that allow to get 
information about the Earth density distribution 
can be obtained in the currently taking data IceCube
experiment \cite{IceCube} and in the future experiments 
PINGU \cite{IceCube-PINGU:2014okk,PINGU}, 
ORCA \cite{KM3Net:2016zxf}, Hyper Kamiokande \cite{HK}  and 
DUNE \cite{DUNE:2018tke}, which are under construction.

The idea of using the absorption method of Earth tomography with 
atmospheric neutrinos was discussed first,  
to our knowledge, in \cite{Gonzalez-Garcia:2007wfs}. 
In 2018 in \cite{Donini:2018tsg} the authors 
used the data of the IceCube experiment on multi-TeV atmospheric  
$\nu_{\mu}$ and $\bar{\nu}_\mu$ with sufficiently long paths in the Earth 
and obtained information about the Earth density distribution, which, although 
not very  precise, broadly agrees with the PREM model. 
More specifically, in \cite{Donini:2018tsg} it is assumed that 
the Earth density distribution is spherically symmetric.
The analysis is performed with a five layer Earth model: 
the inner core, two equal width layers of the outer core 
and two equal width layers of the mantle. 
The densities in each of the five layers are varied independently. 
The external constraints on the Earth total mass 
which is known with a remarkable high precision 
\cite{EarthMI1,EarthMI2,EarthMI3},
was not applied. The results are obtained with the IceCube 
data on the zenith angle dependence of the 
fluxes of up-going atmospheric $\nu_\mu$ and $\bar{\nu}_\mu$ 
producing muons  with energies in the interval  
$E_\mu = (0.4 - 20.0)$ TeV 
\cite{IceCube:2016rnb}. Four different models of the initial fluxes 
of atmospheric $\nu_\mu$ and $\bar{\nu}_\mu$ were used in the analysis. 
The value of the Earth mass found in 
\cite{Donini:2018tsg}, $M^\nu_\oplus = (6.0 ^{+1.6}_{-1.3})\times 10^{24}$ kg, 
is in good agreement  with gravitationally determined value  
\cite{EarthMI1,EarthMI2}, 
$M_\oplus = (5.9722 \pm 0.0006) \times 10^{24}~{\rm kg}$.
Thus, the Earth was ``weighted'' with neutrinos.
The results obtained in \cite{Donini:2018tsg}  
contain evidence at $2\sigma$ C.L. 
that the core is denser than the mantle:
$\bar{\rho}^\nu_{c} ({\rm 3layer}) 
- \bar{\rho}^\nu_{man}({\rm 2layer}) 
= (13.1^{+5.8}_{-6.3})$ g/cm$^3$,
where $\bar{\rho}^\nu_{c} ({\rm 3layer})$ and 
$\bar{\rho}^\nu_{man}({\rm 2layer})$ are the values of the average 
core and mantle densities determined in \cite{Donini:2018tsg}. 
This was the first time 
the study of neutrinos traversing the Earth provided information 
of the Earth interior and marked the beginning of 
real experimental data driven neutrino tomography of the Earth.

The Earth tomography based on the study of the effects of Earth matter 
on the oscillations of atmospheric neutrinos 
with different path-lengths in the Earth is discussed in
\cite{Winter:2015zwx,Bourret:2017tkw,Kumar:2021faw}.
In \cite{Winter:2015zwx} 
the sensitivity of PINGU and ORCA experiments to the radial density 
distribution of the Earth has been investigated. 
The analysis is performed by dividing 
the PREM density distribution in seven layers as a function of the 
radial distance $d$ from the Earth surface: 
1. Crust, $0\lesssim d \lesssim 35$ km; 2. Lower Lithosphere, 
$35~{\rm km} \lesssim d \lesssim 60$ km; 
3. Upper Mesosphere,  $60~{\rm km} \lesssim d \lesssim 410$ km;
4. Transition zone, $410~{\rm km} \lesssim d \lesssim 660$ km;
5. Lower Mesosphere, $660~{\rm km} \lesssim d \lesssim 2860$ km;
6. Outer Core, $2860~{\rm km} \lesssim d \lesssim 5151$ km;
7. Inner Core, $5151~{\rm km} \lesssim d \lesssim 6371$ km.
The layers 2, 3, 4 and 5 correspond to the mantle in PREM.
The oscillation probabilities are evaluated 
by dividing further the layers into certain number of shells 
with constant densities chosen to match the PREM average densities 
in each of the shells and by using the evolution operator method 
(see, e.g., \cite{Ohlsson:1999um}). In the analysis the densities 
in each layer are varied independently.
The constraints of the Earth total mass 
was not taken into account. Thus, some of the independent variations of 
the densities in the layers performed in the analysis violate 
this constraint.
The azimuth-averaged (``solar-minimum'') 
$\nu_{e,\mu}$ and $\bar{\nu}_{e,\mu}$ atmospheric neutrino fluxes 
are taken from  \cite{Honda:2013}
and correspond for PINGU and ORCA 
to the South Pole and Gran Sasso sites. The detection characteristics and the 
simulation of events in PINGU and ORCA are based respectively on 
ref. \cite{IceCube-PINGU:2014okk} and  ref. \cite{Yanez:2015}.
The best fit oscillation parameters and their respective uncertainties 
are taken from \cite{Gonzalez-Garcia:2014bfa}.
The results are obtained by the $\chi^2$-minimisation method 
(for further details of the analysis see  \cite{Winter:2015zwx}).
It is found in \cite{Winter:2015zwx} that using neutrino oscillations 
it is impossible to get information with the PINGU and ORCA set-ups
 about the densities in the Crust, Lower Lithosphere and the Inner Core,
while the information about the densities of the 
Upper Mesosphere and the Transitions zone is very imprecise. 
For example, the density in the Transition zone can be determined 
with ORCA with $1\sigma$ uncertainty of -61.2\%/+35.6\%
(-52.7\%/+45.8\%) for neutrino mass spectrum  
with normal (inverted) ordering (NO (IO) spectrum).
The sensitivity of PINGU and ORCA to the densities of the 
Lower Mesosphere $\tilde{\rho}_{\rm LM}$ and the 
Outer Core $\tilde{\rho}_{\rm OC}$, is found to be significantly higher. 
In the case of NO (IO) spectrum,  $\tilde{\rho}_{\rm LM}$ and 
$\tilde{\rho}_{\rm OC}$ can be determined, e.g., with ORCA, 
according to \cite{Winter:2015zwx}, with 
$1\sigma$ uncertainties respectively of 
$\pm 4\%$ (-4.7\%/+4.8\%) and -5.4\%/+6.0\% 
(-6.5\%/+7.1\%).

In \cite{Bourret:2017tkw} the authors have analysed the 
angular and energy distributions of the events in the ORCA detector 
with the aim of obtaining information on the composition of the Earth core.
They conclude, in particular, that for NO (IO) spectrum, 
after ten years of operation of ORCA the average electron number 
density in the mantle and in the outer core, 
for which radial distribution is assumed, can be determined with 
a precision of $\pm 3.6\%$ ($\pm 4.6\%$) and $\pm 7.4\%$ ($\pm 10.0\%$) 
at $1\sigma$ C.L. 
 These results are obtained accounting only for the statistical errors
of the measurements. In addition, the Earth total mass constraint 
was not taken into account when varying the density of the mantle or 
of the outer core.

The possibility to obtain evidence for the existence 
of the Earth's denser core using the atmospheric neutrino data 
from the future planned Iron Calorimeter (ICAL) detector
at the India-based Neutrino Observatory \cite{ICAL:2015stm}
was studied in \cite{Kumar:2021faw}. The authors assume, 
following the PREM model, that the average core density is larger than 
the average mantle density by a factor $\sim 2.5$. 
The results obtained in  \cite{Kumar:2021faw} show, in particular,
that using prospective 10 year ICAL data, the simple two-layered 
mantle-crust Earth density profile can be disfavored 
with a median $\chi^2$ of 7.45 (4.83) if the case of NO (IO) 
neutrino mass spectrum,
which would provide additional neutrino evidence for the 
 existence of the Earth's denser core. 

 In the somewhat related studies 
\cite{Agarwalla:2012uj,Rott:2015kwa,Bourret:2020zwg} the authors 
have analysed the IceCube sensitivity to the Earth matter effects 
in oscillations of atmospheric neutrinos \cite{Agarwalla:2012uj} 
and the  IceCube \cite{Rott:2015kwa} 
and ORCA \cite{Bourret:2020zwg} sensitivities to the Earth core 
composition. In  \cite{Rott:2015kwa,Bourret:2020zwg}
the PREM density distribution of the Earth core was used as 
input in the corresponding analyses.

 Estimates of the sensitivity to the Earth 
core density of large $\sim$1 Mt 
(SuperKamiokande-like) water Cerenkov and 
$\sim 100$ Kt liquid argon (LAr) detectors 
using atmospheric neutrino oscillation data 
were made in \cite{ChGSTP2011}. 
For the PINGU detector similar estimates was made in 
\cite{ChSTP2014}. A brief account of the studies 
performed in \cite{ChGSTP2011,ChSTP2014} 
and the results obtained therein is given in Appendix A.

  Using PREM as a reference model for the 
Earth density distribution we investigate  in the present article 
the sensitivity of the 
ORCA detector to deviations of the Earth 
i) outer core (OC) density, ii) inner core (IC) density, 
iii) total core density, and iv) mantle density,  
from their respective PREM densities. We consider the case when 
the radial dependence of the densities of the layers of interest, 
$\rho_{i}(r)$, $i= {\rm IC,OC,core,mantle}$, is given by PREM and 
the deviations correspond to an overall scaling factor, i.e., 
have the form $\rho^\prime_{i}(r)  = (1 + \kappa_i)\rho_{i}(r)$, 
where $\kappa_i$ is a real positive or negative constant. 
The change of density in each Earth layer (IC, OC 
and mantle) as described by PREM is taken effectively into account, i.e., 
we do not use the  constant density approximation in the layers and shells. 
The analysis is performed by studying the effects of the Earth matter 
on the oscillations of atmospheric $\nu_{\mu}$, $\nu_e$, $\bar{\nu}_\mu$ 
and $\bar{\nu}_e$. 
For the unoscillated fluxes of the atmospheric neutrinos 
we use the updated azimuth-averaged energy and zenith angle dependent fluxes 
from  \cite{Honda:2015fha}  at the Frejus cite 
\footnote{In \cite{Winter:2015zwx} the atmospheric neutrino fluxes from 
\cite{Honda:2013} at the Gran Sasso cite, which is more distant 
from the ORCA location than the Frejus cite, were used.
}.
The type of light neutrino mass spectrum is assumed to be known 
and we obtain results for both the NO and IO spectra.
The relevant detection characteristics of the ORCA set-up - 
the energy and angular resolutions, the dependence of the effective volumes 
for the different classes of events on the initial neutrino energy,
the prospective systematic uncertainties, etc. are taken from the    
the ORCA proposal \cite{KM3Net:2016zxf}. In our analysis we take into 
account also a number of potential systematic uncertainties identified 
in \cite{Capozzi:2017syc} and
\footnote{The study of ORCA sensitivity to the Earth density structure 
was performed in \cite{Winter:2015zwx} prior to the publication of the 
ORCA proposal \cite{KM3Net:2016zxf} and, as a consequence,
with less systematic error sources than those used   
in the present work.
}
we show the dependence of the results 
on the type of systematic uncertainties used in the respective analysis.
In what concerns the statistical errors, our results correspond to 10 
years of operation of ORCA.

 Determining the sensitivity of ORCA to the densities of the different 
Earth structures (or layers) requires to vary the density of a given 
structure (layer)
with respect to its PREM density. 
Such variation can be incompatible with the total Earth mass value.
In order to avoid this in our analysis we systematically 
implement the total Earth mass constraint. This is done 
by compensating the variation of the density in a given 
 structure or layer by a corresponding change of the density 
in one of the other 
structures or layers. For example, when we vary 
the OC density, 
we compensate it by a corresponding variation of the i) IC density, 
and of the ii) mantle density, so that the total Earth mass constraint 
is always satisfied. In order to assess the effects of this constraint 
we present also results without imposing it
\footnote{ As the author of \cite{Winter:2015zwx} indicates,
the total Earth mass constraint was not implemented in the analysis 
performed  in \cite{Winter:2015zwx}. This implies that at least some
of the cases of independent variation of the densities 
in the six layers considered in \cite{Winter:2015zwx} are unphysical.
}.

The paper is organised as follows.
In Section \ref{sec:basics} we discuss the basic ingredients
of the analysis performed by us, 
including  the PREM input used, the calculation of the relevant 
neutrino oscillation probabilities, the implementation of the 
Earth mass constraint and the simulation of events in ORCA. 
In Section \ref{sec:results} we report our results on the sensitivity 
of ORCA to the IC, OC and mantle densities in the cases of NO and IO 
neutrino mass spectra. A summary and the conclusions of our work 
are presented in Section \ref{sec:summary}. 
Appendix A contains a brief account of the studies 
performed in \cite{ChGSTP2011,ChSTP2014} 
and the results obtained therein. 
%%%%%%%%%%%%%%%%%%%%%%%%%%%%%%%%%%
%
\section{Basics of the Analysis}
\label{sec:basics}
%
%%%%%%%%%%%%%%%%%%%%%%%%%%%%%%%%%%%
% 

\vspace{0.3cm}
\leftline{{\bf A. PREM Input and Calculation of  Oscillation 
Probabilities}}

\vspace{0.3cm}
 We use the PREM model as a reference model of the 
Earth density distribution $\rho_{\rm E}(r)$ 
and  assume that the location of the 
mantle-core and the outer core-inner core  
boundaries are correctly described by the model. 
Compared to seismic waves, which are usually 
reflected or refracted at density jumps, neutrino oscillations are 
essentially not sensitive to changes of density 
in the Earth mantle and core taking place over distances 
which are smaller than the neutrino oscillation length 
\cite{Ohlsson:2001ck} (see also \cite{ParticleDataGroup:2018ovx})
that in the cases we are going to study 
is typically $\sim 1000$ km 
\footnote{The estimate we quote refers to the relevant quantity, 
$L_{\rm osc}/(2\pi)$, $L_{\rm osc}$ being the neutrino oscillation length.
}. 
Thus, they are sensitive, in general, to the average densities of 
the mantle (or layers of the mantle having width $\sim 1000$ km),
and of the inner core  and the outer core
(or possibly of two layers of the outer core each having width $\sim 1000$ km). 
However, through the mantle-core interference 
(or neutrino oscillation length resonance- (NOLR-) like) effect 
\cite{Petcov:1998su,Chizhov:1999az,Chizhov:1999he}
(see also \cite{Chizhov:1998ug,AMS:2006hb,AMS:2005yj} 
and references quoted therein), 
the neutrino oscillations are sensitive to 
the difference of the  densities 
of the mantle and the core, i.e., to the magnitude 
of the density ``jump'' in the mantle-core 
narrow transition zone. They might be sensitive also 
to the difference between the IC and OC (average) densities. 

 Some of our results will be obtained 
by assuming that the density distribution in the mantle, 
which is known with a relatively good precision 
\cite{Masters:2003},
is correctly described by the PREM model.  
We recall that, according to the PREM model, 
the Earth core has a radius of $R_c= 3480$ km, 
the IC has a radius of $R_{\rm IC} = 1221.5$ km and 
OC extends radially from $R_{\rm IC} = 1221.5$ km to $R_{\rm C} = 3480$ km,
so the OC radial width is  $2258.5$ km.
The Earth mantle depth is 2856 km.
The mean densities of the mantle and the core are 
respectively $\bar{\rho}_{man} = 4.45$ g/cm$^3$ and 
$\bar{\rho}_{c} = 10.99$ g/cm$^3$, 
while the mean densities of the IC and OC 
are  $\bar{\rho}_{\rm IC} = 12.89$ g/cm$^3$ and
$\bar{\rho}_{\rm OC} = 10.90$ g/cm$^3$.

For a spherically symmetric Earth density
distribution, the neutrino trajectory
in the Earth is specified by the value
of the nadir angle $\theta_n$ of the trajectory.
For $\theta_n\leq 33.17^{o}$,
or path lengths $L \geq 10665.7$ km,
neutrinos cross the Earth core.
The path length for neutrinos which
cross only the Earth mantle is given
by $L =  2 R_{\oplus}\cos \theta_n$.
If neutrinos cross the Earth core,
the lengths of the paths in the mantle,
$2L^{\rm man}$, and in the core,
$L^{\rm core}$, are determined by:
$L^{\rm man} = R_{\oplus}\cos \theta_n -
(R_c^2 - R^2_{\oplus}\sin^2\theta_n)^{1\over 2}$,
$L^{\rm core} = 2(R_c^2 - R^2_{\oplus}\sin^2\theta_n)^{1\over 2} $.
Correspondingly, the neutrinos cross the 
core, the inner core and the outer core
for $\cos \theta_n$ lying respectively 
in the intervals [0.84,1.00], [0.98,1.00] and [0.98,0.84] 
(the corresponding intervals in $\theta_n$ read [0,33.17$^\circ$], 
[0,10.98$^\circ$] and [10.98$^\circ$,33.17$^\circ$]).

 The Earth matter effects in the neutrino oscillations of interest 
depend on the matter potential 
\cite{MSW1,Barger80,Langa83}
%%%%%%%%%%%%%%%%%%%%%%%%%%%%%%%%%%%%%
\begin{equation}
V = \sqrt{2}\,G_{\rm F}\, N_e\,,
\label{eq:V}
\end{equation}
%%%%%%%%%%%%%%%%%%%%%%%%%%%%%%%%%%%%
%
which involves the electron number density 
$N_e$ along the path of the neutrinos.
The relation between the Earth density and 
electron number density includes the electron 
fraction number $Y_e$ (or $Z/A$ factor) 
of the corresponding Earth 
structure or layer: $N^{(E)}_{e}(r) = \rho_{\rm E}(r)\,Y_e/m_{\rm N}$, 
where $m_{\rm N}$ is the nucleon mass. 
For isotopically symmetric matter 
$Y_e = 0.5$. However, the compositions of the 
Earth mantle and core are not exactly 
isotopically symmetric.
For the outer core, for example, 
different composition models give a value 
of $Y_e$ in the interval $Y_e^{oc} = 0.466 - 0.471$ 
(see, e.g., 
\cite{Bardo:2015,Kaminski:2013,Sakamaki:2009,McDonough:2003,EarthRef}).
The value of $Y_e$ in the mantle is closer to 0.5 
\cite{PREM,McDonough:2003}:
 $Y_e^{man} = 0.490 - 0.496$. 
In this study we will use the following default values of 
$Y_e$ in the mantle and the core: 
$Y_e^{man} = 0.490$ and  $Y_e^{c} = 0.467$~ 
\footnote{The relative density deviations from the PREM 
reference densities to which ORCA may be sensitive 
we are going to derive do not depend on the specific 
choices of  $Y_e^{man}$ and  $Y_e^{c}$ (see further).
}. 

  As is well known, for NO (IO) neutrino mass spectrum, 
the matter effects can lead to strong enhancement of the neutrino 
(antineutrino) transition probabilities of interest 
$P(\nu_\alpha \rightarrow \nu_\beta) \equiv P_{\alpha\beta}$ and 
$P(\bar{\nu}_\alpha \rightarrow \bar{\nu}_\beta) \equiv \bar{P}_{\alpha\beta}$, 
$\alpha\neq \beta = e,\mu$, and $\alpha =e$, $\beta = \tau$,
for neutrino energies  $E\sim (6-10)$ GeV and $\sim (3-5)$ GeV, 
corresponding respectively to the resonance 
in the mantle \cite{MSW1,MSW2} (see also \cite{Barger80})
and to the mantle-core interference 
(NOLR) effect \cite{Petcov:1998su}.
Although most of the Earth density dependent effects are 
contained in the energy interval $E\sim (2-10)$ GeV,
we will perform our analysis in a large energy interval,
$E = (2-100)$ GeV, since, in particular, due to the not very good ORCA 
energy resolution at low energies (see further), the matter effects can 
appear above 10 GeV in the reconstructed neutrino energy.

 For the unoscillated fluxes of atmospheric $\nu_\mu$, $\nu_e$,  
and $\bar{\nu}_\mu$, $\bar{\nu}_e$, 
$\Phi_\alpha(\theta_n,E)$ and $\bar{\Phi}_\alpha(\theta_n,E)$,
we use azimuth-averaged double differential 
$d^2\Phi_\alpha/(d\cos\theta_n dE)$
and $d^2\bar{\Phi}_\alpha/(d\cos\theta_n dE)$ updated fluxes from 
\cite{Honda:2015fha} 
\footnote{More precisely, we use the
``solar minimum, without mountain over the detector'' 
fluxes at the Frejus cite
from the tables given in the web site quoted 
in \cite{Honda:2015fha}.}
at the Frejus cite  which is close to the ORCA site.
The ``source'' of atmospheric neutrinos is assumed to be 
a layer located at 15 km above the Earth surface.

 The energy spectra and nadir angle dependencies of the 
fluxes of atmospheric neutrinos crossing the Earth
before reaching the detector are
modified by the neutrino oscillation probabilities 
$P_{\alpha\beta}$ and $\bar{P}_{\alpha\beta}$, 
$\alpha = e,\mu$, $\beta = e,\mu,\tau$.
The Earth is divided into five principal shells: inner core, 
outer core, lower mantle, transition zone,
and upper mantle. 
The radial distances where there is a transition between 
the lower mantle and transition zone, and the
transition zone and upper mantle are at  5714.8 km 
and 6014.2 km, respectively. 
For sub-horizon trajectories ($\theta_n < 90^\circ$), 
the effects of Earth matter on 
$P_{\alpha\beta}$  and $\bar{P}_{\alpha\beta}$ are calculated 
up to the second order in Magnus expansion in each Earth shell 
as described in \cite{Fogli:2012ua}. 
 The variation of density in each of the five shells is described  
 by a bi-quadratic polynomial  as was done in
 \cite{Lisi:1997yc,Fogli:2012ua,Capozzi:2015bxa}.
 The coefficients in these polynomials 
 are fixed by the requirement to reproduce in the corresponding shell 
 the variation of density in the PREM model.
Thus, the change of density in each shell 
as described by PREM is taken into account, i.e., we do not use 
the constant density approximation in the shells. 

The oscillation probabilities of interest 
$P_{\alpha\beta}$ and $\bar{P}_{\alpha\beta}$
depend on all the oscillation parameters 
$\Delta m^2_{31(23)}$, $\Delta m^2_{21}$ 
$\sin^2\theta_{12}$, $\sin^2\theta_{13}$, 
$\sin^2\theta_{23}$ and $\delta$, where $\Delta m^2_{31(23)}$ 
corresponds to NO (IO) neutrino mass spectrum, and $\delta$ is the 
Dirac CP violation phase (see, e.g., \cite{ParticleDataGroup:2018ovx}).
In our analysis we use the 
following reference (true) input values of 
$\Delta m^2_{31(23)}$, $\Delta m^2_{21}$ 
$\sin^2\theta_{12}$, $\sin^2\theta_{13}$ and $\delta$
\cite{Capozzi:2017ipn}~
\footnote{The reference values we use are compatible within $1\sigma$ C.L. 
with the best fit values obtained in the latest global analysis in 
\cite{Capozzi:2021fjo}.
}:
%%%%%%%%%%%%%%%%%%%%%%%%%%
\begin{eqnarray} 
\label{dm3132}
& \Delta m^2_{31(23)} = 2.50\times 10^{-3}\,{\rm eV^2}\,,\\[0.30cm]
& \Delta m^2_{21} = 7.54\times 10^{-5}{\rm eV^2}\,,~~~\sin^2\theta_{12} = 0.308\,,
\label{dn21th12}
\\[0.30cm]
\label{th13delta}
& \sin^2\theta_{13} = 0.0215\,,~~~\delta = 3\pi/2\,.
\end{eqnarray}
%%%%%%%%%%%%%%%%%%%%%%%%
%
For neutrino energies of interest the probabilities 
$P_{\alpha\beta}$ and $\bar{P}_{\alpha\beta}$ exhibit 
rather strong dependence on $\theta_{23}$ 
\cite{Chizhov:1998ug,ADLS1998,Bernabeu:2003yp,Petcov:2005rv} 
(see also \cite{ParticleDataGroup:2018ovx}).
This parameter and the CP violation (CPV) 
phase $\delta$ are still determined in the global 
analyses of the neutrino oscillation data with 
a relatively large uncertainty 
(see, e.g., \cite{Capozzi:2017ipn,Capozzi:2021fjo}).
In the statistical analysis we perform 
we vary the true value of  $\sin^2\theta_{23}$  
in the range  $\sin^2\theta_{23} \in [0.4,0.6]$.
The test value of $\sin^2\theta_{23}$
is left free in the interval $[0,1]$, without any external prior.
The true value of $\delta$ is fixed at $3\pi/2$, but its test value is
left free in the interval $[0,2\pi]$, also without any external prior.
The effects of  the deviations of the test values of 
$\Delta m^2_{31(23)}$ and $\sin^2\theta_{13}$
from their true values  are taken into account through 
the pull method \cite{Fogli:2002pt}, assuming the $1\sigma$ 
uncertainties reported in \cite{Capozzi:2017ipn}. 
These uncertainties are somewhat larger than 
those reported in the latest global data analysis 
in \cite{Capozzi:2021fjo}, but this does not 
have a significant effect on our results. 
The test values of $\Delta m^2_{21}$ and $\sin^2\theta_{12}$ 
are kept fixed at their true values.
The analysis is performed assuming that 
the type of neutrino mass spectrum 
- with normal ordering or with inverted ordering - is known.  

\vspace{0.3cm}
\leftline{{\bf B.  Total Earth Mass Constraint}}

\vspace{0.3cm}
In order to estimate the sensitivity of the ORCA detector 
to the IC, OC, core (IC + OC) and mantle densities, 
we vary the density in each of these four structures.
We do this variation in one structure at a time and implement 
the total Earth mass constraint by compensating that variation 
by corresponding change of the density in one of the 
other structures. To be more specific,
the total Earth mass in the case of interest is given by:
%%%%%%%%%%%%%%%%%%%%%%%%%%%%
\begin{equation}
M_\oplus = \int^{R_\oplus}_{0} 4\pi \rho_{\rm E}(r) r^2 dr 
= 4\pi \left [ 
\int^{R_{\rm IC}}_{0} \rho_{\rm IC}(r) r^2 dr + 
\int^{R_{\rm C}}_{R_{\rm IC}} \rho_{\rm OC}(r) r^2 dr +
\int^{R_{\oplus}}_{R_{\rm C}} \rho_{\rm man}(r) r^2 dr \right ]\,,
\label{eq:ME2}
\end{equation}
%%%%%%%%%%%%%%%%%%%%%%%%%%%%
%
where  $\rho_{\rm IC}(r)$,  $\rho_{\rm OC}(r)$ and $\rho_{\rm mant}(r)$
are the IC, OC and mantle densities as a function of $r$ and 
$R_{\rm IC}$, $R_{\rm C}$ and $R_\oplus$ were defined and their 
values given earlier. For simplicity we did not indicate 
in eq. (\ref{eq:ME2}) the division of the mantle  
in three layers we employ. As we have discussed, 
the variation of density in each of the five 
layers we consider (IC, OC and the three mantle layers)
is parametrised by bi-quadratic polynomials 
in such a way as to reproduce the change of density 
in the PREM model. When we change the density in a given 
layer  $\rho_{i}(r)$, $i= {\rm IC,OC,man}$, 
by a factor $(1+\kappa_i)$, where $\kappa_i$ is 
$r$-independent real constant, it means that 
we multiply the corresponding density distribution 
by the same factor:  $\rho_{i}(r) \rightarrow (1+\kappa_i) \rho_{i}(r)$.
We will present results on sensitivity of ORCA to  
$\Delta \rho_i = 100\%\,((1+\kappa_i) \rho_{i}(r) - \rho_{i}(r))/\rho_{i}(r) 
= 100\%\,\kappa_i$.
It follows from eq. (\ref{eq:ME2})
that when we increase (decrease) the density in one layer,
in order to keep $M_\oplus$ 
unchanged, we have to decrease (increase) the density in one 
of the two, or in both, other layers. 
The factor by which that has to be done depends on the 
relative volumes of the layers.

 The approach thus described in the preceding paragraph 
is implemented in our analysis.
When we consider the variation of IC density
$\rho_{\rm IC}(r)$, we study two cases: we compensate it by 
the corresponding change of density of 
i) the outer core $\rho_{\rm OC}(r)$, and 
ii) of the mantle $\rho_{\rm man}(r)$. 
We proceed in a similar way when we analyse the sensitivity 
of ORCA to the OC and mantle densities  
 $\rho_{\rm OC}(r)$ and $\rho_{\rm man}(r)$: 
in these two cases we investigate respectively two and three    
ways of compensating the variation of 
the respective densities - 
by the change of $\rho_{\rm IC}(r)$ or of $\rho_{\rm man}(r)$,
and by the change of  $\rho_{\rm IC}(r)$ or of $\rho_{\rm OC}(r)$  
or else of $\rho_{\rm IC}(r) + \rho_{\rm OC}(r) =\rho_{\rm C}(r)$.
We consider also the sensitivity of ORCA to the 
core density $\rho_{\rm C}(r)$. In this case   
the variation of  $\rho_{\rm C}(r)$ 
is compensated by the change of  $\rho_{\rm man}(r)$.
In order to assess the effect of the Earth total 
mass constraint we obtained results on the ORCA's sensitivity 
to the mantle, outer core, inner core and total core densities without 
imposing this external constraint.

 The method of implementing the total Earth mass constraint 
using the average IO, OC, core and mantle densities, although less precise, 
gives quite similar results. We give an example of 
how the method we employ works by varying the average 
OC density and compensating this variation with a change
of the average mantle density. The average densities of the 
mantle and the outer 
core are $\bar{\rho}_{man} = 4.45$ g/cm$^3$ and 
$\bar{\rho}_{\rm OC} = 10.90$ g/cm$^3$, respectively. 
The contribution to the mass of the Earth depends also on the
volume of the layer. The ratio between the volume of the mantle 
and the volume of the outer core is 
approximately 5.3. Therefore, a change 
the outer core density by, e.g., 10\%,
should be compensated with 
a variation of the mantle density by   
10\%$\times$ 10.90/(4.45$\times$ 5.3) = 4.62\%.

 As we have indicated, the 
neutrino oscillation probabilities relevant for our analysis
depend on the Earth electron number density $N^{(E)}_e$ and not 
directly on the Earth matter density $\rho_{\rm E}(r)$:
$N^{(E)}_{e}(r) = \rho_{\rm E}(r)\,Y_e/m_{\rm N}$.
In our analysis we fixed the electron fraction numbers 
(or the Z/A factors) of the mantle and the core to the 
following reference values:
 $Y_e^{man} = 0.490$ and  $Y_e^{c} = 0.467$.
Therefore when we vary 
$\rho_{\rm IC}(r)$, $\rho_{\rm OC}(r)$ $\rho_{\rm C}(r)$ and  $\rho_{\rm man}(r)$ 
we vary the electron number densities in these layers,
$N^{(i)}_{e}(r)$, $i={\rm IC,OC,C,man}$.
For any fixed value of $Y_e^{i}$ of a given layer, 
the quantity $\Delta \rho_i$ 
we will determine statistically from prospective ORCA data 
does not depend on  $Y_e^{i}$. As a consequence we have:
$\Delta \rho_i = 100\%\,((1+\kappa_i)N^{i}_e(r) - N^{i}_e(r))/N^{i}_e(r) 
=\Delta N^{(i)}_e$, i.e., our results on sensitivity of ORCA to 
$\Delta \rho_i$ are also results on sensitivity of ORCA 
to deviations of the electron number densities of the different Earth 
layers from their PREM reference values.

We note also that the uncertainties in the values of 
$Y_e^{man}$ and  $Y_e^{c}$ 
 \cite{Bardo:2015,Kaminski:2013,Sakamaki:2009,McDonough:2003}
induce uncertainties in the sensitivity of ORCA 
to the IC, OC, core and mantle matter densities 
which are much smaller than those due to 
the combination of statistical and systematic errors  
in the ORCA data. For this reason we did not take 
them into account. 

\vspace{0.3cm}
\leftline{{\bf C. Simulation of Events  in ORCA}}

\vspace{0.3cm}
 We calculate the principal observables in ORCA detector - 
the double differential event spectra in the neutrino energy $E$ and 
% nadir (zenith) angle $\theta_n$ ($\theta$) - 
nadir angle  $\theta_n$ using the methods developed and 
described in \cite{Capozzi:2015bxa,Capozzi:2017syc}.
We comment briefly on some of the technical aspects of the calculations.

 The neutrino events in ORCA are 
divided into two classes \cite{KM3Net:2016zxf}: 
``track-like'' and ``cascade-like''.
Track-like events involve an outgoing $\mu^-$ or $\mu^+$
and originate from charged current
(CC) interactions of $\nu_\mu$, $\bar{\nu}_\mu$
and  $\nu_\tau$, $\bar{\nu}_\tau$.  
Cascade-like events result from CC 
interactions of  $\nu_e$, $\bar{\nu}_e$
and $\nu_\tau$, $\bar{\nu}_\tau$, and 
from neutral current interactions and consist
of hadronic and electromagnetic showers.

The ORCA ``Letter of Intent'' \cite{KM3Net:2016zxf} contains  
estimates of the probabilities of flavour-misidentification 
as well as of identifying the $\tau$ and the neutral current 
events as track or cascade events.
We include this information from \cite{KM3Net:2016zxf} 
in our analysis.

 The  relevant detection characteristics of ORCA - 
the energy and angular resolutions and the dependence of the 
effective volumes for the different types/classes of events 
on the initial neutrino energy -  are taken from \cite{KM3Net:2016zxf} 
and correspond to the benchmark (9 m vertical spacing) 
configuration of the ORCA experiment. Note that in \cite{KM3Net:2016zxf}
such characteristics are only provided in graphical form, whereas
the analytical formulas that have been implemented in our code,
as well as the relevant factors,
are obtained through private communication with the 
experimental collaboration, as it also happened in \cite{Capozzi:2017syc},
where one of the authors of this work was involved.

  In our analysis we consider 
$E \in [2,100]$ GeV and $\theta_n/\pi  \in [0,0.5]$.
These two ranges are divided into 20 equally-spaced bins 
(linearly for 
$\theta_n$ and logarithmically for E), 
for a total of 400 bins for cascade events and an equal number 
for track events.

   The statistical analysis of ORCA event distributions is performed 
employing the  $\chi^2$ method described in \cite{Capozzi:2015bxa}.
In the analysis we include, in addition to the statistical uncertainties, 
the following systematic uncertainties \cite{Capozzi:2017syc}:

\begin{enumerate}

\item oscillation and normalization uncertainties, where the latter 
include an overall normalization error (15\%), as well as
the relative $\mu/e$ and $\nu/\bar{\nu}$ 
flux uncertainties (8\% and 5\%, respectively);

\item energy-scale (5\%) and energy-angle resolution uncertainties (10\%),  
independently for cascade and track events;

\item energy-angle spectral shape uncertainties, via quartic polynomials 
in both 
% $\theta$
$\theta_n$  and $E$. These are meant to characterize systematic effects 
including: uncertainties in the primary
cosmic ray fluxes, differential atmospheric neutrino fluxes and 
cross sections and, to
some extent, energy-angle detection efficiencies;

\item residual uncorrelated systematics in each bin, representing 
the presence of unknown uncertainties, like those coming from a finite 
Monte Carlo statistics in experimental simulations.

\end{enumerate}

We define as ``minimal'' set of systematics the one including only  
those described in point 1) in the preceding list. 
When we add the uncertainties at point 2), 3) and 4), assuming a prior of 
0.75\% (1.5\%) on the coefficients of the quartic polynomials and a 
0.75\% (1.5\%) uncorrellated error in each bin, we obtain our  
``optimistic'' set (``default'' set). 
Finally, if we instead consider a 3\% uncertainty on polynomial coefficients 
and uncorrelated errors we get our ``conservative'' set. All the systematics 
mentioned above are implemented using the pull method \cite{Fogli:2002pt}.

%%%%%%%%%%%%%%%%%%%%%%%%%%%%%%%%%%
%
\section{Results}
\label{sec:results}
%
%%%%%%%%%%%%%%%%%%%%%%%%%%%%%%%%%%%
% 

 In the next five subsections
we present the results of our analysis 
on the sensitivity of ORCA to the 
i) OC density $\rho_{\rm OC}(r)$, ii) IC density  $\rho_{\rm OC}(r)$, 
iii) total core density  $\rho_{\rm core}(r)$
and iv) mantle density  $\rho_{\rm man}(r)$.  In each case we impose the 
total Earth mass constraint, compensating the variation of the density 
in a given layer by  corresponding changes of density in one of the 
other layers. In other to assess the effects of this constraint we 
show also results without imposing it.
  
 To illustrate the dependence of the sensitivity of ORCA on the 
systematic uncertainties we obtained results for the four types 
of possible systematic uncertainties, which can affect significantly  
the sensitivity of ORCA and which are still not well determined:    
the ``minimal'', ``optimistic'', ``default'' and ``conservative'' 
sets defined earlier. We choose to present results only for 
the  ``minimal'', ``optimistic'' 
and ``conservative'' sets of uncertainties.

As we have discussed earlier, in the analysis we have performed 
we kept  $\delta$ and $\sin^2\theta_{23}$ fixed to certain values.
We have obtained results for  $\delta = 3\pi/2$ 
and  eleven values of $\sin^2\theta_{23}$ from the interval
[0.40,0.60]. In what follows we  present results for three reference values 
of  $\sin^2\theta_{23} = 0.42$, 0.50 and 0.58~
\footnote{The results for the additional values of 
$\sin^2\theta_{23}$ are used to assess the effects of the 
Earth hydrostatic equilibrium conditions on the results 
derived accounting for the Earth 
mass constraint (see further).}, 
which belong to, and essentially span, the  $3\sigma$ range of 
allowed values of $\sin^2\theta_{23}$ obtained in the latest global 
neutrino data analyses \cite{Capozzi:2021fjo,Esteban:2020cvm}.
All results are obtained assuming 10 years of ORCA operation. 
 In subsections 3.1-3.4 we present results for NO  neutrino mass spectrum. 
Results for IO spectrum are reported in subsection 3.5.

One more comment is in order. It follows from the seismological data 
as well as the condition of hydrostatic equilibrium of the Earth that 
the following inequalities should always hold:
%%%%%%%%%%%%%%%%%%%%%
\begin{equation}
\rho_{man} \leq \rho_{OC} \leq \rho_{IC}\,.
\label{eq:equil}
\end{equation}
%%%%%%%%%%%%%%%%%%%%%%%%%
%
These constraints are not {\it a priori} satisfied 
when we vary the density in a given layer 
and compensate it with a change of density in another layer.
However, we indicate in each specific case 
what are the restrictions they lead to whenever these 
restrictions are relevant.

%%%%%%%%%%%%%%%%%%%%%%%%%%%%%%%
%
\subsection{Sensitivity to the Outer Core Density}
\label{ssec:OC}
%
%%%%%%%%%%%%%%%%%%%%%%%%%%%%%%%%

 In the present subsection we show results on the sensitivity of ORCA 
to the OC density in the cases of mantle and IC being the 
``compensating'' layers and when the total Earth mass constraint is 
not imposed.

\vspace{0.3cm}
\noindent {\bf A. Compensation with Mantle Density}

\begin{figure}[!t]
  \centering
\begin{tabular}{lll}
{\includegraphics[width=0.3\linewidth]{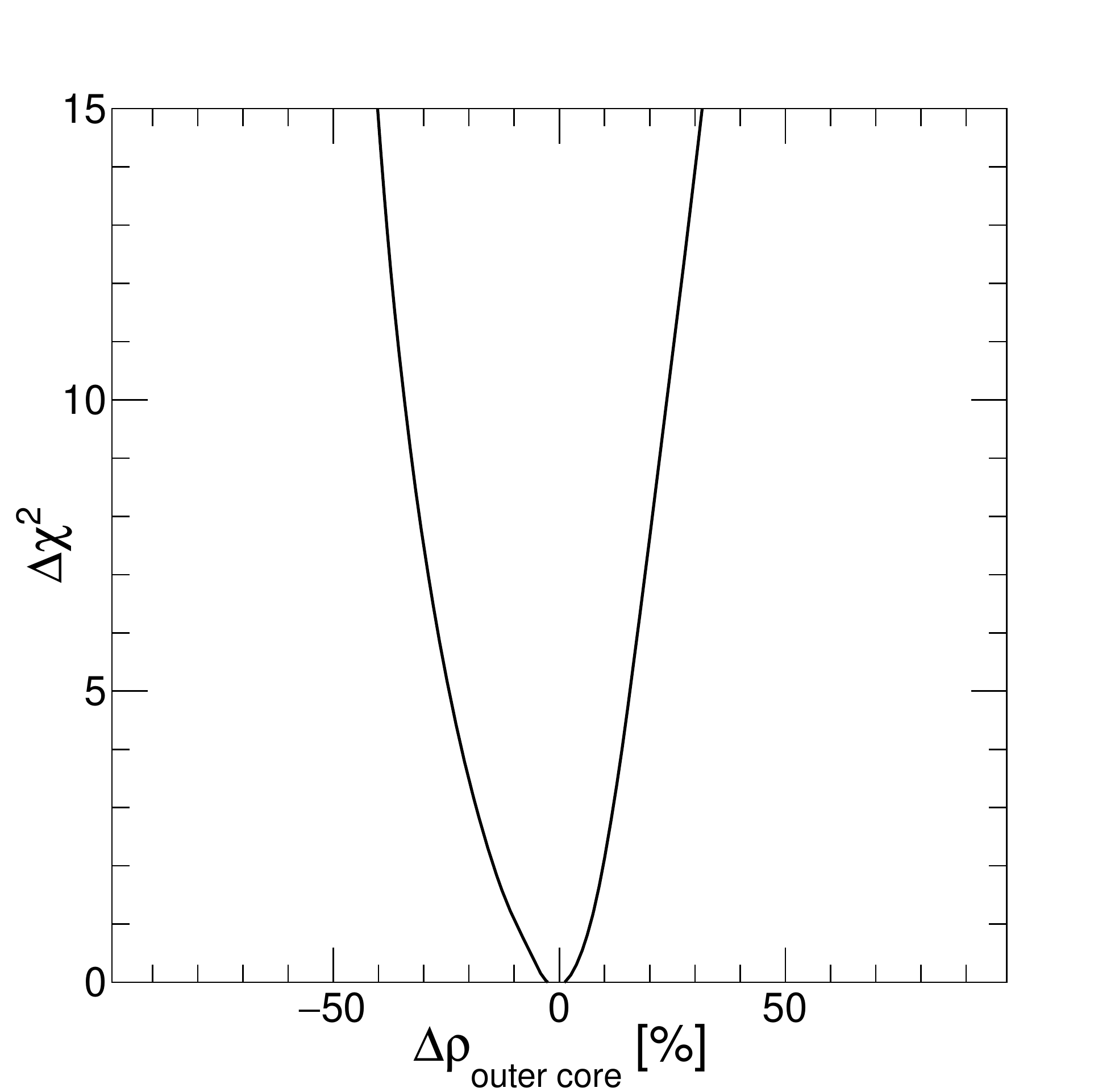}}
    &
   {\includegraphics[width=0.3\linewidth]{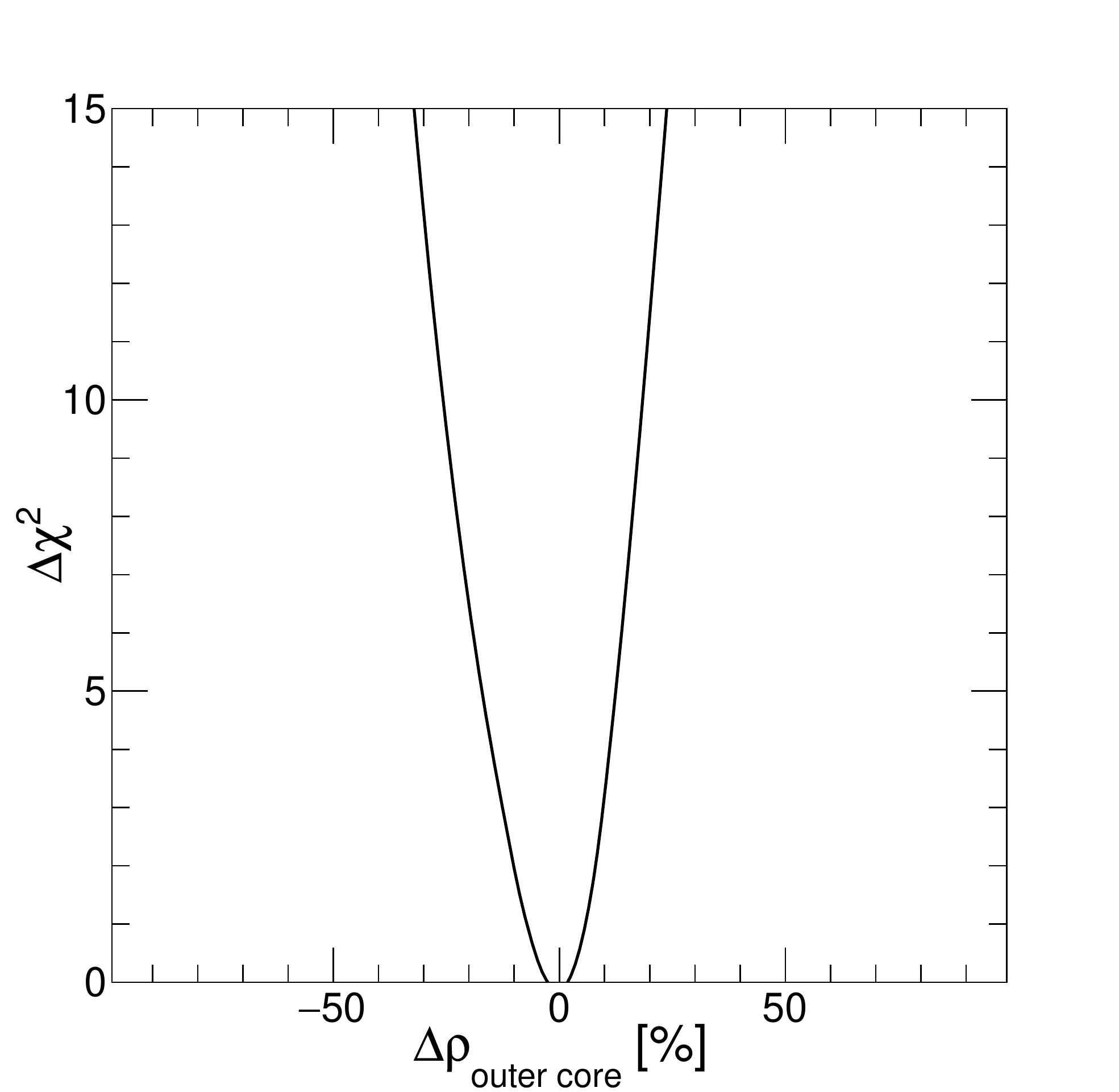}}
    &
    {\includegraphics[width=0.3\linewidth]{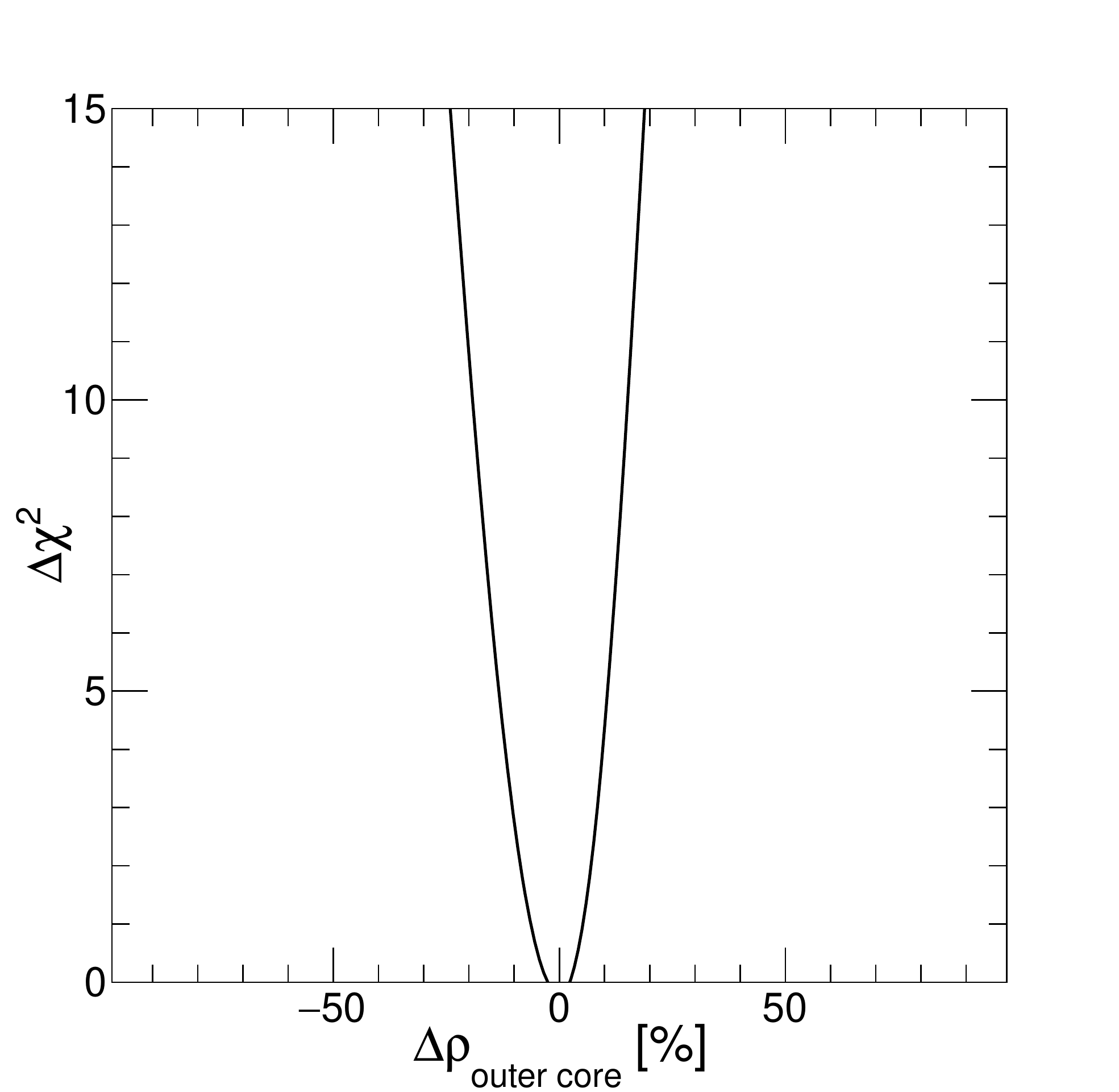}}
    \\
    {\includegraphics[width=0.3\linewidth]{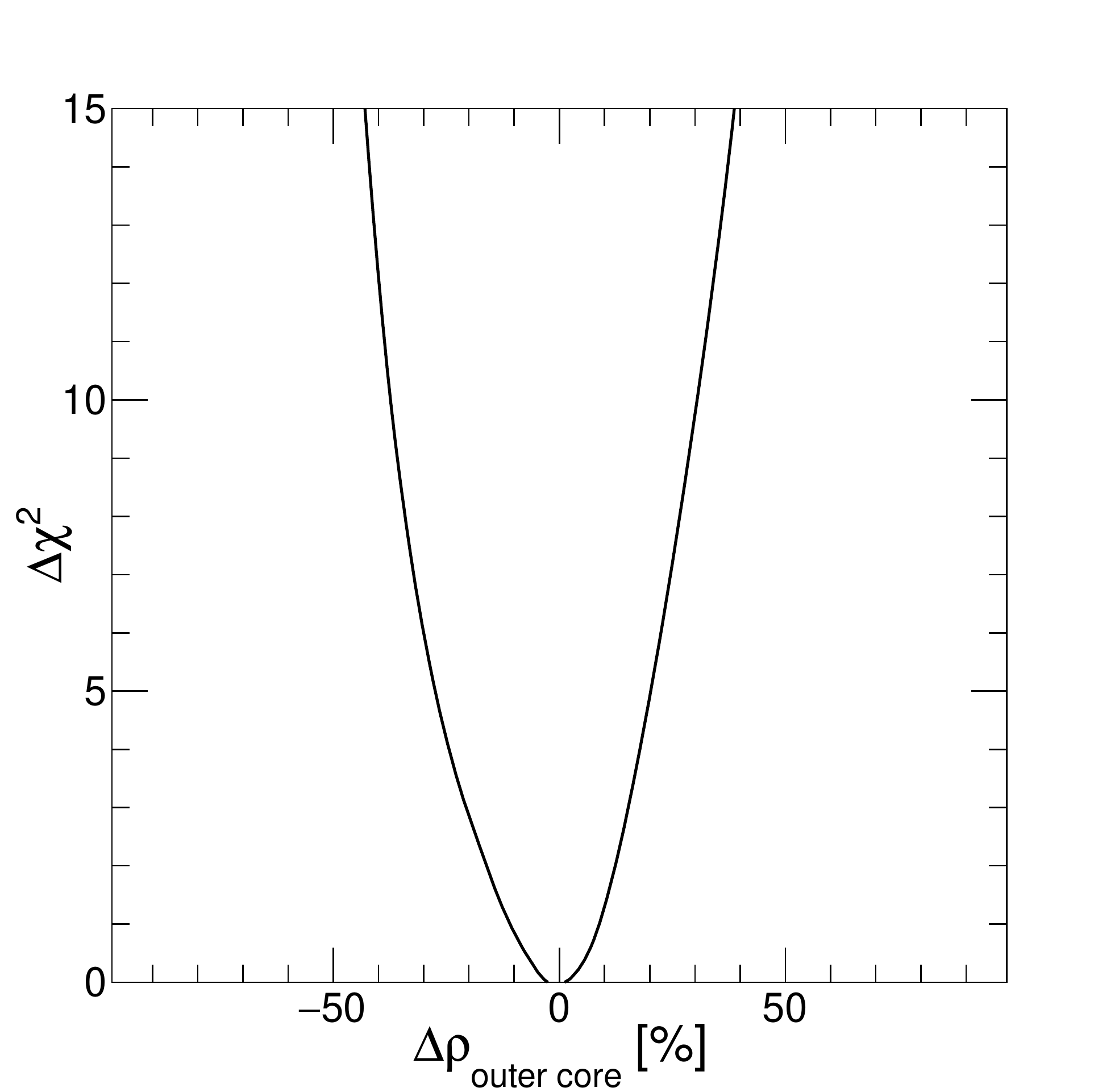}}
    &
    {\includegraphics[width=0.3\linewidth]{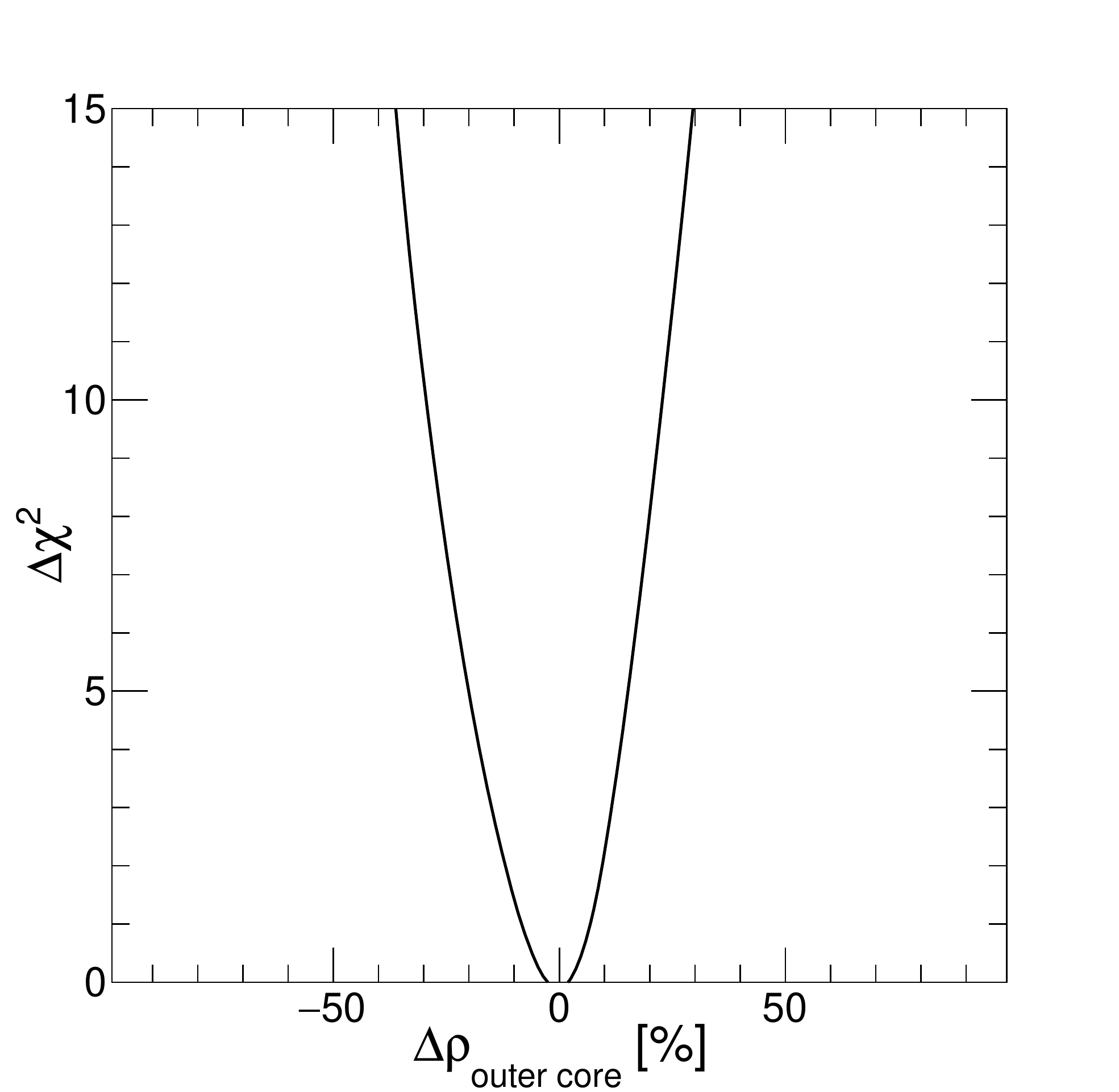}}
    &
   { \includegraphics[width=0.3\linewidth]{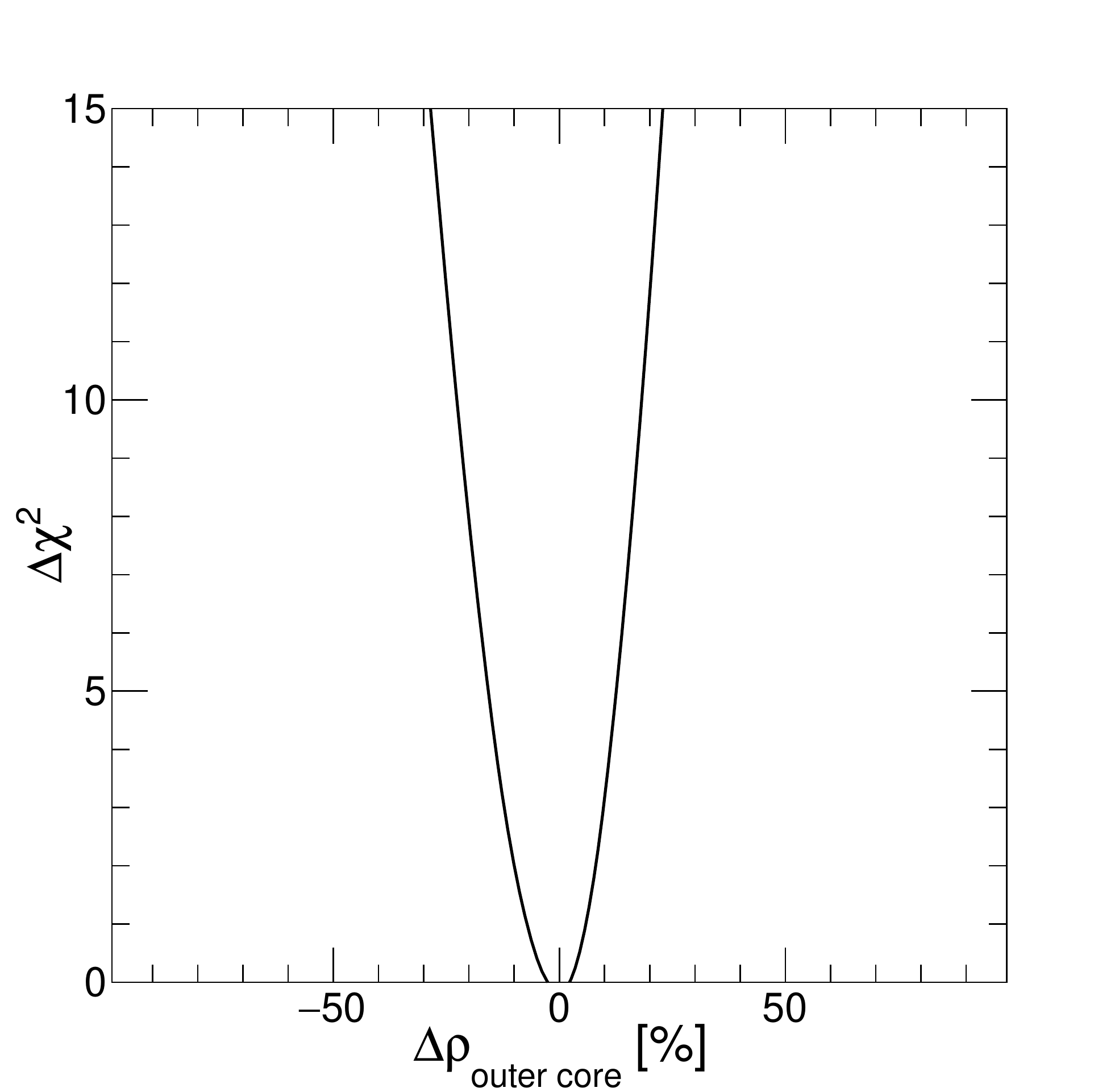}}
    \\
   {\includegraphics[width=0.3\linewidth]{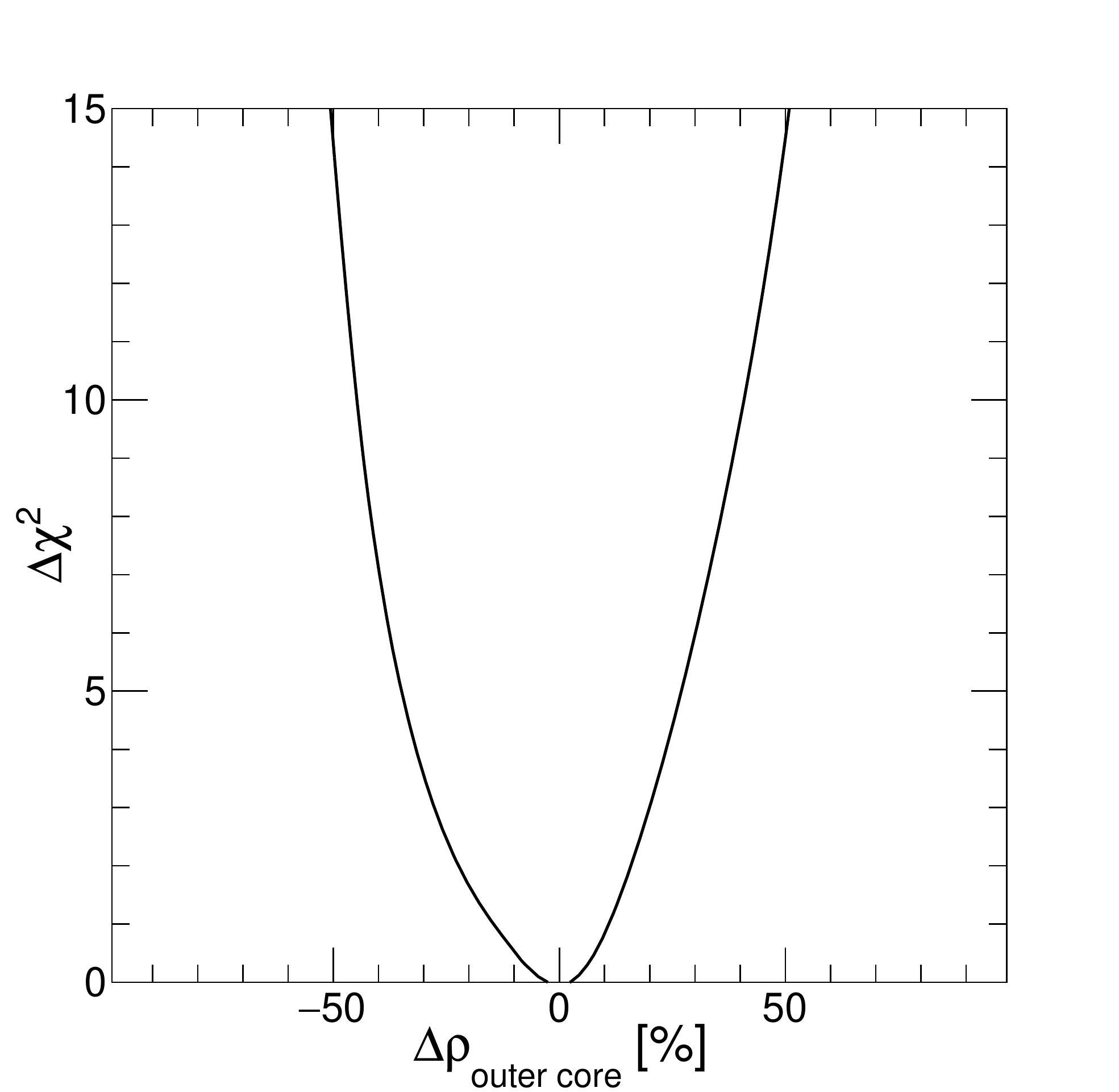}}
    &
    {\includegraphics[width=0.3\linewidth]{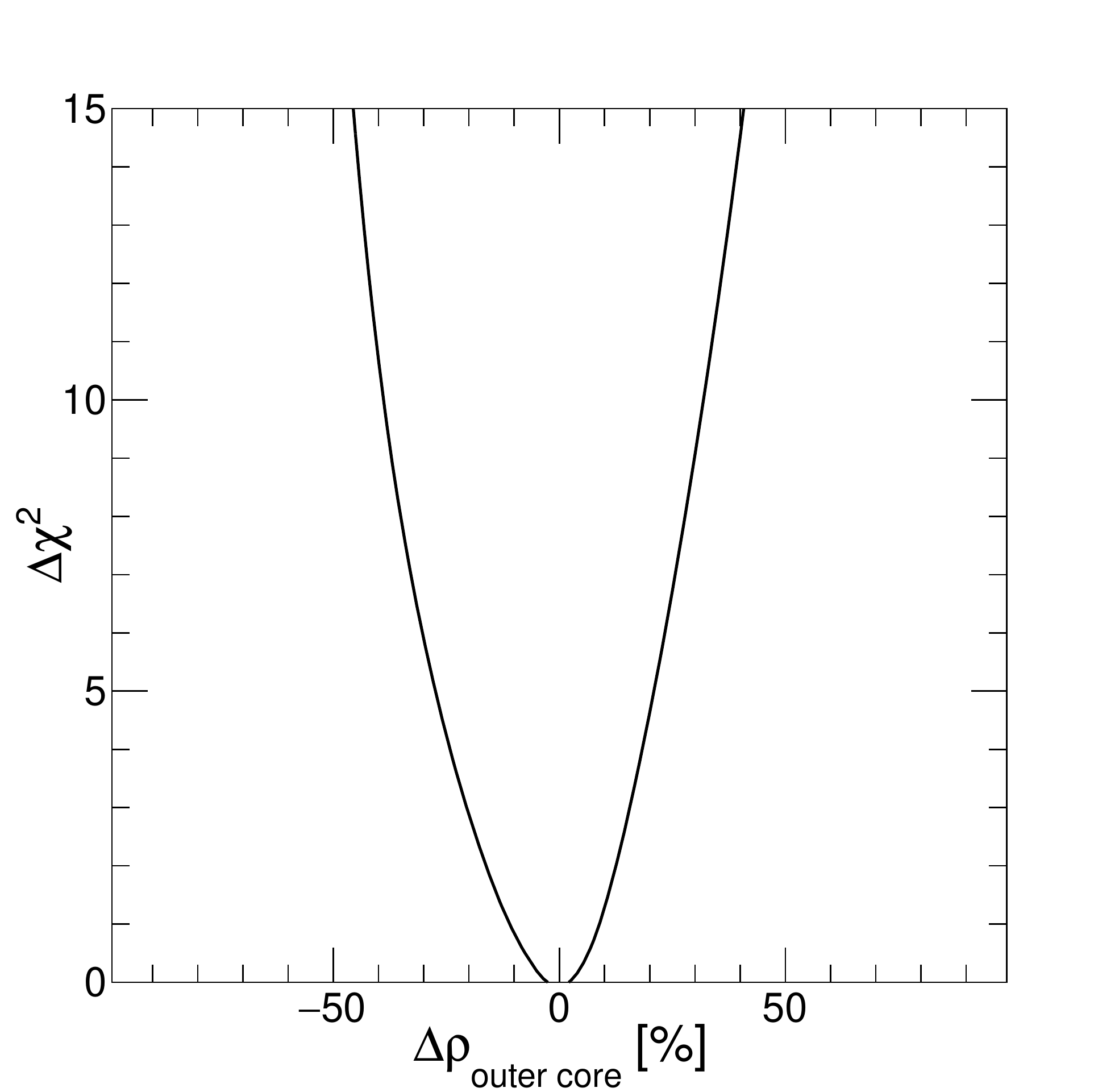}}
    &
   {\includegraphics[width=0.3\linewidth]{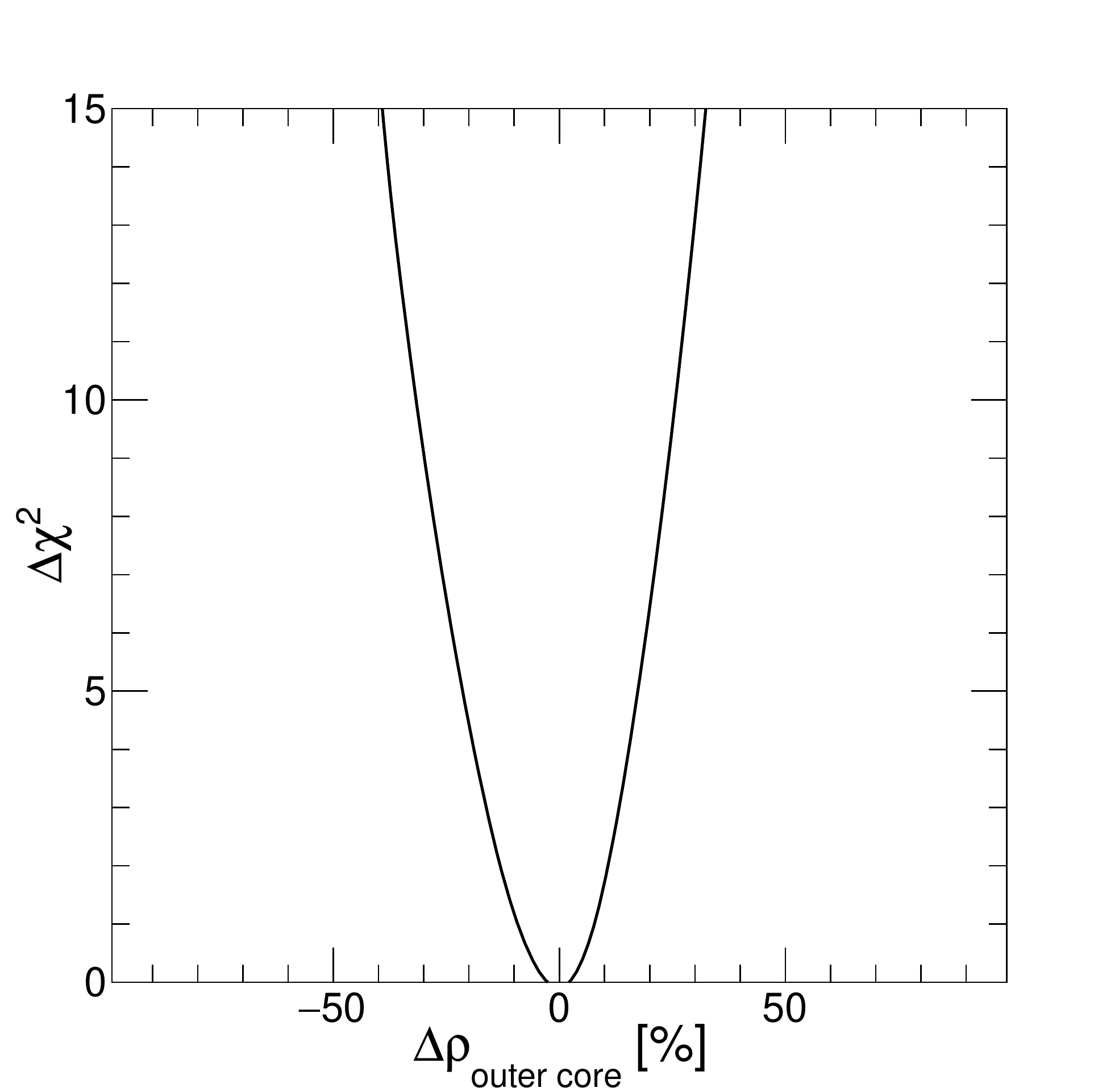}}
  \end{tabular}
 \caption{Sensitivity to the OC density in the case of NO spectrum 
 and 10 years of data. 
The Earth total mass constraint 
 is implemented by compensating the OC density variation with a 
 corresponding mantle density change. The 
 results shown are for $\sin^2\theta_{23} = 0.42$, 0.50, 0.58 
(left, center and right panels) and in the cases of 
``minimal'', ``optimistic'' and ``conservative'' systematic errors 
(top, middle and bottom panels). See text for further details.
}
\label{fig:NHOCvsMantle}
\end{figure}

\vspace{0.3cm}
 Figure \ref{fig:NHOCvsMantle} illustrates the sensitivity of ORCA 
to $\rho_{\rm OC}(r)$ when the Earth total mass constraint 
is implemented and the OC density variation is compensated with a 
corresponding mantle density change. The figure shows 
the $\chi^2$-distribution as a function of the OC relative density 
variation with respect to the PREM value, $\Delta \rho_{\rm outer~core}$. 
The results shown are for $\sin^2\theta_{23} = 0.42$, 0.50, 0.58 
(left, center and right panels) and in the cases of 
``minimal'', ``optimistic'' and ``conservative'' systematic errors 
(top, middle and bottom panels). All $\chi^2$-distributions
in Fig. \ref{fig:NHOCvsMantle}  have a symmetric, or slightly 
asymmetric, Gaussian form. As follows from Fig. \ref{fig:NHOCvsMantle}, 
%%%%%%%%%%%%%%%%%%%%%%%%%
%%%%%%%%%%%%%%%%%%%%%%%%%%%%%%%
%
\noindent 
in the case of 
``minimal'' systematic errors, 
ORCA can determine the OC density at $3\sigma$ with 
an uncertainty of (-33\%)/+23\%,  (-24\%)/+18\% and (-18\%)/+15\% 
respectively for  $\sin^2\theta_{23} = 0.42$, 0.50, 0.58.
The positive (negative) values correspond to 
$\Delta \rho_{\rm outer~core} > 0$ ($\Delta \rho_{\rm outer~core} < 0$). 
In the case of ``conservative'' systematic errors,
the sensitivity is noticeably worse:
(-43\%)/+39\%,  (-37\%)/+30\% and (-30\%)/+24\% 
for  $\sin^2\theta_{23} = 0.42$, 0.50, 0.58, 
respectively. For  $\sin^2\theta_{23} = 0.58$ (0.50),
the sensitivity is worse approximately by a factor of (1.6-1.7) 
(of 1.5 (of 1.7) for negative (positive) $\Delta \rho_{\rm outer~core}$).
As could be expected, in the case of ``optimistic''  systematic errors 
the sensitivity of ORCA under discussion is somewhat worse (better) than 
that obtained with ``minimal'' (``conservative'') systematic errors.
%%%%%%%%%%%%%%%%%%%%%%%%%%%%%%%%%%%%%%%%%%%%%%%
\begin{figure}[!t]
 \centering
% \begin{center}
% \begin{tabular}{ccc}
% \subfigure
% [width=0.65\textwidth]
% \includegraphics[width=8truecm,height=6cm]
\begin{tabular}{lll}
{\includegraphics[width=5truecm]{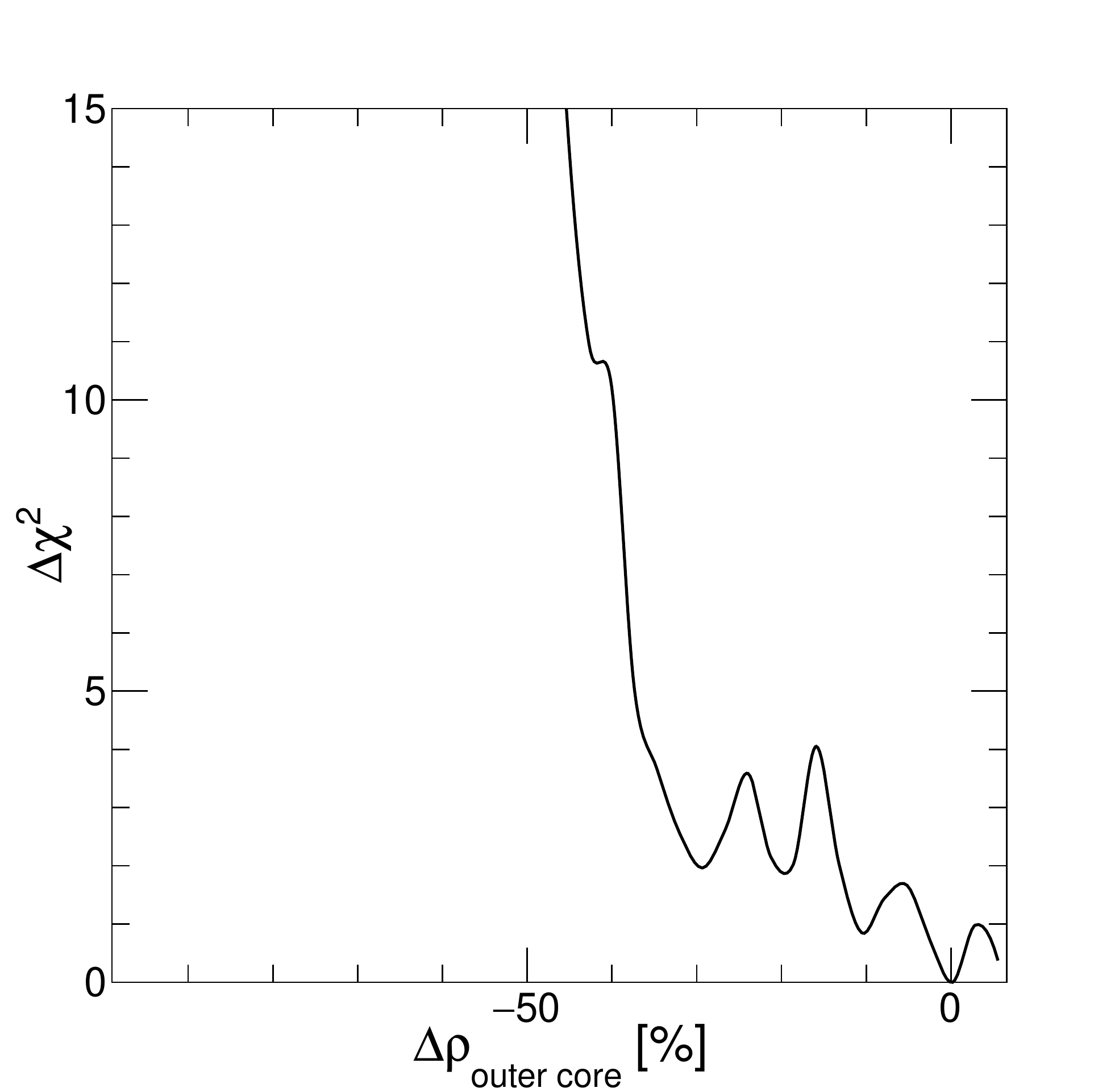}} & 
{\includegraphics[width=5truecm]{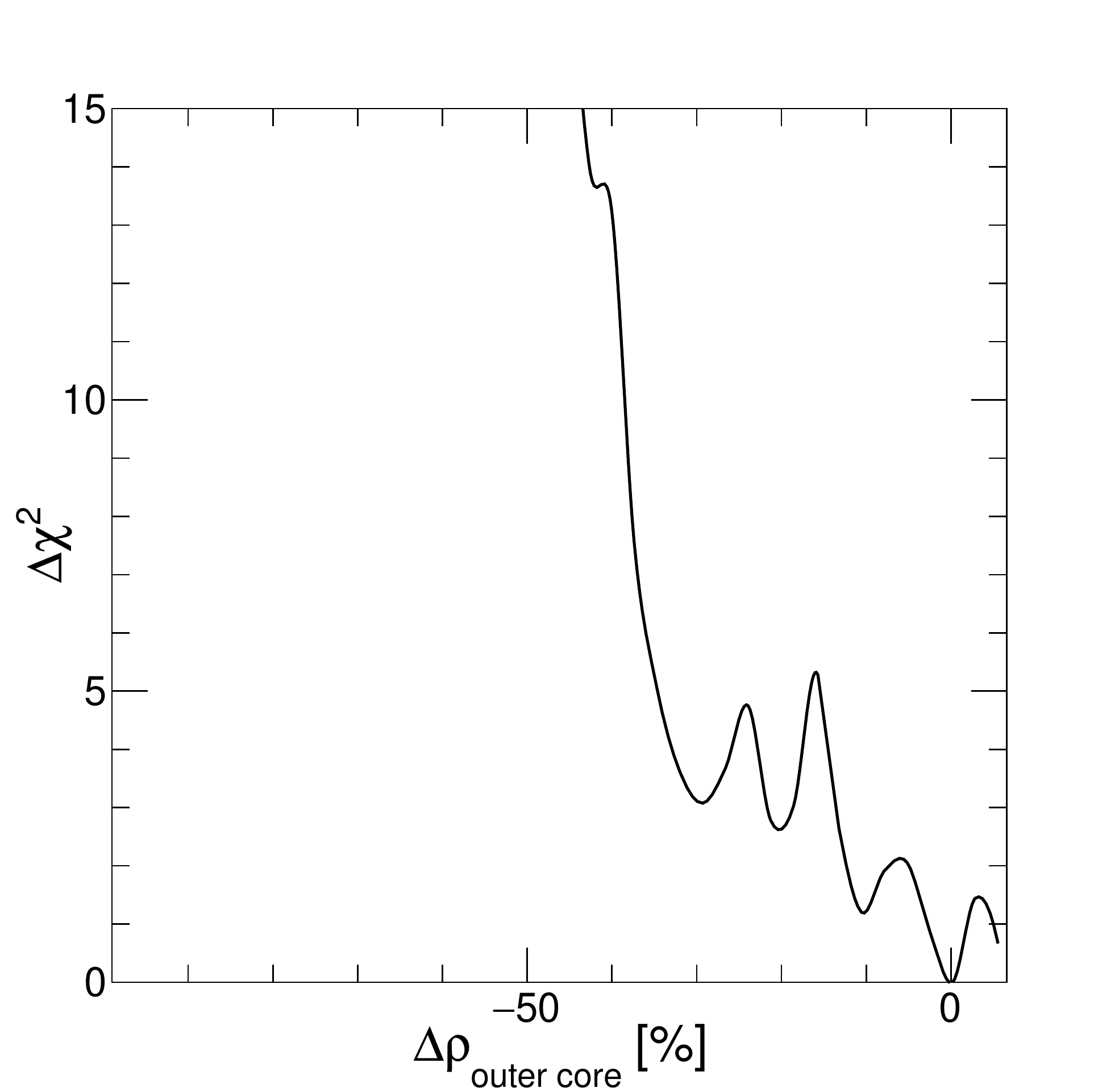}} & % \\
{\includegraphics[width=5truecm]{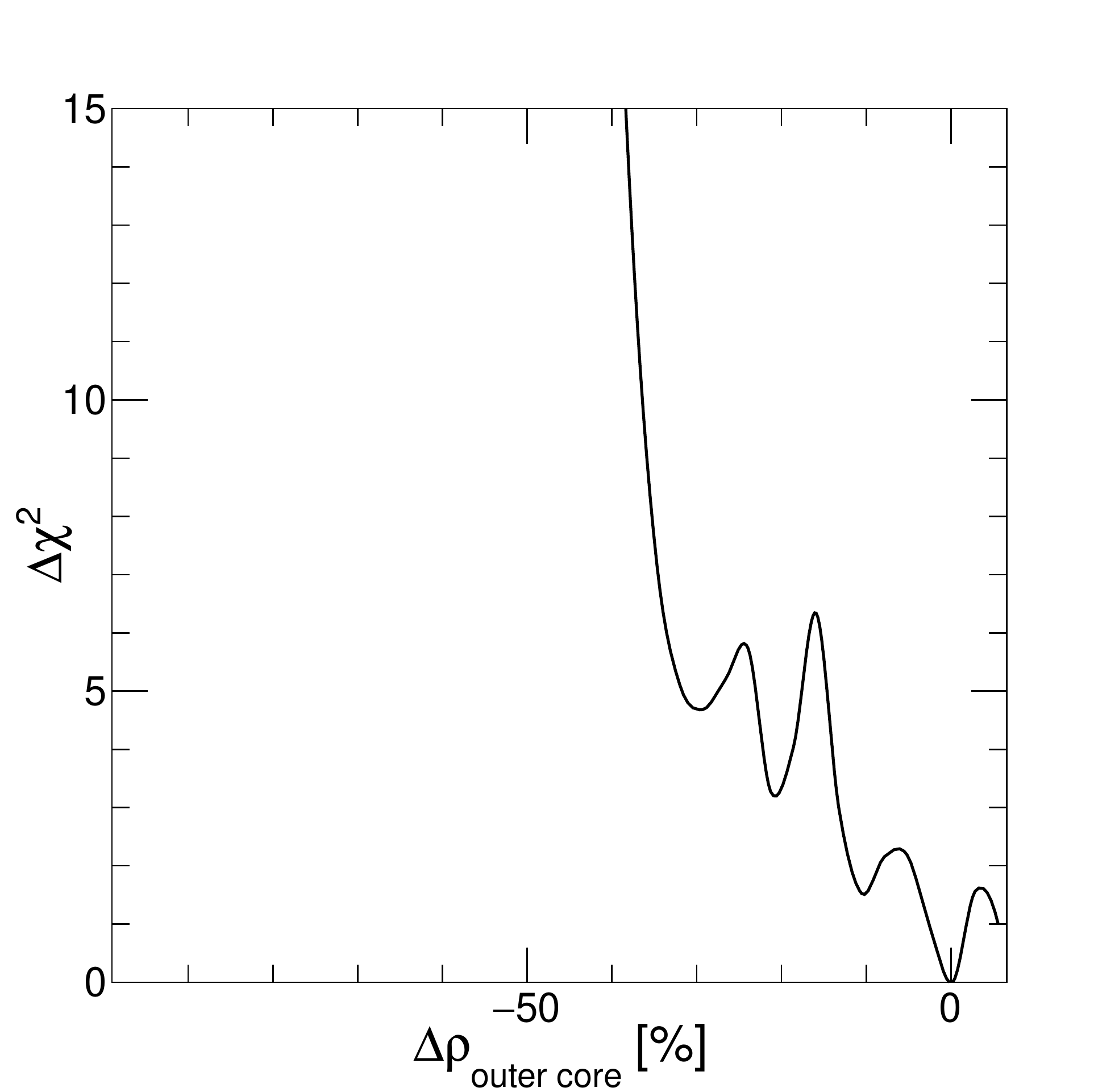}}  \\
% \end{tabular}
% \begin{tabular}{ccc}
% {\includegraphics[width=8truecm,height=6cm]
{\includegraphics[width=5truecm]{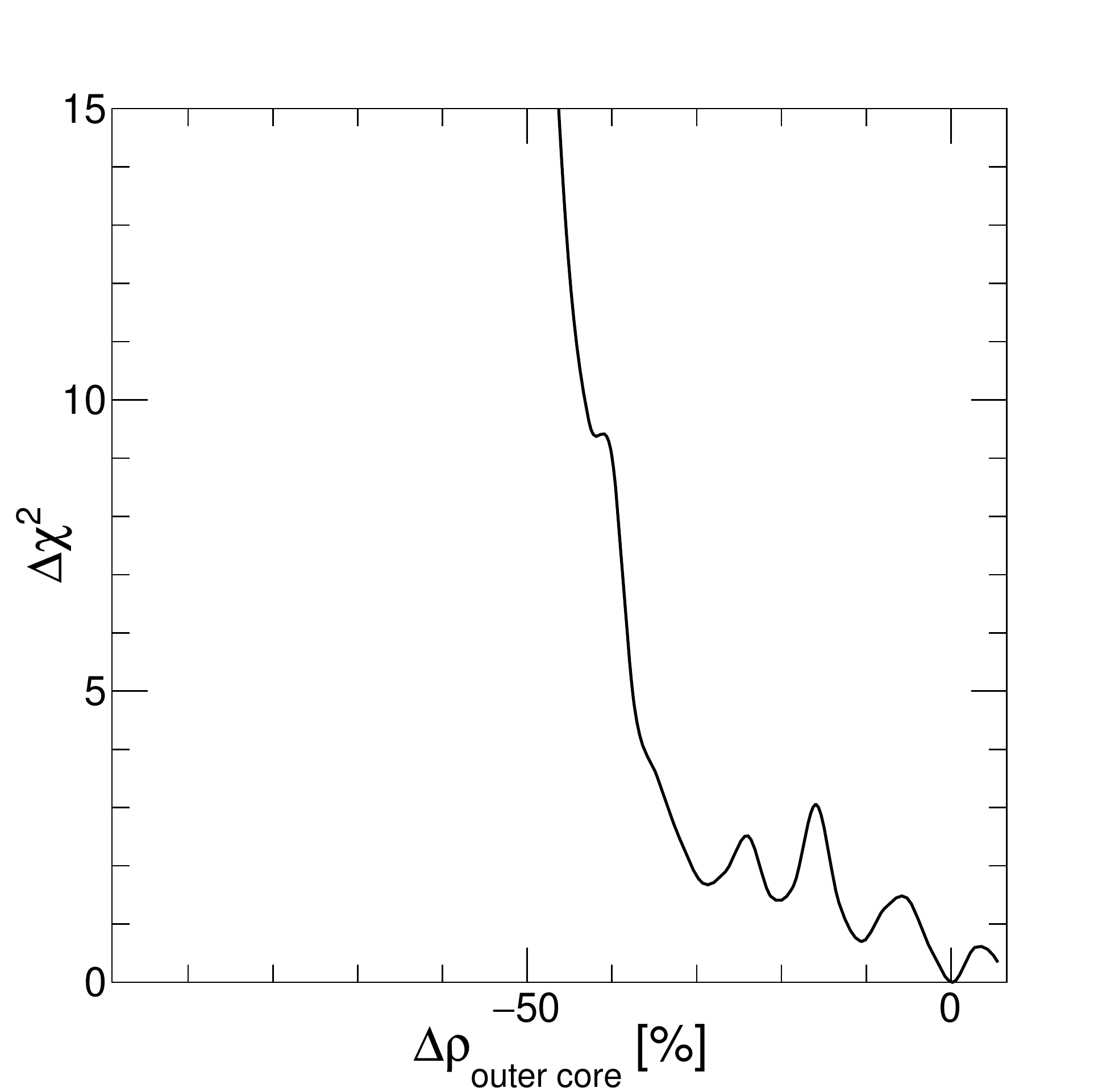}} & % \\
 {\includegraphics[width=5truecm]{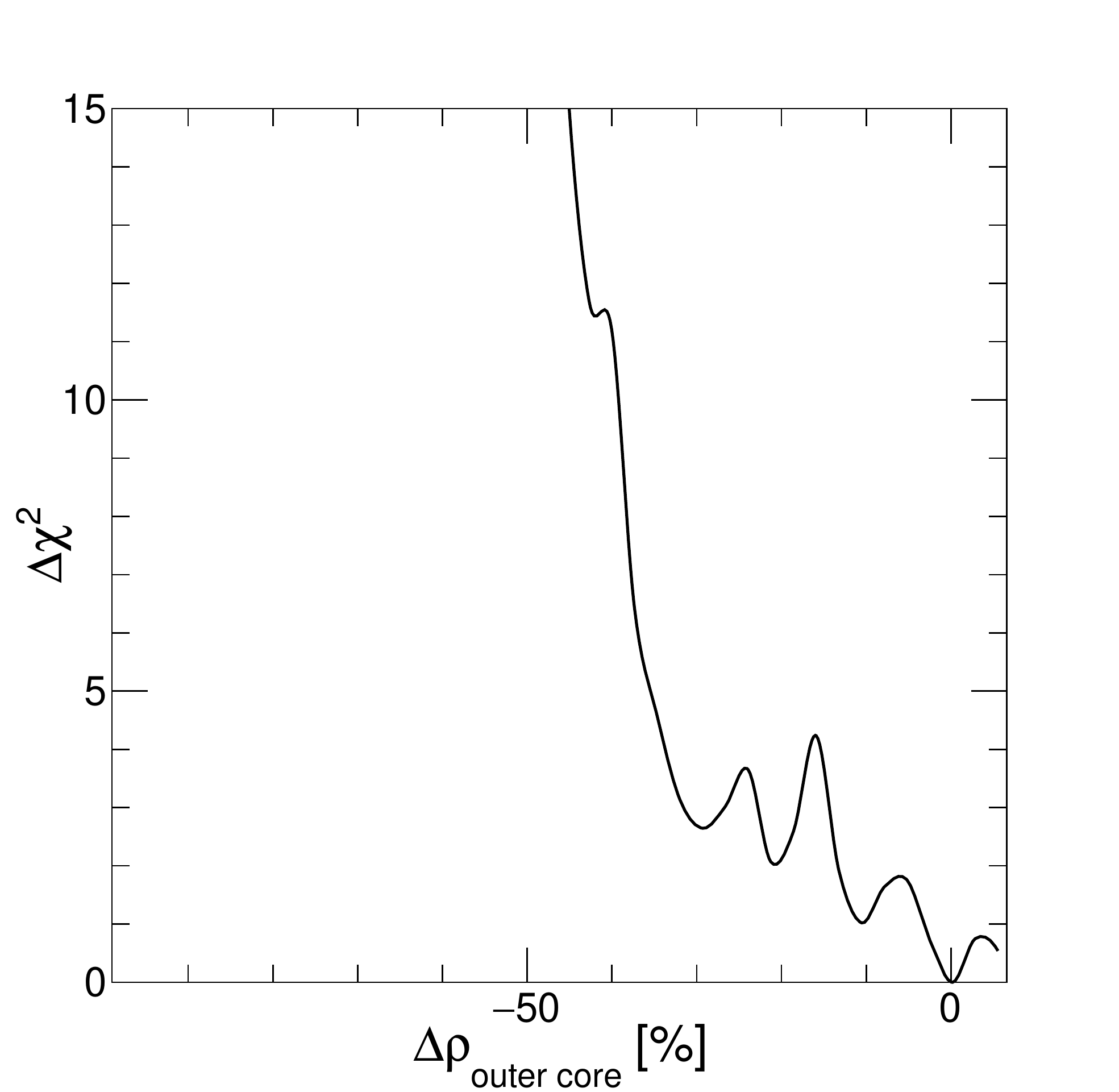}} & %\\
{\includegraphics[width=5truecm]{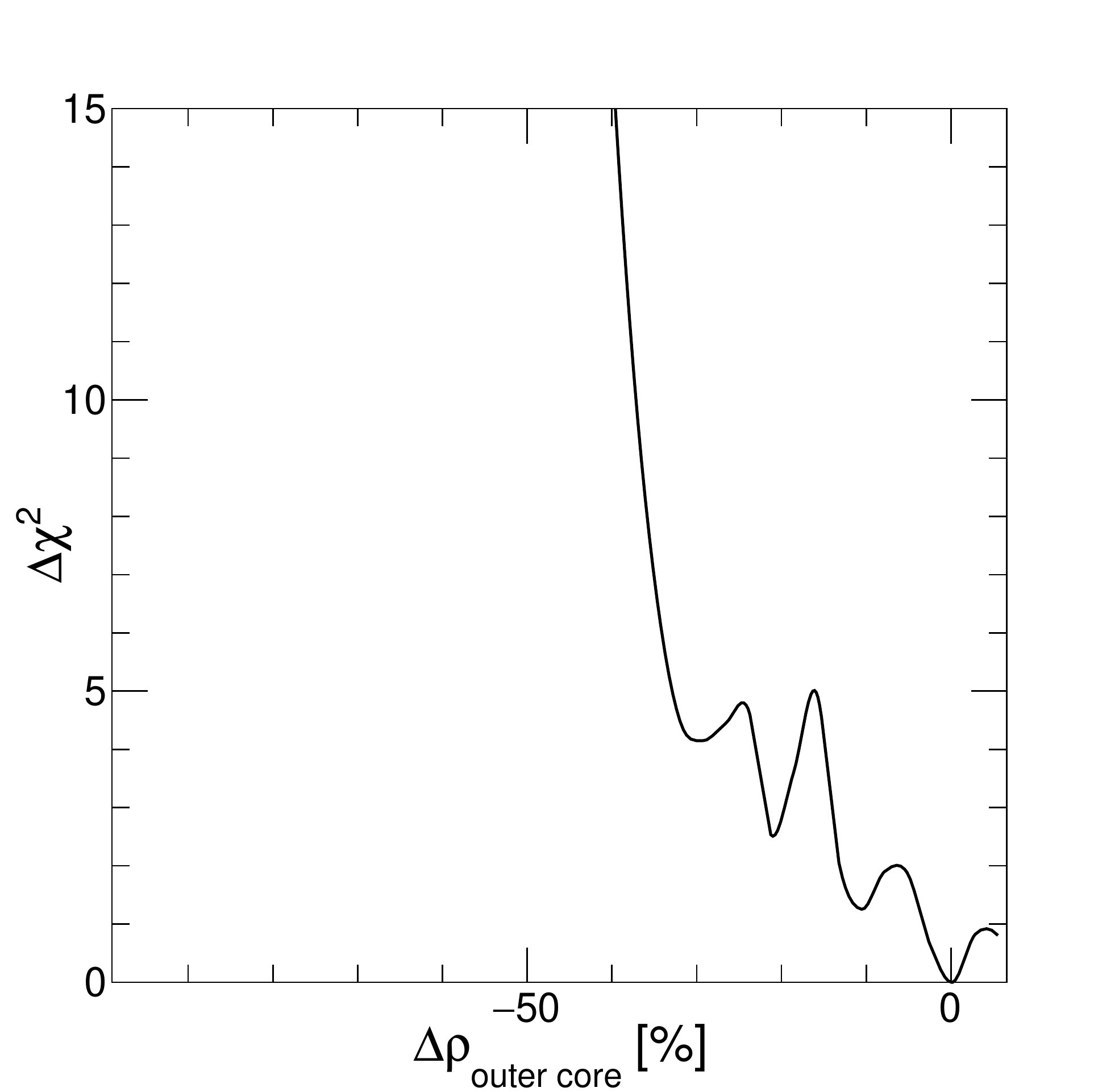}}\\
% \end{tabular}
% \begin{tabular}{ccc}
{\includegraphics[width=5truecm]{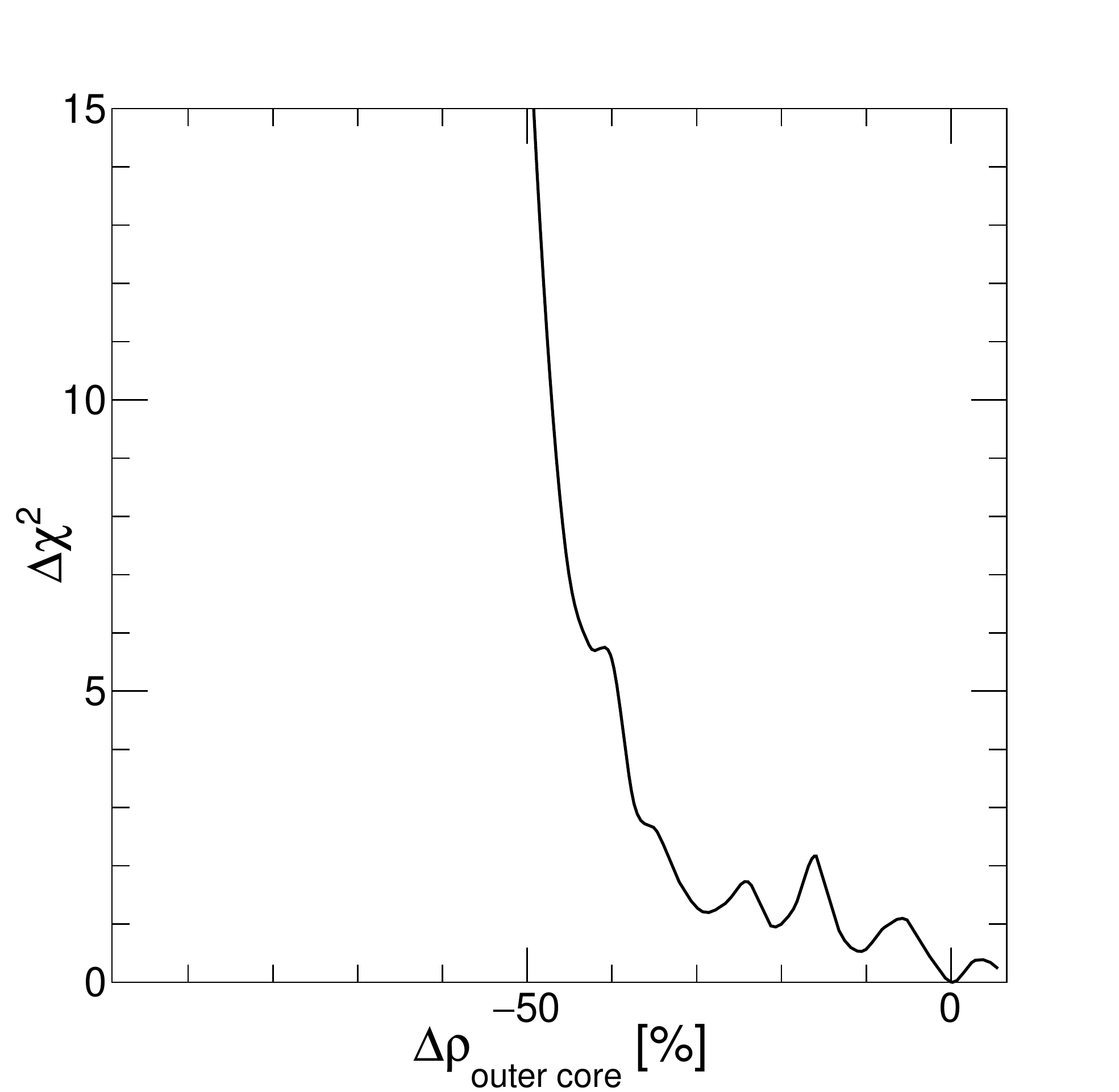}} & % \\
{\includegraphics[width=5truecm]{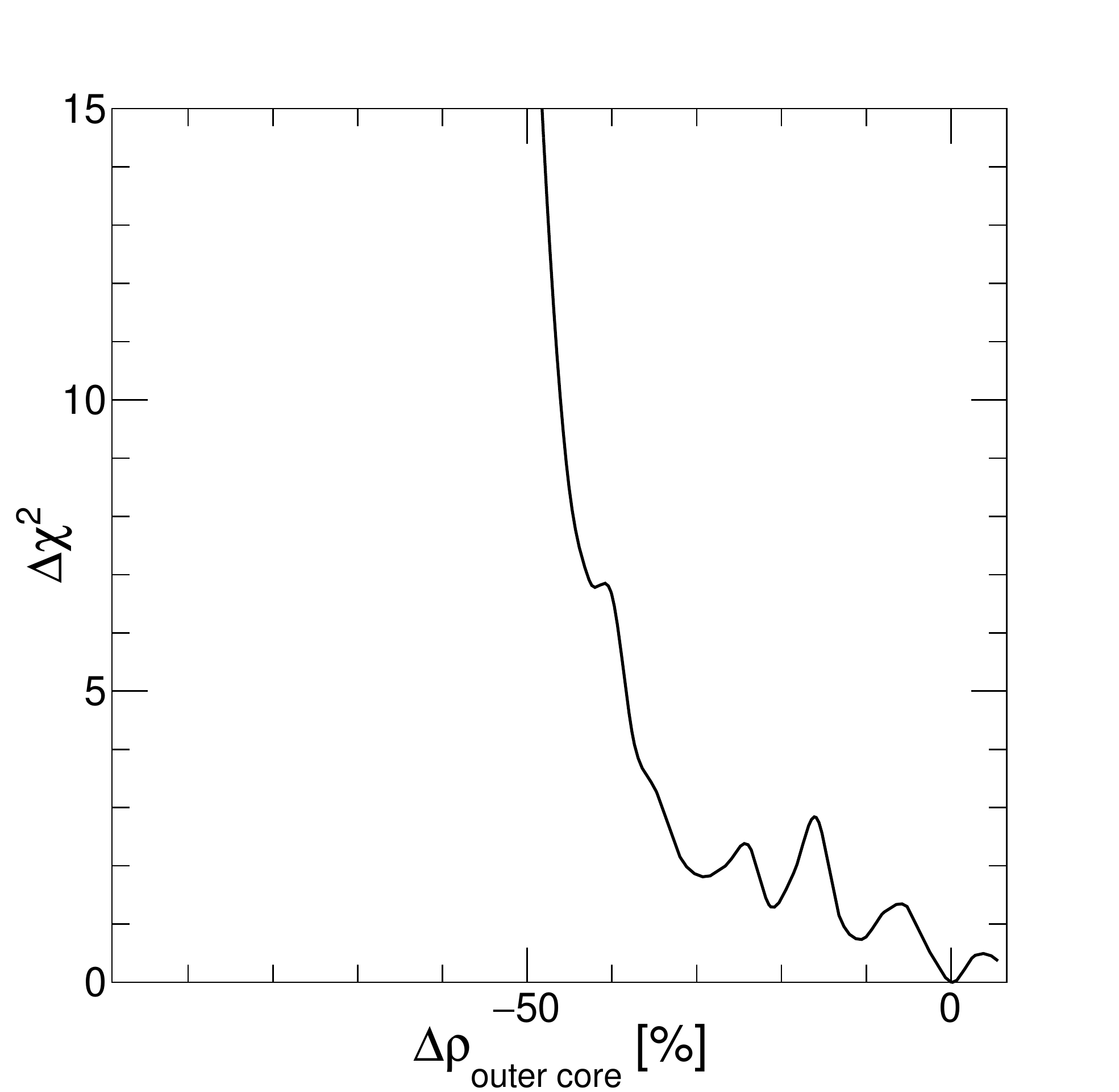}} & % \\
{\includegraphics[width=5truecm]{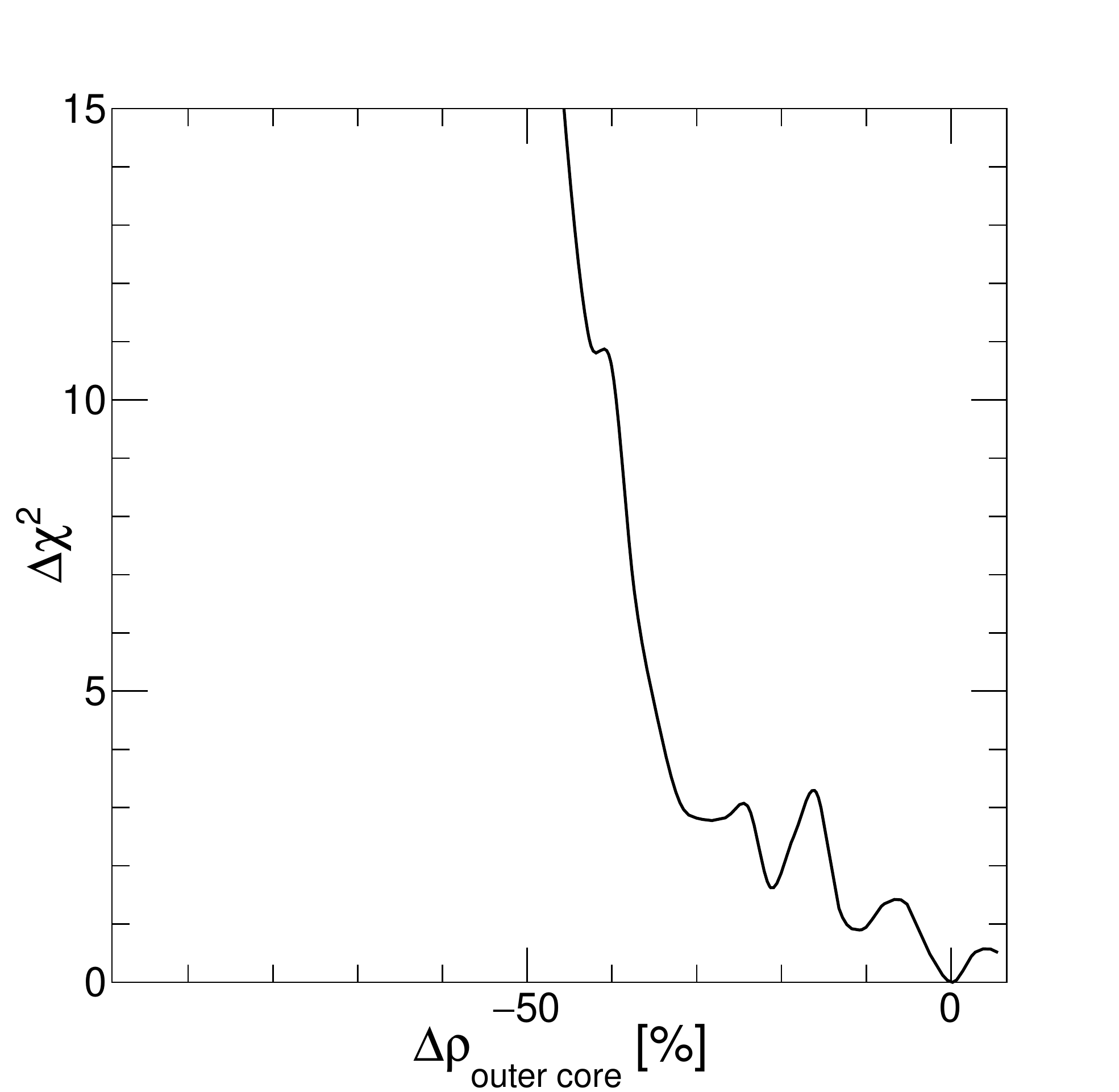}}
\end{tabular}
% \vspace{-0.2cm}
% \end{center}
\caption{
The same as in Fig.  \ref{fig:NHOCvsMantle}, but  
with the Earth total mass constraint 
implemented by compensating the OC density variation with a 
corresponding inner core density change. The 
results shown are for $\sin^2\theta_{23} = 0.42$, 0.50, 0.58 
(left, center and right panels) and in the cases of  
``minimal'', ``optimistic'' and ``conservative'' systematic errors 
(top, middle and bottom panels). See text for further details.
}
\label{fig:NHOCvsIC}
\end{figure}
%%%%%%%%%%%%%%%%%%%%%%%%%%%%%%%%%%%%%%%%%%%%

 Working in the considered case with the average PREM values 
of $\bar{\rho}_{man} = 4.45$ g/cm$^3$, 
$\bar{\rho}_{\rm OC} = 10.90$ g/cm$^3$ and 
$\bar{\rho}_{\rm IC} = 12.89$ g/cm$^3$,
it is not difficult to show that the inequalities in eq. (\ref{eq:equil}) 
imply approximately $(-49.6\%) \ltap \Delta \rho_{\rm outer~core} \ltap 18.3\%$.
In what concerns the derived ORCA $3\sigma$ sensitivity ranges of 
$\Delta \rho_{\rm outer~core}$, 
the lower limit from the external constraint  (\ref{eq:equil}) 
has no effect on them. The effect of the upper limit of 18.3\% 
depends on the type of implemented systematic errors and on 
the value of $\sin^2\theta_{23}$: 
the smaller the systematic errors and/or the larger $\sin^2\theta_{23}$,
the smaller the effect is.
In the case of ``minimal'' systematic errors, the 
indicated upper limit 
has no effect on the ORCA $3\sigma$ sensitivity ranges 
for any $\sin^2\theta_{23}\gtap 0.50$; 
for $\sin^2\theta_{23} =  0.42$ it corresponds 
to the maximal value of the ORCA $2\sigma$ sensitivity range.
For the set of  ``optimistic'' systematic errors,
18.3\% represents approximately the 
maximal value of the $2\sigma$, $2.4\sigma$ and 
$2.6\sigma$ ORCA sensitivity ranges 
for $\sin^2\theta_{23} = 0.50$, 0.54 and 0.58, 
respectively. The effect of the discussed constraint 
is largest for the ranges of interest obtained 
with conservative systematic errors: 
18.3\% correponds, e.g., to the maximal value of 
the ORCA $1.9\sigma$ sensitivity range 
at $\sin^2\theta_{23} = 0.58$.

\vspace{0.3cm}
{\bf B. Compensation with Inner Core Density}

\vspace{0.3cm}
 In Fig. \ref{fig:NHOCvsIC} we present results on the sensitivity of ORCA 
to the OC density when the Earth total mass constraint 
is implemented and the OC density variation is compensated with a 
corresponding IC density change. All $\chi^2$-distributions
shown in Fig. \ref{fig:NHOCvsIC}  have an asymmetric  
%%%%%%%%%%%%%%%%%%%%%%%%%%%%%%%%%%%%%%%%%%%%%%%
\begin{figure}[!t]
\centering
\begin{tabular}{lll}
{\includegraphics[width=5truecm]{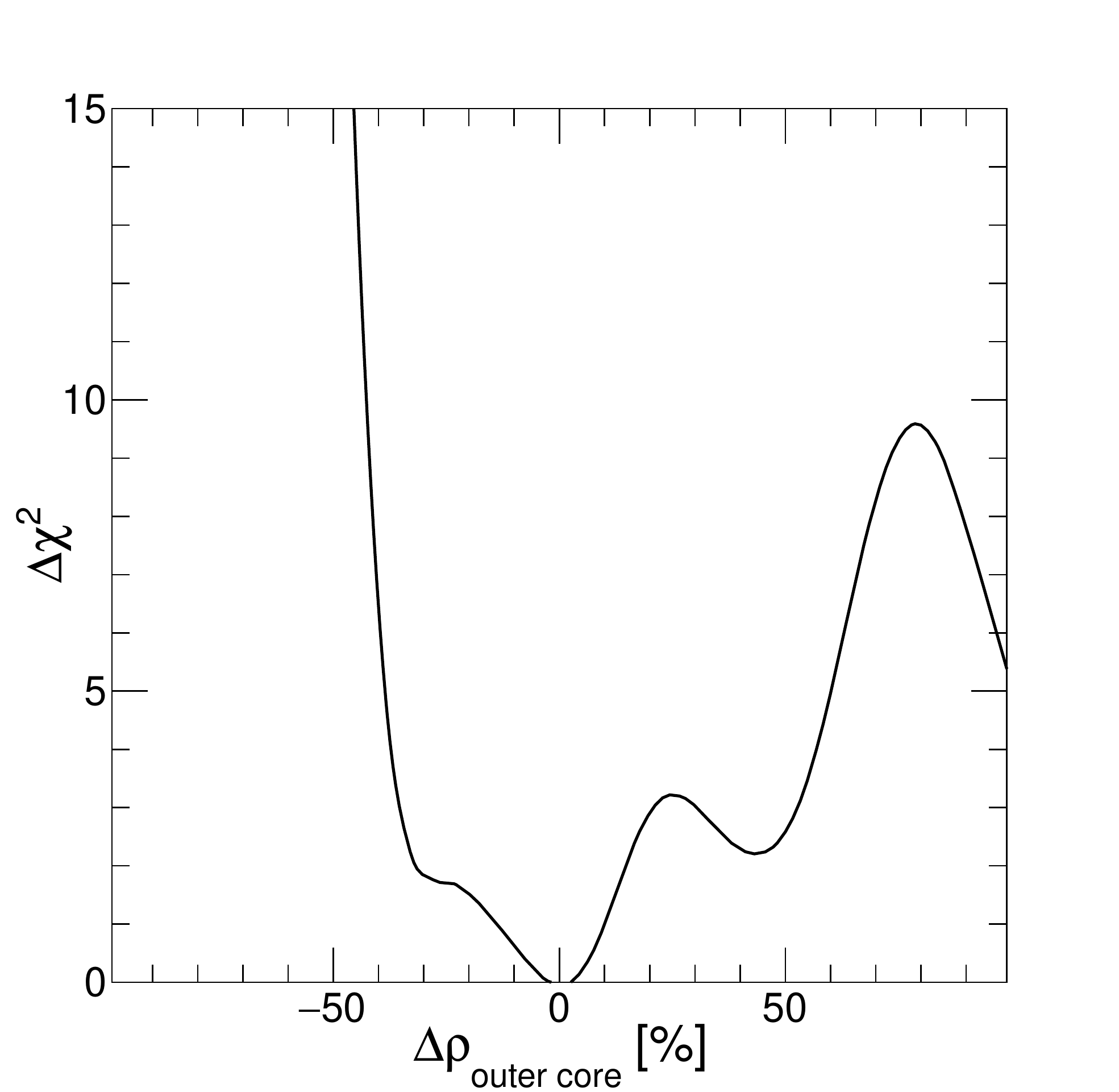}} & % \\
{\includegraphics[width=5truecm]{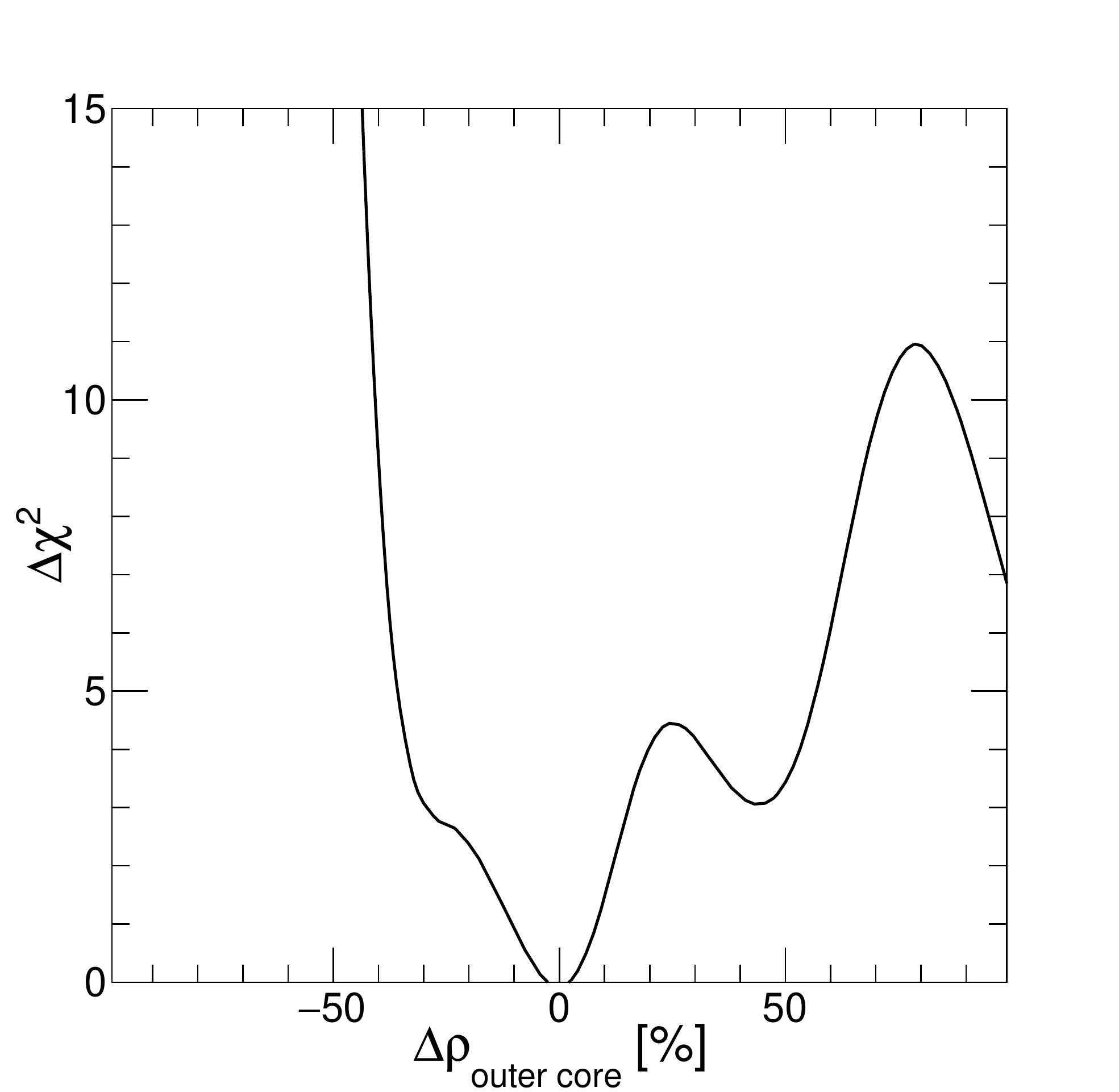}} & % \\
{\includegraphics[width=5truecm]{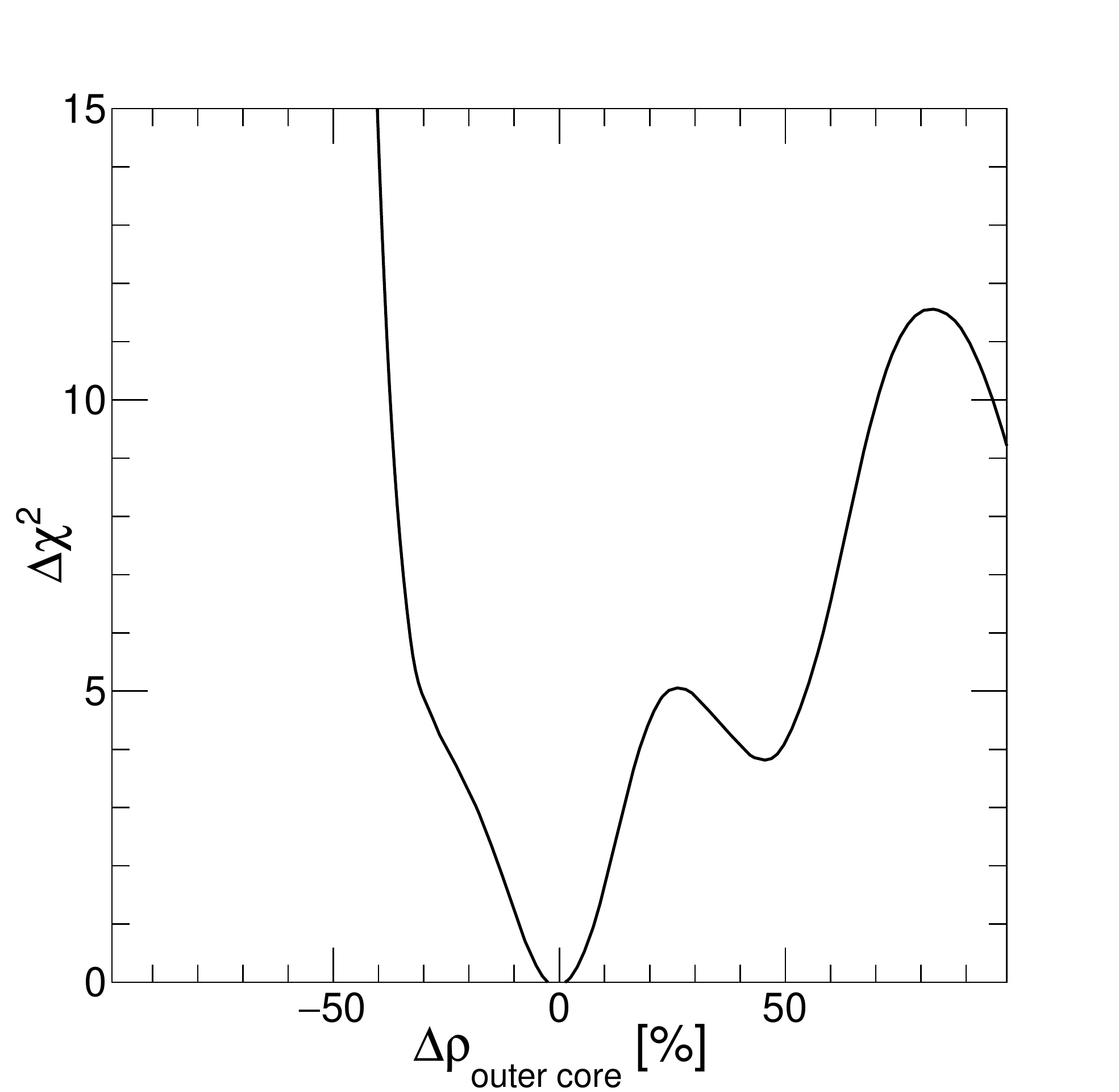}}  \\
{\includegraphics[width=5truecm]{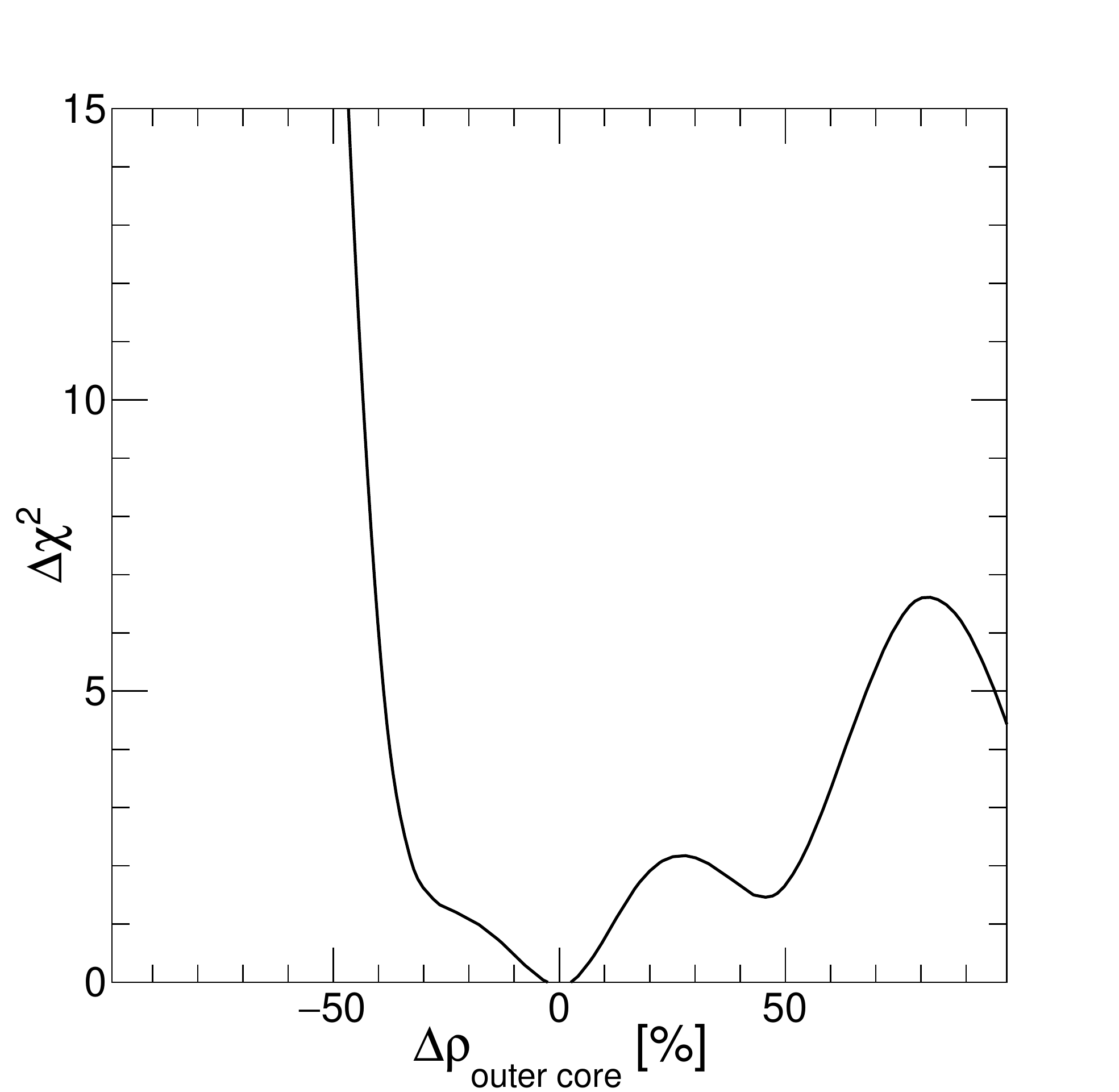}} & % \\
 {\includegraphics[width=5truecm]{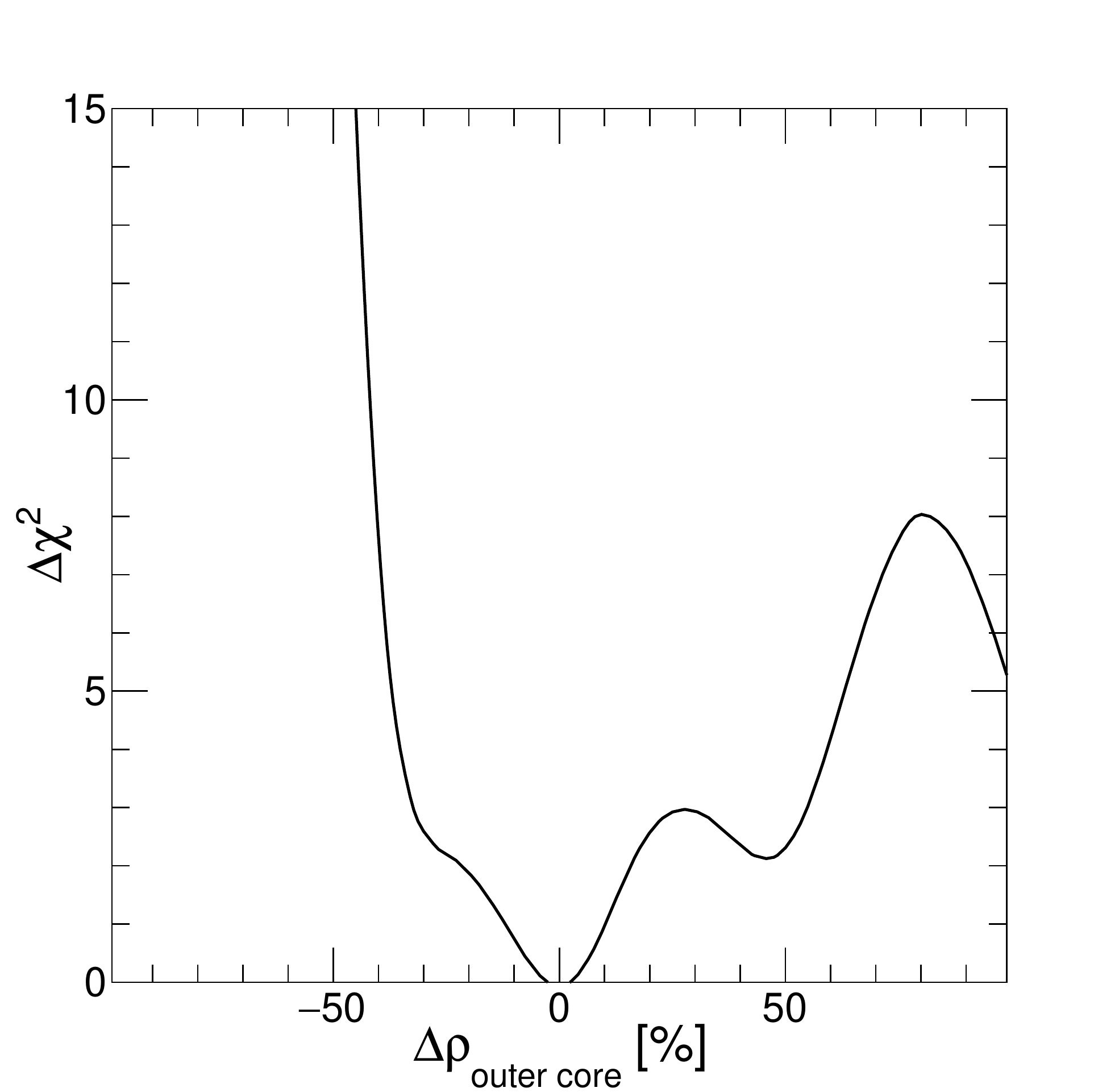}} & %\\
{\includegraphics[width=5truecm]{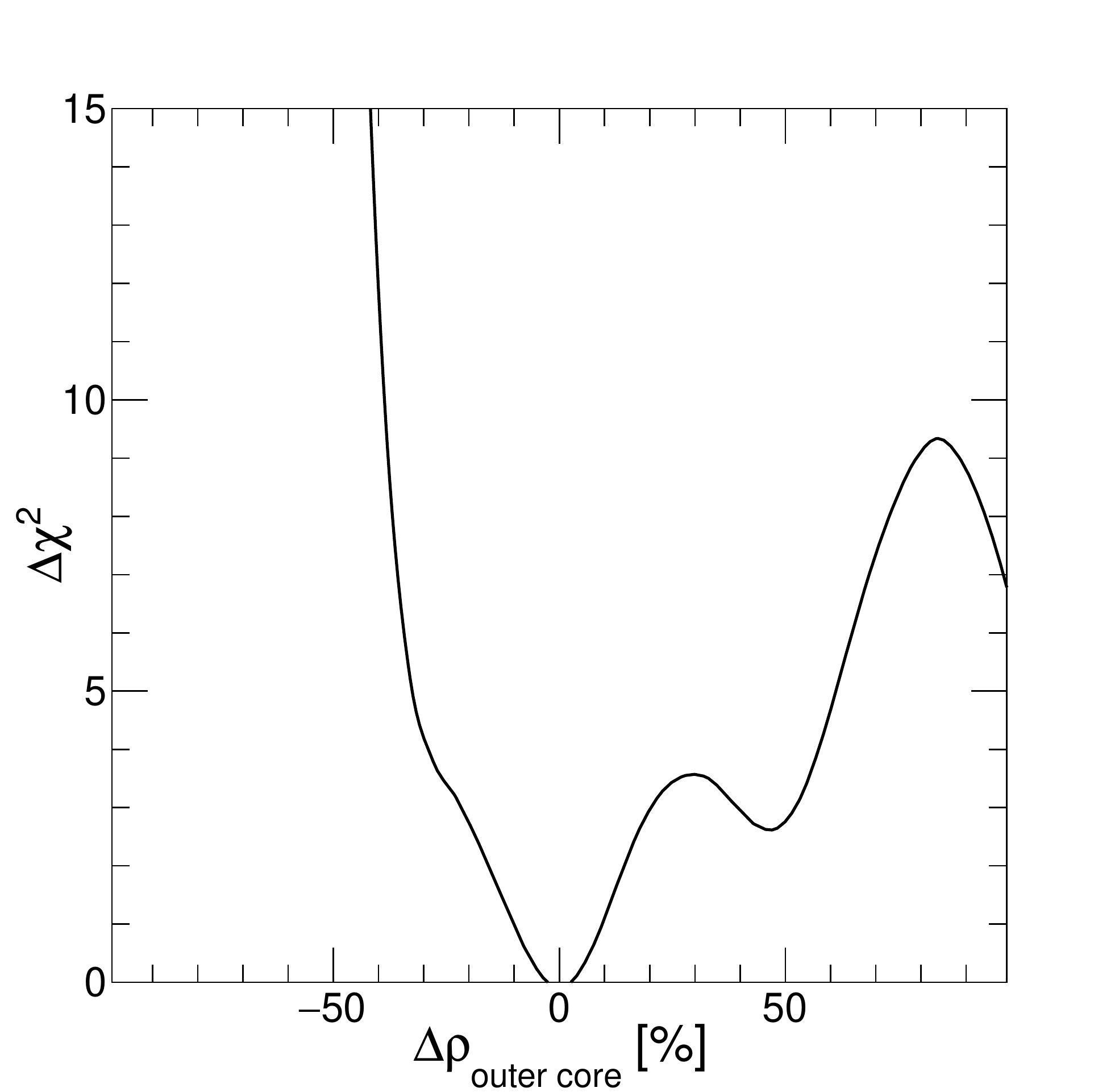}}\\
{\includegraphics[width=5truecm]{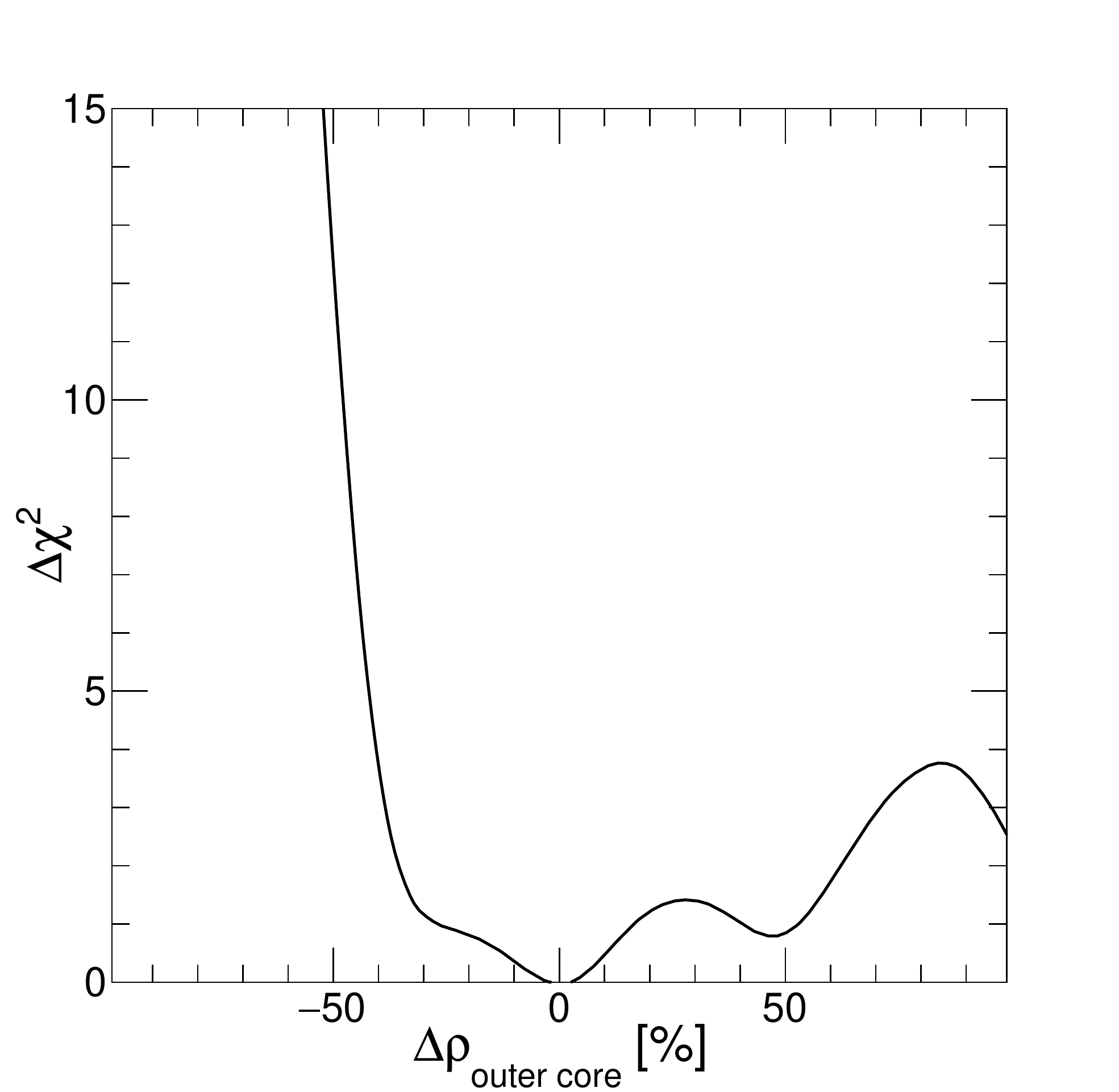}} & % \\
{\includegraphics[width=5truecm]{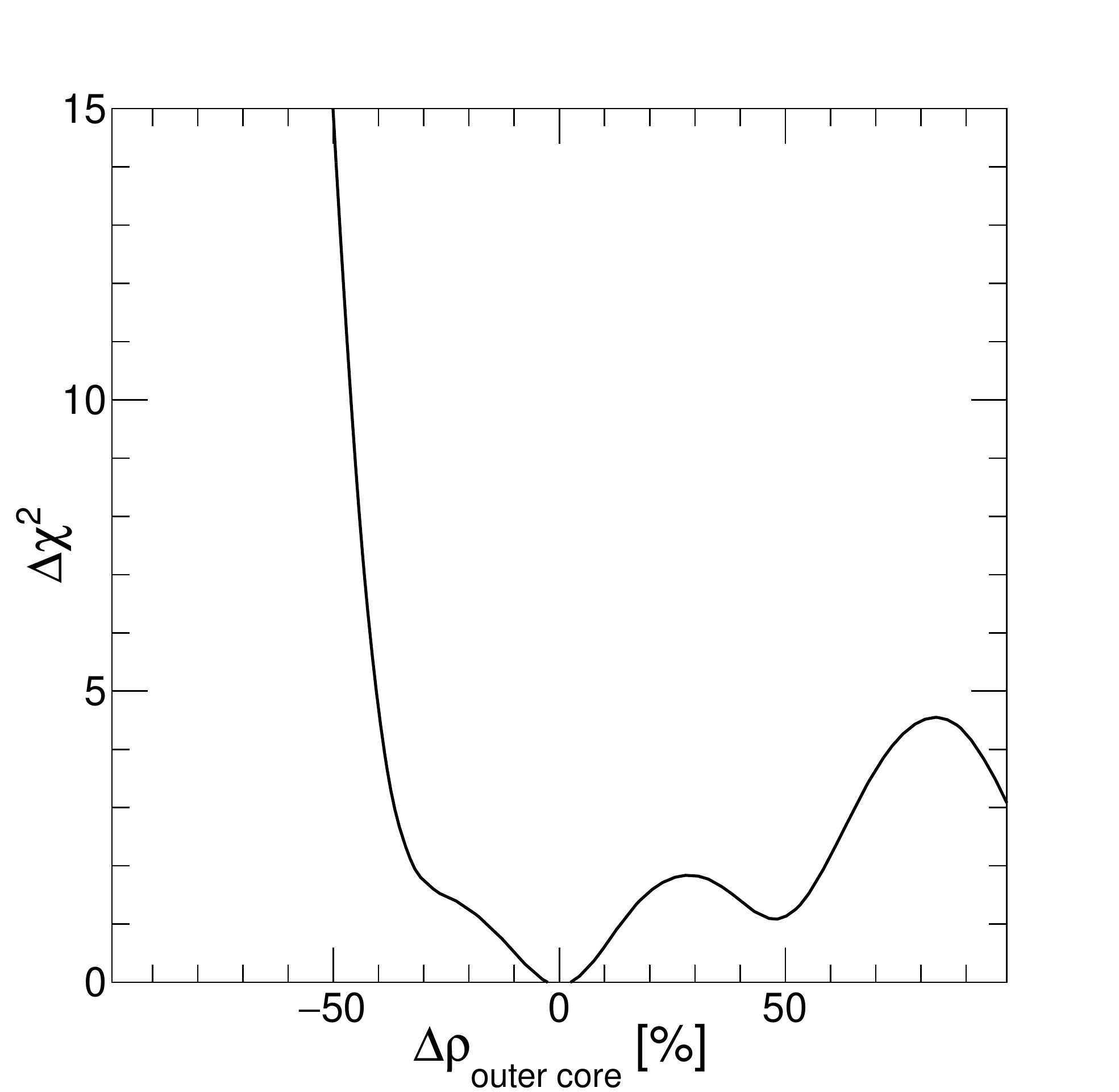}} & % \\
{\includegraphics[width=5truecm]{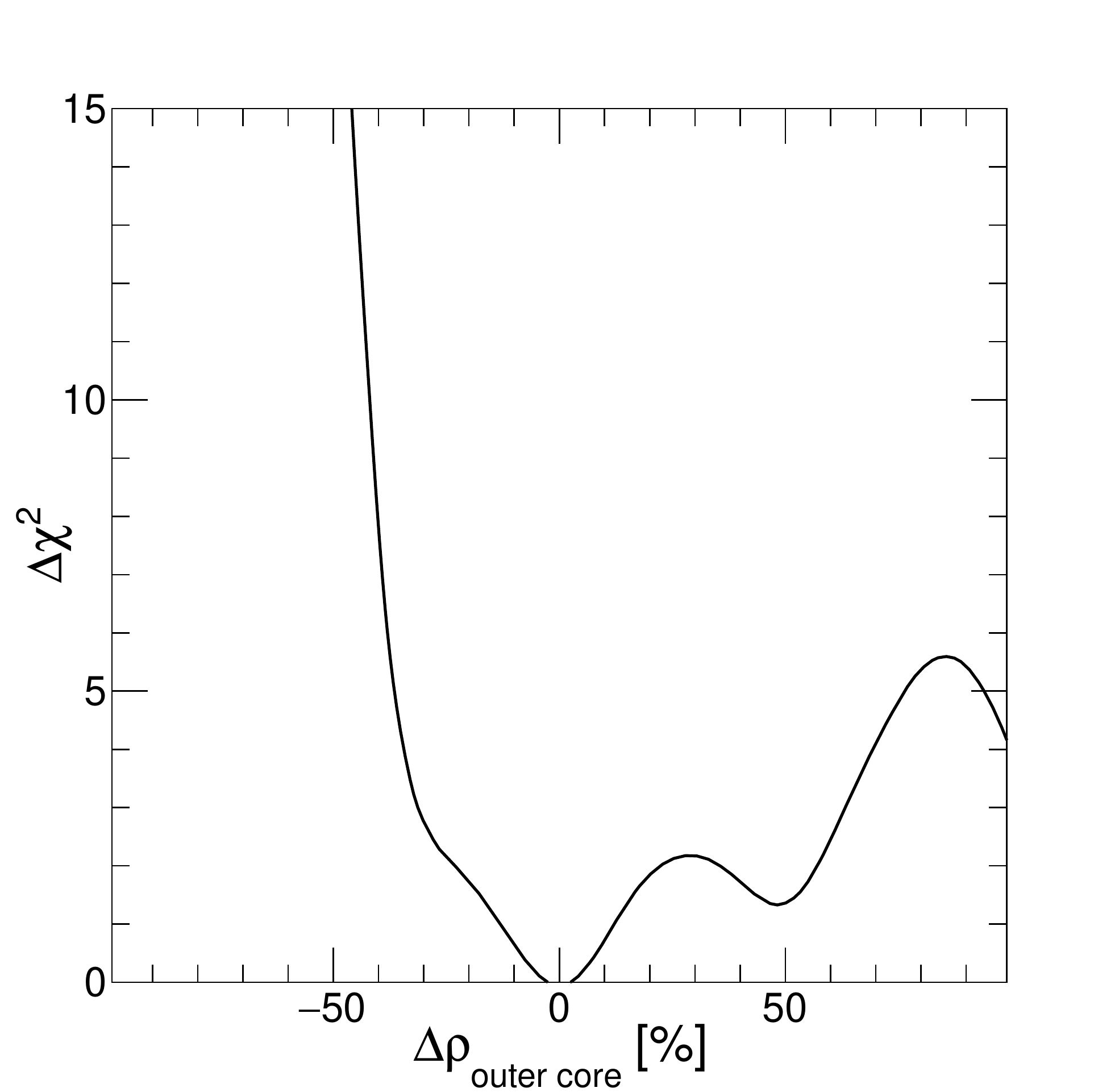}}
\end{tabular}
\caption{
 The same as in Fig.  \ref{fig:NHOCvsMantle}, but  
without implementing the Earth total mass constraint. 
The results shown are for $\sin^2\theta_{23} = 0.42$, 0.50, 0.58 
(left, center and right panels) and in the cases of 
``minimal'', ``optimistic'' and ``conservative'' systematic errors 
(top, middle and bottom panels). See text for further details.
}
\label{fig:NHOCnoComp}
\end{figure}
%%%%%%%%%%%%%%%%%%%%%%%%%%%%%%%%%%%%%%%%%%%%
%
\noindent
non-Gaussian form.
As Fig.  \ref{fig:NHOCvsIC} indicates, in this case ORCA is not sensitive 
to OC densities which are larger than the PREM OC density.
This is due, in particular, to the fact that the IC mass 
is much smaller that the OC mass and only an insignificant increase 
of the OC density can be compensated by a decrease of the IC density.
The wiggles seen in Fig.  \ref{fig:NHOCvsIC} reflect the dependence 
of the relevant oscillation probabilities 
on the correlated modifications of the OC and IC densities.
In the case of 
``minimal'' systematic errors 
and $\Delta \rho_{\rm outer~core} < 0$, ORCA can determine 
the OC density for  $\sin^2\theta_{23} = 0.42$, 0.50, 0.58
with an uncertainty respectively of 
(-39\%),  (-37\%) and (-35\%) at $3\sigma$ C.L.
In the case of ``conservative'' systematic errors,
the sensitivity reads (-46\%),  (-45\%) and (-38\%)
for  $\sin^2\theta_{23} = 0.42$, 0.50, 0.58, 
respectively. All these values are bigger than the external constraint 
 $(-49.6) \ltap \Delta \rho_{\rm outer~core}$ following 
from  eq. (\ref{eq:equil}).  
There is a relatively weak dependence on both  
 $\sin^2\theta_{23}$ and the type of the implemented systematic 
uncertainty within those considered by us.
The sensitivity to OC density in this case is 
much worse than in the case when the mantle is used as a 
``compensating'' layer, considered earlier.

\vspace{0.2cm}
{\bf C. Without Compensation}

\vspace{0.2cm}
We get very different results on the sensitivity of ORCA to the 
OC density when the total Earth mass constraint is 
not implemented. This case is unphysical, except for 
insignificant deviations of the OC density from the PREM value.
The corresponding results are shown in Fig. \ref{fig:NHOCnoComp}.
The $\chi^2$ dependence on  $\Delta \rho_{\rm outer~core}$ 
is asymmetric and non-Gaussian. 
The sensitivity is, in general, much worse 
than in the case of enforcing the total Earth mass constraint 
with mantle being the ``compensating'' layer.
Even in the ``most favorable'' case of 
``minimal'' uncertainties and $\sin^2\theta_{23} = 0.58$,
for example, ORCA can be sensitive at $3\sigma$
only to relatively large deviations of OC density from the PREM value,
which are $\sim 38\%$ for $\Delta \rho_{\rm outer~core} < 0$ and are 
larger for $\Delta \rho_{\rm outer~core} > 0$. 

\vspace{-0.2cm}
%%%%%%%%%%%%%%%%%%%%%%%%%%%%%%%%%%
%
\subsection{Sensitivity to the Inner Core Density}
\label{ssec:IC}
%%%%%%%%%%%%%%%%%%%%%%%%%%%%%%

  Our results show that ORCA  is essentially insensitive to deviations 
of the IC density from the PREM value, corresponding to 
$|\Delta\rho_{\rm inner~core}| \leq 100\%$. This conclusion is valid 
for all values of $\sin^2\theta_{23}$, all possible ``compensating'' 
layers and all possible systematic uncertainties considered.
Although the IC density is largest in the Earth, the IC volume, 
as given by PREM, is much smaller than the OC and mantle volumes. 
As a consequence, the IC contribution to the Earth total mass is also 
much smaller that the OC and mantle contributions. 

\begin{figure}[!t]
  \centering

\begin{tabular}{lll}
{\includegraphics[width=0.3\linewidth]{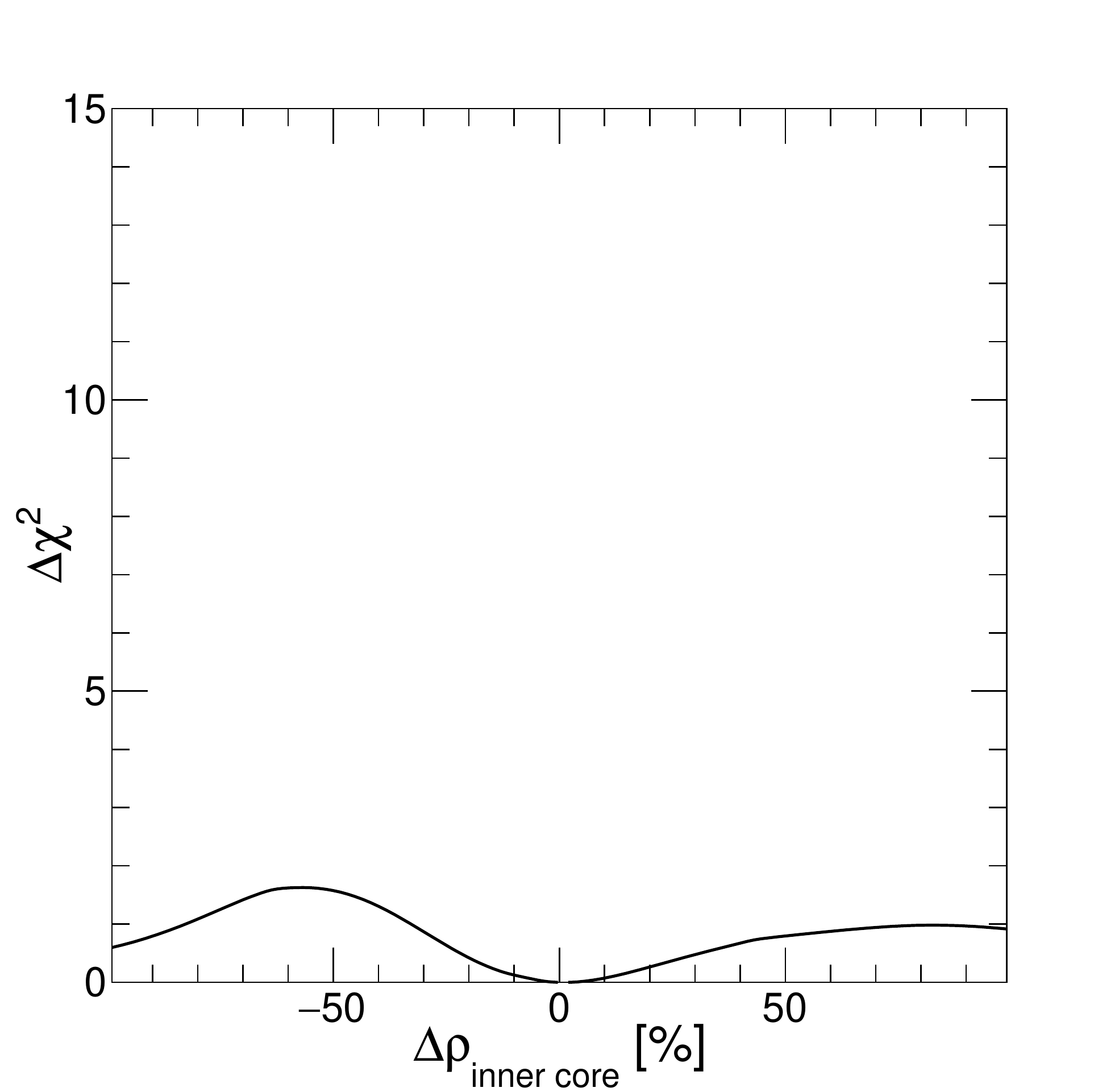}}
    &
   { \includegraphics[width=0.3\linewidth]{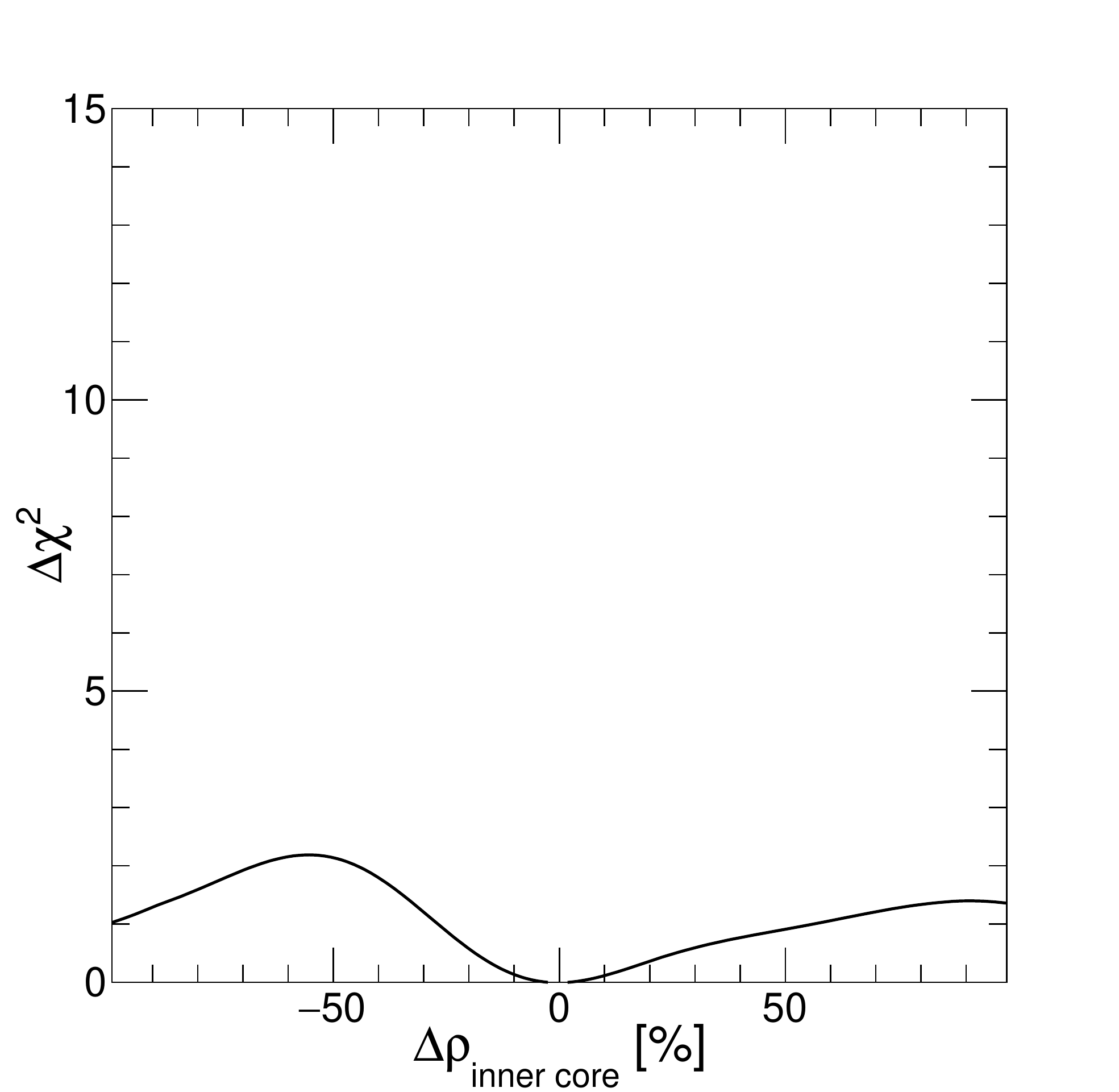}}
    &
    {\includegraphics[width=0.3\linewidth]{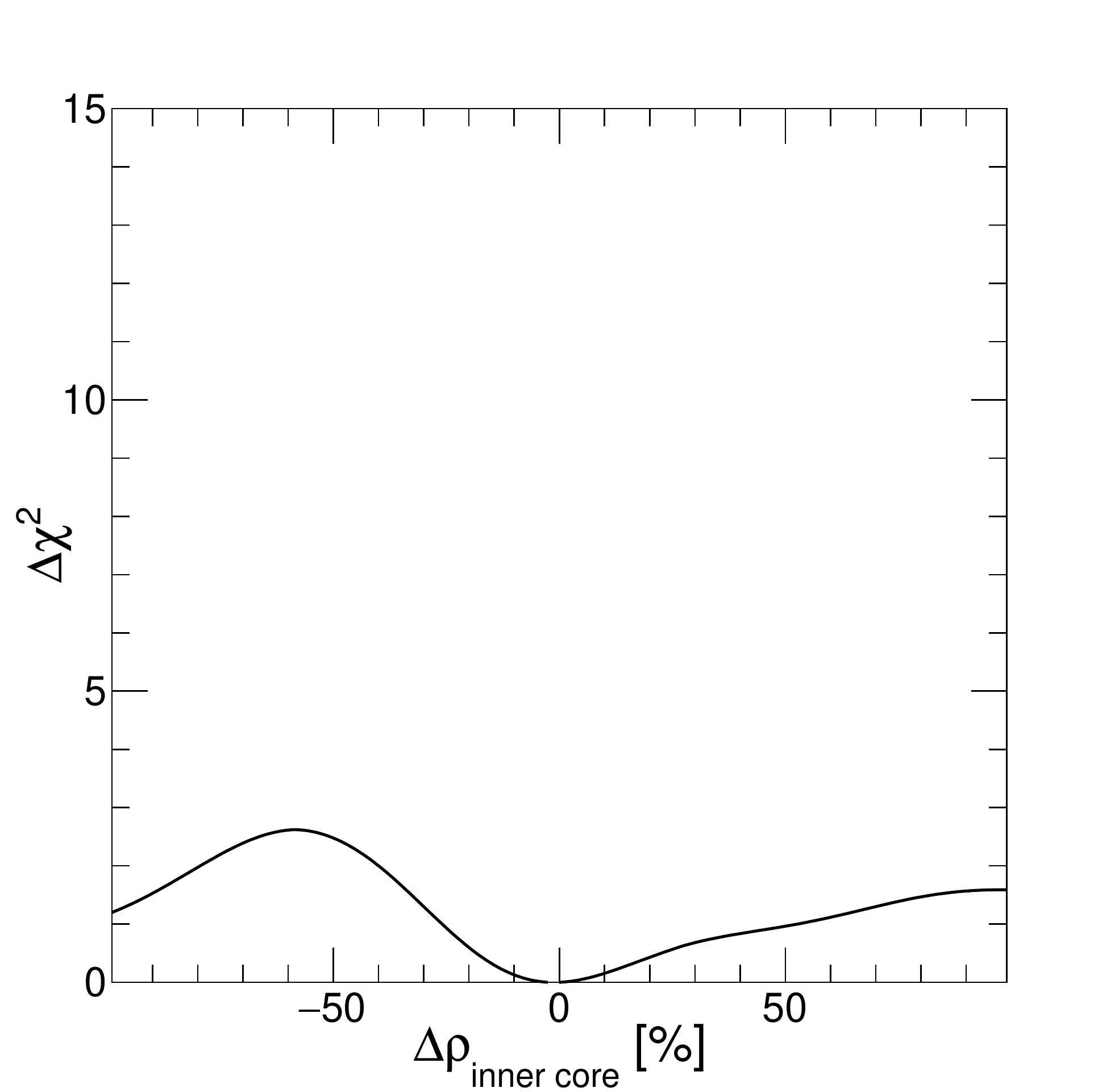}}
   \\
    {\includegraphics[width=0.3\linewidth]{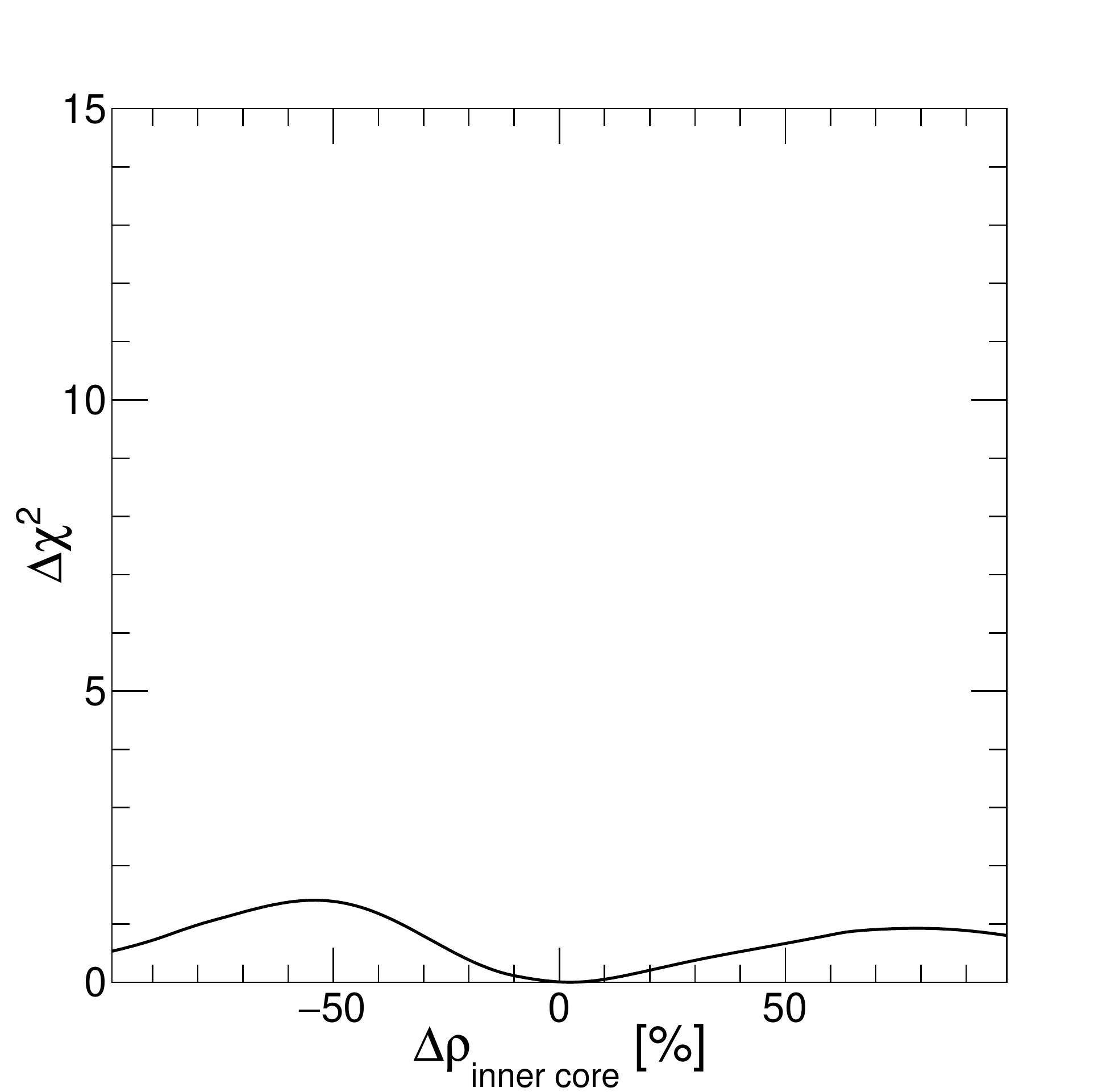}}
    &
    {\includegraphics[width=0.3\linewidth]{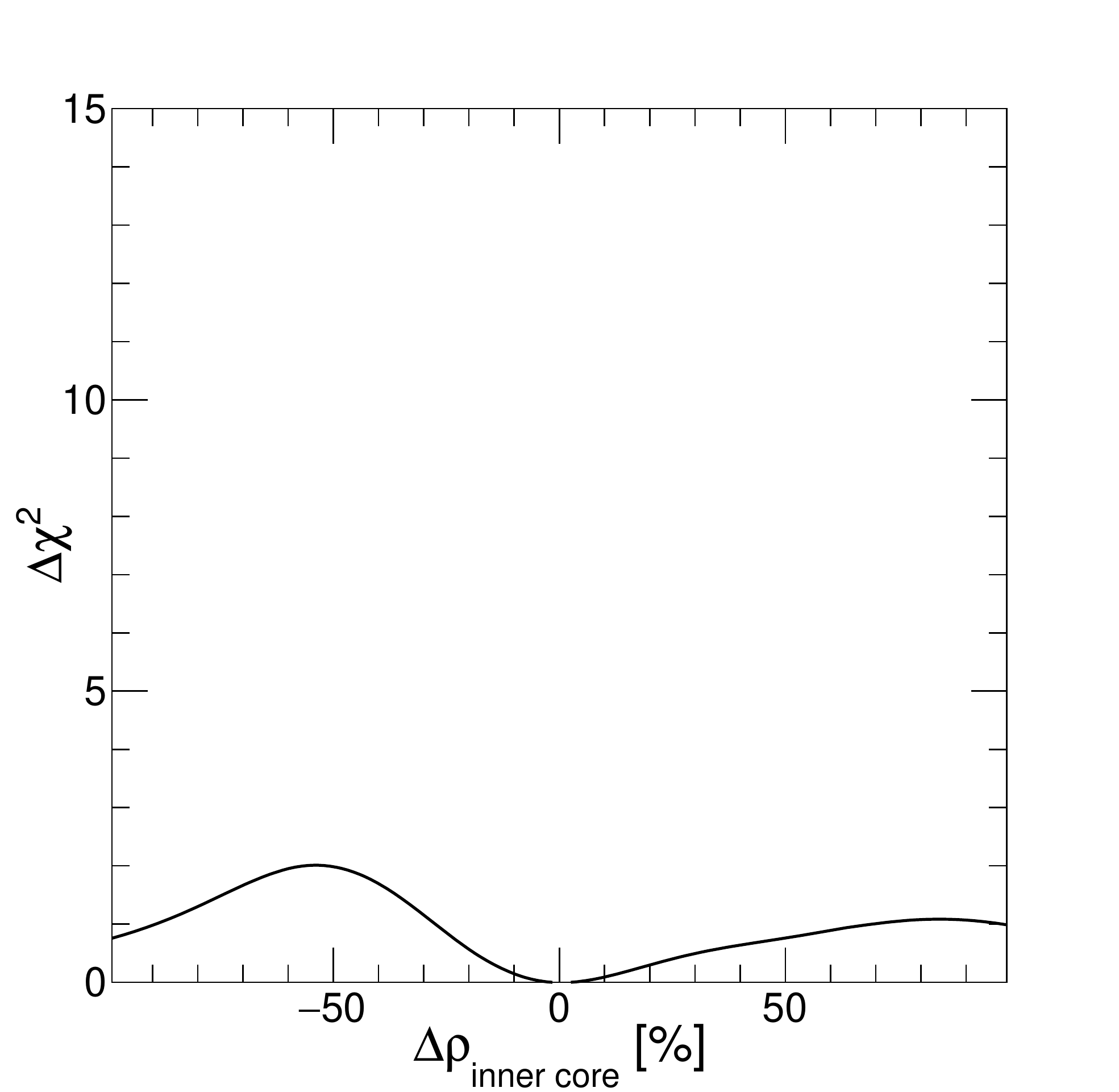}}
    &
   { \includegraphics[width=0.3\linewidth]{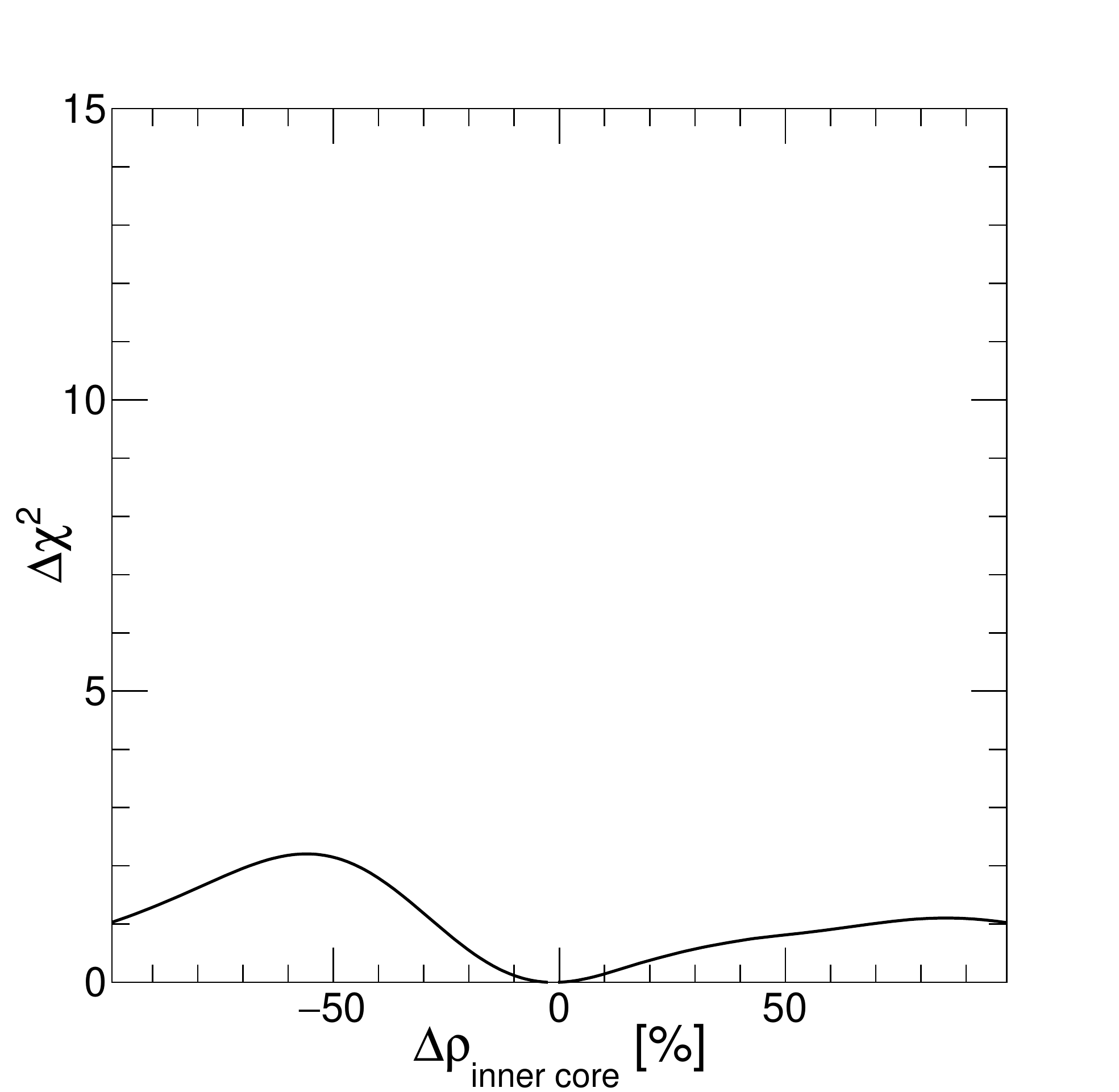}}
\end{tabular}
 \caption{Sensitivity to the IC density. 
The Earth total mass constraint 
i) is implemented by compensating the IC density variation with a 
corresponding mantle density change (top panels), 
ii) is not implemented (bottom panels). The 
results shown are for $\sin^2\theta_{23} = 0.42$, 0.50, 0.58 
(left, center and right panels) and in the case of 
``minimal'' systematic errors. See text for further details.
}
\label{fig:NHICvsMantNoComp}
\end{figure}

 The sensitivity of ORCA to the IC density is illustrated 
in Fig. \ref{fig:NHICvsMantNoComp} in the case of ``minimal'' 
systematic uncertainties. The top panels are obtained 
by imposing the Earth total mass constraint 
and compensating the IC density variation with a 
corresponding mantle density change, while in the bottom panels 
we show results derived without implementing this constraint. 

%%%%%%%%%%%%%%%%%%%%%%%%%%%%%
%
\subsection{Sensitivity to the Core Density}
\label{ssec:OC}
%
%%%%%%%%%%%%%%%%%%%%%%%%%%%%%

 In the present subsection we present results on the ORCA prospective 
sensitivity to the total core density. They are obtained by 
considering deviations of the IC and OC densities by the same 
constant scale factor from their respective PREM densities. 
We consider two cases: i) compensating the variation of the core 
density by corresponding change of the mantle density so as to satisfy 
the total Earth mass constraint, and 
ii) not imposing the total Earth mass constraint 
when varying the total core density. 

\vspace{0.3cm}
{\bf A. Compensation with Mantle Density}

\begin{figure}[!t]
  \centering
\begin{tabular}{lll}
{\includegraphics[width=0.3\linewidth]{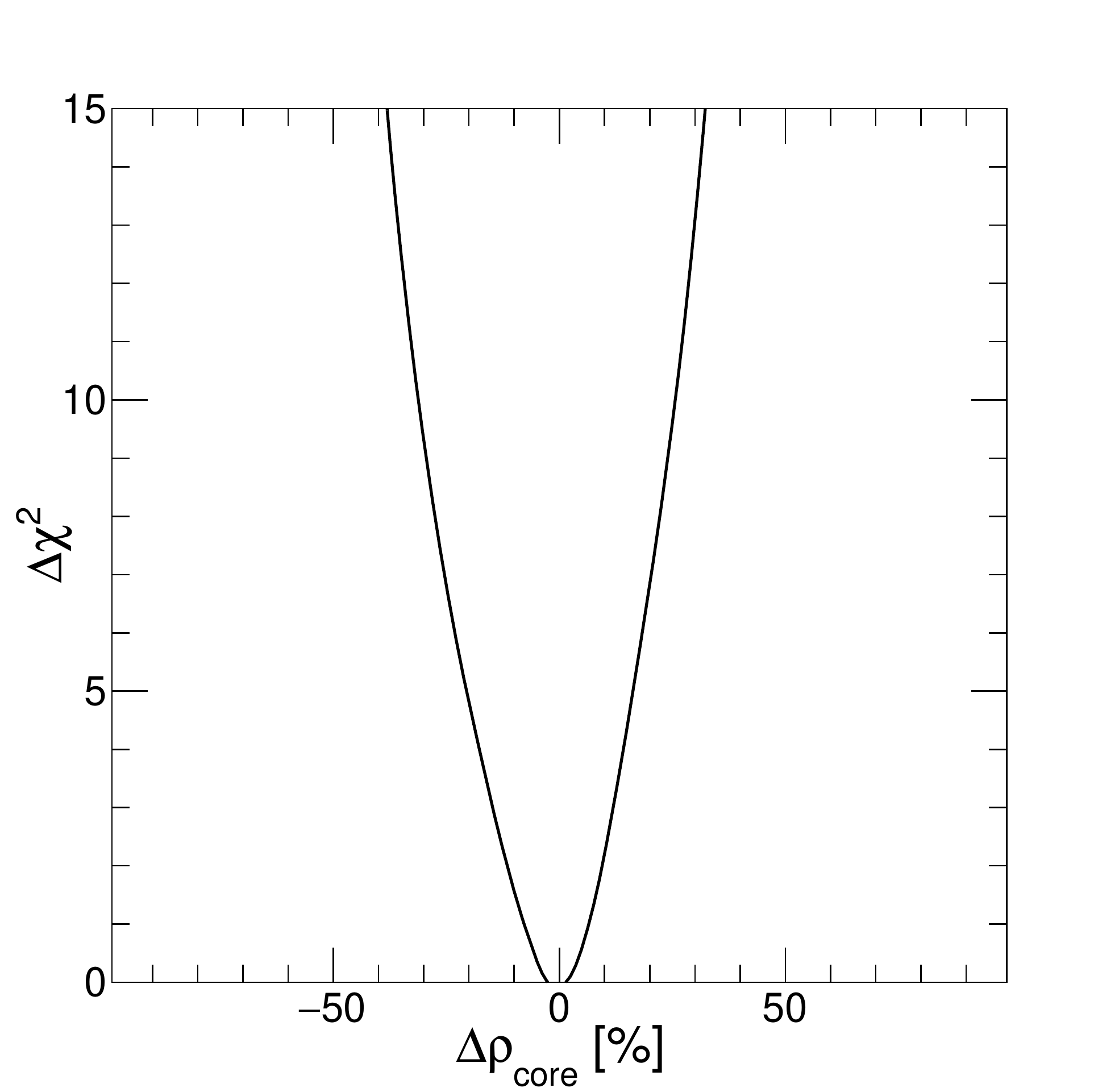}}
    &
   { \includegraphics[width=0.3\linewidth]{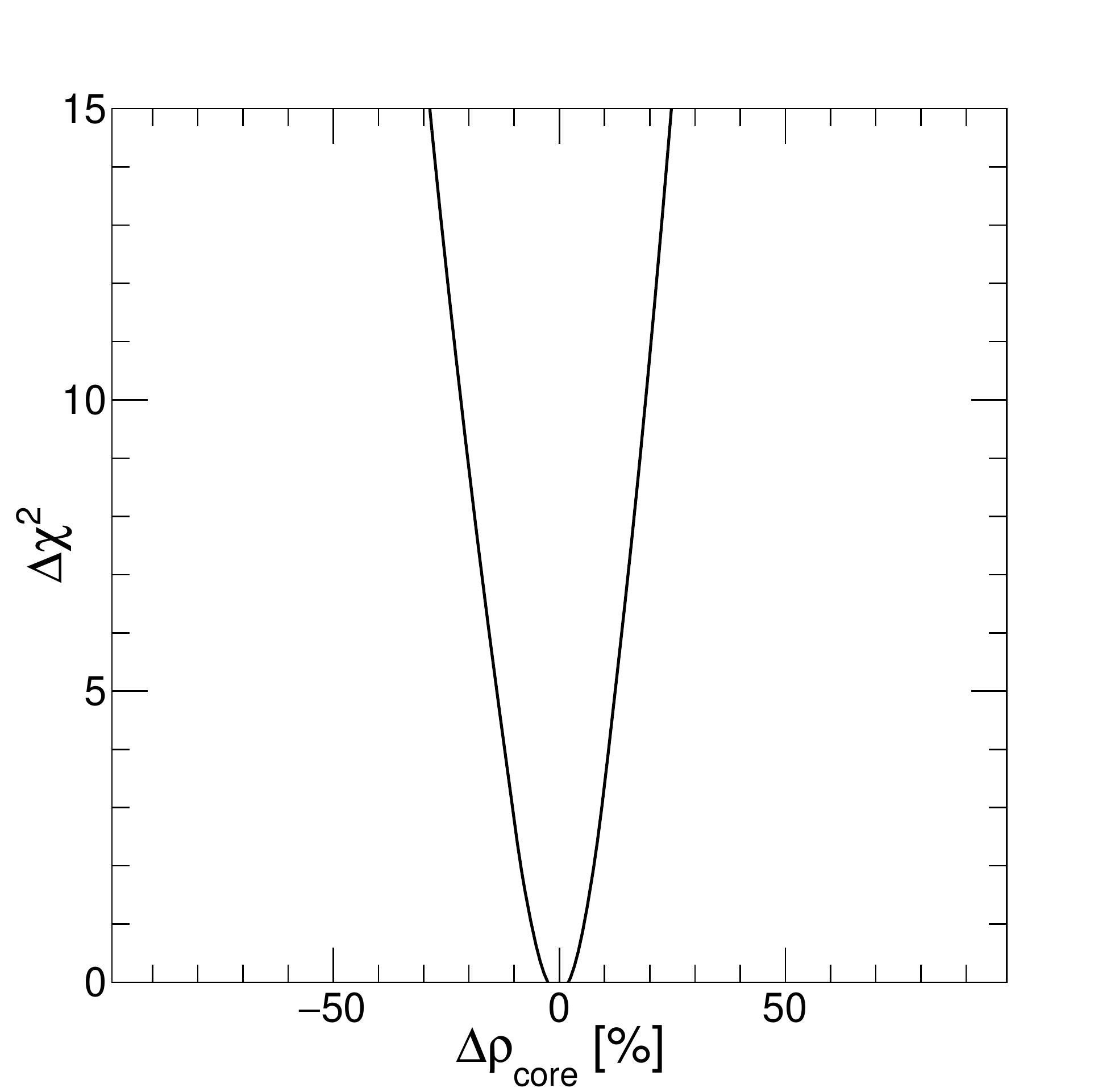}}
    &
    {\includegraphics[width=0.3\linewidth]{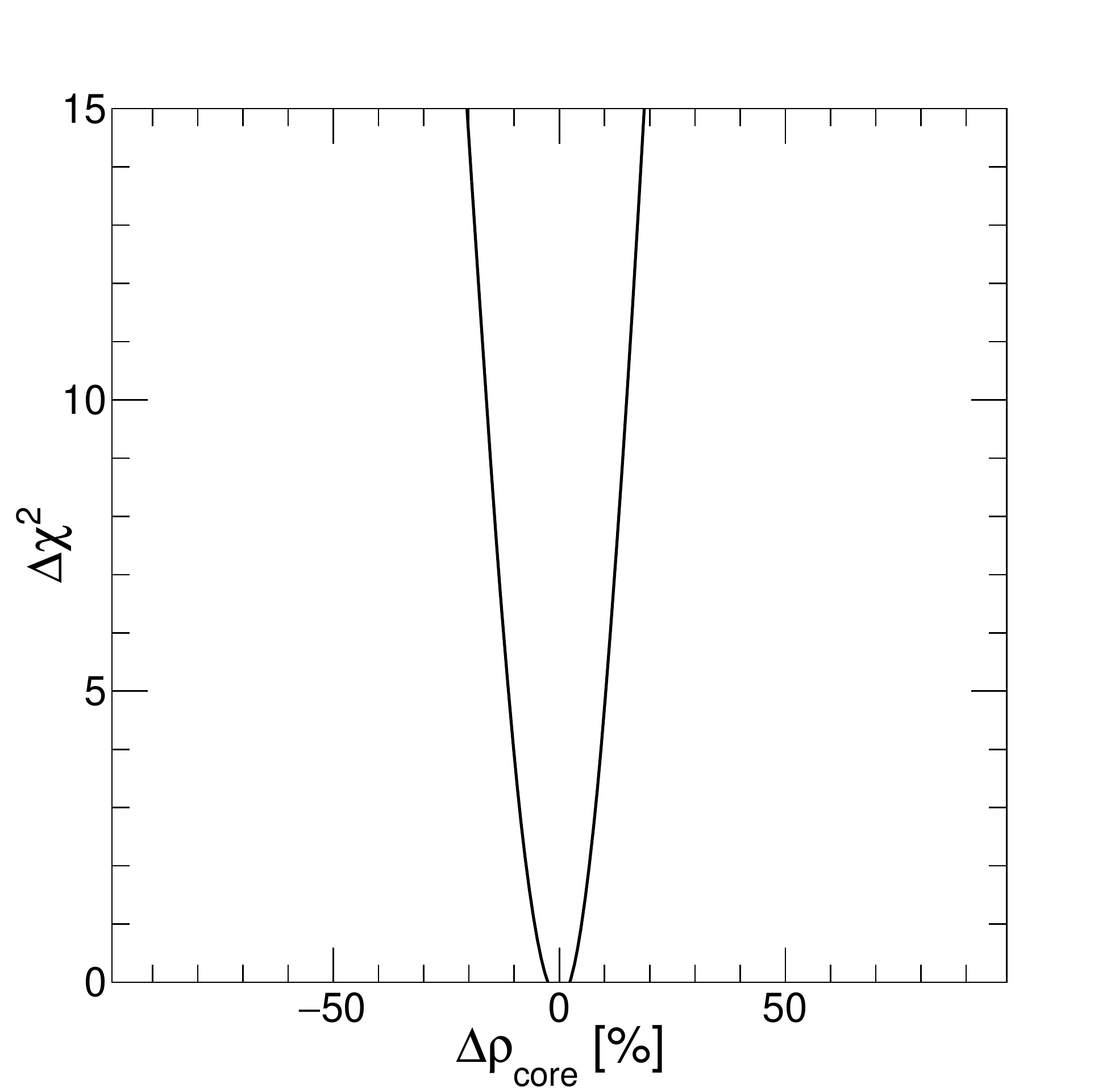}}
    \\
    {\includegraphics[width=0.3\linewidth]{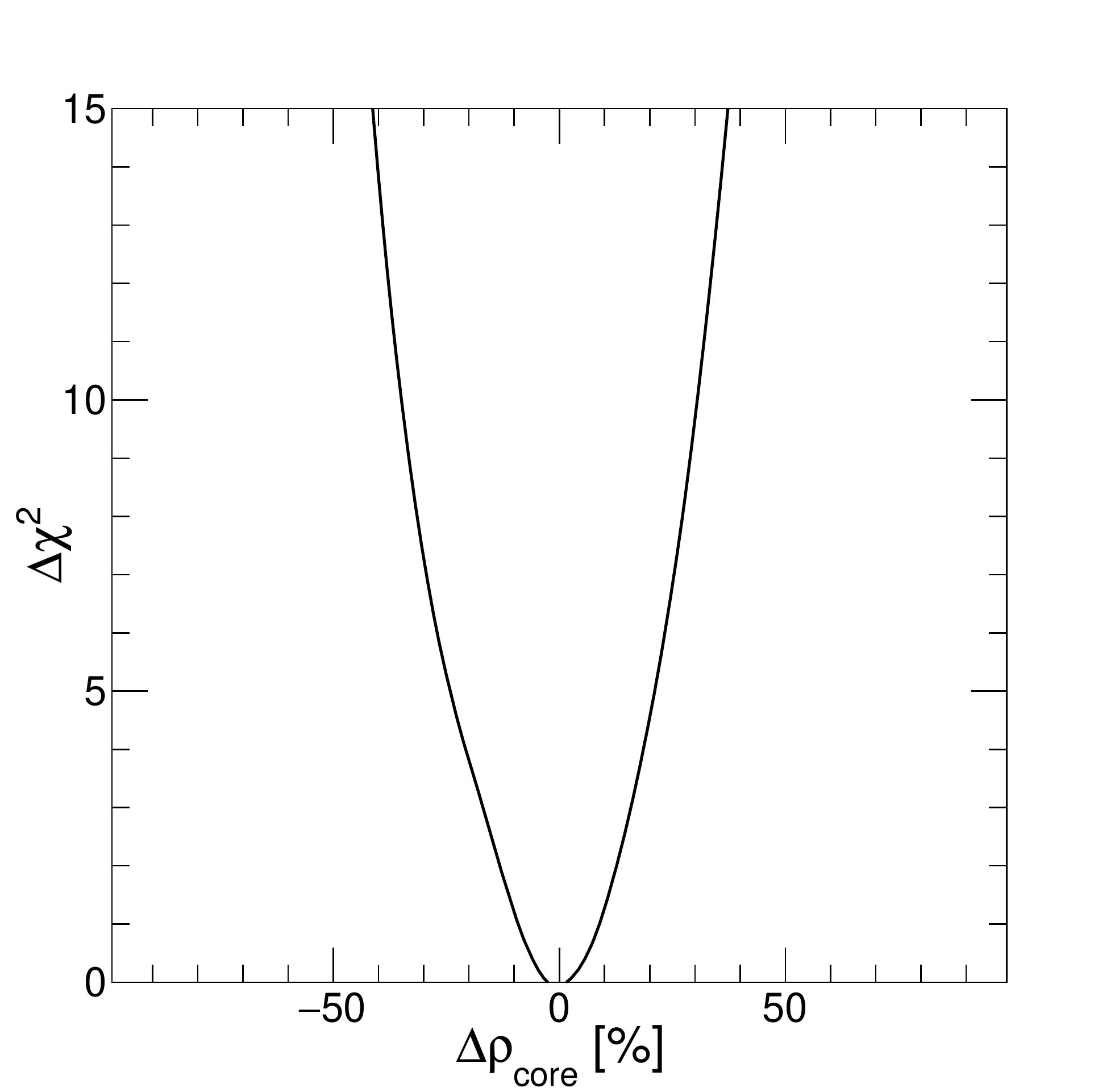}}
    &
    {\includegraphics[width=0.3\linewidth]{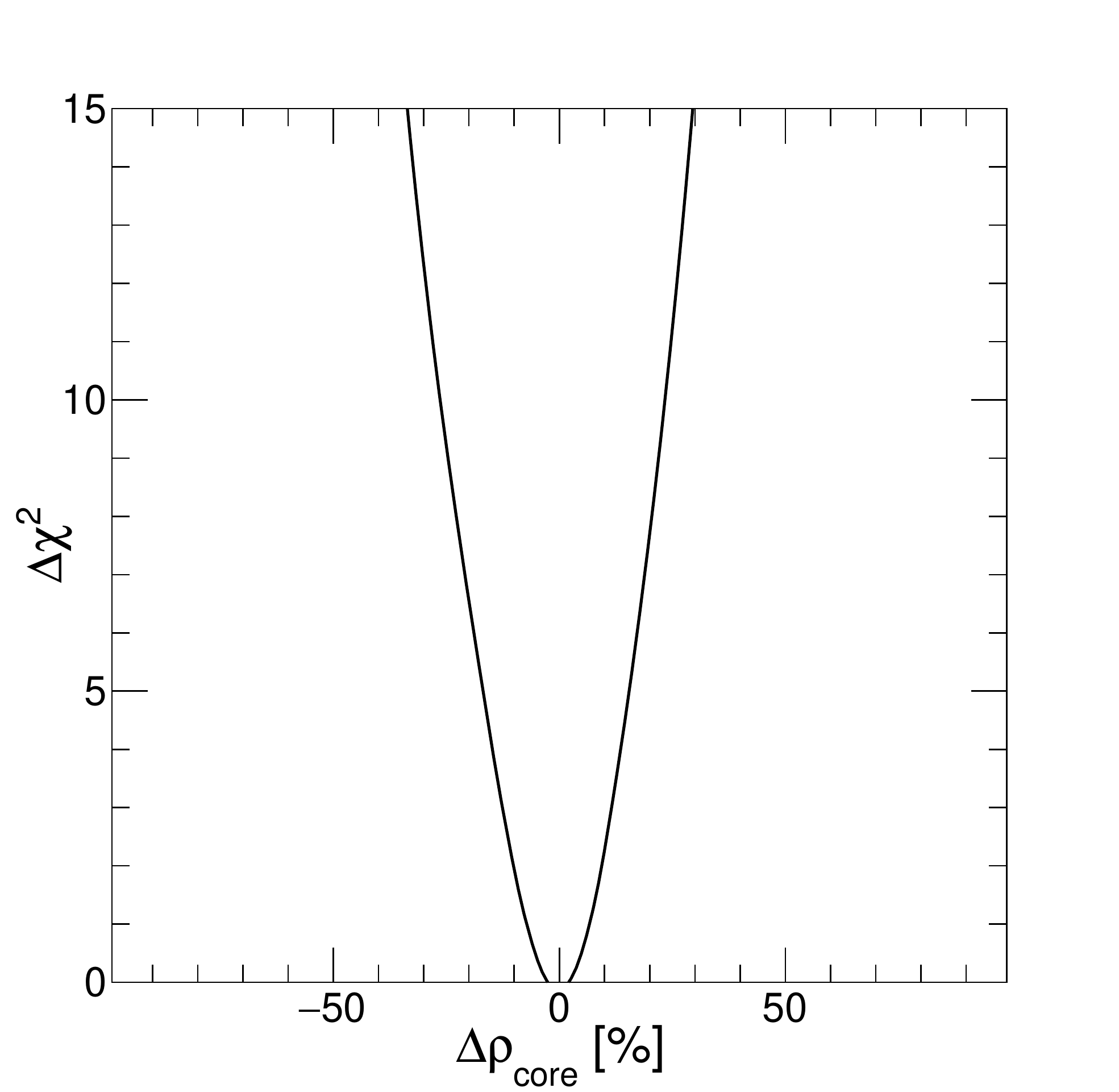}}
    &
   { \includegraphics[width=0.3\linewidth]{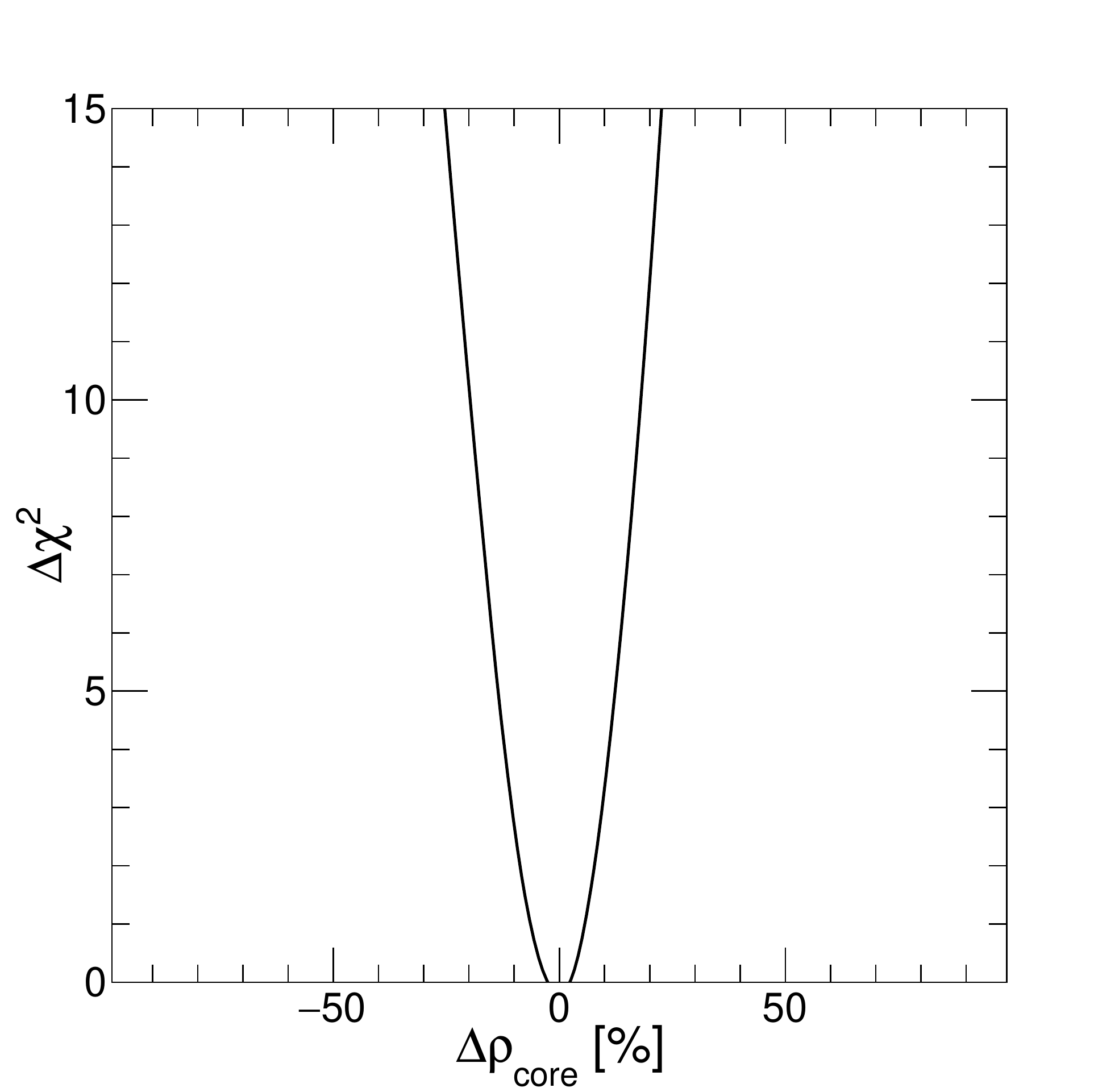}}
    \\
   { \includegraphics[width=0.3\linewidth]{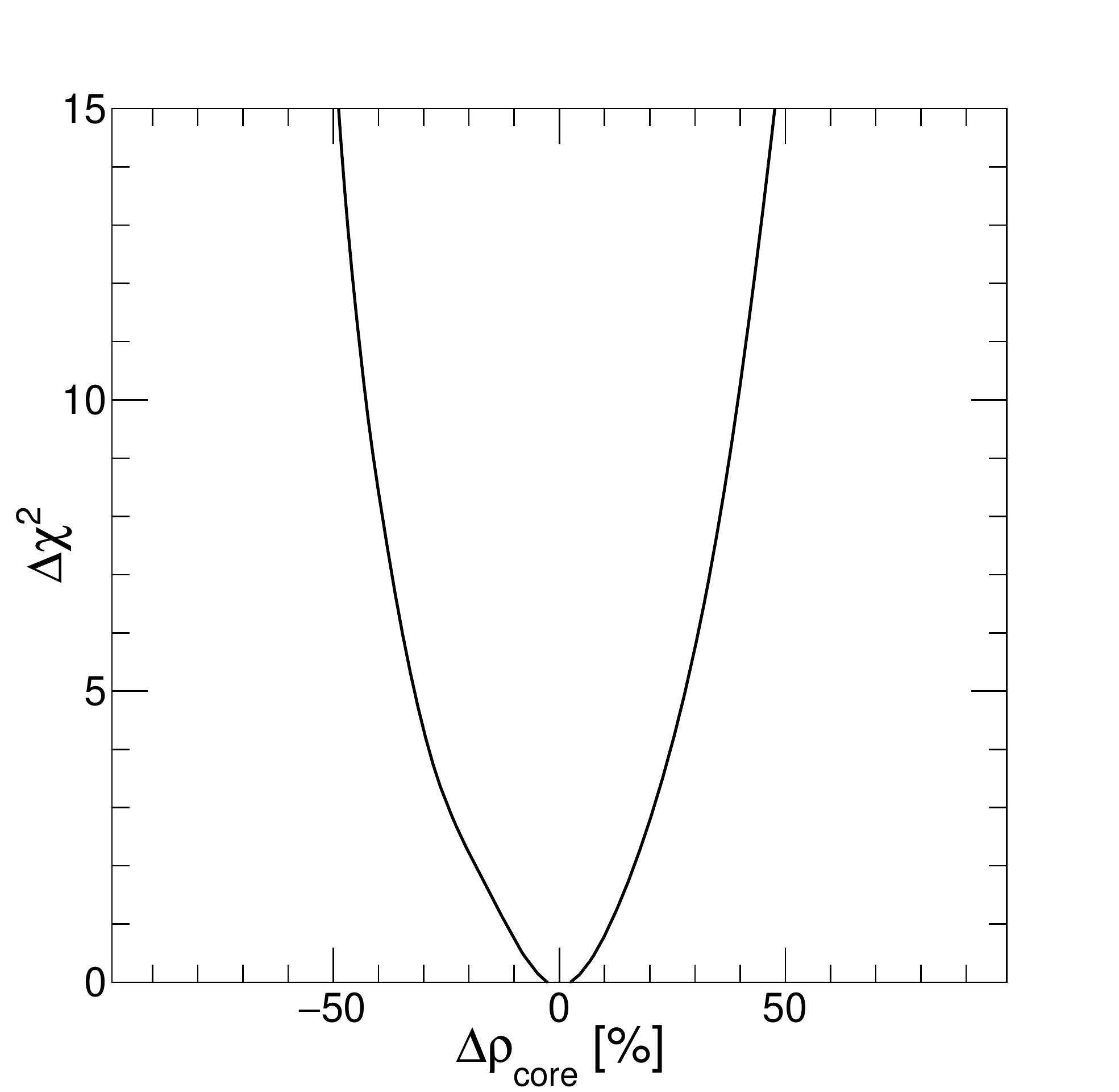}}
    &
    {\includegraphics[width=0.3\linewidth]{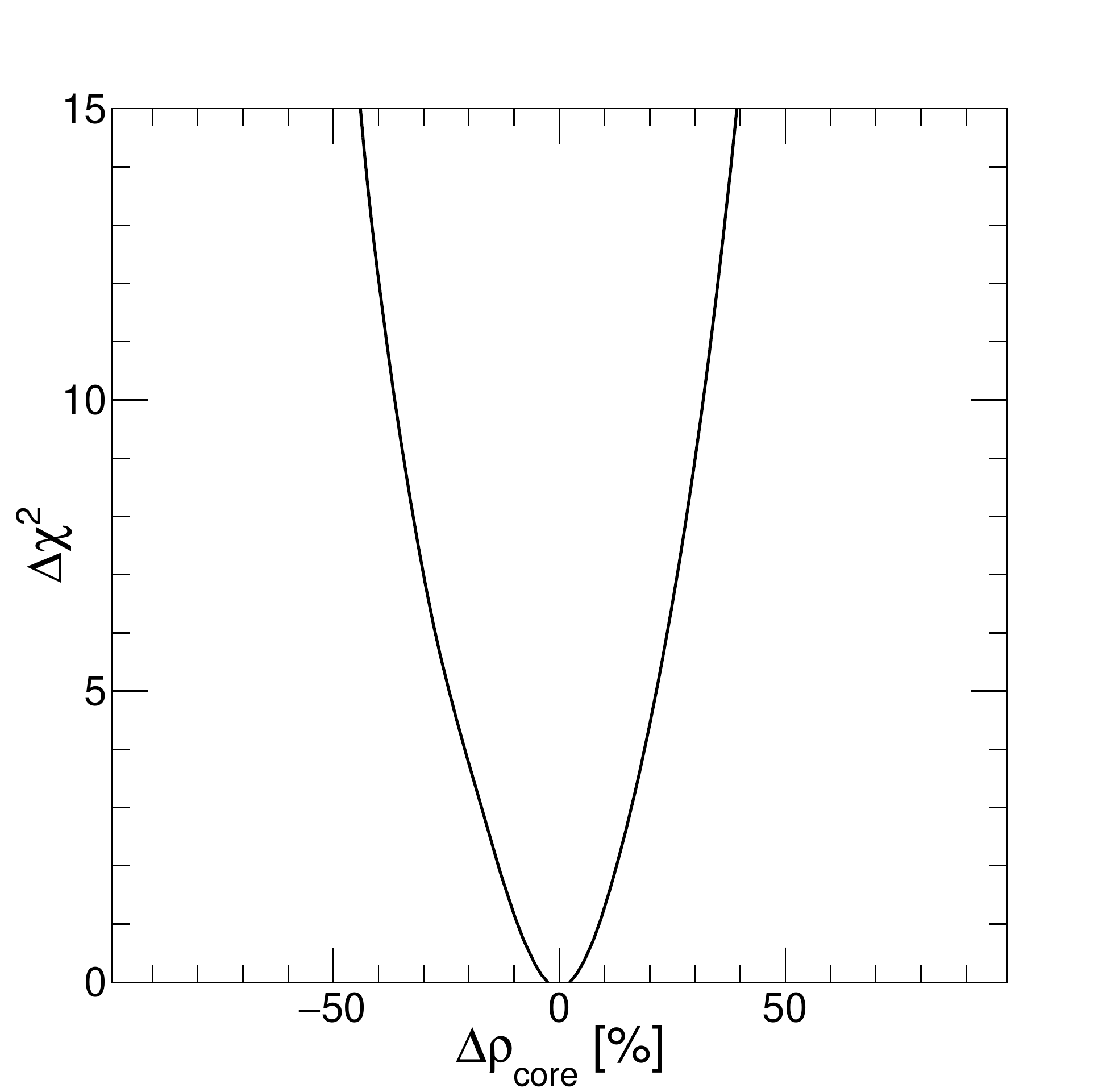}}
    &
   {\includegraphics[width=0.3\linewidth]{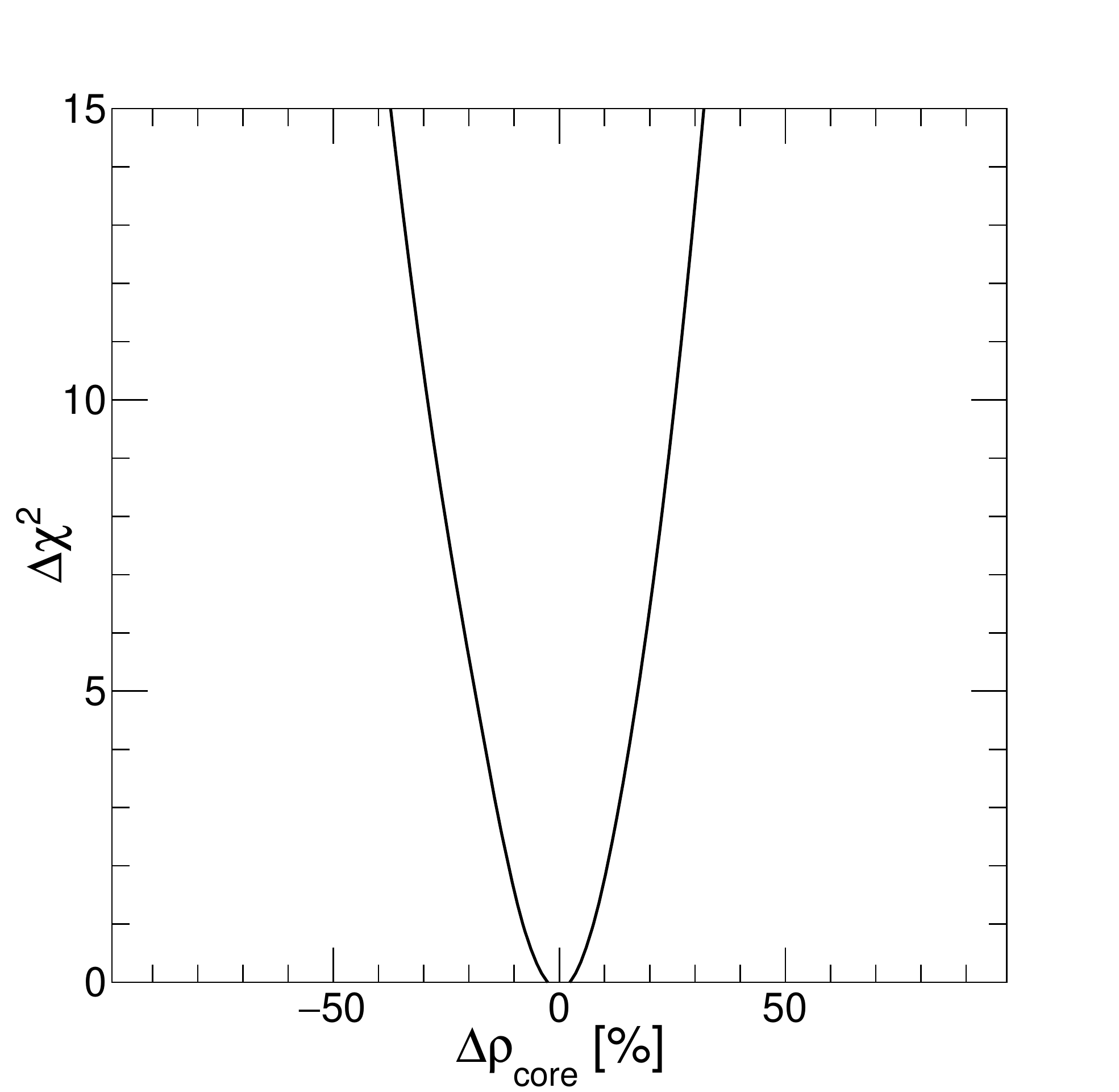}}
  \end{tabular}
 \caption{Sensitivity to the core density in the case of NO spectrum 
and 10 years of data. 
The Earth total mass constraint 
is implemented by compensating the core density variation with a 
corresponding mantle density change. The 
results shown are for $\sin^2\theta_{23} = 0.42$, 0.50, 0.58 
(left, center and right panels) and in the cases of 
``minimal'', ``optimistic'' and ``conservative'' systematic errors 
(top, middle and bottom panels). See text for further details.
}
\label{fig:NHCvsMant}
\end{figure}

\vspace{0.3cm}
  Figure \ref{fig:NHCvsMant} illustrates the sensitivity of ORCA 
to the total core density $\rho_{\rm C}(r)$ when 
the ``compensating'' layer is the mantle. 
In the figure, the $\chi^2$-distribution is shown 
as a function of the core relative density 
variation with respect to the PREM value, $\Delta \rho_{\rm core}$. 
As in the other considered cases, the reported results are 
for $\sin^2\theta_{23} = 0.42$, 0.50, 0.58 
(left, center and right panels) and in the cases of 
``minimal'', ``optimistic'' and ``conservative'' systematic errors 
(top, middle and bottom panels). 

 According to our results, the sensitivity of ORCA to $\Delta \rho_{\rm core}$ 
for the three types of systematic 
errors considered are similar to, or somewhat better than,
the results on ORCA sensitivity to $\Delta \rho_{\rm outer~core}$
when OC density change is compensated by mantle density change,
reported in Fig. \ref{fig:NHOCvsMantle}.
All $\chi^2$-distributions in Fig. \ref{fig:NHCvsMant} have a symmetric 
or slightly asymmetric Gaussian form. 
In the case of 
``minimal'' systematic errors, 
ORCA can determined the core density at $3\sigma$ with 
an uncertainty of (-29\%)/+24\%,  (-20\%)/+18\% and (-15\%)/+15\% 
respectively for  $\sin^2\theta_{23} = 0.42$, 0.50, 0.58.
In the case of ``conservative'' systematic errors,
the sensitivity reads:
(-40\%)/+38\%,  (-35\%/+30\% and (-28\%)/+24\% 
for  $\sin^2\theta_{23} = 0.42$, 0.50, 0.58, respectively. 
For  $\sin^2\theta_{23} = 0.58$ (0.50),
it is worse by a factor of $\sim (1.6 - 1.9)$  
than the sensitivity corresponding to 
``minimal''  systematic errors.
%%%%%%%%%%%%%%%%%%%%%%%%%
%%%%%%%%%%%%%%%%%%%%%%%%%%%%%%%
%

%%%%%%%%%%%%%%%%%%%%%%%%%
%%%%%%%%%%%%%%%%%%%%%%%%%%%%%%%

The constraint  $\rho_{man} \leq \rho_{core}$ 
implies maximal negative $\Delta\rho_{core}$ of (-57\%),
which has no effect on the $3\sigma$  maximal negative 
variation of $\rho_{OC}$ to which ORCA might be sensitive.
\footnote{
 Adding the Earth moment of inertia constraint, 
would fix (up to the uncertaities in the constraints) 
the values of  $\rho_{man}$ and $\rho_{core}$. This case will 
be considered elsewhere.
}

\vspace{0.3cm}
{\bf B. Without Compensation}

\begin{figure}[!t]
  \centering
\begin{tabular}{lll}
{\includegraphics[width=0.3\linewidth]{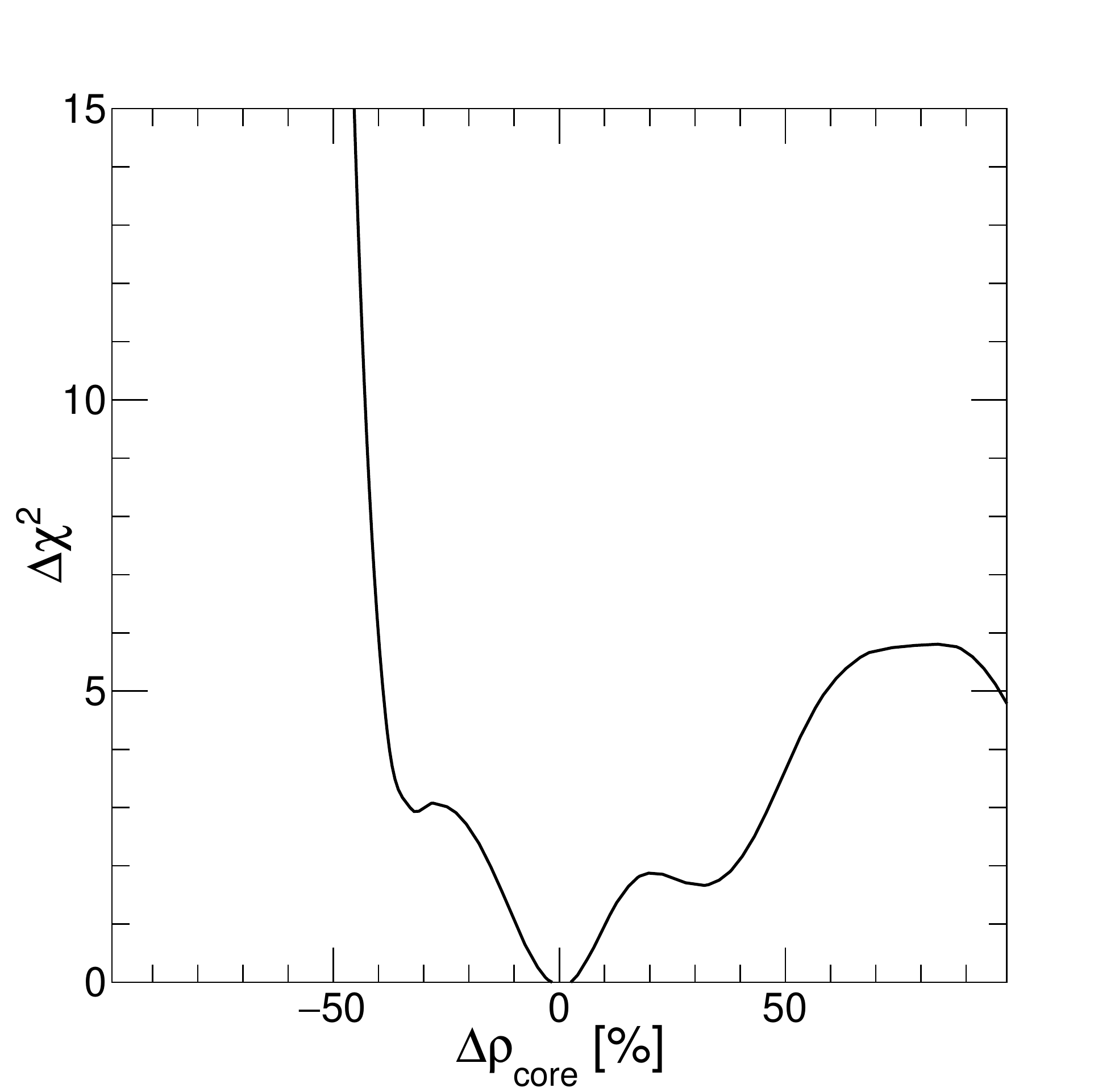}}
    &
   {\includegraphics[width=0.3\linewidth]{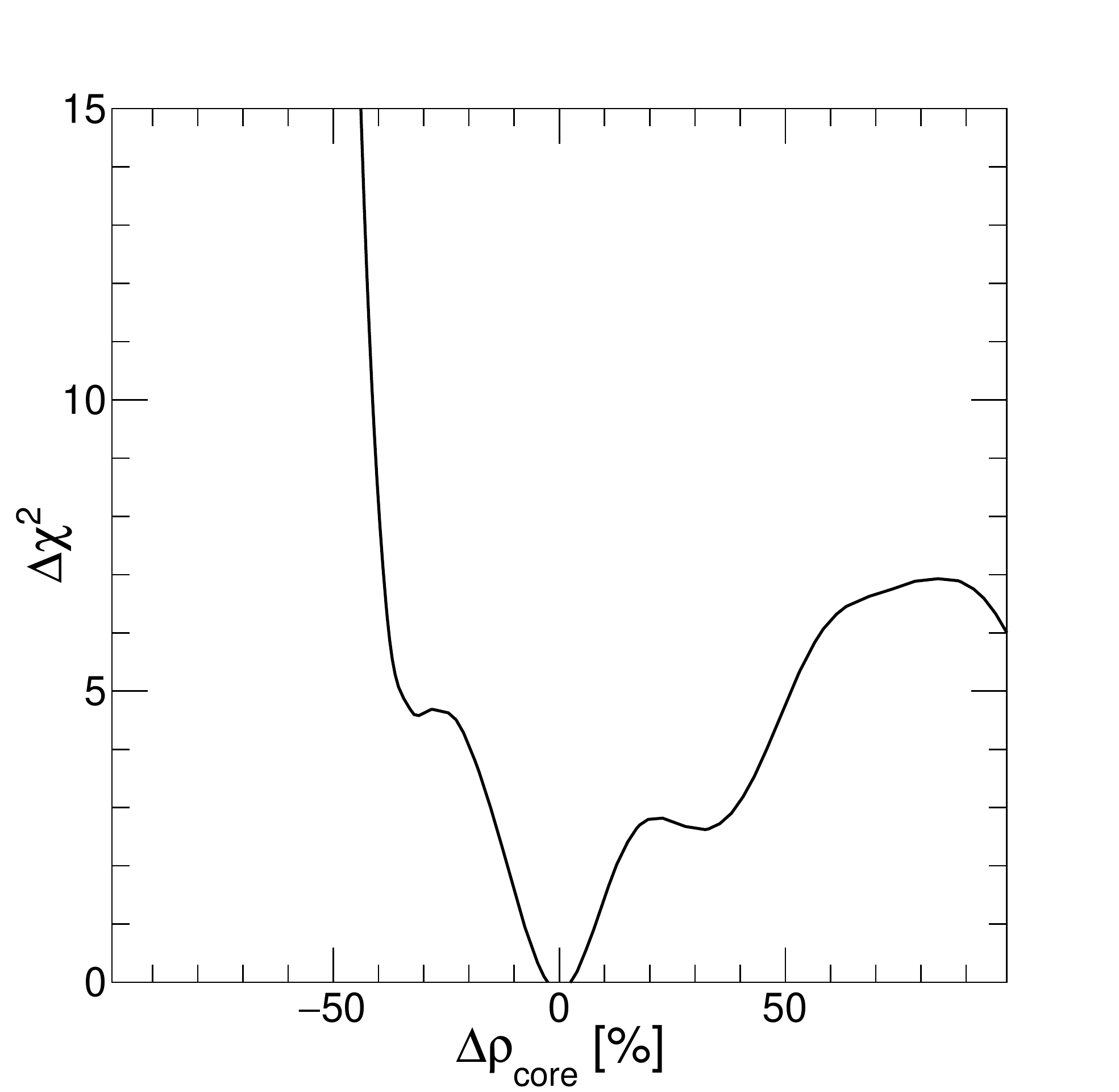}}
    &
    {\includegraphics[width=0.3\linewidth]{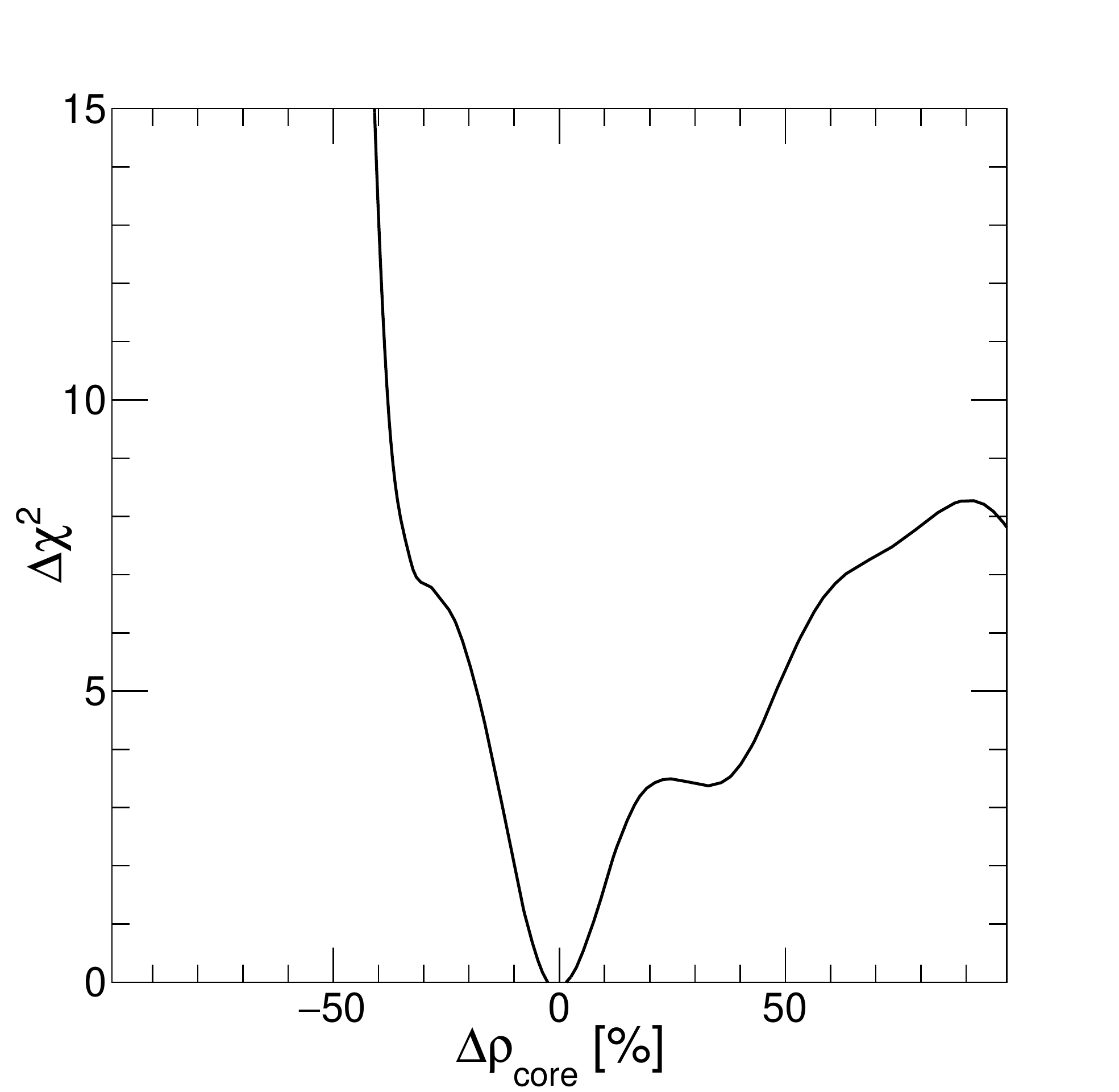}}
    \\
    {\includegraphics[width=0.3\linewidth]{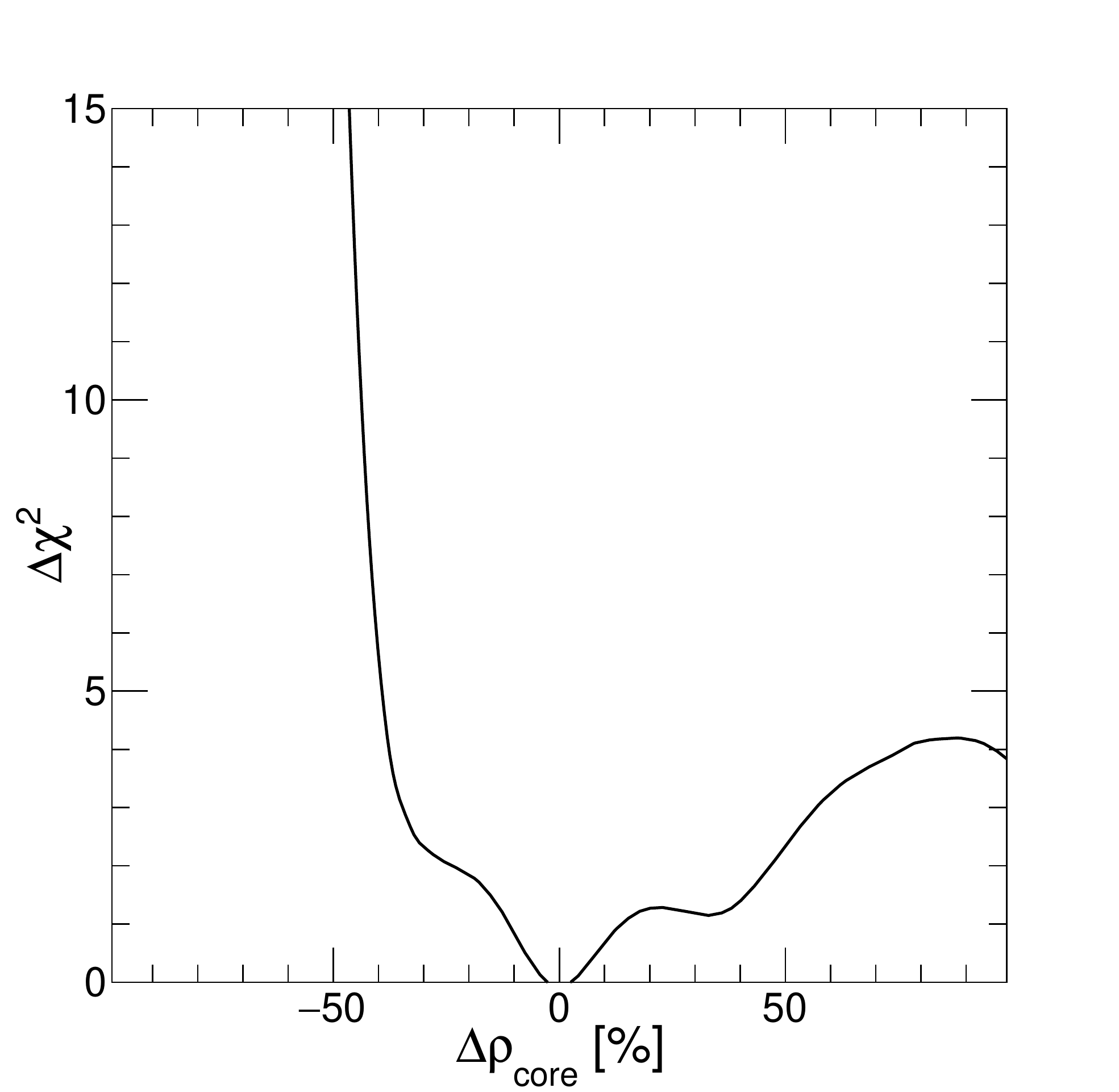}}
    &
    {\includegraphics[width=0.3\linewidth]{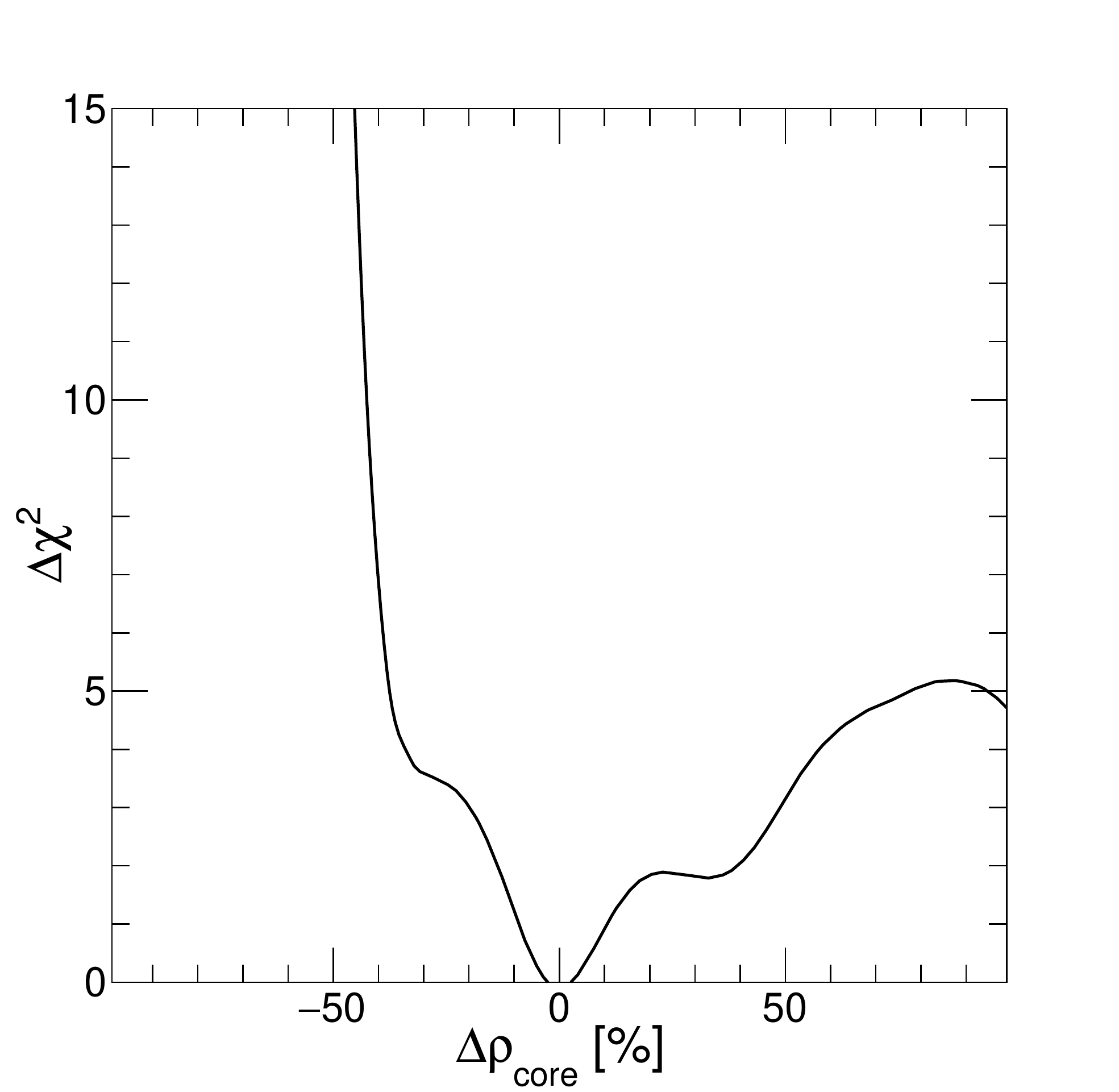}}
    &
   { \includegraphics[width=0.3\linewidth]{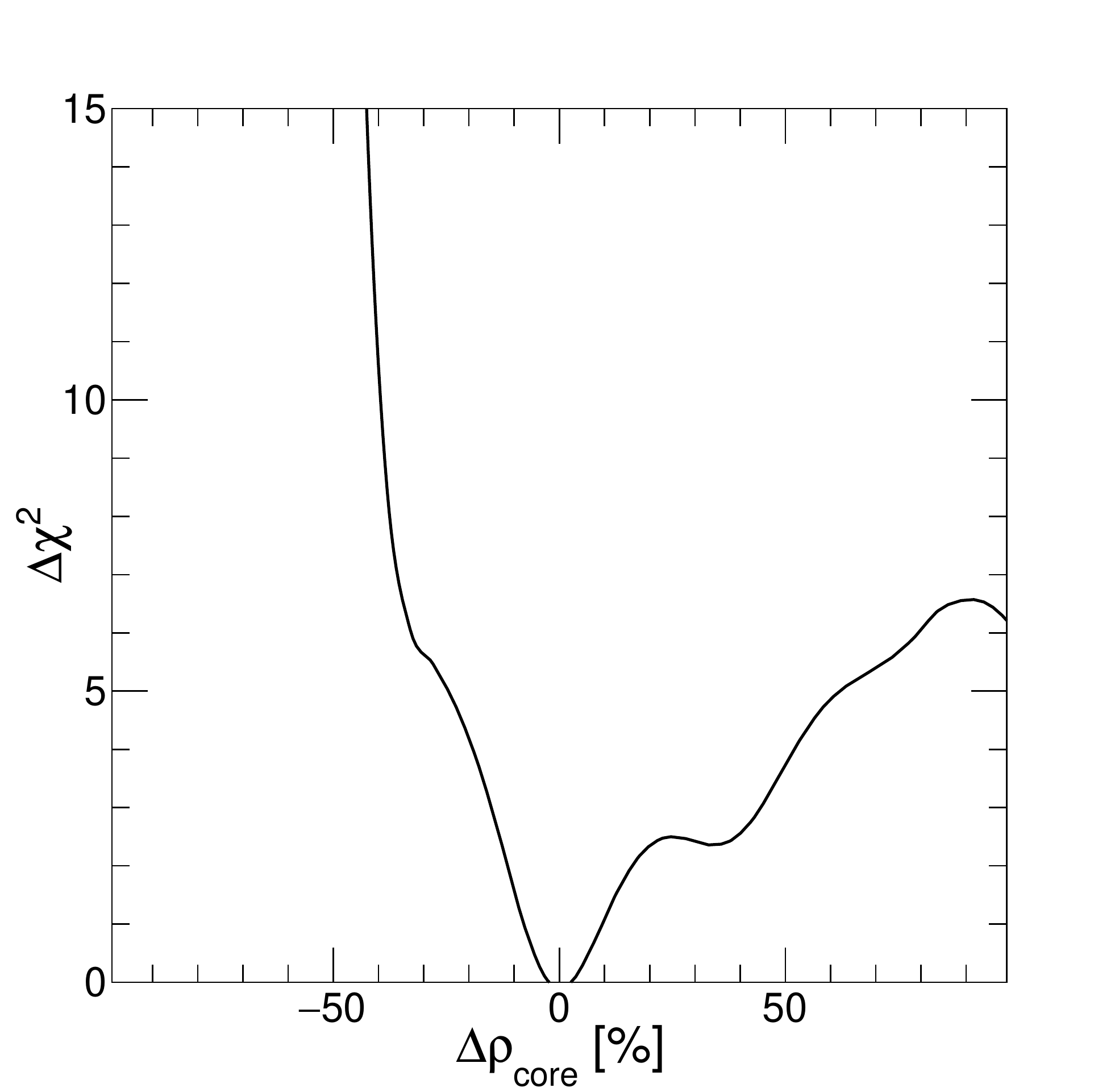}}
    \\
   {\includegraphics[width=0.3\linewidth]{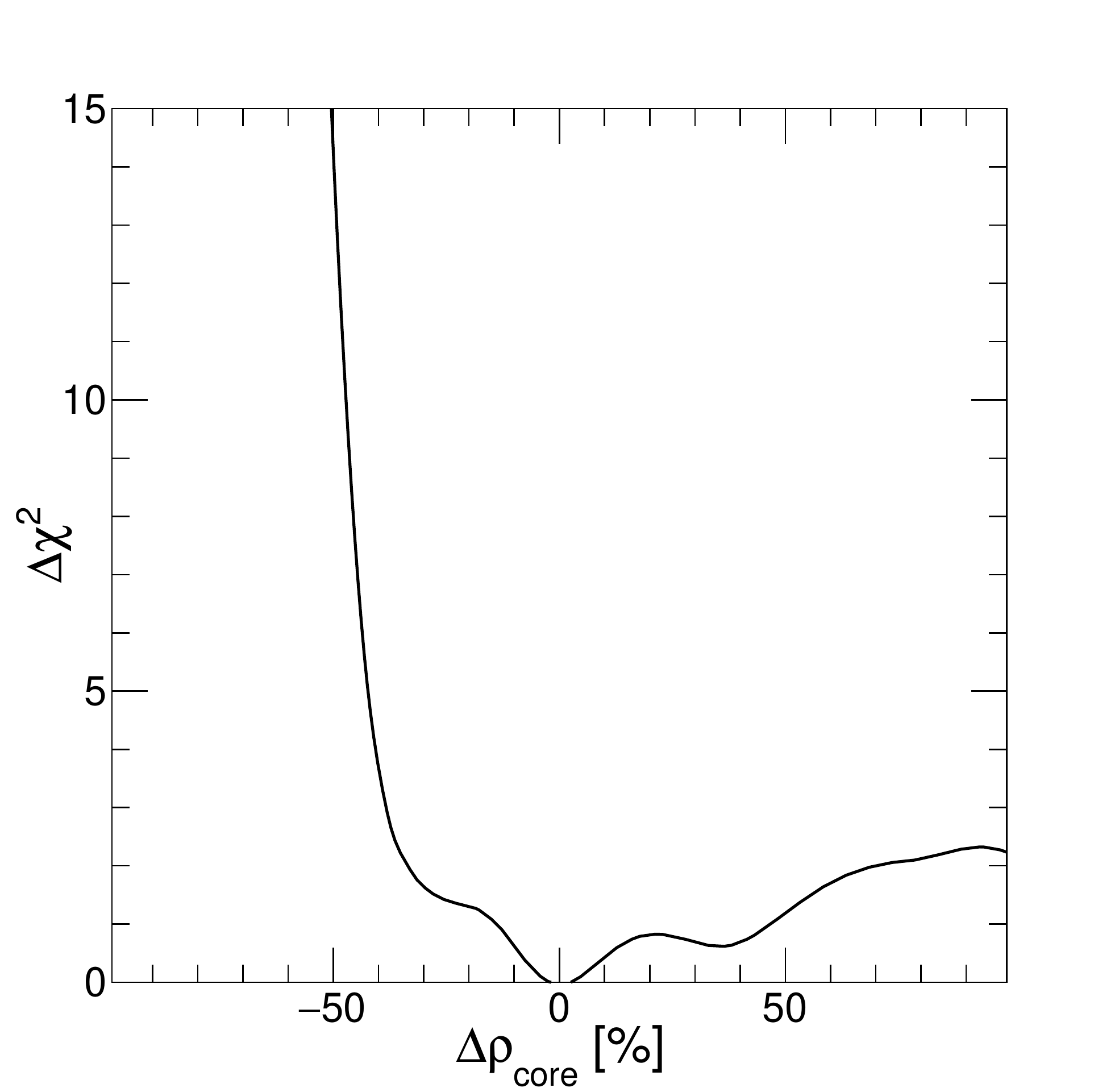}}
    &
    {\includegraphics[width=0.3\linewidth]{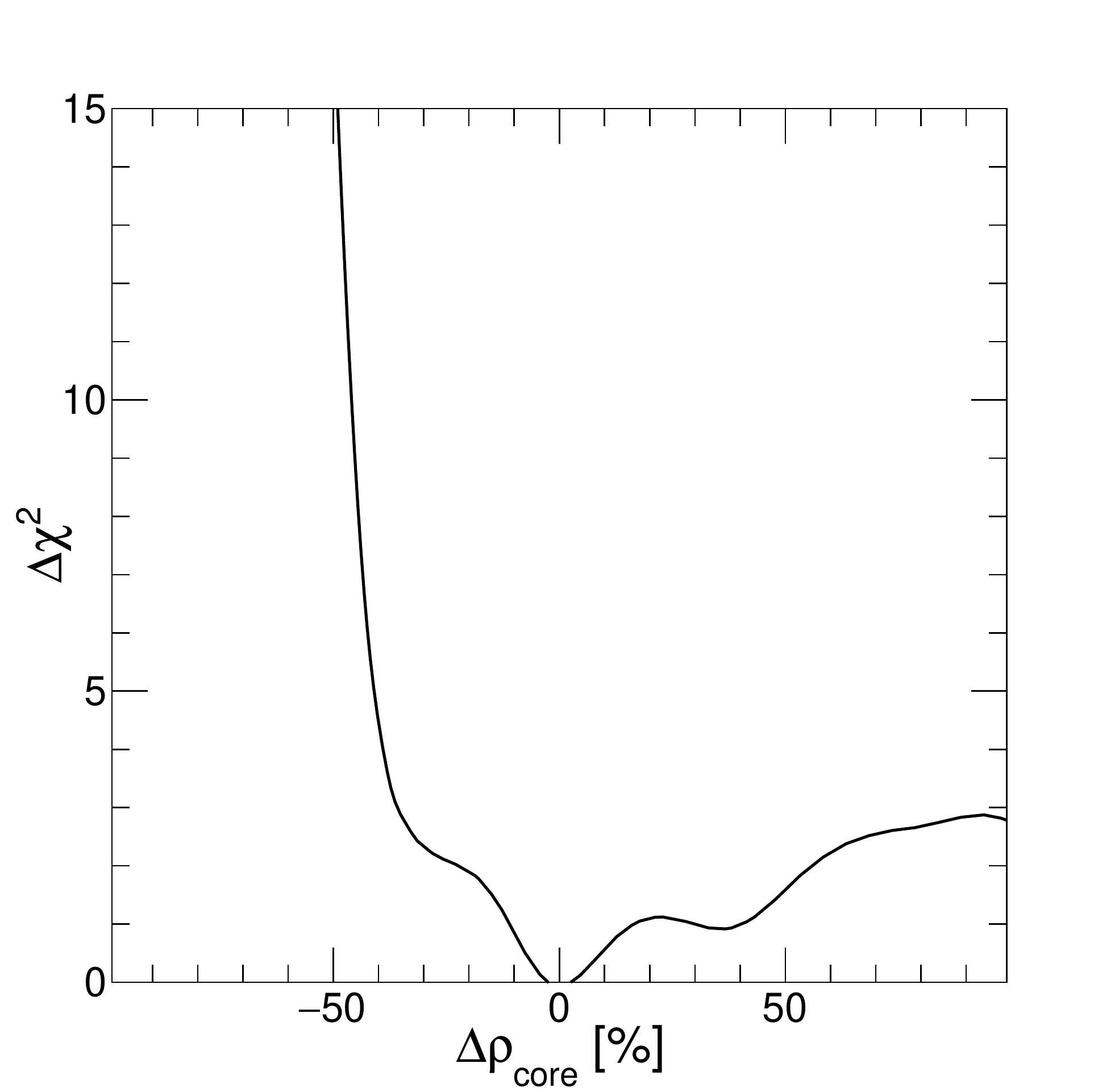}}
    &
   {\includegraphics[width=0.3\linewidth]{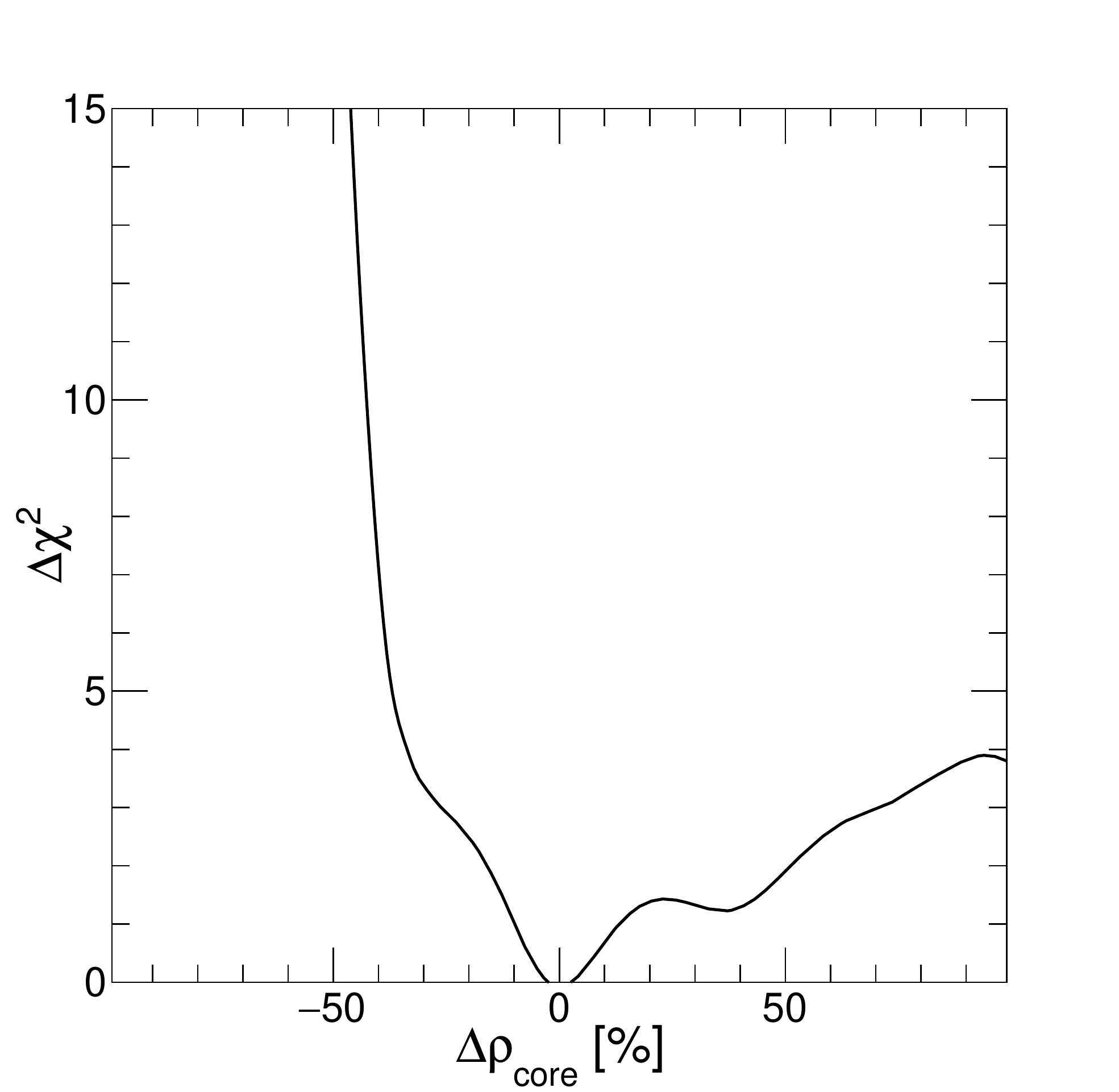}}
  \end{tabular}
 \caption{
The same as in Fig. \ref{fig:NHCvsMant}, but 
without implementing the
Earth total mass constraint. 
The results shown are for $\sin^2\theta_{23} = 0.42$, 0.50, 0.58 
(left, center and right panels) and in the cases of 
``minimal'', ``optimistic'' and ``conservative'' systematic errors 
(top, middle and bottom panels). See text for further details.
}
\label{fig:NHCnoComp}
\end{figure}

\vspace{0.3cm}
The sensitivity of ORCA to the 
core density when the total Earth mass constraint is 
not implemented is similar to that of the OC density under 
the same conditions (see Fig. \ref{fig:NHOCnoComp}).
The results corresponding to this case are presented 
in Fig. \ref{fig:NHCnoComp}.
The $\chi^2$ distributions shown in Fig. \ref{fig:NHCnoComp} 
as a function of $\Delta \rho_{\rm core}$ are asymmetric and non-Gaussian. 
Compared to the case of enforcing the total Earth mass constraint 
with mantle being the ``compensating'' layer, 
the sensitivity is much worse. 
It follows from Fig. \ref{fig:NHCnoComp}, in particular,
that for  $\sin^2\theta_{23} = 0.58$ and 
``minimal'' systematic uncertainties, which are the ``most favorable'' 
conditions ensuring the highest sensitivity,
ORCA can be sensitive at $3\sigma$ C.L. 
only to relatively large deviations 
of the core density from that given by PREM, namely, 
to $\sim 38\%$ for $\Delta \rho_{\rm core} < 0$ and 
to much larger deviations for $\Delta \rho_{\rm core} > 0$. 
%%%%%%%%%%%%%%%%%%%%%%%%%
%%%%%%%%%%%%%%%%%%%%%%%%%%%%%%%

\subsection{Sensitivity to the Mantle Density}
\label{ssec:Mantle}

{\bf A. Compensation with Outer Core Density}

\begin{figure}[!t]
  \centering

\begin{tabular}{lll}
{\includegraphics[width=0.3\linewidth]{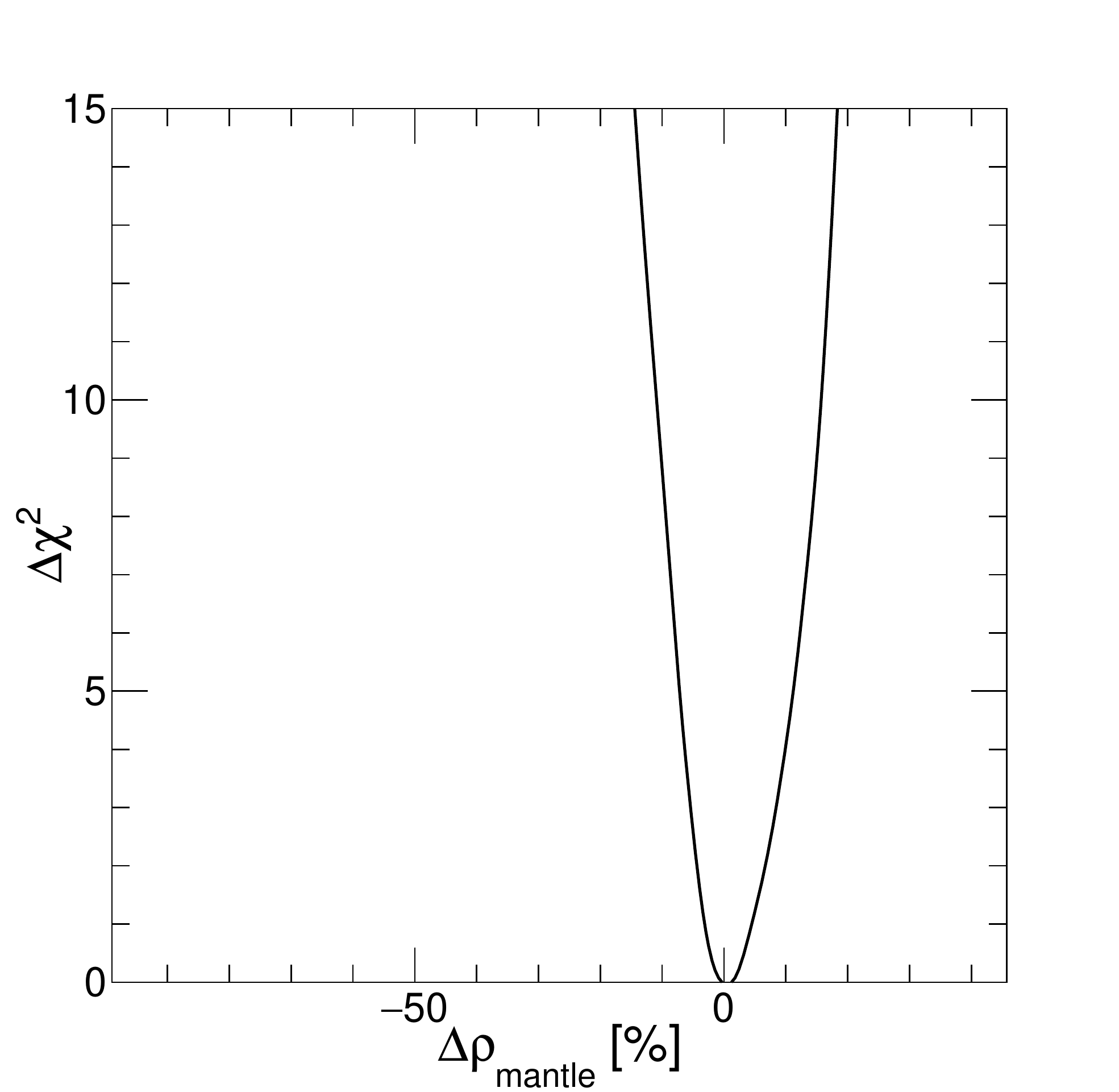}}
    &
   { \includegraphics[width=0.3\linewidth]{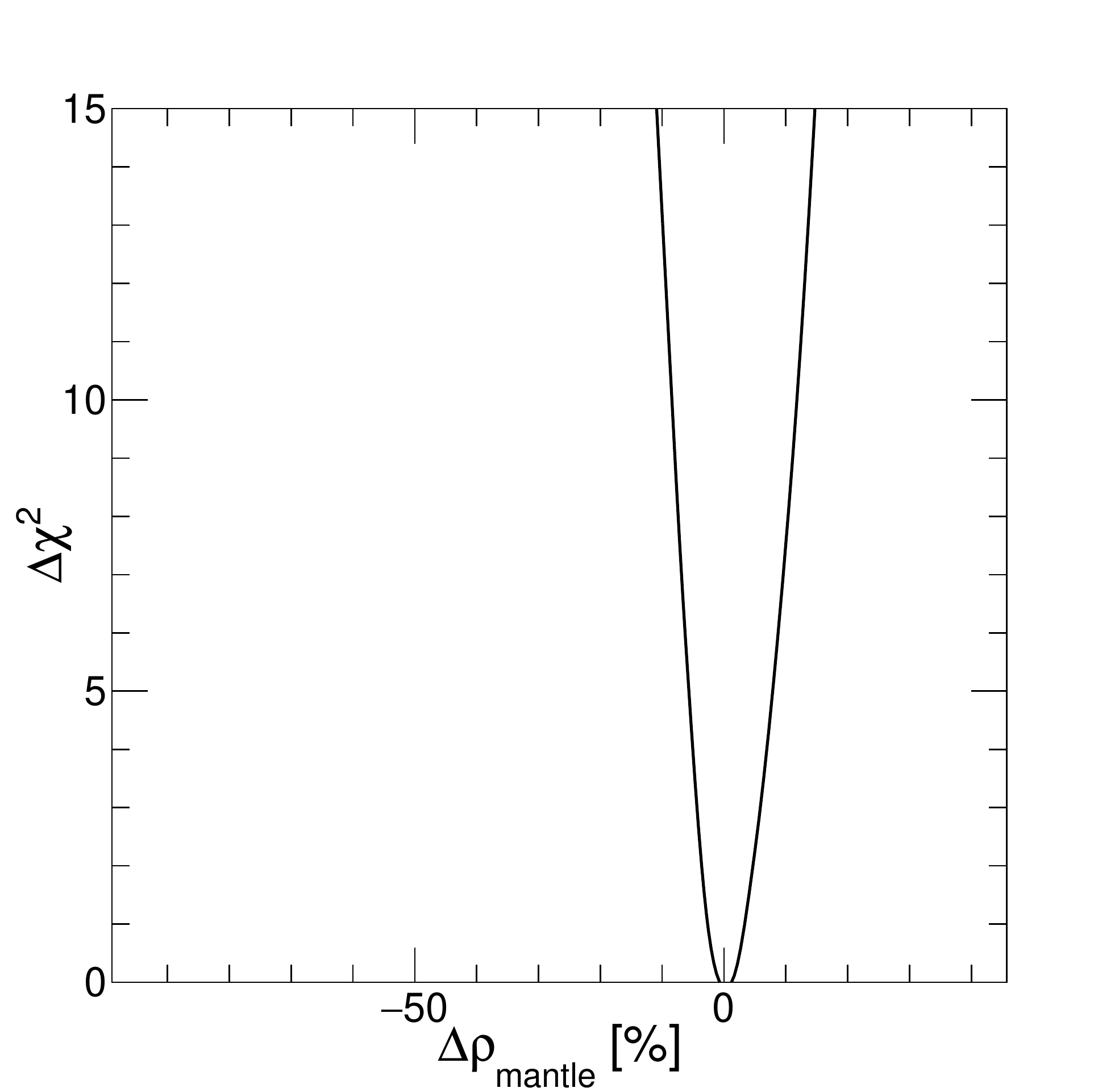}}
    &
    {\includegraphics[width=0.3\linewidth]{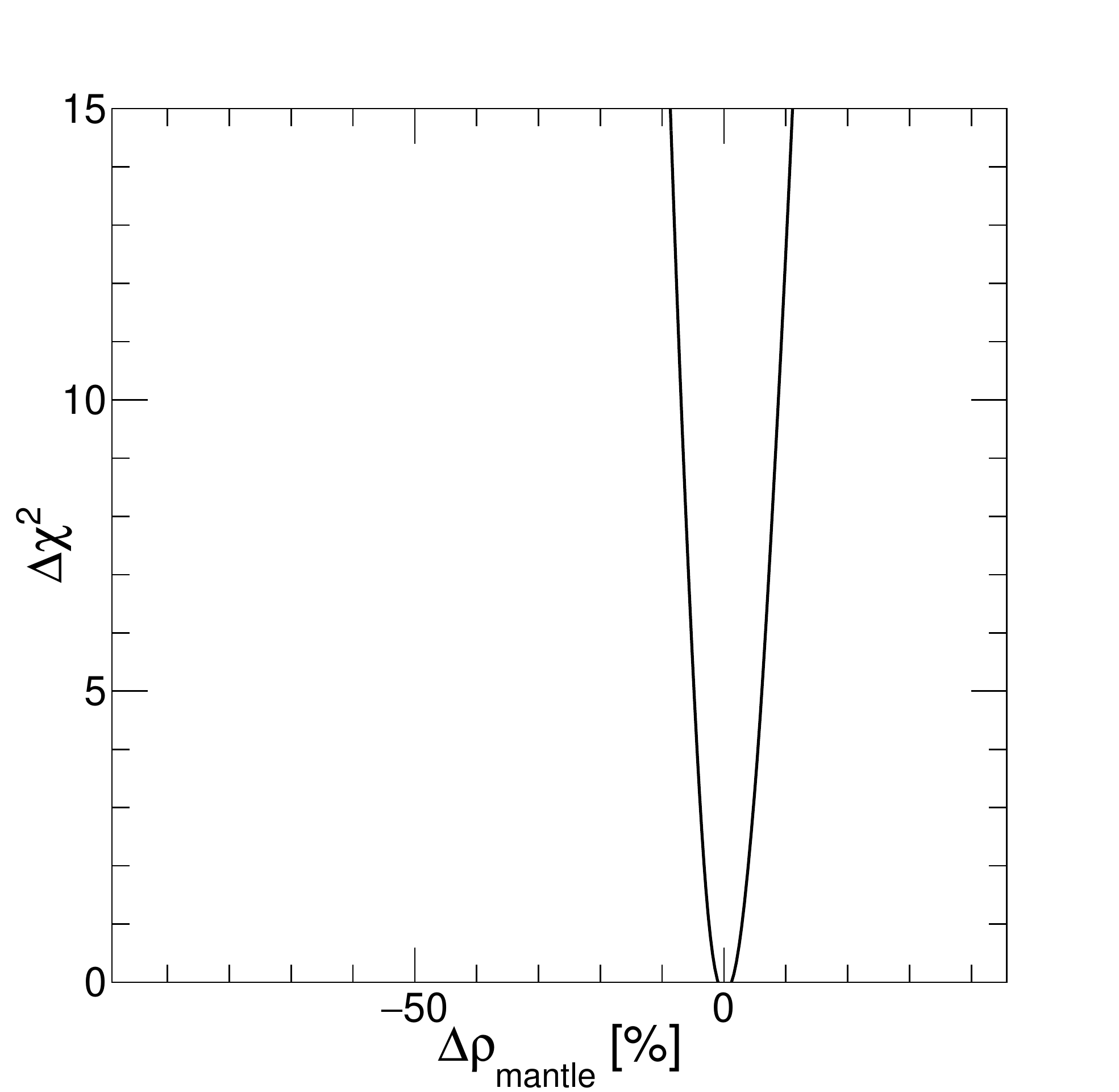}}
    \\
    {\includegraphics[width=0.3\linewidth]{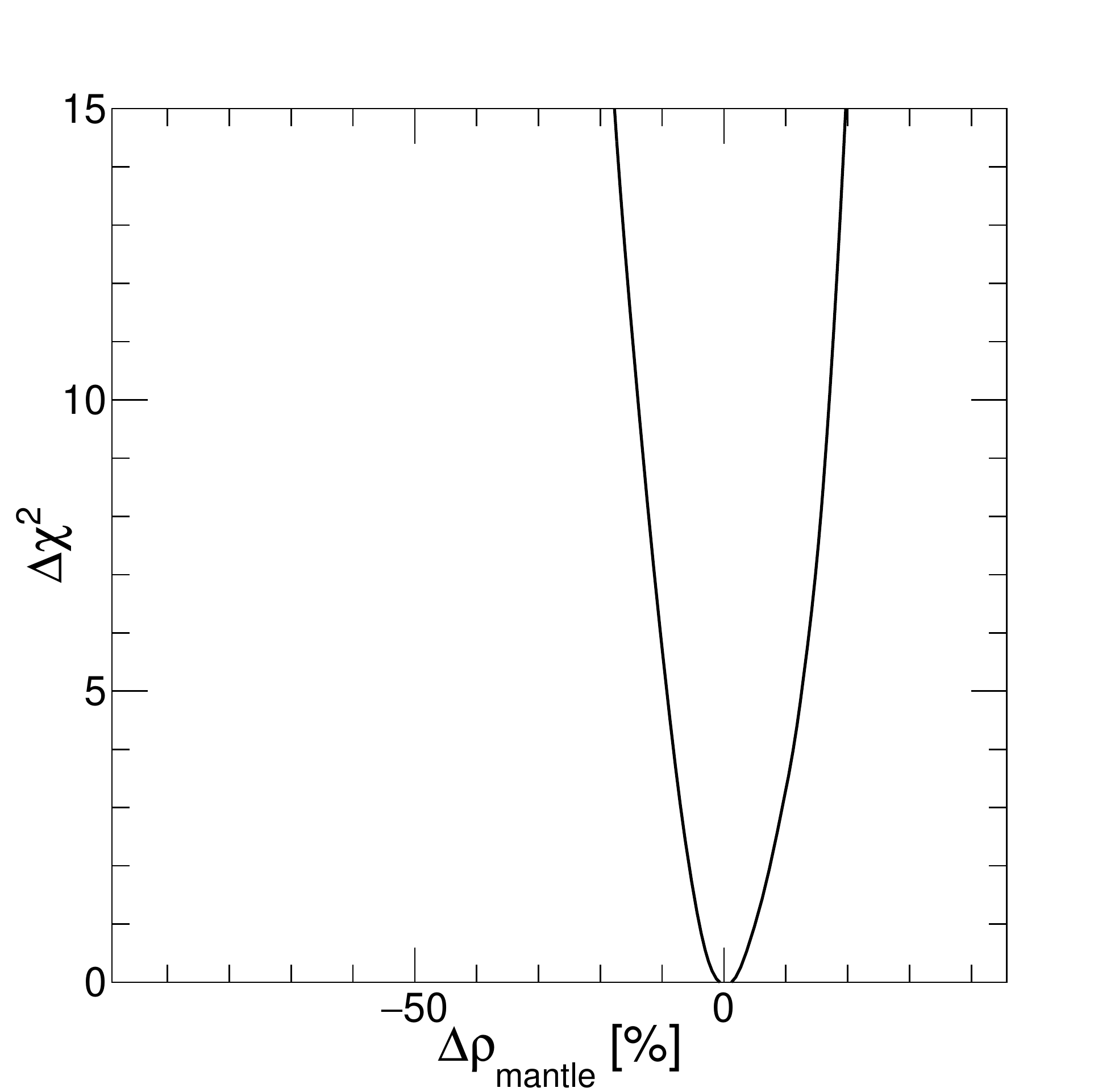}}
    &
    {\includegraphics[width=0.3\linewidth]{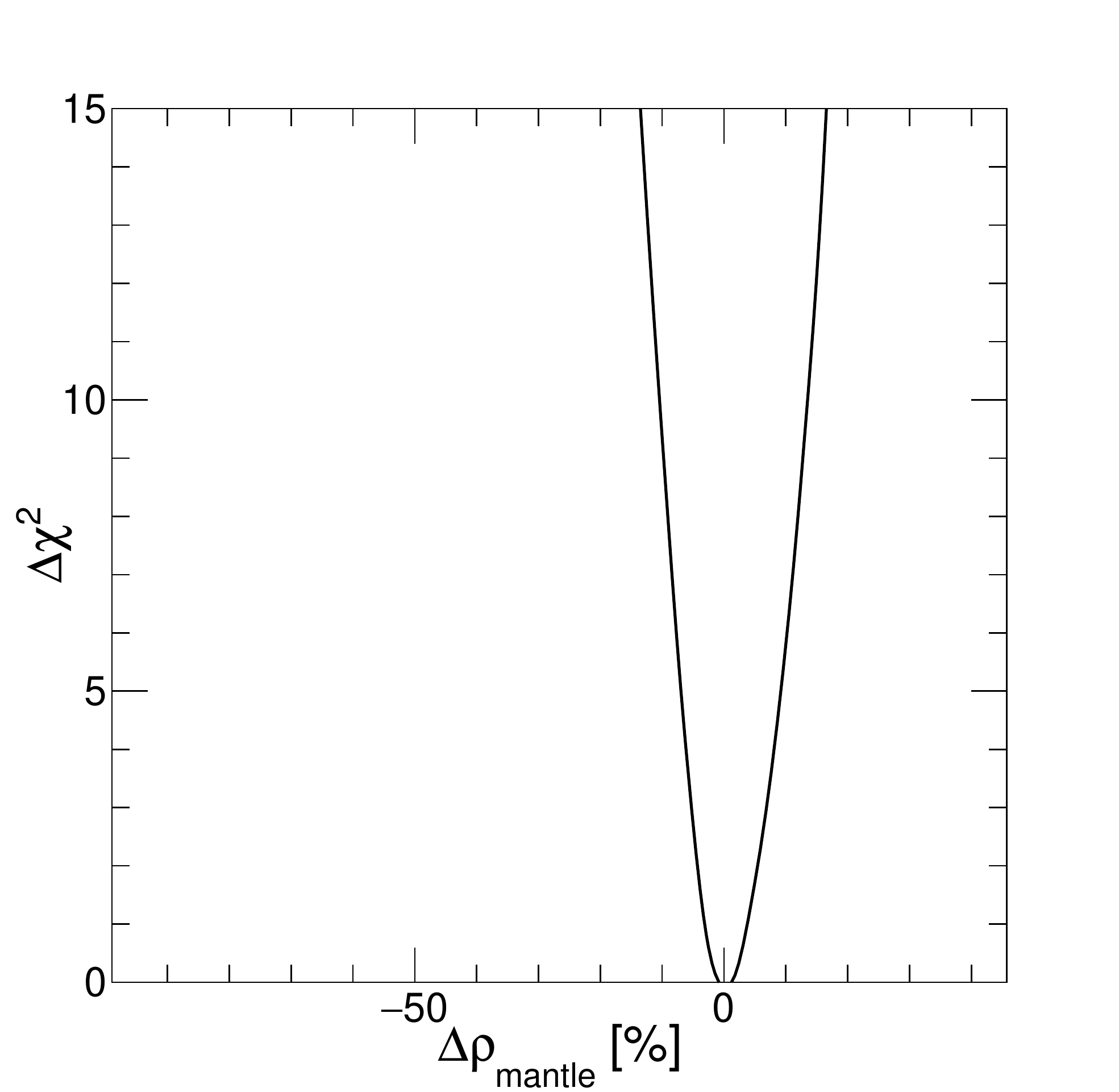}}
    &
   { \includegraphics[width=0.3\linewidth]{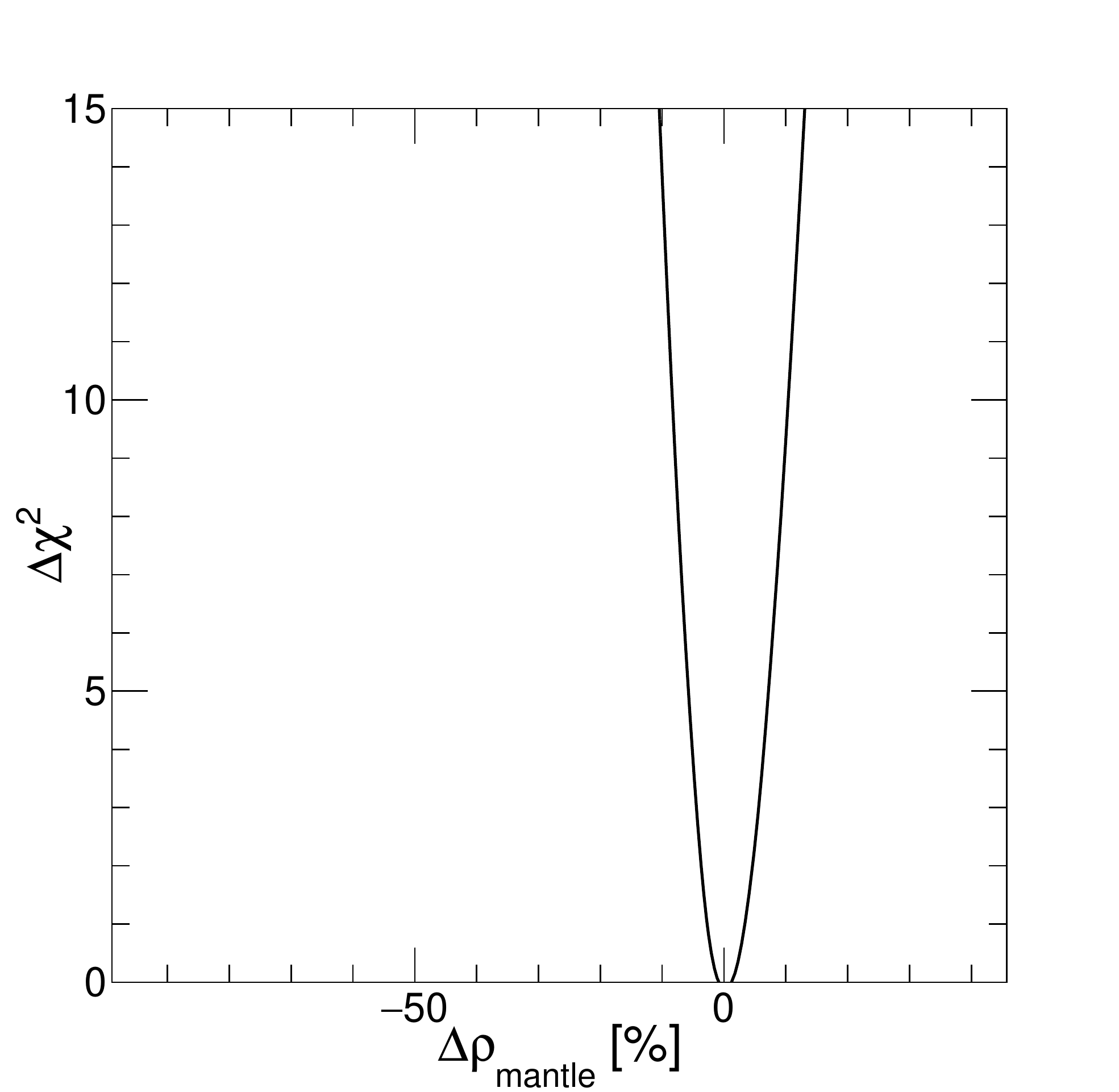}}
    \\
   { \includegraphics[width=0.3\linewidth]{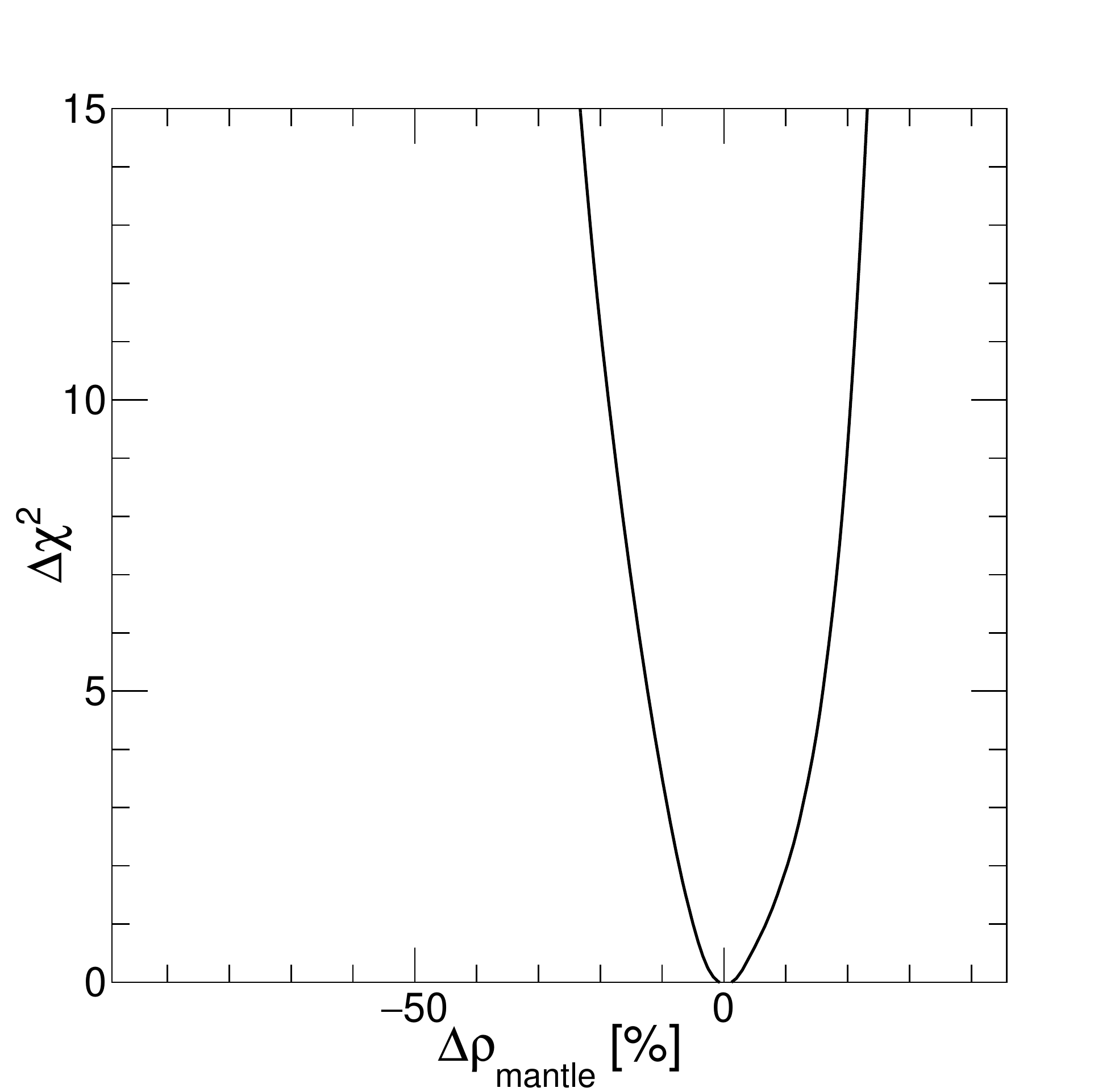}}
    &
    {\includegraphics[width=0.3\linewidth]{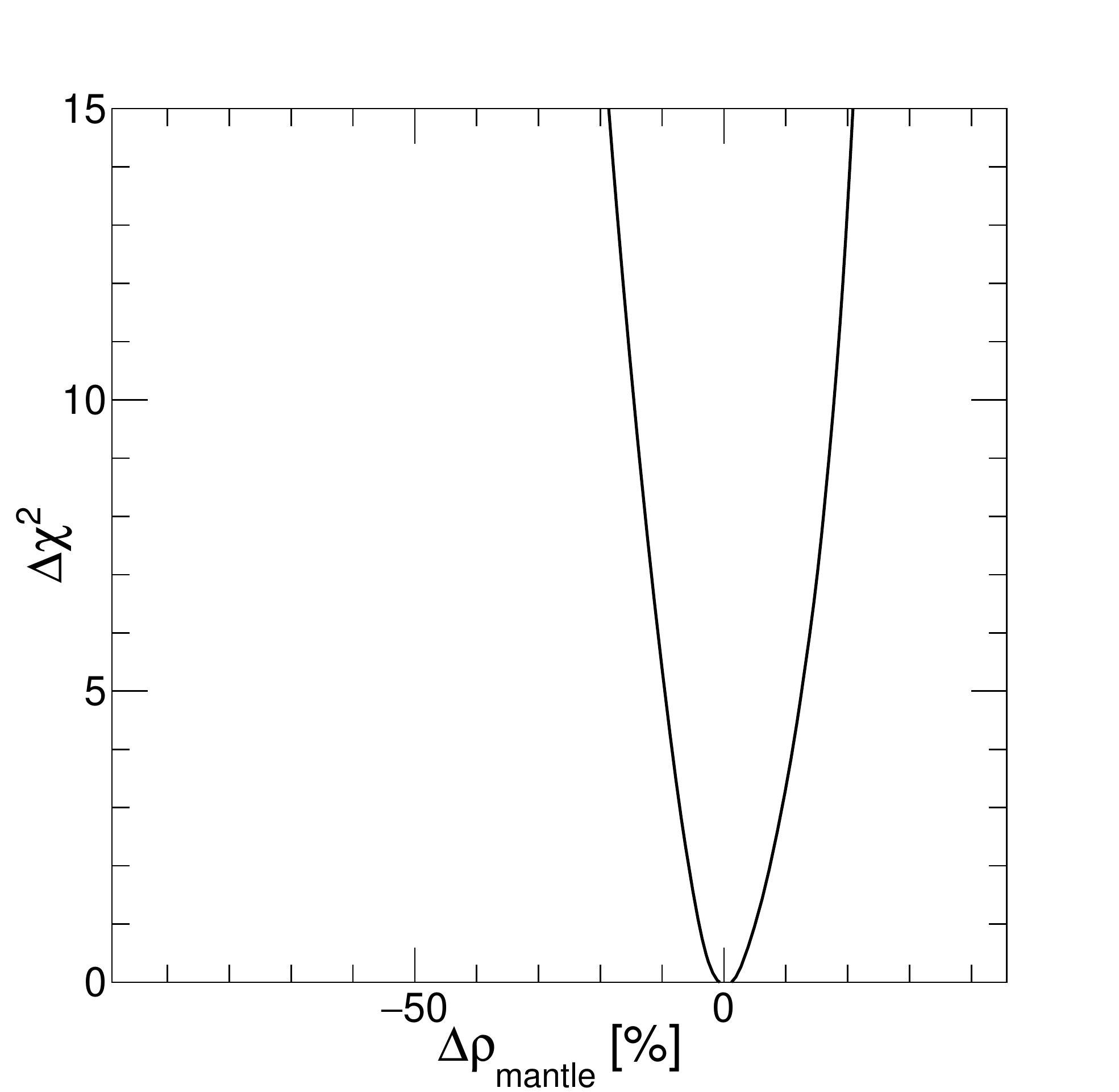}}
    &
   {\includegraphics[width=0.3\linewidth]{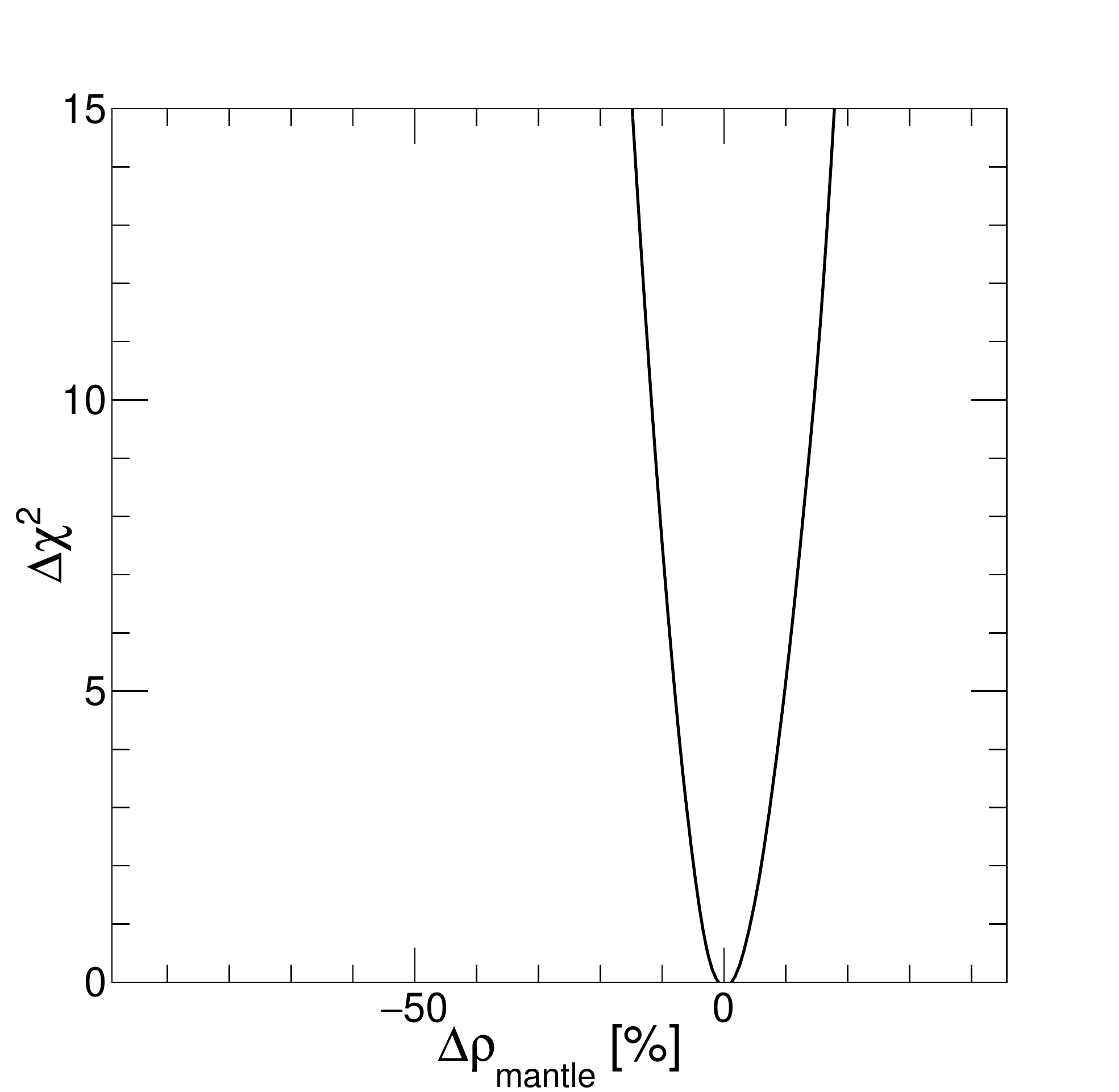}}
  \end{tabular}
 \caption{Sensitivity to the mantle density in the case of NO spectrum 
and 10 years of data.
The Earth total mass constraint 
is implemented by compensating the mantle density variation with a 
corresponding OC density change. The 
results shown are for $\sin^2\theta_{23} = 0.42$, 0.50, 0.58 
(left, center and right panels) and in the cases of 
``minimal'', ``optimistic'' and ``conservative'' systematic errors 
(top, middle and bottom panels). See text for further details.
}
\label{fig:NHMantvsOC}
\end{figure}

\vspace{0.3cm}
Our results show that ORCA will have rather high sensitivity to the 
Earth mantle density if OC (or total Earth core) is used as a 
compensation layer when imposing the total Earth mass constraint. 
They are presented graphically in Fig. \ref{fig:NHMantvsOC}
in which we show  $\chi^2$ versus $\Delta \rho_{\rm mantle}$ 
with OC being the compensating layer. 
The results correspond, as in the previous cases, 
to  $\sin^2\theta_{23} = 0.42$, 0.50, 0.58 
(left, center and right panels) and  
``minimal'', ``optimistic'' and ``conservative'' 
systematic errors (top, middle and bottom panels).
%%%%%%%%%%%%%%%%%%%%%%%%%
%%%%%%%%%%%%%%%%%%%%%%%%%%%%%%%
%
The $\chi^2$ dependence on $\Delta \rho_{\rm mantle}$ 
in Fig. \ref{fig:NHMantvsOC} has symmetric or slightly asymmetric 
Gaussian form. As follows from Fig. \ref{fig:NHMantvsOC},
ORCA can determine the mantle density 
$3\sigma$ with an uncertainty of 
(-11\%)/+13\%,  (-9\%)/+11\% and (-6\%)/+8\% 
for  $\sin^2\theta_{23} = 0.42$, 0.50, 0.58,
respectively, and  ``minimal'' systematic errors.
The sensitivity of ORCA is somewhat worse 
in the case of   ``conservative'' systematic errors:  
for $\sin^2\theta_{23} = 0.42$, 0.50, 0.58 the uncertainties 
of interest are given at $3\sigma$ C.L. respectively by 
(-17\%)/+20\%,  (-13\%/+17\% and (-11\%)/+14\%,
implying still a relatively high ORCA sensitivity to 
$\rho_{\rm man}$. Depending on the value of $\sin^2\theta_{23}$,
the uncertainties are larger than those corresponding to 
``minimal'' systematic errors approximately 
by factors of $\sim (1.4 - 1.8)$.
We find that for ``optimistic'' systematic errors 
the uncertainty under discussion for 
$\sin^2\theta_{23} = 0.42$, 0.50, 0.58 at $3\sigma$ C.L. 
reads, respectively: (-14\%)/+17\%, 
(-10\%)/+13\% and (-8\%)/+10\%.
As can be expected, it is somewhat larger (smaller) than the 
ORCA uncertainty corresponding to 
``minimal'' (``conservative'') 
systematic errors. 

As can be shown, the conditions in eq. (\ref{eq:equil}) 
imply  $(-8.3\%)\ltap \Delta \rho_{mantle} \ltap 22.8\%$.  
These conditions do not restrict from above the 
the ORCA $3\sigma$ sensitivity range of 
$\Delta \rho_{\rm mantle} > 0$ even in the case of ``conservative'' 
systematic errors. The maximal negative variation 
of $\Delta \rho_{\rm mantle}$ of (-8.3\%) 
restricts from below the ORCA sensitivity 
ranges derived employing the Earth mass constraint,
the smaller the systematic erors and/or the larger 
$\sin^2\theta_{23}$, the smaller the effect 
of the external condition  
$\Delta \rho_{\rm mantle} \gtap -\,8.3\%$ is.
In the case of ``minimal'' systematic errors, 
$\Delta \rho_{\rm mantle} = -8.3\%$ 
corresponds to the minimal values of the 
ORCA $2\sigma$, $2.4\sigma$, $2.8\sigma$ and $3.0\sigma$ 
sensitivity ranges derived 
for $\sin^2\theta_{23} = 0.46$, 0.50, 0.54 and 0.58, 
respectively. With implemented ``optimistic'' systematic errors, 
it coincides with the minimal values of 
the ORCA $2.0\sigma$, $2.4\sigma$ and $2.7\sigma$  
sensitivity ranges  for  
$\sin^2\theta_{23} = 0.48$, 0.54 and 0.58. 
And in the case of ``conservative'' systematic errors, 
it is equal to the minimal value of the 
$2.0\sigma$ sensitivity range for 
$\sin^2\theta_{23} = 0.58$.

With maximally reduced systematic errors 
and certain further improvements 
(e.g., the discussed ``favorable'' 6 m vertical spacing 
configuration of ORCA experiment \cite{KM3Net:2016zxf} 
or the Super-ORCA version of the detector
\cite{Hofestadt:2019whx}) 
the ORCA sensitivity to negative 
$\Delta \rho_{\rm mantle}$ might 
%%%%%%%%%%%%%%%%%%%%%%%%%
\begin{figure}[!t]
  \centering

\begin{tabular}{lll}
{\includegraphics[width=0.3\linewidth]{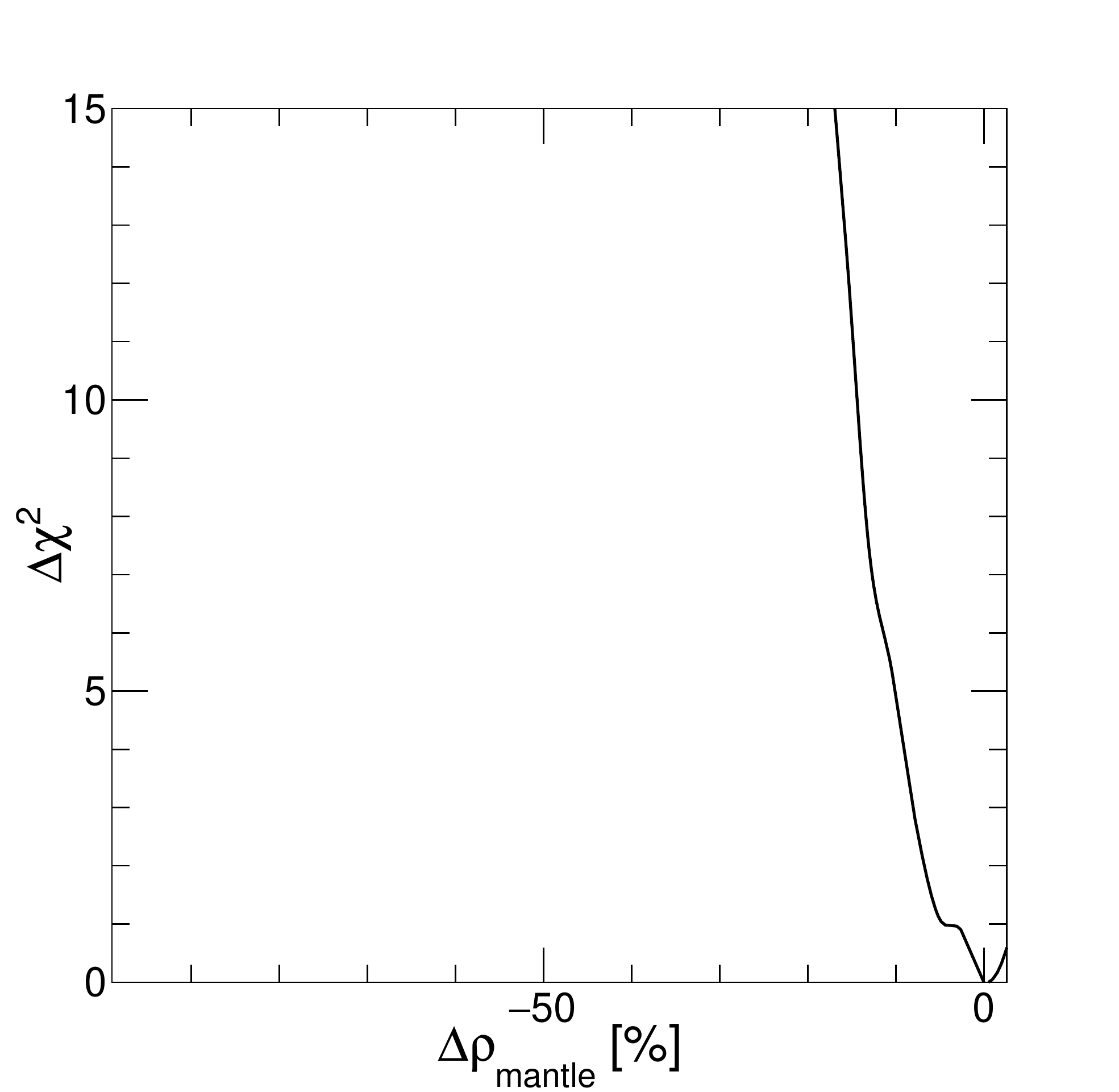}}
    &
   { \includegraphics[width=0.3\linewidth]{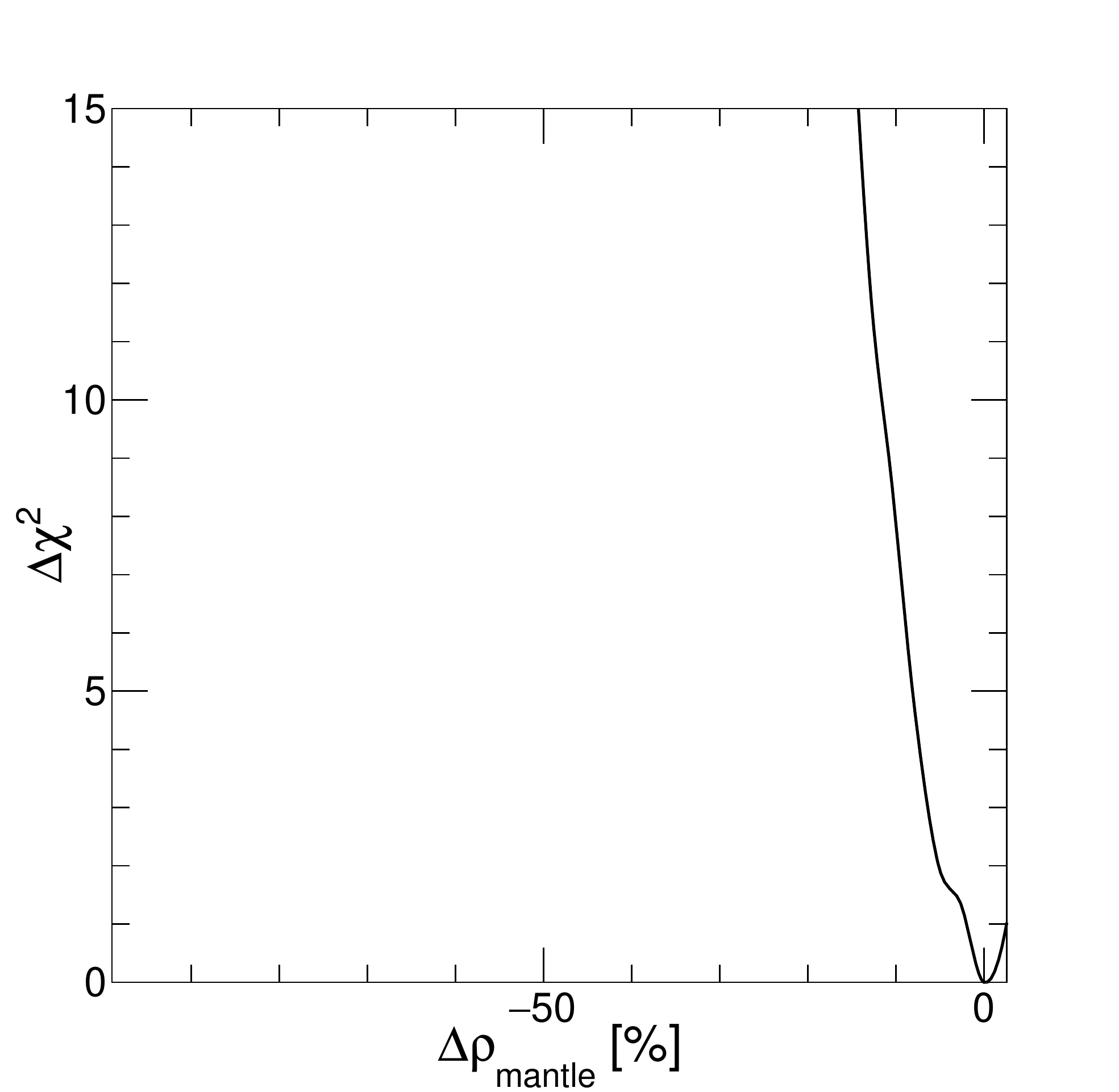}}
    &
    {\includegraphics[width=0.3\linewidth]{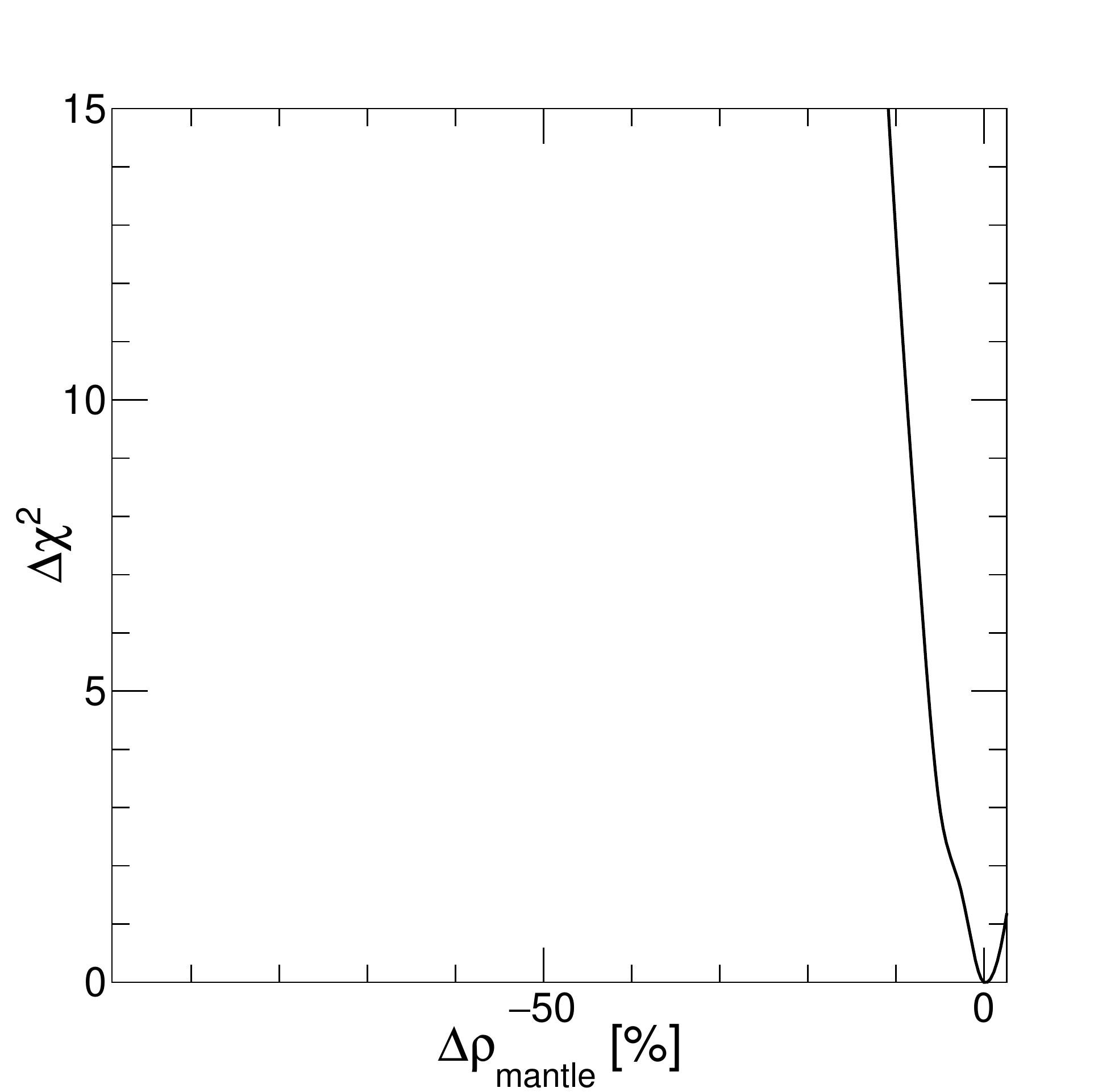}}
    \\
    {\includegraphics[width=0.3\linewidth]{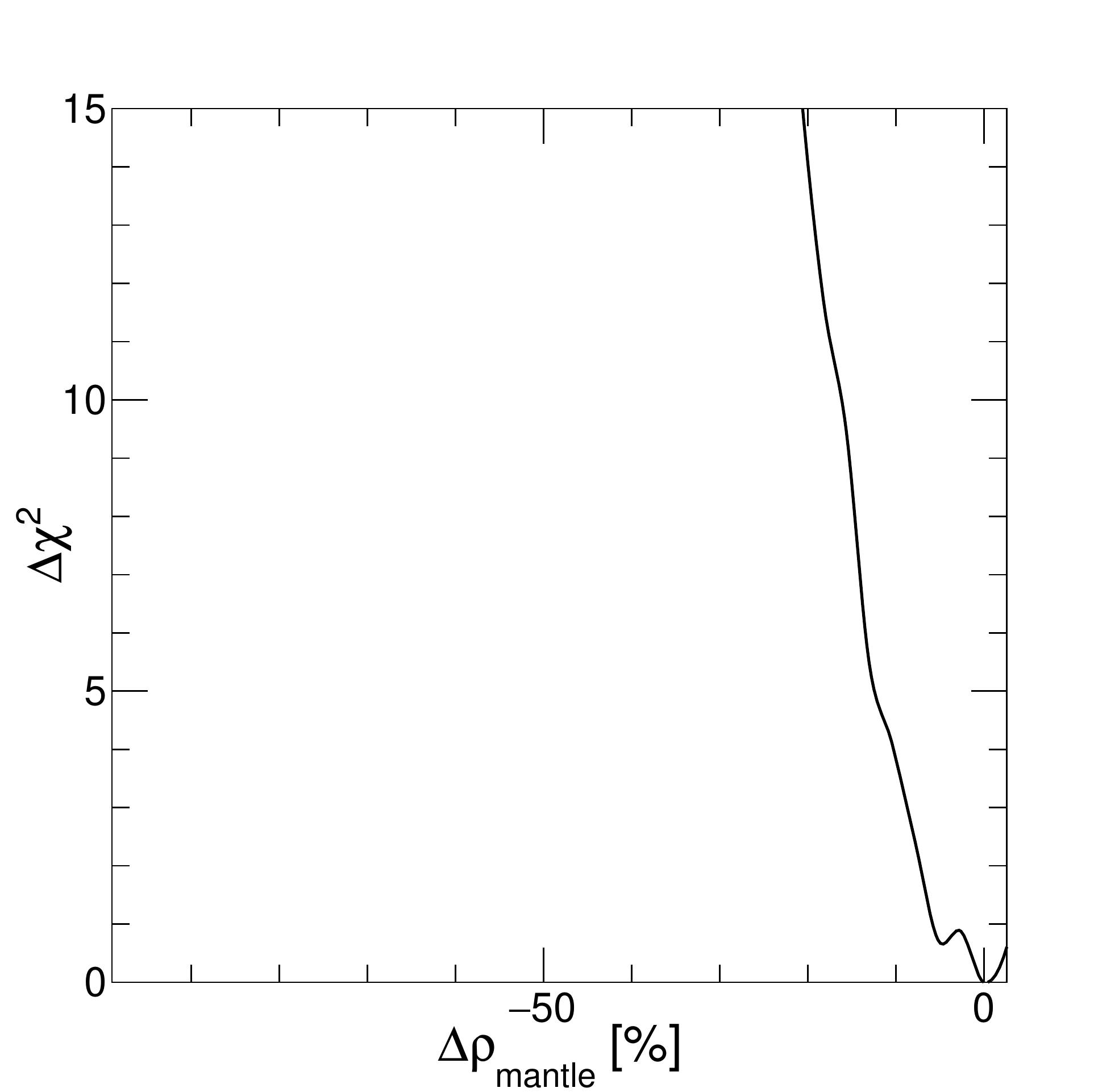}}
    &
    {\includegraphics[width=0.3\linewidth]{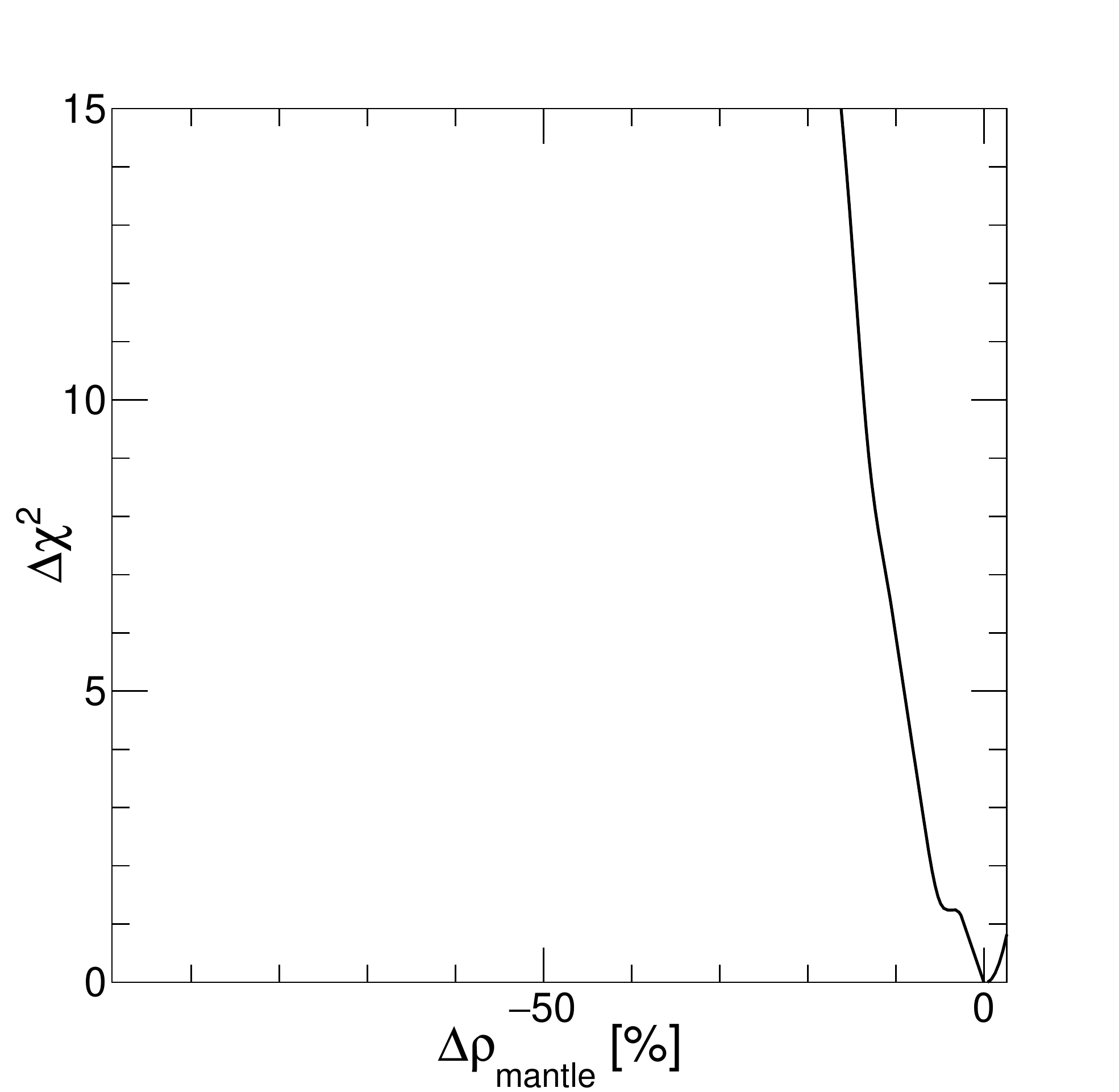}}
    &
   { \includegraphics[width=0.3\linewidth]{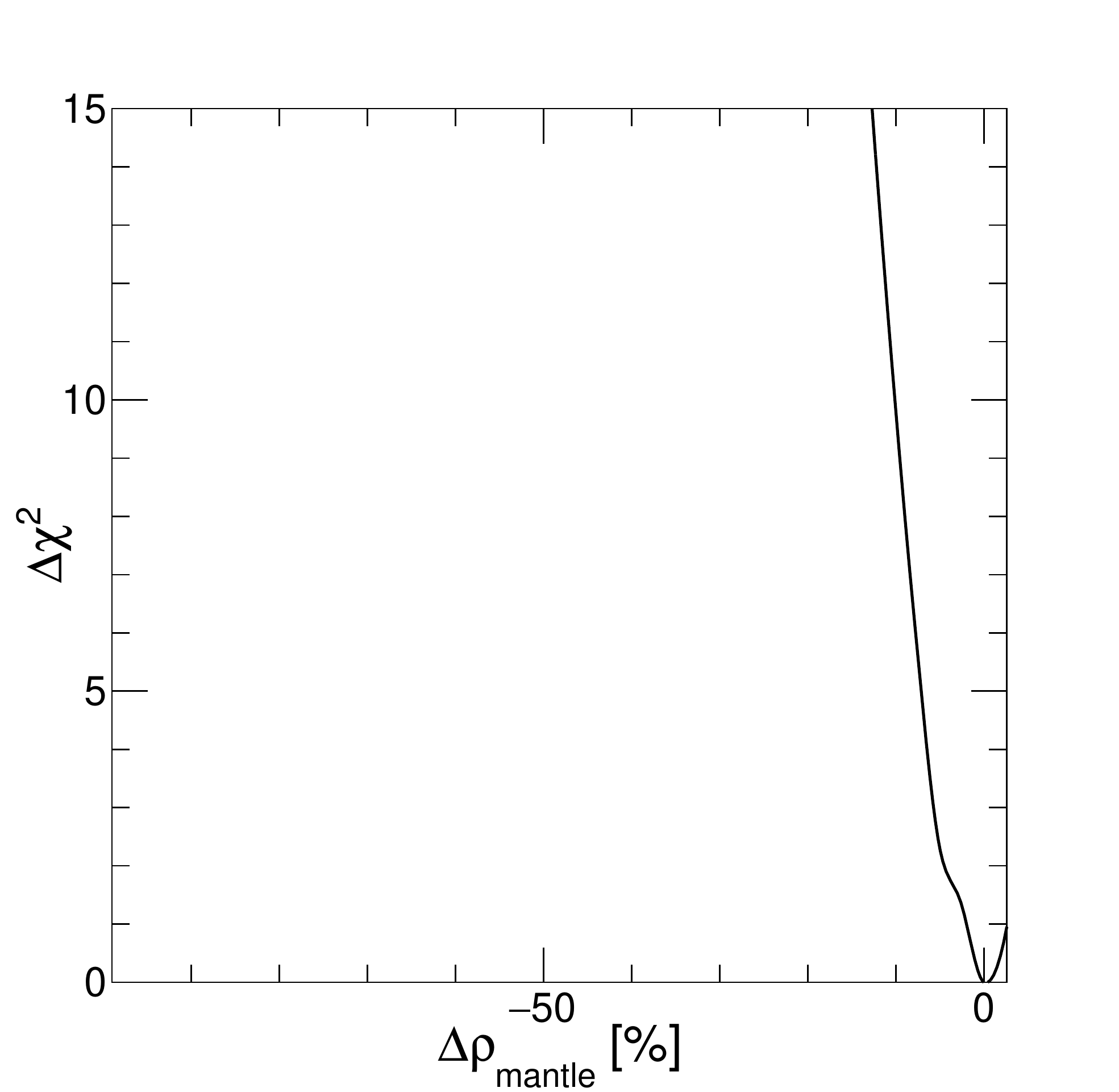}}
    \\
   { \includegraphics[width=0.3\linewidth]{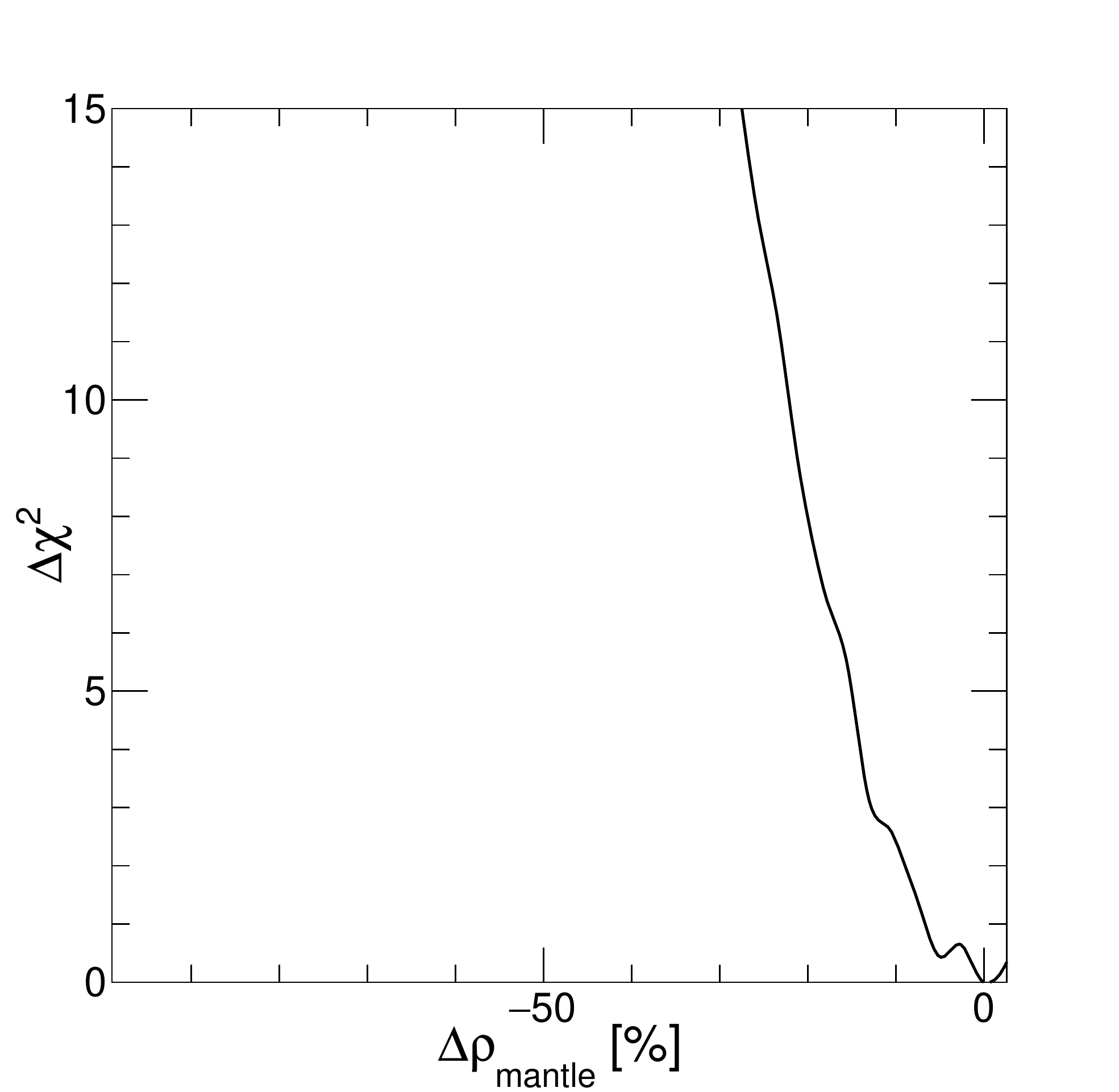}}
    &
    {\includegraphics[width=0.3\linewidth]{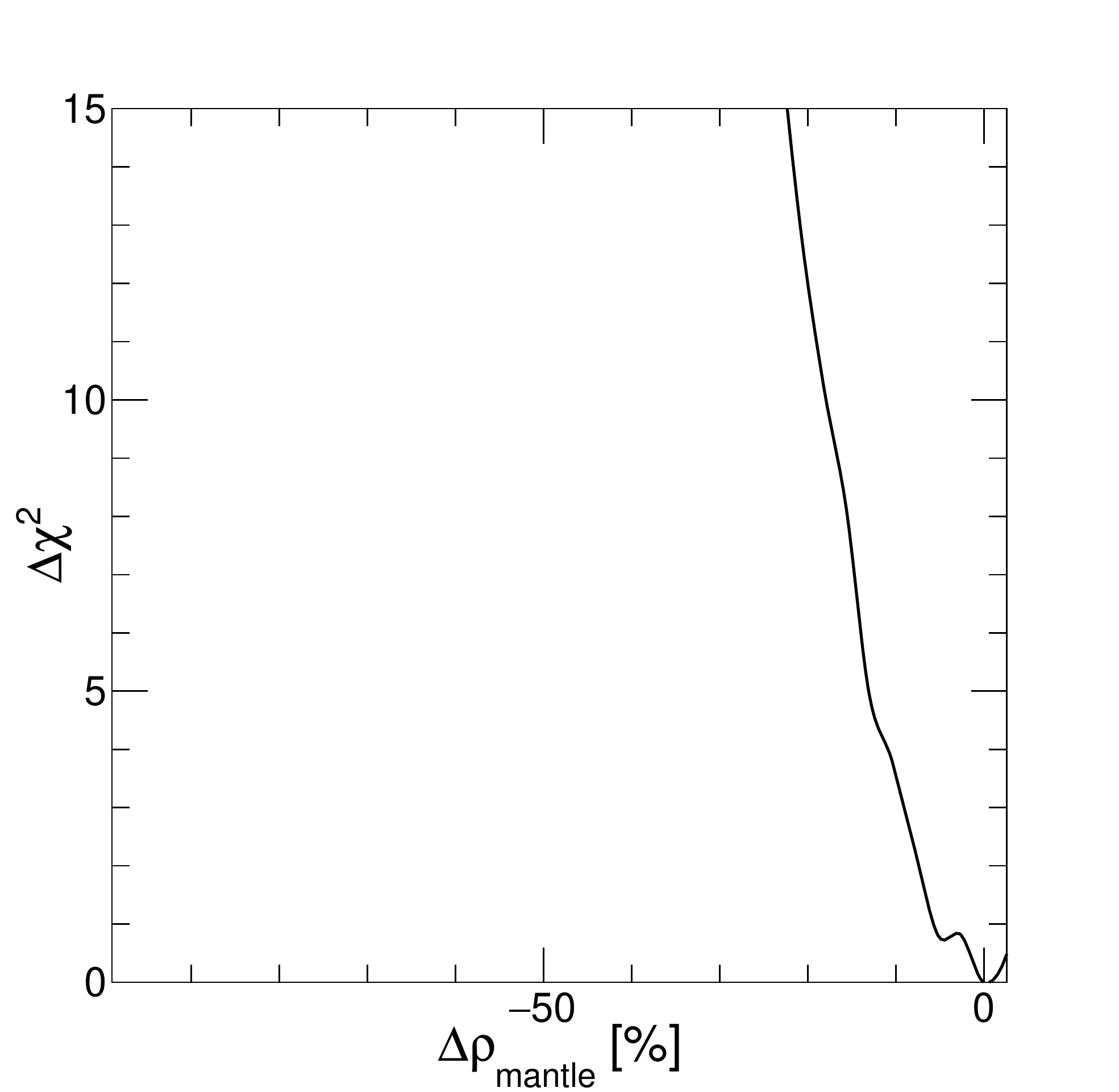}}
    &
   {\includegraphics[width=0.3\linewidth]{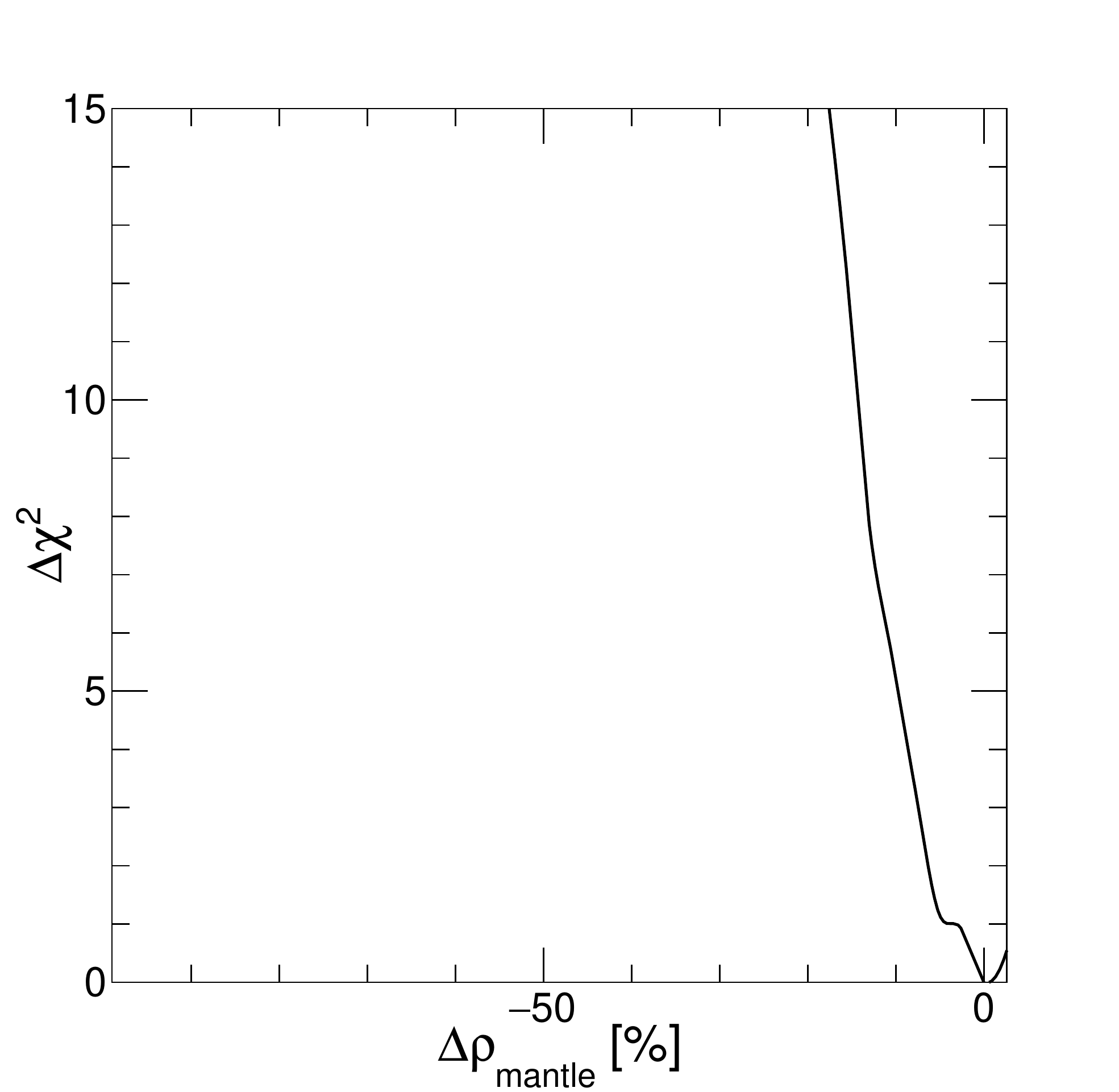}}
  \end{tabular}
 \caption{
The same as in Fig. \ref{fig:NHMantvsOC}, but 
with implementing the 
Earth total mass constraint 
by compensating the mantle density variation with a 
corresponding IC density change. The 
results shown are for $\sin^2\theta_{23} = 0.42$, 0.50, 0.58 
(left, center and right panels) and in the cases of 
``minimal'', ``optimistic'' and ``conservative'' systematic errors 
(top, middle and bottom panels). See text for further details.
}
\label{fig:NHMantvsIC}
\end{figure}
%%%%%%%%%%%%%%%%%%%%%%%%%%%%%%%
%
\noindent
increase sufficiently so that the external constraints 
$(-8.3\%) \ltap \Delta \rho_{\rm mantle}$ 
would have no effect on the ORCA $3\sigma$ 
sensitivity to negative $\Delta \rho_{\rm mantle}$.

\vspace{0.2cm}
{\bf B. Compensation with Inner Core Density}

\vspace{0.2cm}
We get quite different results when IC is used as a ``compensating'' layer.
They are presented in Fig. \ref{fig:NHMantvsIC}.
The sensitivity of ORCA to $\rho_{\rm man}$ is still very high for 
$\Delta \rho_{\rm mantle} < 0$. For  $\Delta \rho_{\rm mantle} > 0$ 
ORCA practically is not sensitive to  $\rho_{\rm man}$.
As in the similar case of variation of 
%%%%%%%%%%%%%%%%%%%%%%%%%
\begin{figure}[!t]
  \centering

\begin{tabular}{lll}
{\includegraphics[width=0.3\linewidth]{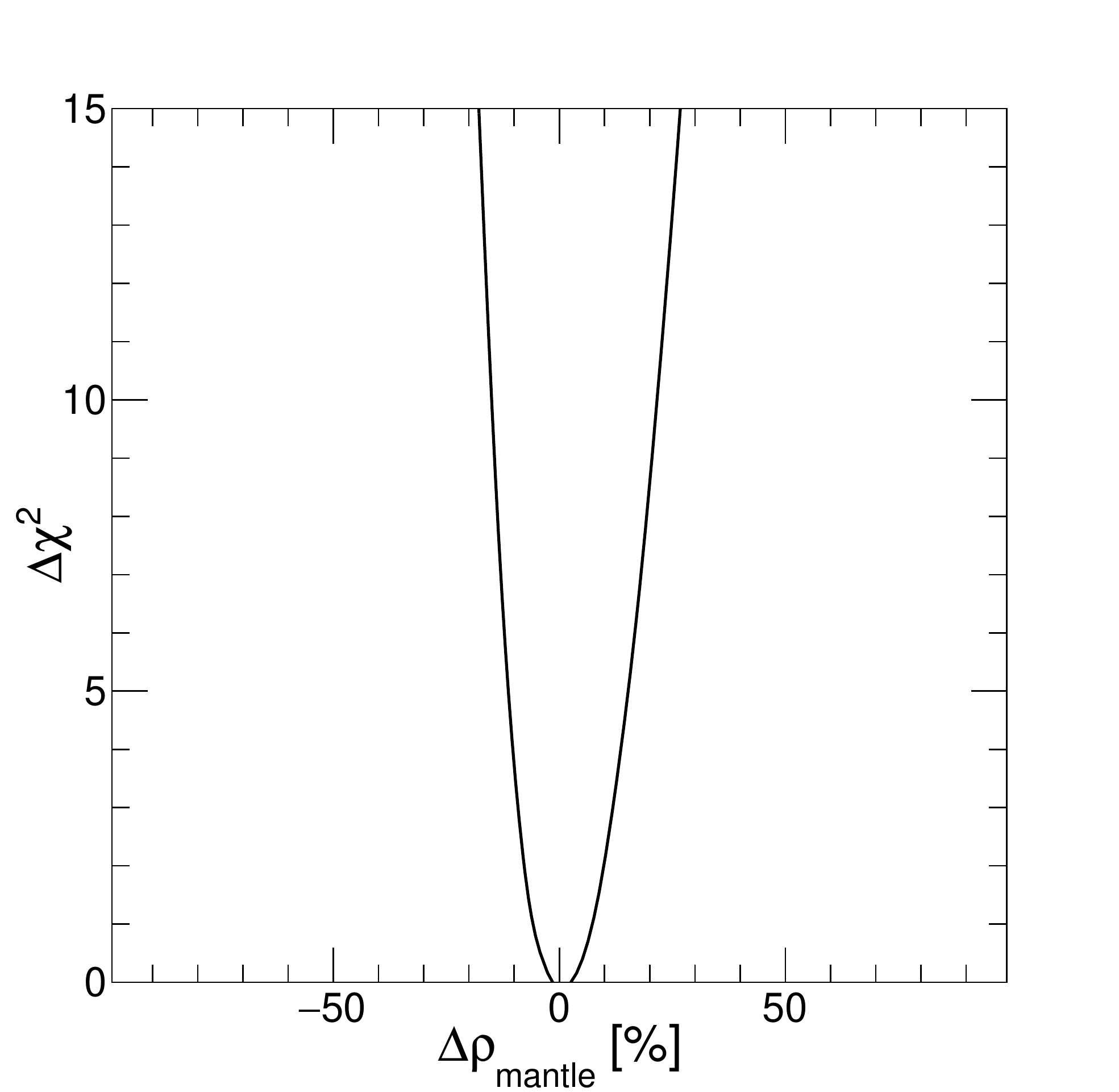}}
    &
   { \includegraphics[width=0.3\linewidth]{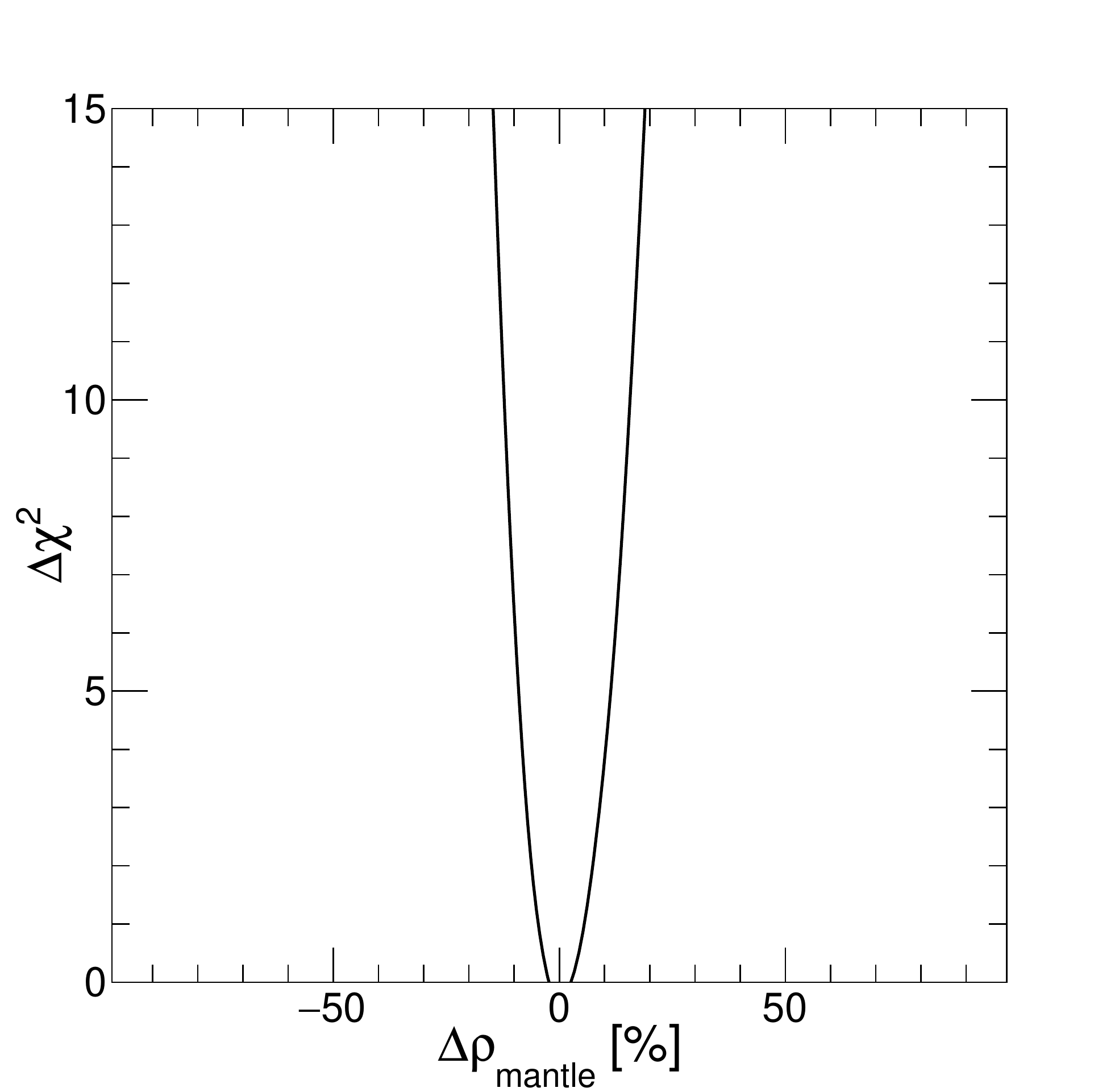}}
    &
    {\includegraphics[width=0.3\linewidth]{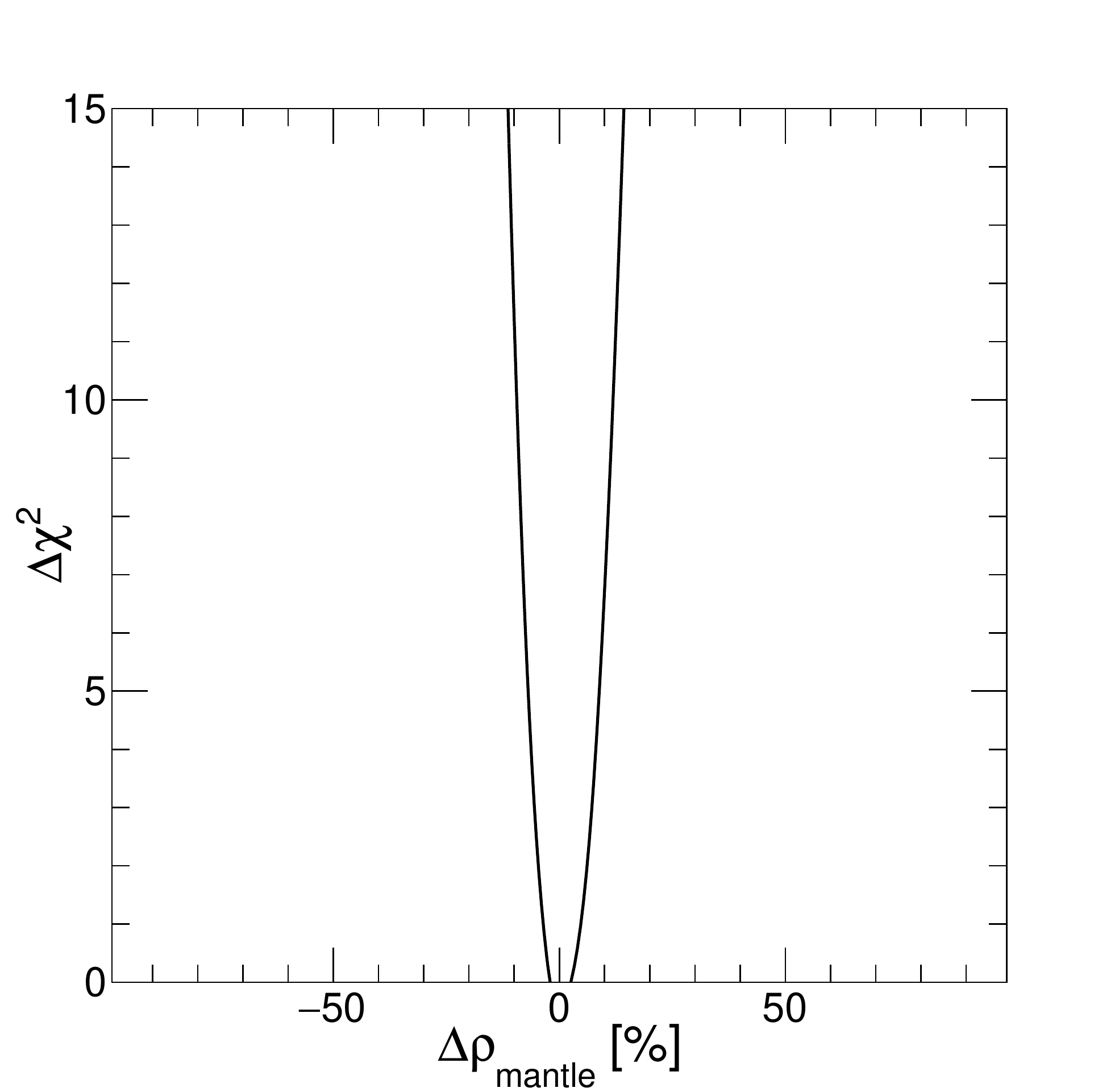}}
    \\
    {\includegraphics[width=0.3\linewidth]{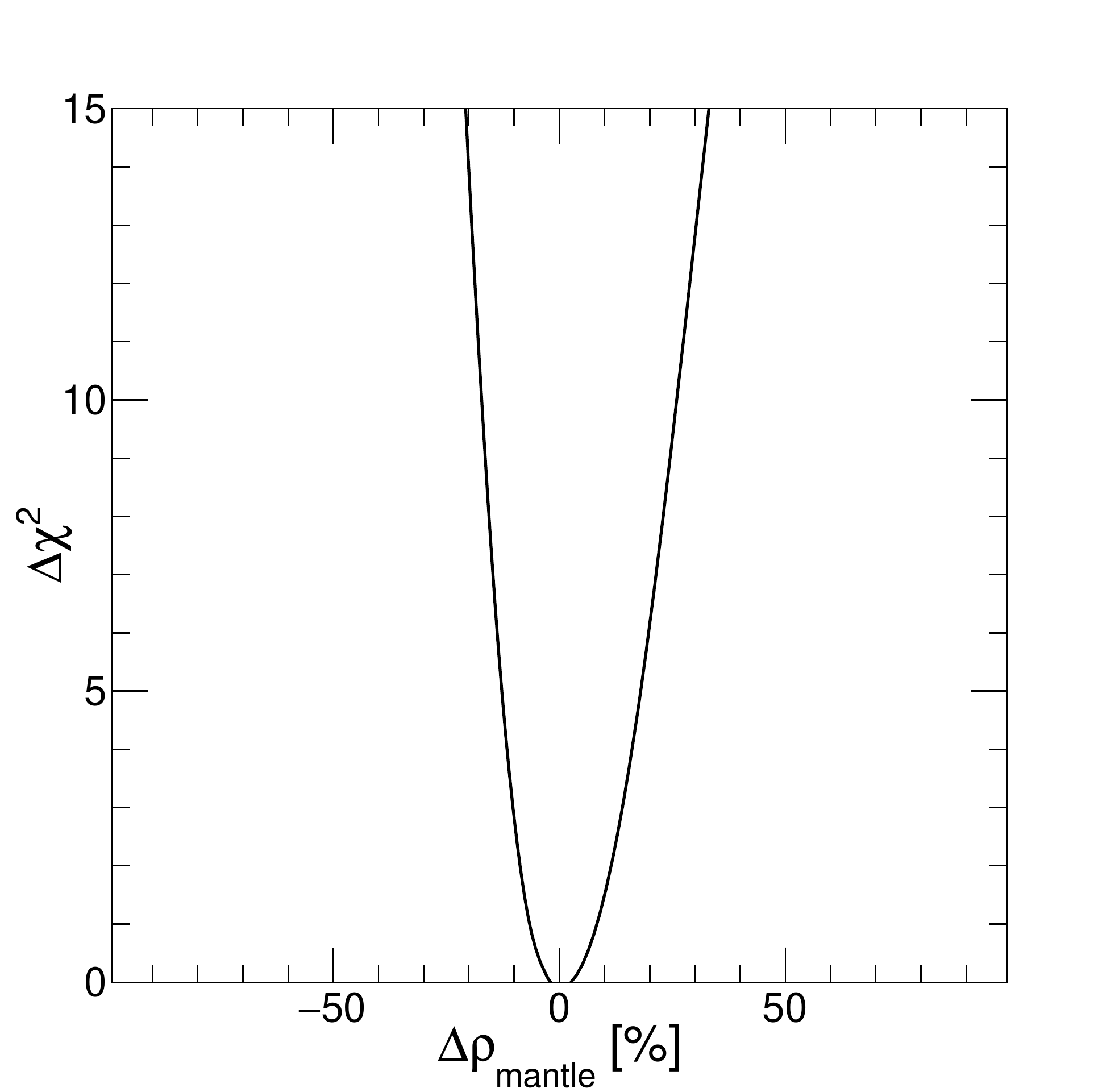}}
    &
    {\includegraphics[width=0.3\linewidth]{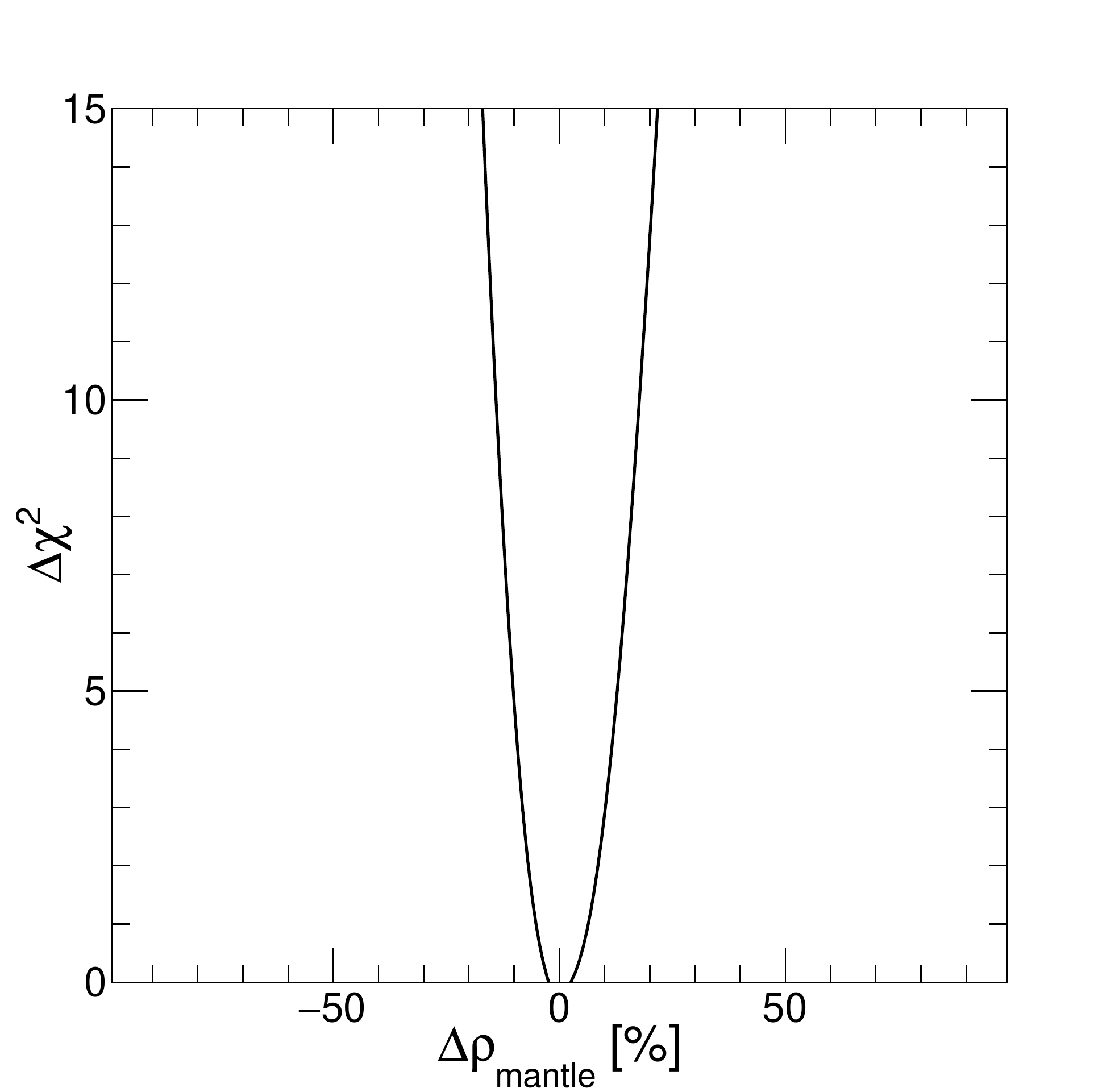}}
    &
   { \includegraphics[width=0.3\linewidth]{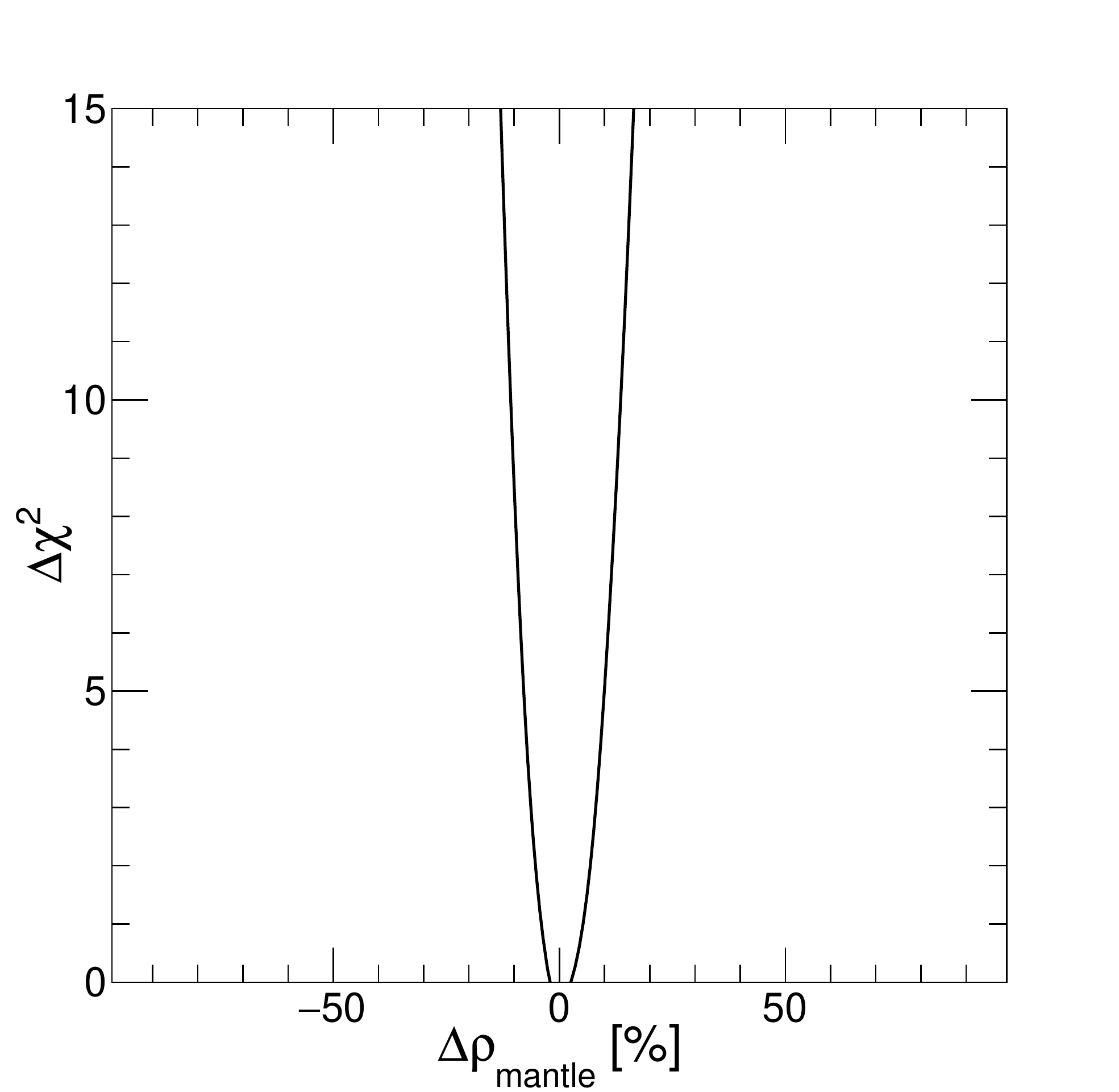}}
    \\
   { \includegraphics[width=0.3\linewidth]{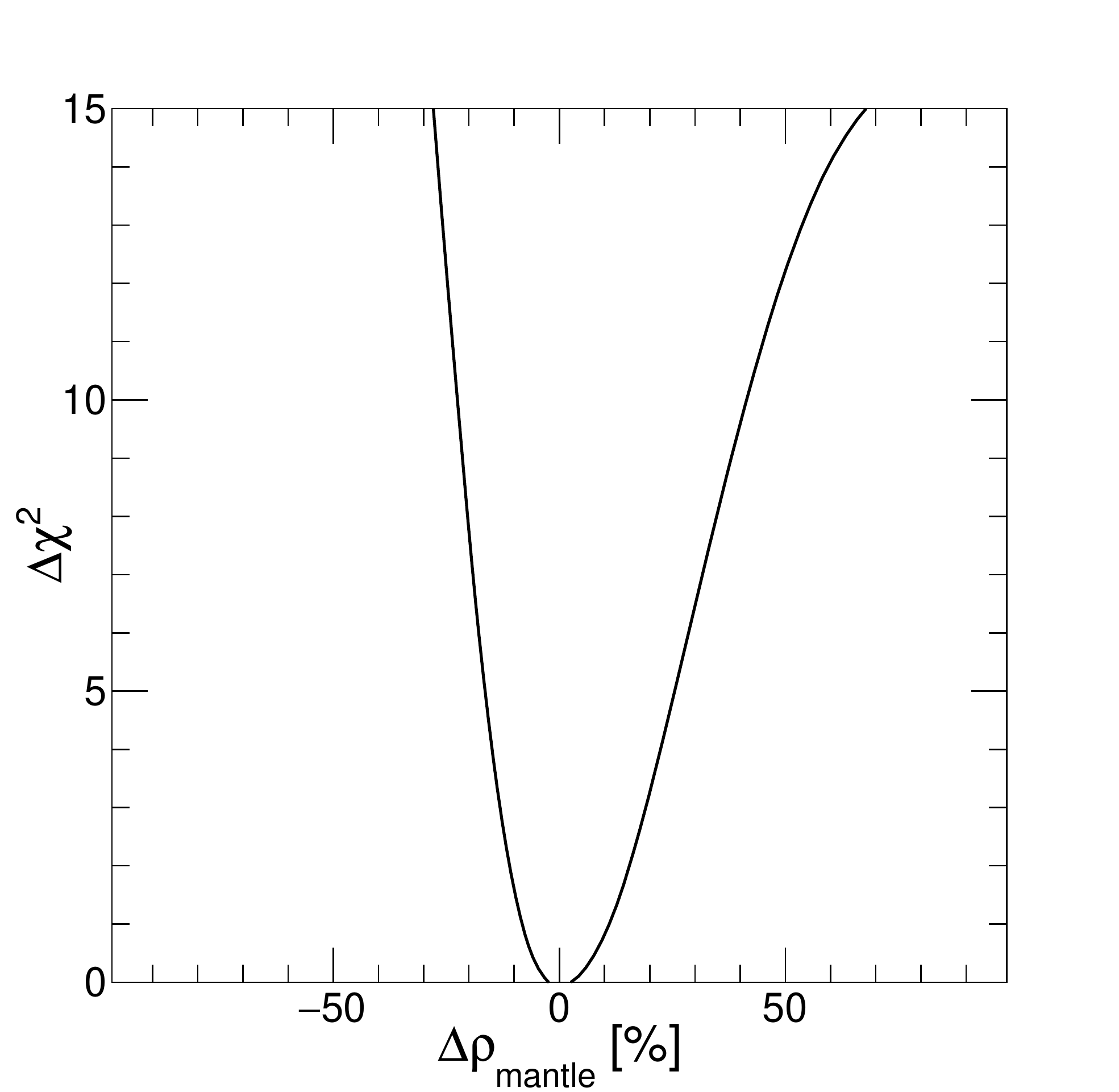}}
    &
    {\includegraphics[width=0.3\linewidth]{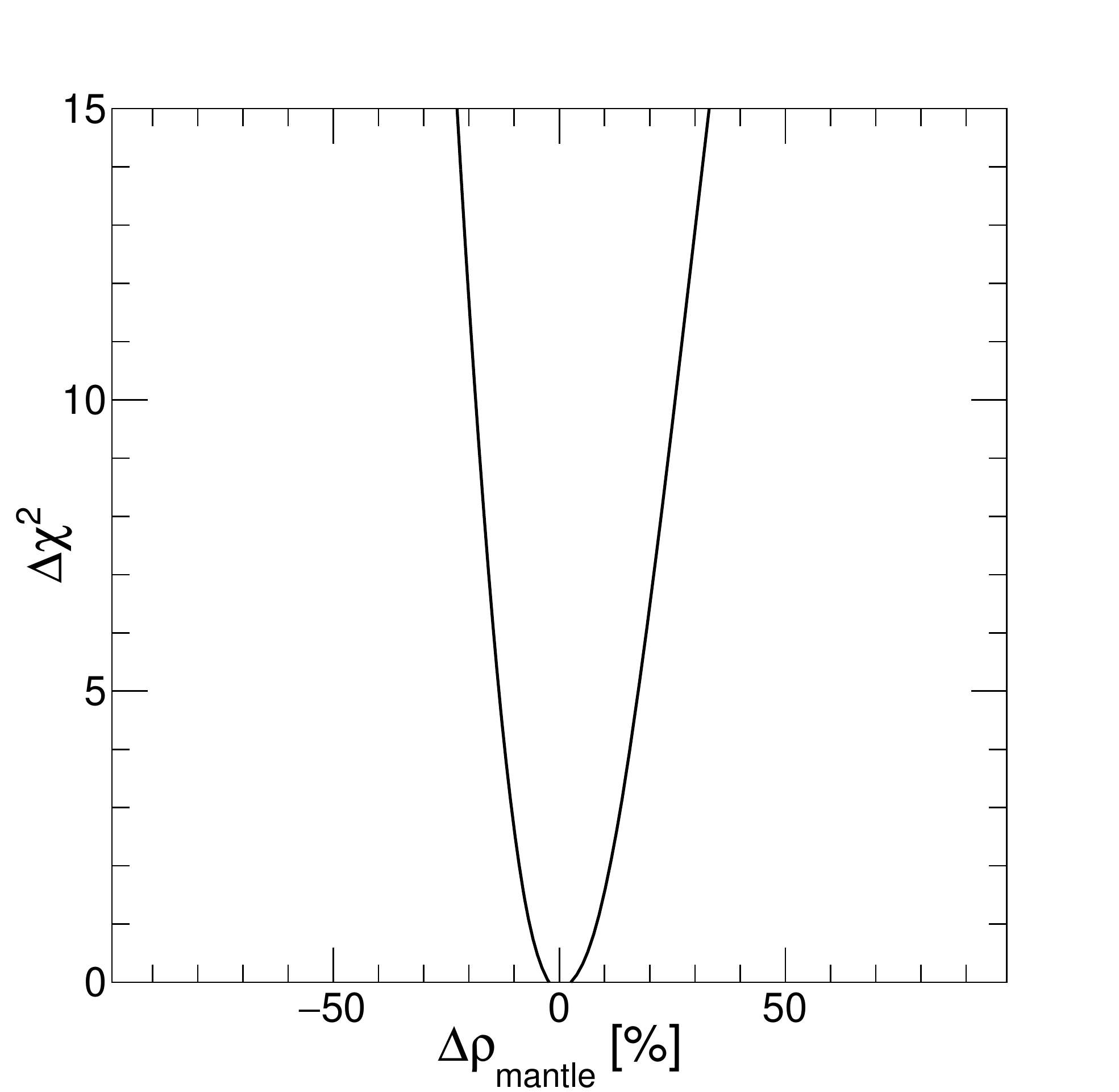}}
    &
   {\includegraphics[width=0.3\linewidth]{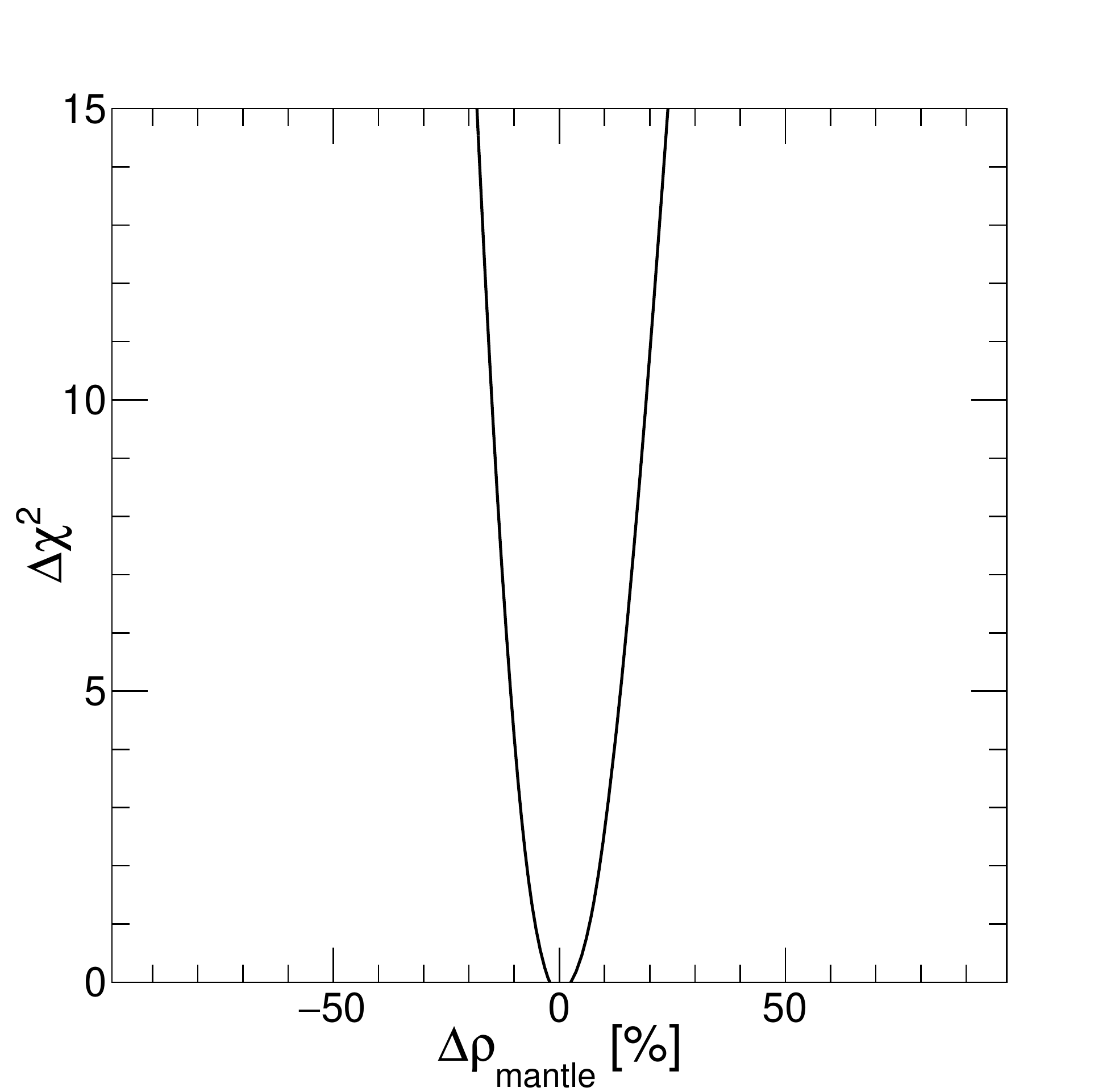}}
  \end{tabular}
 \caption{
 The same as in Fig. \ref{fig:NHMantvsOC}, but 
without implementing the Earth total mass constraint. 
The results shown are for $\sin^2\theta_{23} = 0.42$, 0.50, 0.58 
(left, center and right panels) and in the cases of 
``minimal'', ``optimistic'' and ``conservative'' systematic errors 
(top, middle and bottom panels). See text for further details.
}
\label{fig:NHMantleNoComp}
\end{figure}
%%%%%%%%%%%%%%%%%%%%%%%%%%%%%%%
%
\noindent
 $\rho_{\rm OC}$ compensated with 
a change of  $\rho_{\rm IC}$, this is a consequence of the fact 
that the IC mass is much smaller than the mantle mass and only 
insignificant increase of the mantle mass can be compensated by 
decreasing the IC mass. 

As  Fig. \ref{fig:NHMantvsIC} indicates,
in the considered case ORCA can determine the mantle density 
at $3\sigma$ C.L. with an uncertainty  $\Delta \rho_{\rm mantle}$ 
of (-14\%),  (-12\%) and (-8\%)
for  $\sin^2\theta_{23} = 0.42$, 0.50, 0.58,
respectively, and  ``minimal'' systematic errors.
The sensitivity of ORCA in the case of   ``conservative'' 
systematic errors and  $\sin^2\theta_{23} = 0.42$, 0.50, 0.58  
corresponds at $3\sigma$ C.L. respectively to 
$\Delta \rho_{\rm mantle}$ of (-22\%),  (-18\%) and (-14\%).
We find that for ``optimistic'' 
systematic errors 
the uncertainty under discussion for 
$\sin^2\theta_{23} = 0.42$, 0.50, 0.58 at $3\sigma$ C.L. 
reads, respectively: (-16\%),  (-14\%) and (-9\%).

\vspace{0.25cm}
{\bf D. Without Compensation}

\vspace{0.25cm}
The sensitivity of ORCA to the mantle density  $\rho_{\rm man}$ 
when the total Earth mass constraint is 
not imposed is significantly worse than in the 
case of imposing it and OC is used as a compensating layer.
The results corresponding to this case are reported 
graphically in Fig. \ref{fig:NHMantleNoComp}.
The $\chi^2$ dependence on $\Delta \rho_{\rm mantle}$ 
in Fig. \ref{fig:NHMantleNoComp} has somewhat asymmetric 
Gaussian form. According to Fig. \ref{fig:NHMantleNoComp},
ORCA can determine the mantle density at $3\sigma$ C.L. 
when the total Earth mass constraint is not implemented  
with an uncertainty  $\Delta \rho_{\rm mantle}$ 
of (-14\%)/+20\%,  (-11\%)/+17\% and (-9\%)/+12\%
for  $\sin^2\theta_{23} = 0.42$, 0.50, 0.58, respectively, and  
``minimal'' systematic errors.
In case of ``optimistic'' and ``conservative'' 
systematic errors we get for  $\sin^2\theta_{23} = 0.42$, 0.50, 0.58 
respectively (-17\%)/+24\%,  (-13\%)/+18\%, (-10\%)/+13\%
and  (-22\%)/+36\%,  (-18\%)/+25\%, (-14\%)/+18\%.
Depending on the value of  $\sin^2\theta_{23}$, 
the $3\sigma$ uncertainty in  $\rho_{\rm mantle}$ in 
the case of ``conservative'' systematic errors is larger than 
that for ``minimal'' systematic errors by a factor
of $\sim (1.5 - 1.8)$.

%%%%%%%%%%%%%%%%%%%%%%%%%%%%%
%
\subsection{Results for IO Neutrino Mass Spectrum}
\label{ssec:IO}
%
%%%%%%%%%%%%%%%%%%%%%%%%%%%%%

 We have performed exactly the same analysis 
(``minimal'', ``optimistic'', ``default'' and ``conservative'' 
sets of systematic errors, eleven values of 
$\sin^2\theta_{23}$ from the interval [0.40, 060], 
$\delta = 3\pi/2$, implementing the total Earth mass constraint 
and not imposing it) assuming IO neutrino mass spectrum.
In all cases considered we find that the sensitivity of ORCA detector 
to deviations of the densities of the different Earth 
structures (IC, OC, core, mantle)  from their respective PREM 
reference values is significantly worse than that in the case of 
%%%%%%%%%%%%%%%%%%%%%%%%%
\begin{figure}[H]%[!t]
  \centering
\begin{tabular}{lll}
{\includegraphics[width=0.26\linewidth]{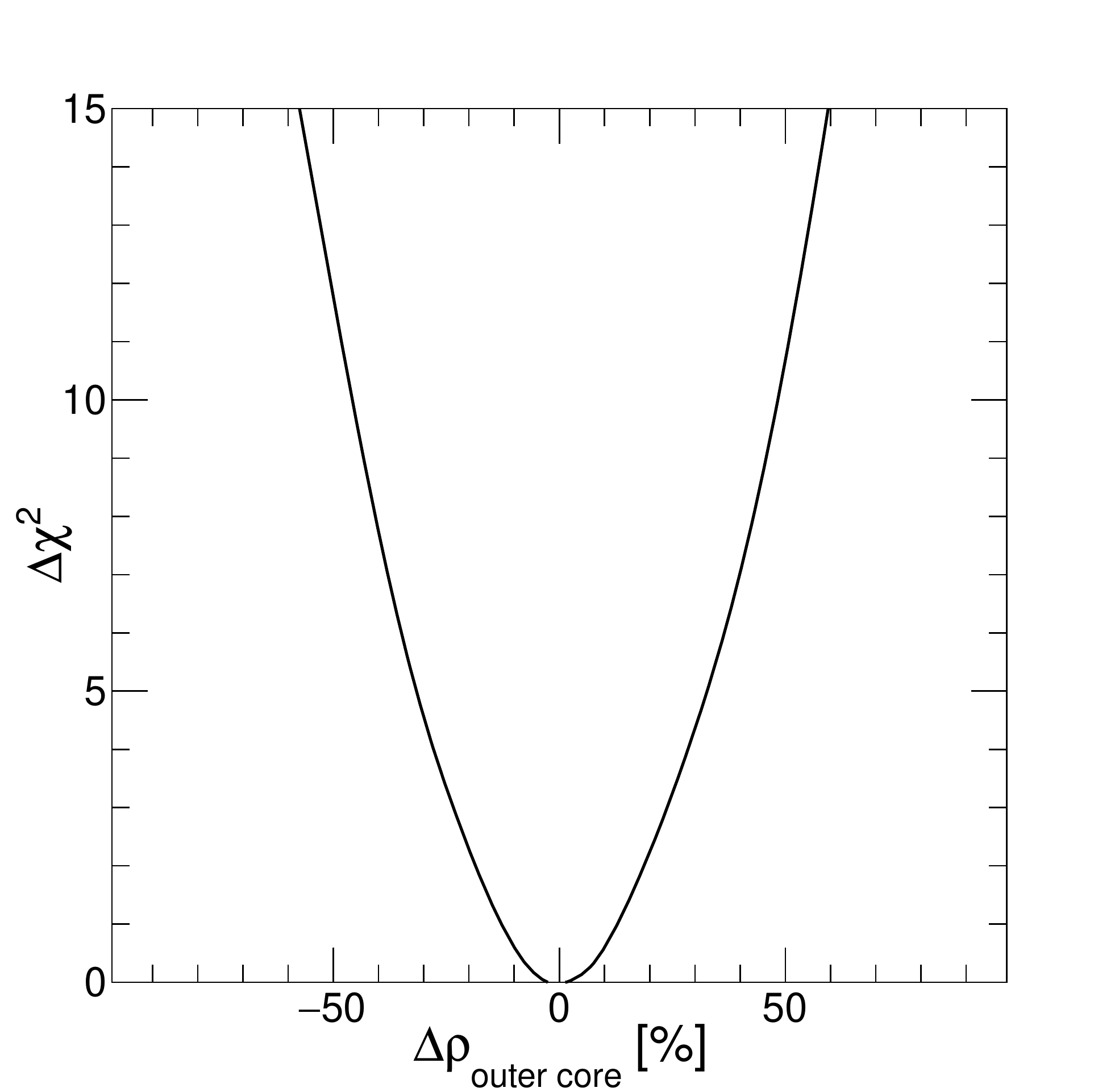}}
    &
   {\includegraphics[width=0.26\linewidth]{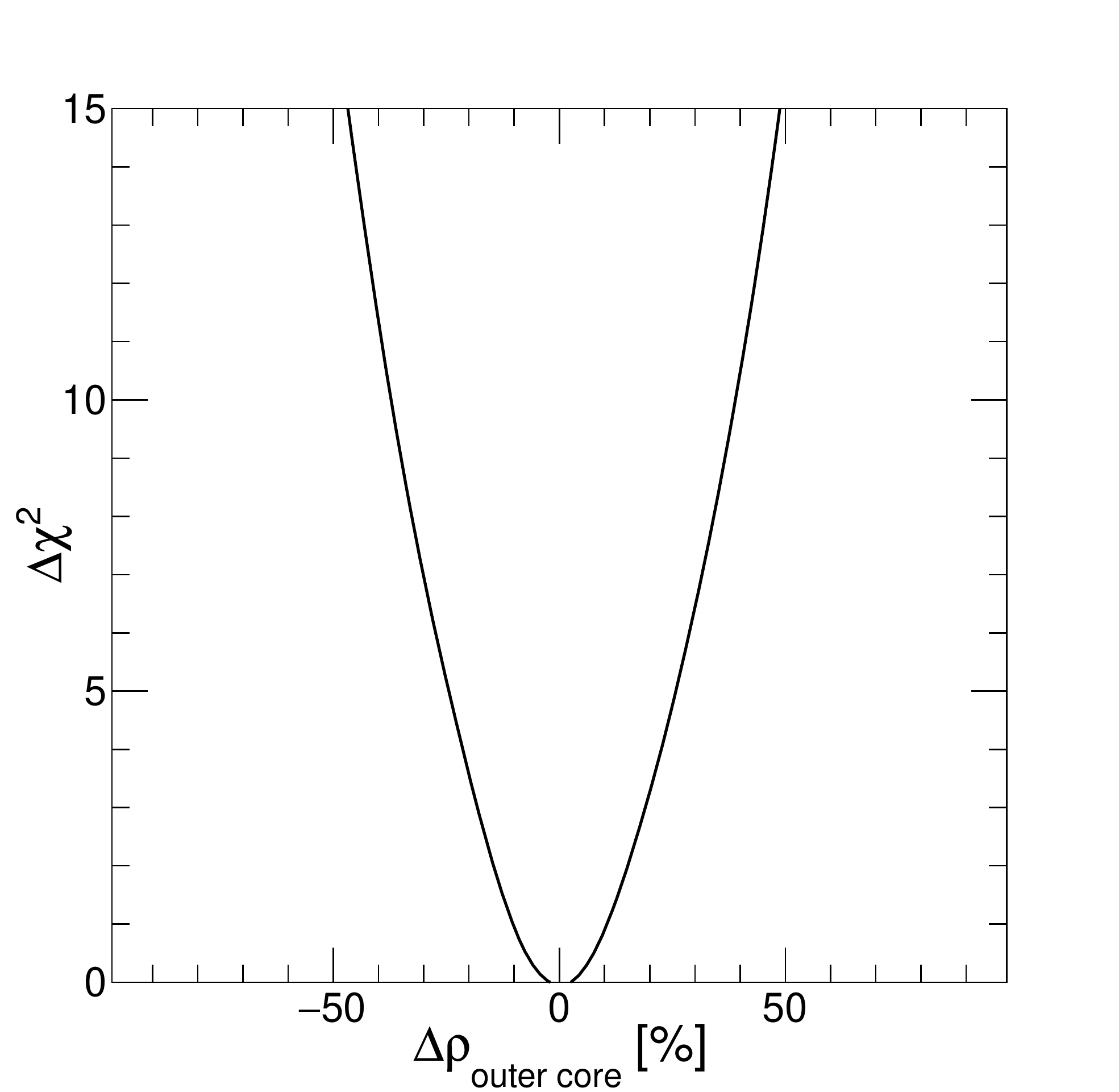}}
    &
    {\includegraphics[width=0.26\linewidth]{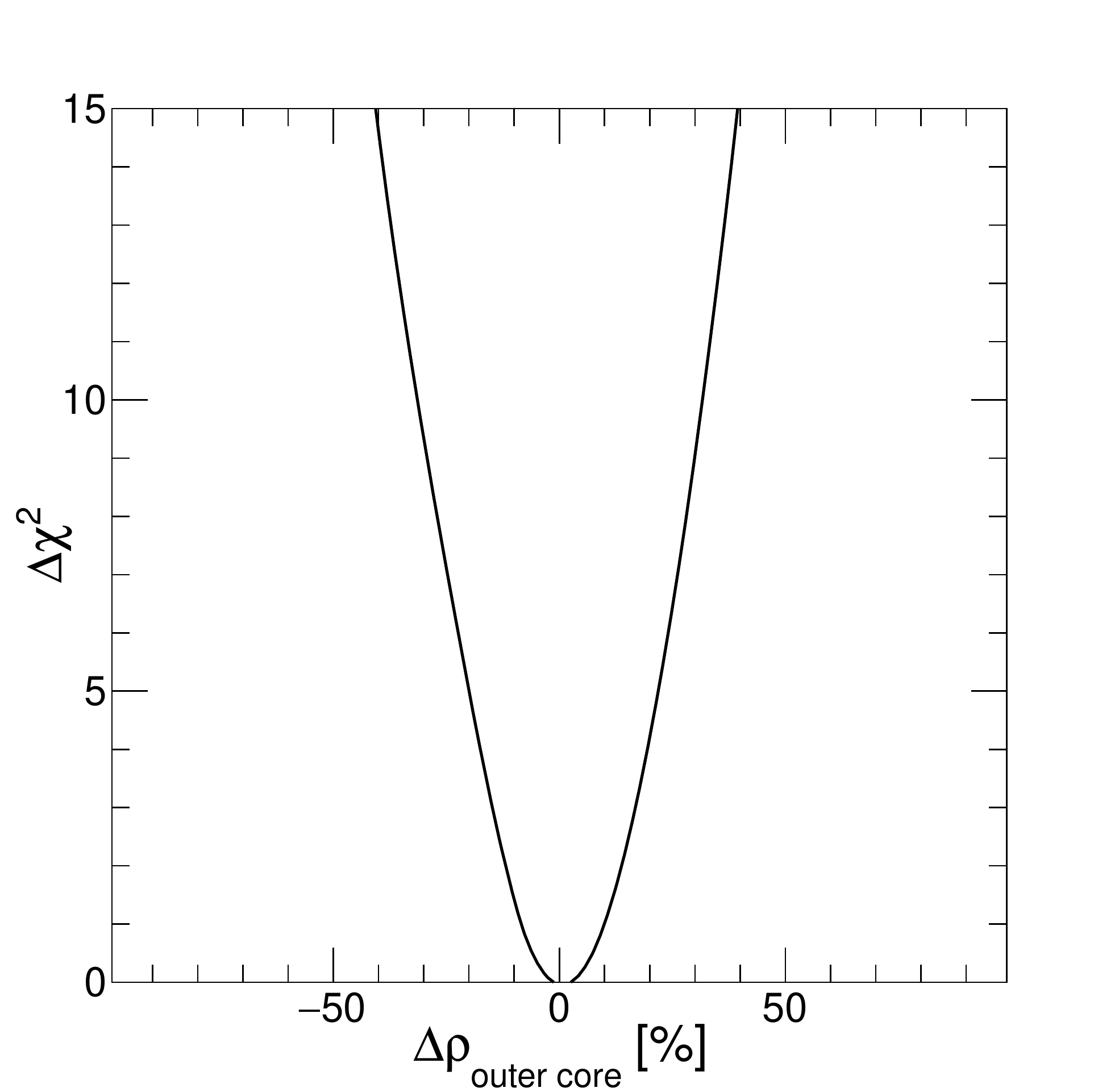}}
    \\
    {\includegraphics[width=0.26\linewidth]{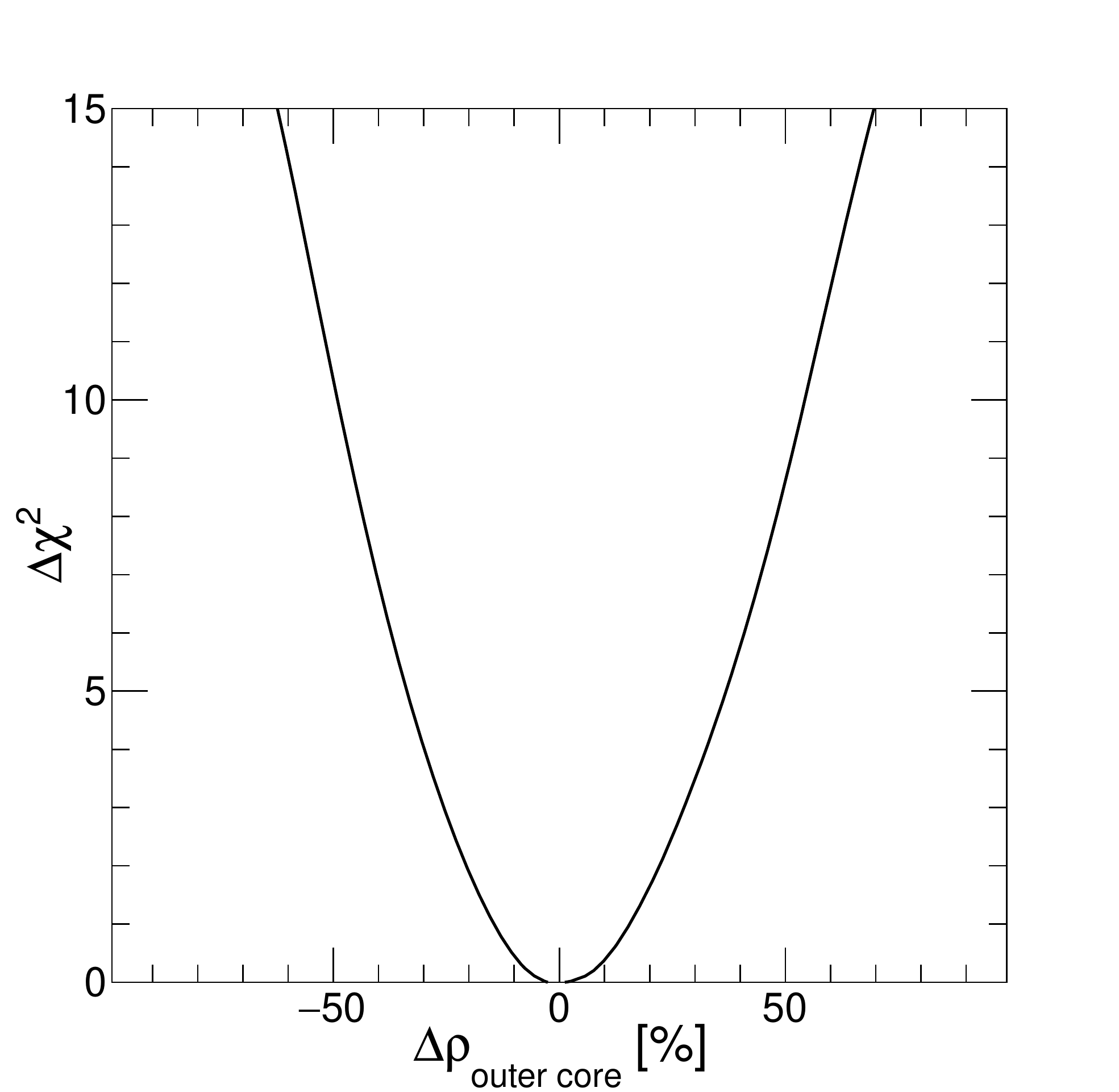}}
    &
    {\includegraphics[width=0.26\linewidth]{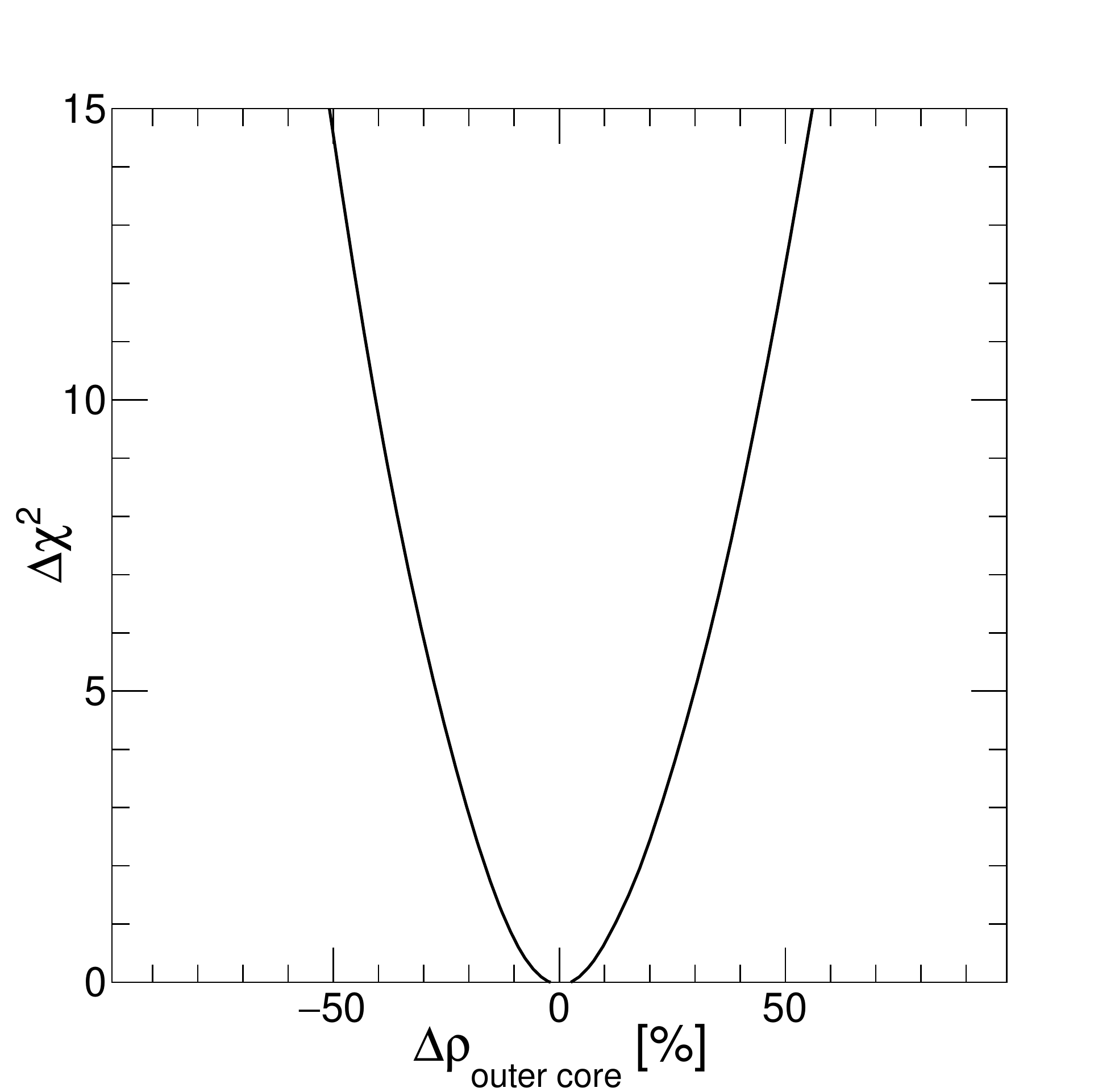}}
    &
   { \includegraphics[width=0.26\linewidth]{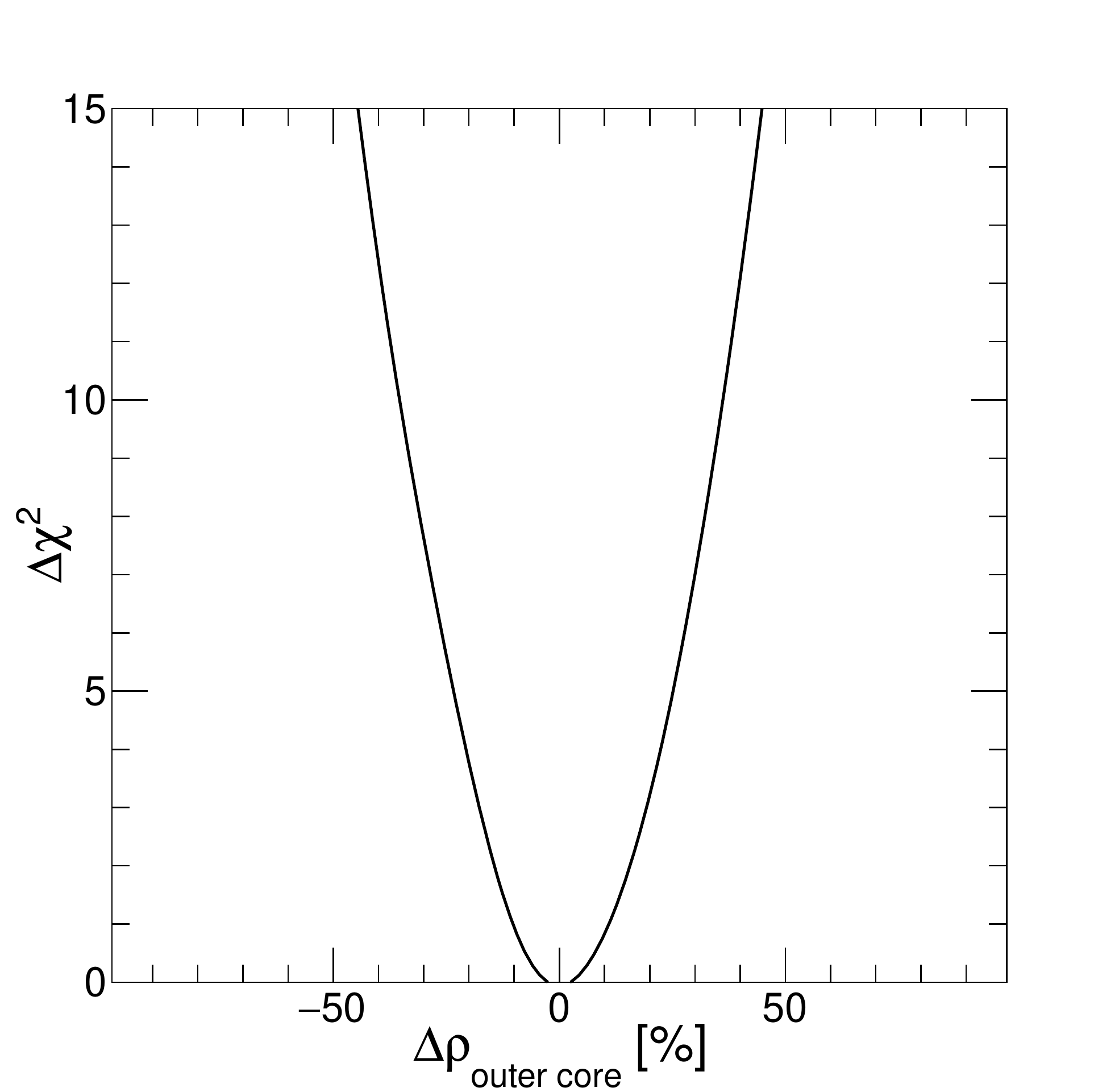}}
\\    
{\includegraphics[width=0.26\linewidth]{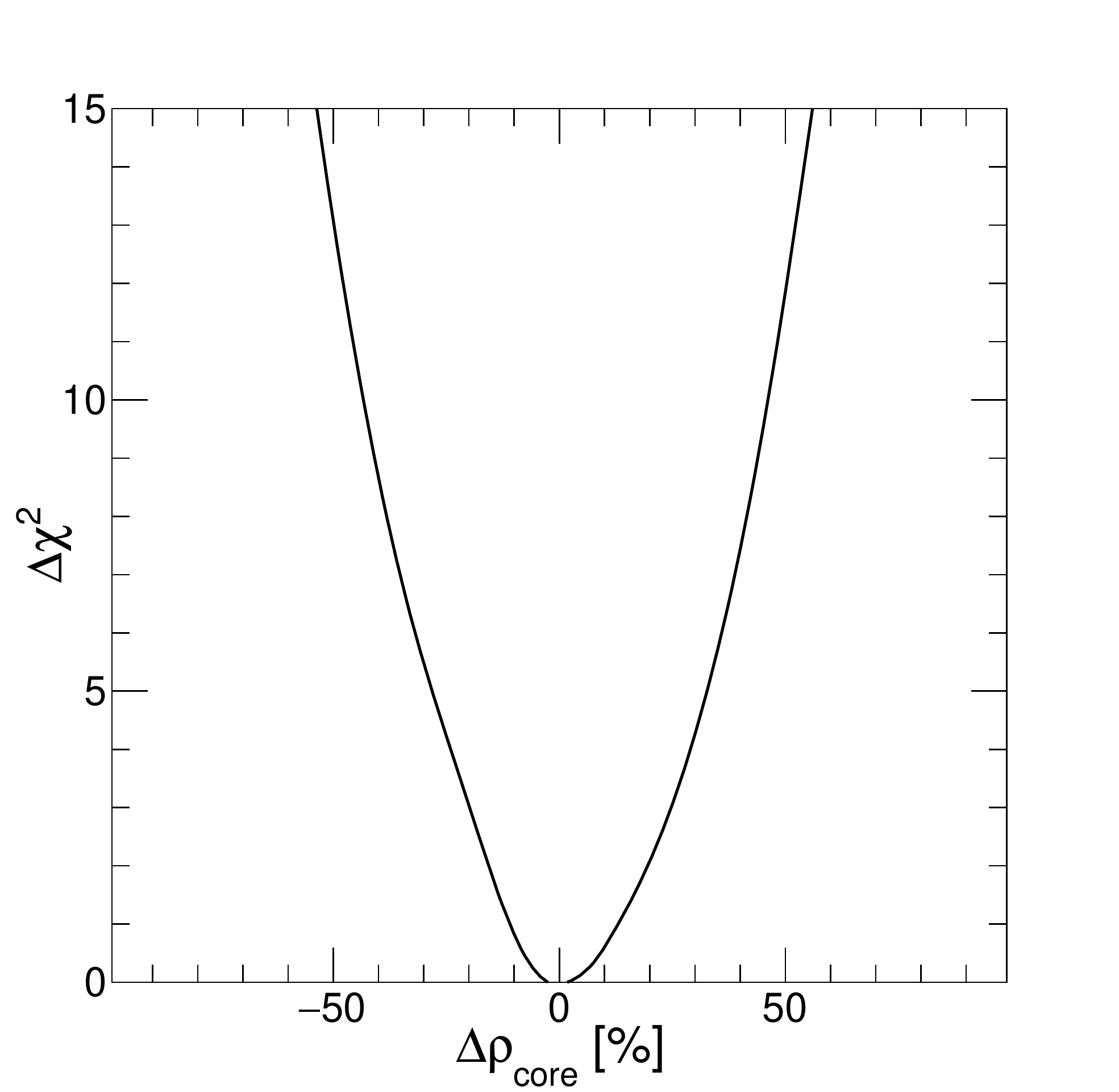}}
    &
   { \includegraphics[width=0.26\linewidth]{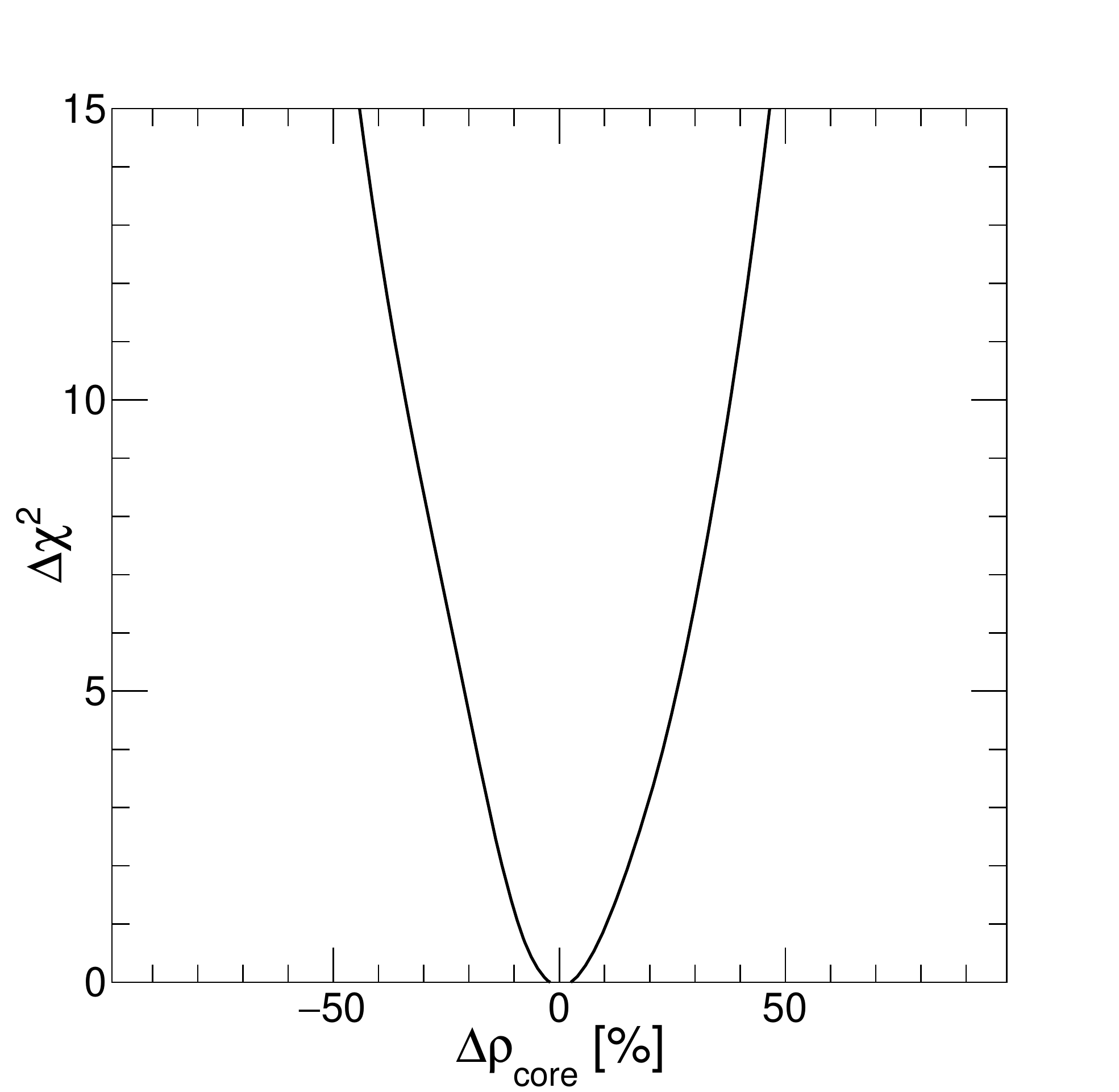}}
    &
    {\includegraphics[width=0.26\linewidth]{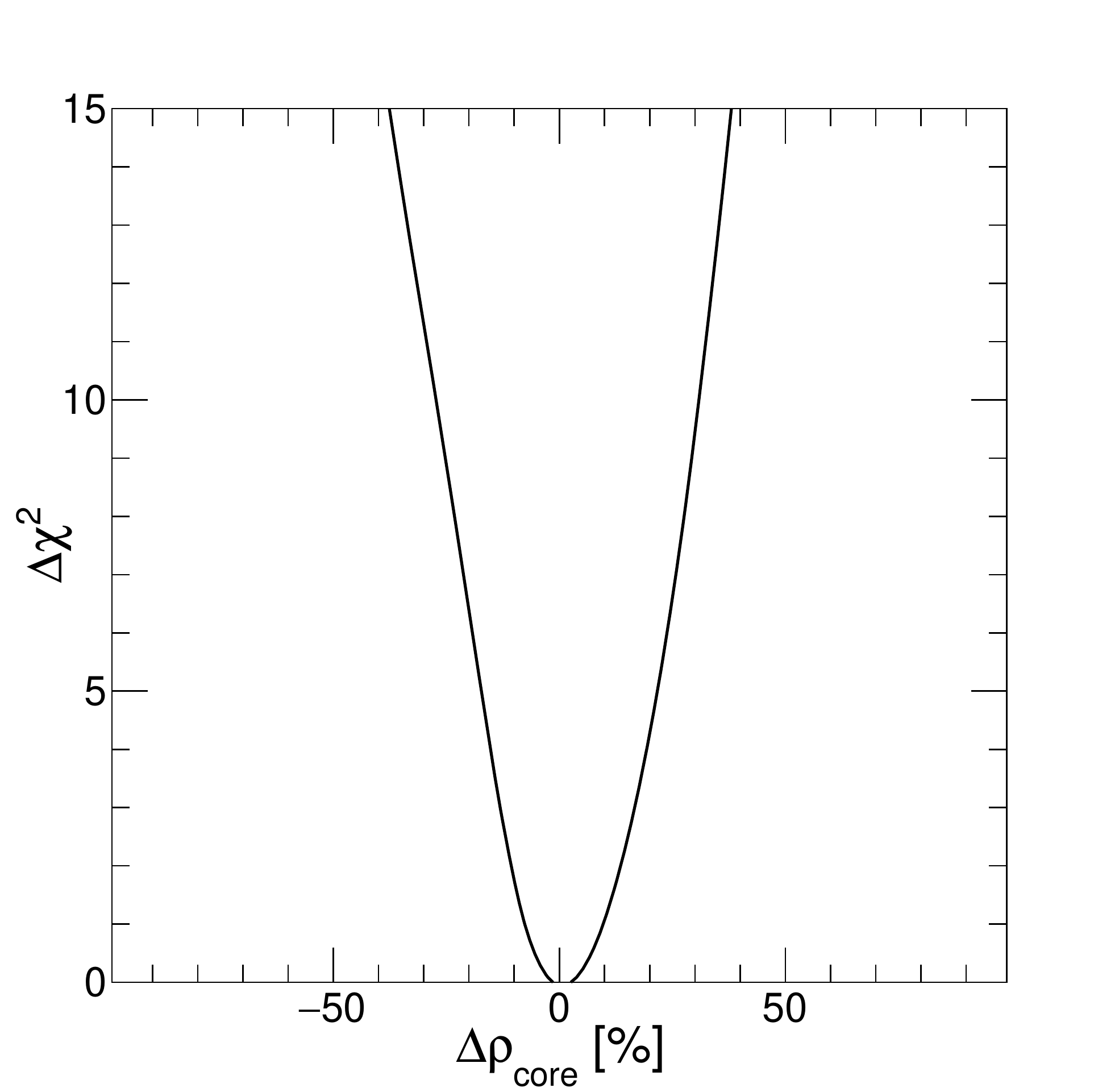}}
\\
   { \includegraphics[width=0.26\linewidth]{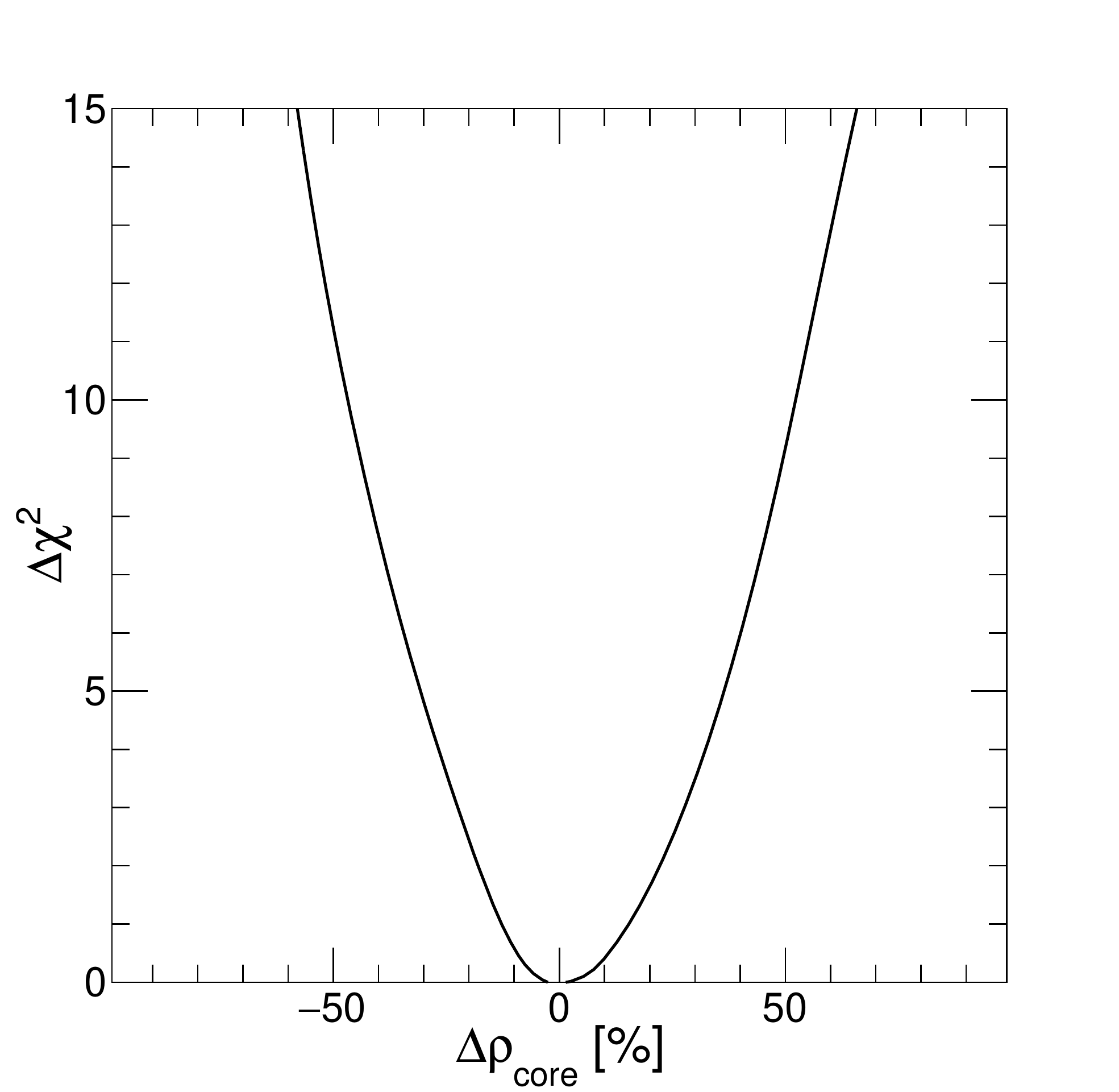}}
    &
    {\includegraphics[width=0.26\linewidth]{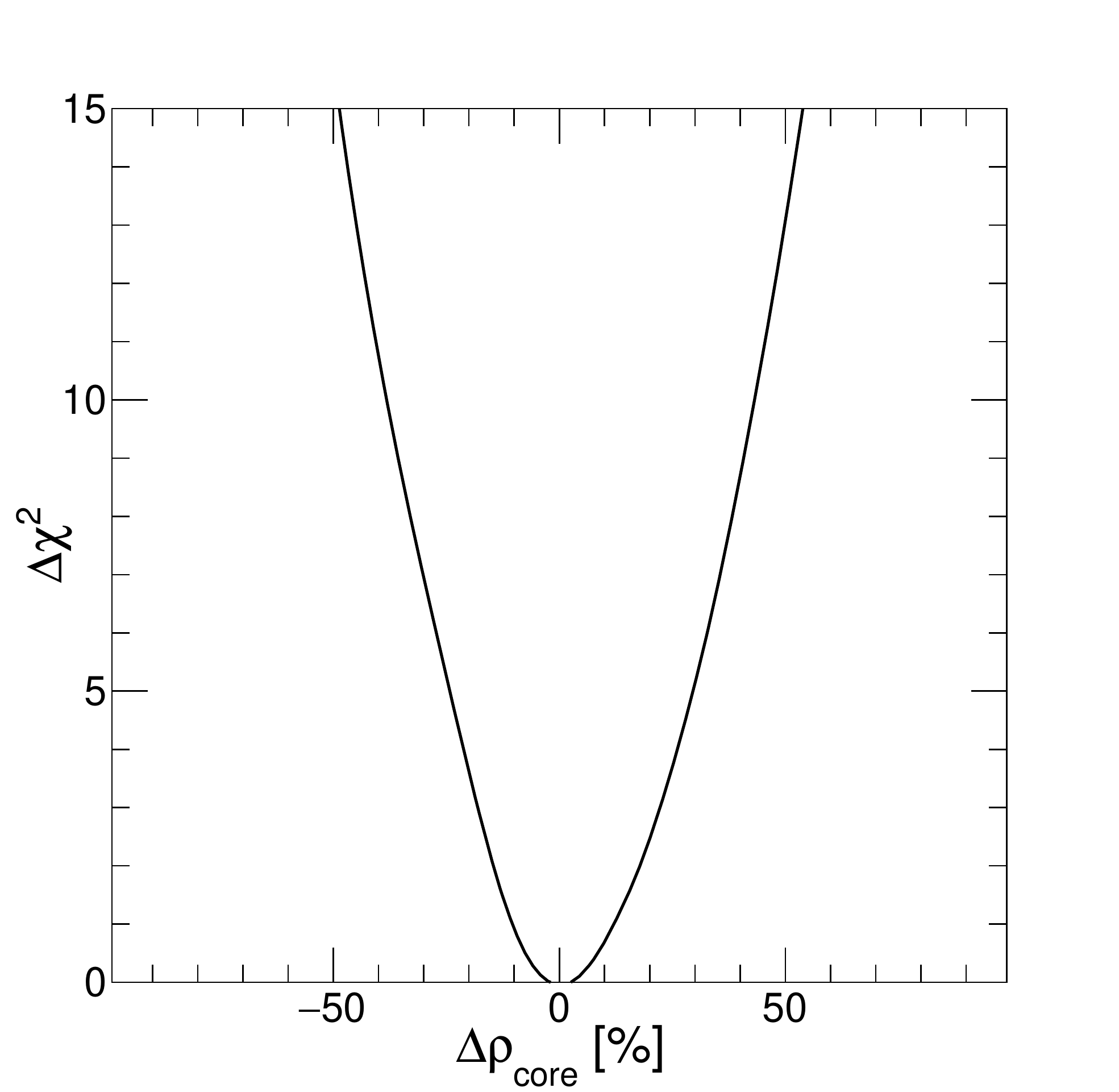}}
    &
   {\includegraphics[width=0.26\linewidth]{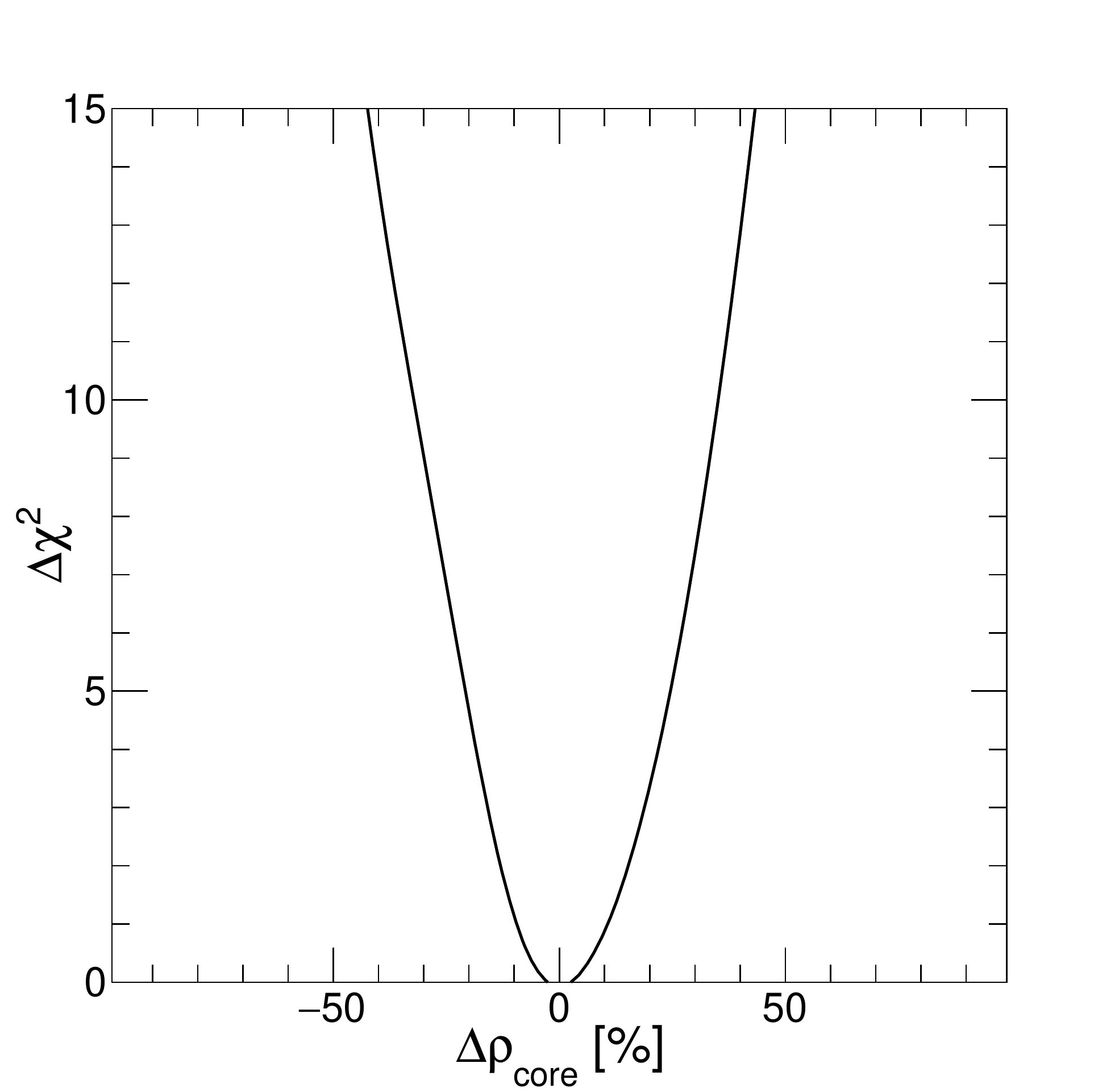}}
  \end{tabular}
 \caption{Sensitivity to the OC (1st and 2nd row panels)
and core (3rd and 4th row panels)
densities in the case of IO spectrum 
and 10 years of data.
The Earth total mass constraint 
is implemented by compensating the variations of the 
OC and  core densities with  
corresponding mantle density changes. The 
results shown are for $\sin^2\theta_{23} = 0.42$, 0.50, 0.58 
(left, center and right panels) and in the cases of 
``minimal'' (1st and 3rd row panels) and ``optimistic''
(2nd and 4th row panels) 
sets of errors. See text for further details.
}
\label{fig:IHOCCvsMant}
\end{figure}
%%%%%%%%%%%%%%%%%%%%%%%%%%%%%%%
%
\noindent 
NO neutrino mass spectrum.
This is essentially due to the fact that for the 
IO spectrum only the antineutrino oscillation probabilities 
$\bar{P}_{\alpha\beta}$, $\alpha = e,\mu$, $\beta = e,\mu,\tau$, 
can be amplified by the matter effects, while for the energies 
of interest the anti-neutrino cross sections are approximately 
by a factor of two smaller than the neutrino cross sections. 
In view of the above we present below only selected illustrative 
minimal subset of our results for the IO spectrum.

In Fig. \ref{fig:IHOCCvsMant} we present results on sensitivity of ORCA 
to the OC and total core densities with mantle being the 
``compensating'' layer. The results shown are for 
$\sin^2\theta_{23} = 0.42$, 0.50, 0.58 
(left, center and right panels). The panels in the first two rows 
(last two rows) in Fig. \ref{fig:IHOCCvsMant} 
illustrate the sensitivity to the OC (core) density and are obtained 
using ``minimal'' (1st and 3rd row panels) and 
``optimistic'' (2nd and 4th row panels) sets of systematic errors.
The $\chi^2$-distributions in  in Fig. \ref{fig:IHOCCvsMant} 
have symmetric or slightly asymmetric Gaussian form.

 In the case of most favorable ``minimal'' systematic error set,
our results show that for
  $\sin^2\theta_{23} = 0.42$, 0.50, 0.58,
ORCA can determine the OC (core) density at $3\sigma$ C.L. with 
uncertainties of 
$\mp 45\%$, $\mp 37\%$ and $\mp 30\%$ 
((-40\%)/+44\%, (-32\%)/+36\% and (-25\%)/+30\%), respectively.
For the ``optimistic'' and ``conservative'' sets of errors 
we get respectively for the indicated three values of 
$\sin^2\theta_{23}$:
(-47\%)/+52\%, (-38\%)/+42\%, (-33\%)/+34\%, and 
(-63\%)/+70\%, (-52\%)/+57\%, (-47\%)/+48\%
((-44\%)/+50\%, (-36\%)/+40\%, (-30\%)/+33\%),
and ((-58\%)/+57\%, (-50\%)/+53\% and (-43\%)/+45\%))
\footnote{The results for the ``conservative'' set of errors 
are not shown in  Fig. \ref{fig:IHOCCvsMant}. 
We quote them here for completeness.
}.
They should be compared with the corresponding  
results for the NO spectrum reported in Figs. \ref{fig:NHOCvsMantle} 
and \ref{fig:NHCvsMant}.
It follows form this comparison, in particular, that in the case 
of IO spectrum the ORCA sensitivity to the OC and core densities 
is worse than the sensitivity in the case of NO spectrum
by factors that can be as large as $\sim 2.5$.

%%%%%%%%%%%%%%%%%%%%%%%%%%%%%%%%%%
%
\section{Summary and Conclusions}
\label{sec:summary}
%
%%%%%%%%%%%%%%%%%%%%%%%%%%%%%%%%%%%

 In the present article we have investigated 
the sensitivity of the ORCA detector to deviations of the Earth 
i) outer core (OC) density, ii) inner core (IC) density, 
iii) total core density, and iv) mantle density,  
from their respective PREM reference densities. 
We have considered the case when 
the radial dependence of the densities of the layers of interest, 
$\rho_{i}(r)$, $i= {\rm IC,OC,core,mantle}$, is given by PREM and 
the deviations correspond to an overall scaling factor, i.e., 
have the form $\rho^\prime_{i}(r)  = (1 + \kappa_i)\rho_{i}(r)$, 
where $\kappa_i$ is a real positive or negative constant. 
The results we present on sensitivity of ORCA 
are for the relative deviations of the densities 
$\rho_{i}(r)$ from their PREM reference values, i.e., 
for $100\% (\rho^\prime_{i}(r) - \rho_{i}(r))/\rho_{i}(r) 
= 100\% \kappa_i$.      
The change of density in each Earth layer (IC, OC 
and mantle) as described by PREM was taken effectively into account, i.e., 
we did not use the  constant density approximation in the 
considered layers and shells.
The analysis was performed by studying the effects of the Earth matter 
on the oscillations of atmospheric $\nu_{\mu}$, $\nu_e$, $\bar{\nu}_\mu$ 
and $\bar{\nu}_e$. 
 The relevant neutrino oscillation probabilities depend on the 
electron number densities of the Earth layers of interest, $N^{(i)}_e$:
$N^{(i)}_e(r) = Y^{(i)}_e \rho_{i}(r)/m_{\rm N}$, where  
$Y^{(i)}_e$ is the electron fraction number (or the Z/A factor) 
of the layer $i= {\rm IC,OC,core,mantle}$.
In the analysis we have performed we have 
set $Y_e$ of the mantle and the core to fixed values,
$Y_e^{man} = 0.490$ and  $Y_e^{c} = 0.467$.
Thus, when we vary the matter density $\rho_{\rm i}(r)$, $i={\rm IC,OC,C,man}$, 
we actually vary the electron number density of the 
corresponding layer, $N^{(i)}_{e}(r)$, $i={\rm IC,OC,C,man}$.
As a consequence the results on sensitivity of ORCA to the matter densities 
of the different Earth structures 
are results on sensitivity of ORCA 
to the electron number densities of these Earth structures.
They do not depend on the chosen fixed values of 
$Y_e^{man}$ and  $Y_e^{c}$.

For the unoscillated fluxes of the atmospheric neutrinos 
we used the azimuth-averaged energy and zenith angle dependent fluxes 
from  \cite{Honda:2015fha}  at the Frejus cite.
 We assumed that the type of light neutrino mass spectrum 
is known and we obtain results for both the NO and IO spectra.
The relevant detection characteristics of the ORCA set-up - 
the energy and angular resolutions, the dependence of the effective volumes 
for the different classes of events on the initial neutrino energy,
the prospective systematic uncertainties, etc. were taken form the    
the ORCA proposal \cite{KM3Net:2016zxf}. We took into 
account also a number of potential systematic uncertainties identified 
in \cite{Capozzi:2017syc} and studied the dependence of the results 
on the type of systematic uncertainties used. 
The statistical errors employed in our analysis correspond to 10 
years of operation of ORCA.

 In determining the sensitivity of ORCA to the densities 
of the different Earth layers, which requires to consider 
deviations of the density of a given 
layer from its PREM reference density, 
we systematically implemented the constraint 
following the precise knowledge of 
the total Earth mass. 
% In this way unphysical variations of a given layer's density were avoided.  
This was done by compensating the variation of the density in 
the considered layer by a corresponding change 
of the density in one of the other layers.

 More specifically,
when we consider the variation of IC density
$\rho_{\rm IC}(r)$, we studied two cases: compensating it by 
the corresponding change of density of 
i) the outer core $\rho_{\rm OC}(r)$, and 
ii) of the mantle $\rho_{\rm man}(r)$. 
We proceeded in a similar way when we analyse the sensitivity 
of ORCA to the OC and mantle densities  
 $\rho_{\rm OC}(r)$ and $\rho_{\rm man}(r)$:
in these two cases we investigate respectively two and three    
ways of compensating the variation of 
the respective densities - 
by the change of $\rho_{\rm IC}(r)$ or of $\rho_{\rm man}(r)$,
and by the change of  $\rho_{\rm IC}(r)$ or of $\rho_{\rm OC}(r)$  
or else of $\rho_{\rm IC}(r) + \rho_{\rm OC}(r) =\rho_{\rm C}(r)$.
We considered also the sensitivity of ORCA to the 
core density $\rho_{\rm C}(r)$. In this case   
the variation of  $\rho_{\rm C}(r)$ 
is compensated by the change of  $\rho_{\rm man}(r)$.
In order to asses the effects of the total Earth mass constraint 
we presented also results without imposing it.

 Following the described procedure 
we have derived and reported results 
on the sensitivity of ORCA to the 
i) OC density $\rho_{\rm OC}(r)$, ii) IC density  $\rho_{\rm OC}(r)$, 
iii) total core density  $\rho_{\rm core}(r)$
and iv) mantle density  $\rho_{\rm man}(r)$.  
We have derived results for four different sets of systematic 
uncertainties which can affect significantly  
the sensitivity of ORCA, ``minimal'', ``optimistic'', 
``default'' and ``conservative'' 
(defined in Section \ref{sec:basics}).
To illustrate the dependence of the sensitivity of ORCA on the 
systematic uncertainties and not burden the 
presentation, we presented results only for 
``minimal'', ``optimistic'' and ``conservative''
sets of uncertainties.

In the analysis we have performed 
we kept the Dirac phase  $\delta$ and 
$\sin^2\theta_{23}$ fixed to certain values and assumed 
that mass spectrum of light neutrinos, which can be with normal ordering 
(NO) or inverted ordering (IO) is known.
We have obtained results for $\delta = 3\pi/2$,     
eleven values of $\sin^2\theta_{23}$ from the interval
[0.40,0.60] and the two types of neutrino mass spectrum.
 Rather detailed results were presented only for three 
reference values of  $\sin^2\theta_{23} = 0.42$, 0.50 and 0.58, 
which belong to, and essentially span, the  $3\sigma$ range of 
allowed values of $\sin^2\theta_{23}$ obtained in the latest global 
neutrino data analyses \cite{Capozzi:2021fjo,Esteban:2020cvm},
and for NO spectrum (subsections 3.1-3.4). We reported also 
results obtained assuming IO neutrino mass spectrum 
(subsection 3.5).

 In our analysis the Earth hydrostatic equilibrium constraints 
given in Eq. (\ref{eq:equil}) are not {\it a priori} satisfied 
when we vary the dnsity in a given layer 
and compensate it with a change of density in another layer.
However, we indicated in each specific case 
what are the restrictions they lead to whenever these 
restrictions are relevant. In general, in the cases when they are relevant 
the smaller the systematic errors and/or the larger $\sin^2\theta_{23}$,
the smaller the effect of the constraints is.

 Our results show that the ORCA detector in the configuration 
considered in KM3NeT 2.0 LoI for ORCA \cite{KM3Net:2016zxf},
which we used in our analysis, is practically not sensitive 
to the IC density. We find further that that the sensitivity of ORCA 
to the densities of the outer core, core and mantle 
depend strongly \\ 
1. on the value of $\sin^2\theta_{23}$,\\
2. on the type of systematic errors employed in the analysis,\\
3. on whether the total Earth mass constraint is implemented or not, and\\
4. on the way the compensation of the density variation in a 
given layer by a change of density in another layer 
is implemented, i.e., on the choice of the ``compensating'' layer, 
when the total Earth mass constraint is imposed.\\
It depends also strongly on the type of neutrino mass spectrum.   

The sensitivity of ORCA to the outer core (core) and 
mantle densities is found in our analysis to be highest/maximal\\
i) for the largest value of $\sin^2\theta_{23}$ 
allowed by the data;\\
ii) in the case when the total Earth mass constraint is implemented 
and the variation of the outer core (total core) and 
mantle densities is compensated by changes respectively of the 
mantle and outer core or total core densities;\\
iii) and, obviously, for the ``minimal'' or ``optimistic'' 
set of systematic errors.\\
We have shown , in particular, that in the ``most favorable'' NO case
of ``minimal'' systematic errors, $\sin^2\theta_{23}=0.58$ 
and implemented Earth mass constraint as in point ii),
ORCA can determine, e.g., the OC (mantle) density at $3\sigma$ C.L.
after 10 years of operation with an uncertainty of 
% (-15\%)/+12\%  (of (-6\%)/+8\%).  
 (-18\%)/+12\%  (of (-6\%)/+8\%).  
In the  ``most disfavorable'' NO 
with implemented Earth mass constraint as in point ii) 
but ``conservative'' systematic errors
and  $\sin^2\theta_{23}=0.42$,  
the uncertainty reads (-43\%)/+39\% ((-17\%/+20\%), while for  
for $\sin^2\theta_{23} = 0.50$ and 0.58 it  is noticeably smaller:
(-37)\%/+30\% and (-30\%)/+24\% 
%  ((-13\%)/+16\%  
((-13\%)/+17\% and  (-11\%/+14\%)).

 In the case of OC (mantle) density variation compensated 
by a change of the mantle (OC) density, the  
Earth hydrostatic equilibrium constraints, Eq.  (\ref{eq:equil}), 
imply approximately $(-49.6\%) \ltap \Delta \rho_{\rm outer~core} \ltap 18.3\%$
($(-8.3\%)\ltap \Delta \rho_{mantle} \ltap 22.8\%$). 
In what concerns the drived ORCA $3\sigma$ sensitivity 
ranges of $\Delta \rho_{\rm outer~core}$ reported above, 
i) the lower limit from the external constraints in Eq. (\ref{eq:equil}) 
has no effect on them, 
ii) the effects of the upper limit  of 18.3\%, if any, depends 
on $\sin^2\theta_{23}$ and on the type of implemented systematic errors.
% the larger $\sin^2\theta_{23}$ and the smaller the systematic errors,
% the smaller the effect of the indicated upper limit.
More specifically, the upper limit of 18.3\% has no effect on the ORCA 
$3\sigma$ sensitivity ranges in the case of ``minimal'' systematic errors 
for any $\sin^2\theta_{23} \gtap 0.50$; 
for $\sin^2\theta_{23} =  0.42$, it corresponds 
to the maximal value of the ORCA $2\sigma$ sensitivity range.
Fot the set of  ``optimistic'' systematic errors,
18.3\% represents approximately the 
maximal value of the $2\sigma$, $2.4\sigma$ and 
$2.6\sigma$ ORCA sensitivity ranges 
for $\sin^2\theta_{23} = 0.50$, 0.54 and 0.58, 
respectively. The effect of the discussed constraint 
is largest for the ORCA sensitivity ranges obtained 
with conservative systematic errors: 
18.3\% correponds, e.g., to the maximal value of 
the ORCA $1.9\sigma$ sensitivity range 
at $\sin^2\theta_{23} = 0.58$. 

In a similar way, the external constraints 
in Eq. (\ref{eq:equil}) do not restrict from above the reported 
ORCA $3\sigma$ sensitivity to  $\Delta \rho_{mantle}$
even in the case of ``conservative'' systematic errors.
The maximal allowed negative variation of (-8.3\%) restricts 
from below the ORCA sensitivity ranges. More specifically, 
in the case of ``minimal'' systematic errors, 
$\Delta \rho_{\rm mantle} = -8.3\%$ 
corresponds to the minimal values of the 
ORCA $2\sigma$, $2.4\sigma$, $2.8\sigma$ and $3.0\sigma$ 
sensitivity ranges derived 
for $\sin^2\theta_{23} = 0.46$, 0.50, 0.54 and 0.58, 
respectively. With implemented ``optimistic'' systematic errors, 
it coincides with the minimal values of 
the ORCA $2.0\sigma$, $2.4\sigma$ and $2.7\sigma$  
sensitivity ranges  for  
$\sin^2\theta_{23} = 0.48$, 0.54 and 0.58. 
And in the case of ``conservative'' systematic errors, 
it is equal to the minimal value of the 
$2.0\sigma$ sensitivity range for 
$\sin^2\theta_{23} = 0.58$. 

 With maximally reduced systematic errors 
and certain further improvements, e.g., 
the discussed ``favorable'' 6 m vertical spacing 
configuration of ORCA experiment \cite{KM3Net:2016zxf} 
or the Super-ORCA version of the detector
\cite{Hofestadt:2019whx}, the ORCA sensitivities  
to  positive $\Delta \rho_{\rm outer~core}$ 
(negative $\Delta \rho_{\rm mantle}$) might 
increase sufficiently so that the 
external constraint 
$\Delta \rho_{\rm outer~core} \ltap 18.3\%$
($(-8.3\%) \ltap \Delta \rho_{\rm mantle}$) 
would have no effect on the ORCA $3\sigma$ 
sensitivity to positive variations of $\rho_{\rm OC}$
compensated by changes of $\rho_{\rm mantle}$
(negative variations of $\rho_{\rm mantle}$   
compensated by changes of  $\rho_{\rm OC}$).

The uncertainties in the determination of the 
outer core, total core and mantle densities by ORCA 
in the case of NO spectrum, according to our results,
are considerably larger if the total Earth mass constraint 
is not implemented in the analysis, or if it is implemented 
but the inner core is used as a ``compensation'' layer.  

We find also that the sensitivity 
of ORCA to the outer core, core and mantle densities 
is significantly worse for the IO neutrino mass spectrum 
than for the NO spectrum (Fig. \ref{fig:IHOCCvsMant}, subsection 3.5).
In the case of most favorable ``minimal'' systematic error set 
our results show, for example, that for
$\sin^2\theta_{23} = 0.42$, 0.50, 0.58,
ORCA can determine the OC density 
at $3\sigma$ C.L. if mantle is used as a ``compensating'' layer 
with uncertainties of 
$\mp 45\%$, $\mp 37\%$ and $\mp 30\%$,  
respectively. For the ``optimistic'' and ``conservative'' 
sets of errors we get for the indicated three values of $\sin^2\theta_{23}$:
(-47\%)/+52\%, (-38\%)/+42\%, (-33\%)/+34\%, and 
(-63\%)/+70\%, (-52\%)/+57\%, (-47\%)/+48\%.
They should be compared with the corresponding  
results for the NO spectrum reported in 
Fig. \ref{fig:NHOCvsMantle}.
It follows form this comparison, in particular, that in the case 
of IO spectrum the ORCA sensitivity to the OC 
density is worse than the sensitivity in the case of NO spectrum
by factors that can be as large as $\sim 2.5$.
We find similarly that the sensitivity of ORCA to 
the core and mantle densities for IO neutrino mass spectrum
are also worse than that in the case of NO spectrum.

 On the basis of the results of these study we can conclude that 
the ORCA experiment has the potential of  making  
unique pioneering contributions to the studies of the Earth interior 
with atmospheric neutrinos, i.e., to the neutrino tomography of the Earth.

\vspace{0.3cm} 
\noindent {\it Note Added.}
While the text of this article was being written, 
refs. \cite{Kelly:2021jfs,Denton:2021rgt} appeared on the arXiv.
In \cite{Kelly:2021jfs}
the authors investigated the sensitivity of the 
DUNE experiment to the Earth core, lower mantle and upper 
mantle densities by using prospective DUNE data on oscillations 
of atmospheric neutrinos, while in  \cite{Denton:2021rgt}
prospective atmospheric neutrino DUNE data were used with 
the aim of determining the  radius of the Earth core.

%%%%%%%%%%%%%%%%%%%%%%%
\section*{Acknowledgments}
%%%%%%%%%%%%%%%%%%%%%%%
  
S.T.P. would like to thank S. Choubey 
for many discussions over the years 
of various aspects of neutrino tomography of the Earth. 
This project was supported in part by
the European Union's Horizon 2020 research and innovation programme under 
the Marie Sklodowska-Curie grant agreement No.~860881-HIDDeN,
by the Italian INFN program on Theoretical Astroparticle Physics 
and by the  World Premier International Research Center
Initiative (WPI Initiative, MEXT), Japan (S.T.P.).
The work of F.C. at Virginia Tech 
is supported by the U.S. Department of Energy under the award 
grant number DE-SC0020250. 
The work of F.C. at IFIC (Valencia, Spain) is supported by 
GVA Grant No. CDEIGENT/2020/003.

%%%%%%%%%%%%%%%%%%%%%%%
\section*{Appendix}
%%%%%%%%%%%%%%%%%%%%%%%

\appendix
%%%%%%%%%%%%%%%%%%%%%%%%%%%%%%%%%%%%%%%%%%%%%%%%%%%%%%%%%%%%%%%%%%%%%
%
\section{Early Preliminary Estimates}
\label{Early}
%
%%%%%%%%%%%%%%%%%%%%%%%%%%%%%%%%
In \cite{ChGSTP2011} the sensitivities of $\sim 1$ Mt 
SuperKamiokande-like water Cerenkov detector 
and of $\sim 100$ Kt liquid argon (LAr) detector 
to the Earth (outer) core density were estimated  
using prospective atmospheric neutrino oscillation data. 
The characteristics of the detectors and the methods 
developed in \cite{Gandhi:2007td,Gandhi:2008zs} 
for studies of neutrino mass ordering determination 
with these detectors were employed in \cite{ChGSTP2011}. 

 In the case of the water Cerenkov detector, both 
$e$-like and  $\mu$-like events produced by atmospheric neutrinos 
with energy $E$ in the multi-GeV range (2-10) GeV 
were taken into account. This range was divided into eight bins of equal 
width of 1 GeV. Three Earth core bins in $\cos\theta_z$, 
$\theta_z$ being the zenith angle, were used: 
one corresponding to the inner core and two equal width bins 
corresponding to the outer core. The atmospheric neutrino 
fluxes from \cite{Honda:2004yz} were utilised in the analysis.
Following \cite{Gandhi:2007td}, the energy and zenith angle 
resolution functions were assumed to have Gaussian forms 
with quite optimistic widths:
$\sigma_{\rm E} = 0.10\,E$ (0.05E) and $\Delta \theta_z = 7^\circ$ 
($5^\circ$). Results for the somewhat more realistic values 
$\sigma_{\rm E} = 0.15E$ and $\Delta \theta_z = 10^\circ$ 
were also obtained. The neutrino oscillation parameters were fixed 
to  $|\Delta m^2_{31(23)}| = 2.50\times 10^{-3}\,{\rm eV^2}$,
$\Delta m^2_{21} = 8.0\times 10^{-5}\,{\rm eV^2}$
$\sin^2\theta_{12} = 0.31$, $\sin^2\theta_{23} = 0.50$ and $\delta = 0$.
In what concerns $\sin^22\theta_{13}$, 
it was found that maximal sensitivity is obtained for 
$\sin^22\theta_{13} = 0.05$ and this value was 
used in the study. In the $\chi^2$-analysis only 
statistical errors corresponding to exposure of 
10 Mty were included. The results, e.g., on the outer core density  
sensitivity were derived by changing the PREM outer core density 
by an overall fixed scale factor. The inner core and mantle densities 
were kept fixed at their PREM values.
It was found in \cite{ChGSTP2011}, in particular, 
that in the case of NO neutrino mass spectrum and 
$\sigma_{\rm E} = 0.10\,E$ (0.05E) and $\Delta \theta_z = 7^\circ$ 
($5^\circ$), a water Cerenkov detector may have a $2\sigma$ C.L. 
sensitivity to a 20\% (15\%) outer core density deviation from 
its PREM density. These results were quite discouraging.

 In \cite{ChGSTP2011} aspects of atmospheric neutrino oscillation 
tomography of the Earth with a 100 Kt prototype of LAr 
detector \cite{Rubbia:2004tz,Bueno:2007um,Cline:2006st} 
with magnetisation over the detector's volume 
\cite{Ereditato:2005yx} have been investigated using the 
characteristics of the detector considered in 
\cite{Gandhi:2008zs}. The results of this study were not published 
since the technical characteristics of the proposed LAr detector 
were evolving continuously (the mass of the detector was reduced, 
the magnetisation option was abandoned, etc.) and eventually 
the ``realistic'' design of the detector 
differed significantly from that used in the study. 
%%%%%%%%%%%%%%%%%%%%%%%%%
\begin{figure}[H]  
  \centering
\begin{tabular}{lll}
% \begin{tabular}{l}
{\includegraphics[width=0.3\linewidth]{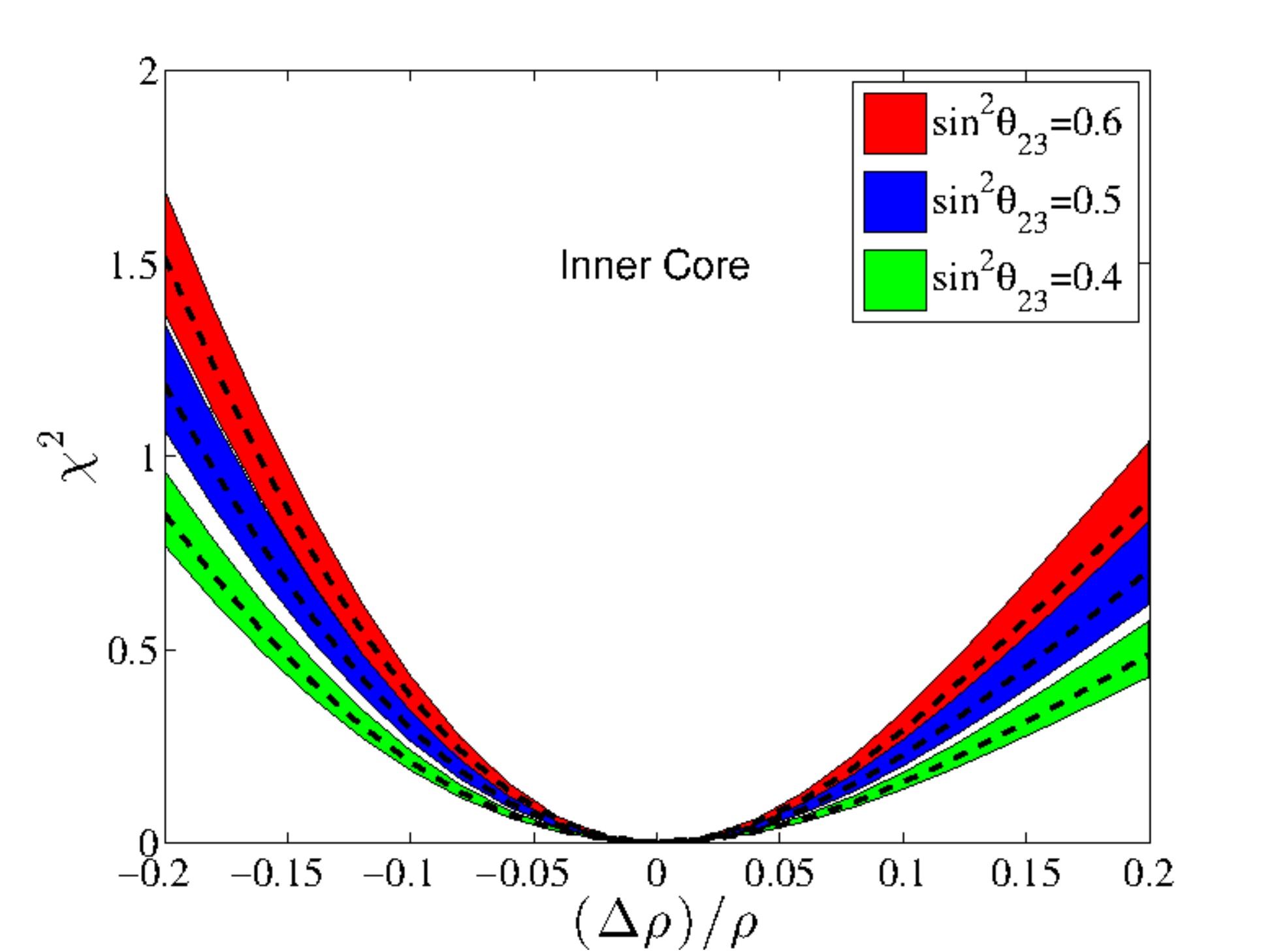}}
    &
   { \includegraphics[width=0.3\linewidth]{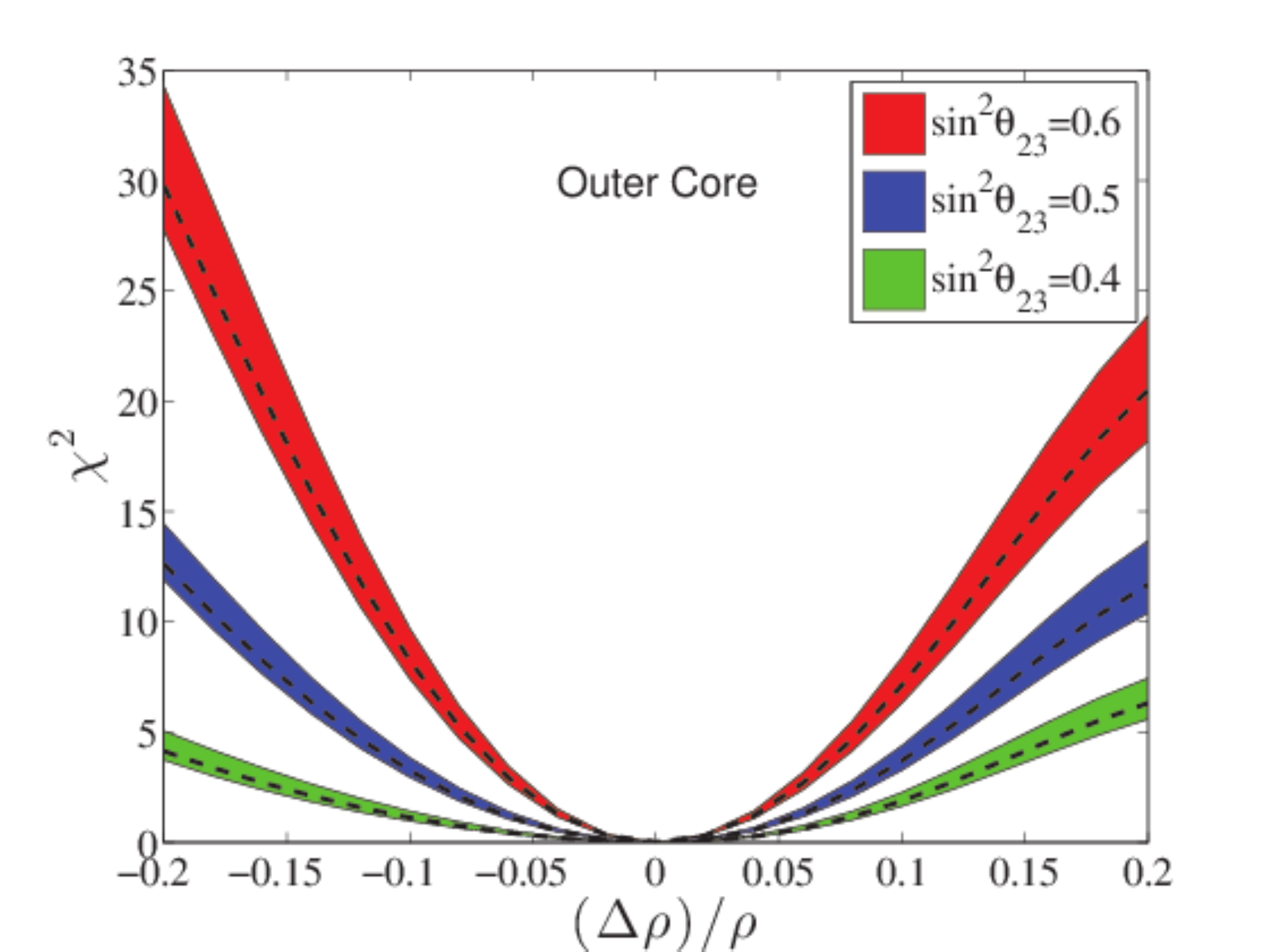}}
    &
    {\includegraphics[width=0.3\linewidth]{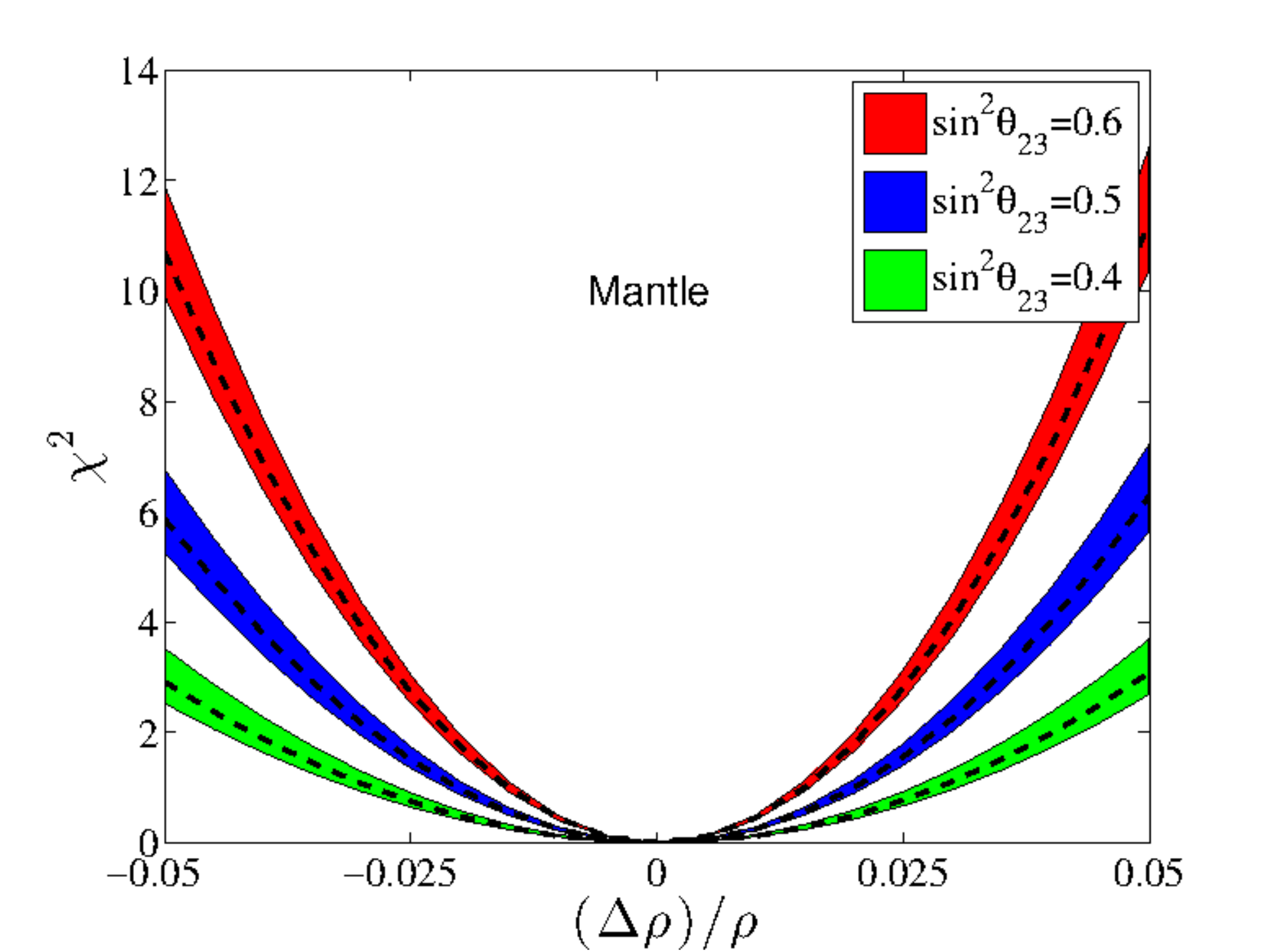}}
 \end{tabular}
 \caption{PINGU prospective sensitivity to the inner core 
(left panel), outer core (middle panel)
and mantle (right panel) densities in the case of NO spectrum 
and 10 years of data. 
The ratios $\Delta \rho/\rho$ in the three panels correspond 
respectively to $\Delta \rho_{\rm inner~core}$, $\Delta \rho_{\rm outer~core}$
and  $\Delta \rho_{\rm mantle}$ defined in the main text.
The results shown are for 
$\sin^2\theta_{23} = 0.40$, 0.50, 0.60. The bands correspond to 
variation of $\sin^2\theta_{13}$ in the interval 0.020 - 0.025 and 
marginalisation over $\theta_{23}$, $\Delta m^2_{31}$ and $\theta_{13}$. 
See text for further details. (Figures from \cite{ChSTP2014}.)
}
\label{fig:PINGU2014}
\end{figure}
%%%%%%%%%%%%%%%%%%%%%%%%%%%%%%%
%
%\noindent
 We show graphically in Fig. \ref{fig:PINGU2014} the results obtained 
in \cite{ChSTP2014} on PINGU sensitivity to the 
inner core, outer core and mantle densities for NO spectrum and after 
10 years of operation. The PINGU characteristics used in the analysis 
were taken from \cite{IceCube-PINGU:2014okk}, while the atmospheric 
neutrino fluxes are from \cite{Honda:2013}. Both $e$-like and $\mu$-like 
events were included in the study.
 The total Earth mass constraint was not implemented 
and only the statistical uncertainties were accounted for 
in the analysis. The results shown in 
Fig. \ref{fig:PINGU2014} were obtained for the following (true) 
values of the neutrino oscillation parameters:
  $\Delta m^2_{31(23)} = 2.46\times 10^{-3}\,{\rm eV^2}$,
$\Delta m^2_{21} = 7.6\times 10^{-5}\,{\rm eV^2}$
$\sin^2\theta_{12} = 0.32$, $\sin^2\theta_{23} = 0.50$ and $\delta = 0$.
The parameter $\sin^22\theta_{13}$ was varied in the interval 
(0.020 - 0.025). The sensitivities of interest were 
calculated with and without marginalisation 
over $\Delta m^2_{31(23)}$,  $\theta_{23}$ and $\theta_{13}$.
The differences between the two sets of results 
depend on the deviations from the PREM densities considered 
but are typically smaller than one unit of the relevant 
$\chi^2$ function. According to the results 
obtained in \cite{ChSTP2014} and illustrated graphically 
in Fig. \ref{fig:PINGU2014}, in the case of NO neutrino mass spectrum, 
PINGU in the configuration proposed in 
\cite{IceCube-PINGU:2014okk} and after 10 years 
of data-taking will have a rather good sensitivity
to the outer core and mantle densities 
and essentially no sensitivity to the inner core density.
More specifically, it follows from Fig. \ref{fig:PINGU2014}, in particular,  
that for, e.g., $\sin^2\theta_{23} = 0.5$, PINGU may 
be sensitive at $3\sigma$ C.L. to  
$\Delta \rho_{\rm outer~core} \cong \pm 17\%$ and 
at $2.5\sigma$ C.L. to  $\Delta \rho_{\rm mantle}\cong \pm 5\%$.

%%%%%%%%%%%%%%%%%%%%%%%%%%%%%%%%%%%%%%%%%%%%%%%%%%%%%%%%%%%%%%%%%%%%%
%


\begin{thebibliography}{99}


\bibitem{Bolt:1991}     
B. A. Bolt, Q. J. R. Astron. Soc. {\bf 32}, 367 (1991).
%
\bibitem{Yoder:1995} C. F. Yoder, ``Global Earth Physics'', vol. 1, 
(ed. T. J. Ahrens, American Geophysical Union, Washington DC, 1995), p. 1.
%
\bibitem{McDonough:2003}
W.F. McDonough, ``Treatise on Geochemistry: 
The Mantle and Core'', vol. 2 (ed. R. W. Carlson, 
Elsevier-Pergamon, Oxford, 2003), p. 547.
%
%\cite{McDonough:2008zz}
\bibitem{McDonough:2008zz}
W.~F.~McDonough and R.~Arevalo,
%``Uncertainties in the composition of Earth, its core and silicate sphere,''
J. Phys. Conf. Ser. \textbf{136} (2008), 022006
(doi:10.1088/1742-6596/136/2/022006).
%
\bibitem{Baffet:1991} B.A. Baffet, H.E. Huppert, J.R. Lister 
and A.W. Woods, Submitted to Nature 1991.
%
\bibitem{Kennett:1998}
B. L. N. Kennett, Geophys. J. Int. {\bf 132}, 374 (1998).
%
\bibitem{Masters:2003}
G. Masters and D. Gubbins, Phys. Earth Planet. Inter. {\bf 140}, 159 (2003).
%
\bibitem{PREM}
  A.~M.~Dziewonski and D.~L.~Anderson,
  %``Preliminary reference earth model,''
  Phys.\ Earth Planet.\ Interiors {\bf 25} (1981) 297.


%%%%%%%%%%%%%%% Attenuation Method %%%%%%%%%%%%%%%%%%%%

\bibitem{PlaZavatt1973} A. Placci and E. Zavattini, 
``On the possibility of using high-energy neutrinos to study the 
Earth's interior", https://cds.cern.ch/record/2258764 (1973), CERN Report.

\bibitem{VolkovaZatsepin1974}
 L. V. Volkova and G. T. Zatsepin, Izv. Akad. Nauk Ser. Fiz. 
{\bf 38N5} (1974) 1060.

\bibitem{Nedyalkov:1981}
 I.~P.  Nedyalkov, ``Notes on neutrino tomography'', 
      Acad. Bulgarian Sci. {\bf 34} (1981) 177. % 177–180.

\bibitem{Nedyalkov:1981yy}
I.~P. Nedyalkov, preprint JINR (Dubna), JINR-P2-81-645, 1981.

\bibitem{Nedyalkov:1982} I.~P. Nedyalkov,  
      ``On the study of the Earth composition by means of neutrino 
      experiments'', 
 Balatonfuered 1982, Proc. Neutrino ’82 {\bf 1} (1982) 300.

\bibitem{Nedyalkov:1983} I.~P. Nedialkov, ``Measurement of projected mass 
density - A basic problem 
 of neutrinogeophysics'',
 Acad. Bulgarian Sci. {\bf 36} (1983) 1515. % 1515-1518. 


\bibitem{DeRujula:1983ya}
A.~De~Rujula, S.~L. Glashow, R.~R. Wilson, G.~Charpak, 
Phys. Rept. {\bf  99} (1983) 341.


\bibitem{Wilson:1983an}
T.~L. Wilson, Nature {\bf 309} (1984) 38.

\bibitem{Askar:1984}
G.~A. Askar'yan,  Usp. Fiz. Nauk {\bf 144} (1984) 523 
[Sov. Phys. Usp. {\bf 27} (1984) 896].

\bibitem{Borisov:1986sm}
A.~B. Borisov, B.~A. Dolgoshein, A.~N. Kalinovsky,  Yad. Fiz. {\bf 44} 
(1986) 681.

\bibitem{Borisov:1989kh}
A.~B. Borisov, B.~A. Dolgoshein,  Phys. Atom. Nucl. {\bf 56} (1993) 755.

\bibitem{Winter:2006vg}
W.~Winter, Earth Moon Planets {\bf 99} (2006) 285.


\bibitem{Kuo95}
C.~Kuo, {\it et~al.},  Earth Plan. Sci. Lett. {\bf 133} (1995) 95.

\bibitem{Jain:1999kp}
P.~Jain, J.~P. Ralston, G.~M. Frichter, Astropart. Phys. {\bf 12}  
(1999) 193.

\bibitem{Reynoso:2004dt}
M.~M. Reynoso, O.~A. Sampayo,  Astropart. Phys. {\bf 21} (2004) 315.

%\cite{Gonzalez-Garcia:2007wfs}
\bibitem{Gonzalez-Garcia:2007wfs}
M.~C.~Gonzalez-Garcia, F.~Halzen, M.~Maltoni and H.~K.~M.~Tanaka,
%``Radiography of earth's core and mantle with atmospheric neutrinos,''
Phys. Rev. Lett. \textbf{100} (2008) 061802
% (doi:10.1103/PhysRevLett.100.061802)
[arXiv:0711.0745 [hep-ph]].

%\cite{ParticleDataGroup:2018ovx}
\bibitem{ParticleDataGroup:2018ovx}
M.~Tanabashi \textit{et al.} [Particle Data Group],
%``Review of Particle Physics,''
Phys. Rev. D \textbf{98} (2018) 030001
(doi:10.1103/PhysRevD.98.030001).
See therein the review ``Neutrino Masses, Mixing
and Oscillations'' by K. Nakamura and S.T. Petcov.

%\cite{Gaisser:2002jj}
\bibitem{Gaisser:2002jj}
T.~K.~Gaisser and M.~Honda,
%``Flux of atmospheric neutrinos,''
Ann. Rev. Nucl. Part. Sci. \textbf{52} (2002), 153-199
%(doi:10.1146/annurev.nucl.52.050102.090645)
[arXiv:hep-ph/0203272].


\bibitem{IceCube} R. Abbasi et al. [IceCube Collab.], 
Astropart. Phys. {\bf 35} (2012) 615;\\
M.G. Aartsen et al. [IceCube Collab.], 
Phys. Rev. Lett. {\bf 120} (2018) 071801;\\
 S. Blot [for the IceCube-PINGU Collab.], 
talk given at the XXIX Int. Conference on Neutrino
Physics and Astrophysics, Chicago, June 22 - July 2, 2020 
(virtual conference), DOI:10.5281/zenodo.3959546.

%\cite{IceCube-PINGU:2014okk} [12]
\bibitem{IceCube-PINGU:2014okk}
M.~G.~Aartsen \textit{et al.} [IceCube-PINGU],
``Letter of Intent: The Precision IceCube Next Generation Upgrade (PINGU),''
[arXiv:1401.2046 [physics.ins-det]].


\bibitem{PINGU} 
%\cite{IceCube:2016xxt}
% \bibitem{IceCube:2016xxt}
M.~G.~Aartsen \textit{et al.} [IceCube],
%``PINGU: A Vision for Neutrino and Particle Physics at the South Pole,''
J. Phys. G \textbf{44} (2017) 054006
% doi:10.1088/1361-6471/44/5/054006
[arXiv:1607.02671 [hep-ex]].


% \bibitem{ORCA} 
%\cite{KM3Net:2016zxf}
\bibitem{KM3Net:2016zxf}
S.~Adrian-Martinez \textit{et al.} [KM3Net],
``Letter of intent for KM3NeT 2.0,''
J. Phys. G \textbf{43} (2016) 084001
% (doi:10.1088/0954-3899/43/8/084001)
[arXiv:1601.07459 [astro-ph.IM]].
% ORCA: J. Brunner, PoS ICRC2015 1140 (2015).

\bibitem{HK} 
%\cite{Hyper-Kamiokande:2018ofw}
% \bibitem{Hyper-Kamiokande:2018ofw}
K.~Abe \textit{et al.} [Hyper-Kamiokande],
%``Hyper-Kamiokande Design Report,''
[arXiv:1805.04163 [physics.ins-det]];\\
M. Ishitsuka [on behalf of the Hyper-Kamiokande Proto-Collab.], 
talk given at the XXIX Int. Conference on Neutrino Physics and Astrophysics, 
Chicago, June 22 - July 2, 2020 (virtual conference), 
DOI:10.5281/zenodo.3959585.

%\cite{DUNE:2018tke}
\bibitem{DUNE:2018tke}
B.~Abi \textit{et al.} [DUNE],
%``The DUNE Far Detector Interim Design Report Volume 1: Physics, Technology and Strategies,''
[arXiv:1807.10334 [physics.ins-det]].

%\cite{Donini:2018tsg}
\bibitem{Donini:2018tsg}
A.~Donini, S.~Palomares-Ruiz and J.~Salvado,
%``Neutrino tomography of Earth,''
Nature Phys. \textbf{15} (2019) 37 % no.1, 37-40
(doi:10.1038/s41567-018-0319-1)
[arXiv:1803.05901 [hep-ph]].

%%%%%%%%%%% Start details of Donini et al, 1803 %%%

\bibitem{EarthMI1} B. Luzum {\it et al.}, Celest. Mech. Phys. {\bf 110} 
(2011) 110.

\bibitem{EarthMI2} H.USAO, USNO, UKHO, ``The Astronomical Almanac'', 
http://asa.usno.navy.mil/.

\bibitem{EarthMI3} W. Chen, J. Ray, W.B. Shen and C.L. Huang, 
J. Geod. {\bf 89} (2015) 179.

%\cite{IceCube:2016rnb}
\bibitem{IceCube:2016rnb}
M.~G.~Aartsen \textit{et al.} [IceCube],
%``Searches for Sterile Neutrinos with the IceCube Detector,''
Phys. Rev. Lett. \textbf{117} (2016) 071801
% (doi:10.1103/PhysRevLett.117.071801)
[arXiv:1605.01990 [hep-ex]].

%%%%%%%%%%%%%%End absorbtion method and details of Donini et al.;%%%%%%%%%%%%%% 
%%%%%%%%%%%%% Start Oscillation method  %%


%\cite{Winter:2015zwx}
\bibitem{Winter:2015zwx}
W.~Winter,
%``Atmospheric Neutrino Oscillations for Earth Tomography,''
Nucl. Phys. B \textbf{908} (2016), 250-267
% doi:10.1016/j.nuclphysb.2016.03.033
[arXiv:1511.05154 [hep-ph]].

\cite{Bourret:2017tkw}
\bibitem{Bourret:2017tkw}
S.~Bourret \textit{et al.} [KM3NeT],
%``Neutrino oscillation tomography of the Earth with KM3NeT-ORCA,''
J. Phys. Conf. Ser. \textbf{888} (2017) 012114  %no.1, 012114
% doi:10.1088/1742-6596/888/1/012114
[arXiv:1702.03723 [physics.ins-det]].

%\cite{Kumar:2021faw}
\bibitem{Kumar:2021faw}
A.~Kumar and S.~K.~Agarwalla,
%``Validating the Earth\textquoteright{}s core using atmospheric 
%  neutrinos with ICAL at INO,''
JHEP \textbf{08} (2021) 139
% doi:10.1007/JHEP08(2021)139
[arXiv:2104.11740 [hep-ph]].

%\cite{Ohlsson:1999um}
\bibitem{Ohlsson:1999um}
T.~Ohlsson and H.~Snellman,
%``Neutrino oscillations with three flavors in matter: Applications to neutrinos traversing the Earth,''
Phys. Lett. B \textbf{474} (2000)  153% 153-162
[erratum: Phys. Lett. B \textbf{480} (2000), 419-419]
%(doi:10.1016/S0370-2693(00)00008-3)
[arXiv:hep-ph/9912295 [hep-ph]].

\bibitem{Honda:2013}
%\cite{SajjadAthar:2012dji}
% \bibitem{SajjadAthar:2012dji}
M.~Sajjad Athar, M.~Honda, T.~Kajita, K.~Kasahara and S.~Midorikawa,
%``Atmospheric neutrino flux at INO, South Pole and Pyhasalmi,''
Phys. Lett. B \textbf{718} (2013), 1375-1380
% (doi:10.1016/j.physletb.2012.12.016)
[arXiv:1210.5154 [hep-ph]].
%
\bibitem{Yanez:2015} P. Yanez and A. Kouchner, 
Adv. High Energy Phys. 271968 (2015);\\
%\cite{Brunner:2015ltd}
% \bibitem{Brunner:2015ltd}
J.~Brunner [KM3NeT],
%``KM3NeT - ORCA: Measuring neutrino oscillations and the mass hierarchy in the Mediterranean Sea,''
PoS \textbf{ICRC2015} (2016) 1140
(doi:10.22323/1.236.1140).

%\cite{Gonzalez-Garcia:2014bfa}
\bibitem{Gonzalez-Garcia:2014bfa}
M.~C.~Gonzalez-Garcia, M.~Maltoni and T.~Schwetz,
%``Updated fit to three neutrino mixing: status of leptonic CP violation,''
JHEP \textbf{11} (2014) 052
% (doi:10.1007/JHEP11(2014)052)
[arXiv:1409.5439 [hep-ph]].

%\cite{ICAL:2015stm}
\bibitem{ICAL:2015stm}
S.~Ahmed \textit{et al.} [ICAL],
%``Physics Potential of the ICAL detector at the 
% India-based Neutrino Observatory (INO),''
Pramana \textbf{88} (2017) 79
% (doi:10.1007/s12043-017-1373-4)
[arXiv:1505.07380 [physics.ins-det]].

% \cite{Agarwalla:2012uj}
\bibitem{Agarwalla:2012uj}
S.~K.~Agarwalla, T.~Li, O.~Mena and S.~Palomares-Ruiz,
%``Exploring the Earth matter effect with atmospheric neutrinos in ice,''
[arXiv:1212.2238 [hep-ph]].

%\cite{Rott:2015kwa}
\bibitem{Rott:2015kwa}
C.~Rott, A.~Taketa and D.~Bose,
%``Spectrometry of the Earth using Neutrino Oscillations,''
Sci. Rep. \textbf{5} (2015), 15225
doi:10.1038/srep15225
[arXiv:1502.04930 [physics.geo-ph]].

%\cite{Bourret:2020zwg}
\bibitem{Bourret:2020zwg}
S.~Bourret, J.~Coelho, E.~Kaminski and V.~Van Elewyck,
%``Probing the Earth Core Composition with Neutrino Oscillation Tomography,''
PoS \textbf{ICRC2019} (2020), 1024
doi:10.22323/1.358.1024
%
\bibitem{ChGSTP2011} S. Choubey, P. Ghoshal and S.T. Petcov, 
studies performed in the period 2008 - 2011, 
unpublished.
%
\bibitem{ChSTP2014} S. Choubey and S.T. Petcov, 
studies performed in 2014, unpublished.
%
%\cite{Honda:2015fha}
\bibitem{Honda:2015fha}
M.~Honda, M.~Sajjad Athar, T.~Kajita, K.~Kasahara and S.~Midorikawa,
%``Atmospheric neutrino flux calculation using the NRLMSISE-00 
% atmospheric model,''
Phys. Rev. D \textbf{92} (2015)  023004
%(doi:10.1103/PhysRevD.92.023004)
[arXiv:1502.03916 [astro-ph.HE]]. 
The tables for the fluxes and production height calculated
for the Frejus/Gran Sasso cite are available in the
web page, http://www.icrr.u-tokyo.ac.jp/~mhonda 
cited in this paper.


%\cite{Capozzi:2017syc}
\bibitem{Capozzi:2017syc}
F.~Capozzi, E.~Lisi and A.~Marrone,
%``Probing the neutrino mass ordering with KM3NeT-ORCA: 
% Analysis and perspectives,''
J. Phys. G \textbf{45} (2018) 024003
% (doi:10.1088/1361-6471/aa9503)
[arXiv:1708.03022 [hep-ph]].

%\cite{Ohlsson:2001ck}
\bibitem{Ohlsson:2001ck}
T.~Ohlsson and W.~Winter,
%``Reconstruction of the earth's matter density profile using a 
% single neutrino baseline,''
Phys. Lett. B \textbf{512} (2001) 357 % -364
% (doi:10.1016/S0370-2693(01)00731-6}
[arXiv:hep-ph/0105293 [hep-ph]].

%\cite{Petcov:1998su}
\bibitem{Petcov:1998su}
S.~T.~Petcov,
%``Diffractive - like (or parametric resonance - like?) enhancement of the earth (day - night) effect for solar neutrinos crossing the earth core,''
Phys. Lett. B \textbf{434} (1998) 321 % 321-332
(doi:10.1016/S0370-2693(98)00742-4)
[arXiv:hep-ph/9805262 [hep-ph]];
Phys. Lett. B \textbf{444} (1998) 584 (Erratum).

%\cite{Chizhov:1999az}
\bibitem{Chizhov:1999az}
M.~V.~Chizhov and S.~T.~Petcov,
%``New conditions for a total neutrino conversion in a medium,''
Phys. Rev. Lett. \textbf{83} (1999) 1096 %, 1096-1099
% (doi:10.1103/PhysRevLett.83.1096)
[arXiv:hep-ph/9903399 [hep-ph]].

%\cite{Chizhov:1999he}
\bibitem{Chizhov:1999he}
M.~V.~Chizhov and S.~T.~Petcov,
%``Enhancing mechanisms of neutrino transitions in a medium of nonperiodic constant density layers and in the earth,''
Phys. Rev. D \textbf{63} (2001) 073003
% (doi:10.1103/PhysRevD.63.073003)
[arXiv:hep-ph/9903424 [hep-ph]].

%\cite{Chizhov:1998ug}
\bibitem{Chizhov:1998ug}
M.~Chizhov, M.~Maris and S.~T.~Petcov,
%``On the oscillation length resonance in the transitions of solar and atmospheric neutrinos crossing the earth core,''
[arXiv:hep-ph/9810501 [hep-ph]].


%\cite{Akhmedov:2006hb}
% \bibitem{Akhmedov:2006hb}
\bibitem{AMS:2006hb}
E.~K.~Akhmedov, M.~Maltoni and A.~Y.~Smirnov,
%``1-3 leptonic mixing and the neutrino oscillograms of the Earth,''
JHEP \textbf{05} (2007) 077
% (doi:10.1088/1126-6708/2007/05/077)
[arXiv:hep-ph/0612285 [hep-ph]].
%
% \cite{Akhmedov:2005yj}
% \bibitem{Akhmedov:2005yj}
\bibitem{AMS:2005yj}
E.~K.~Akhmedov, M.~Maltoni and A.~Y.~Smirnov,
%``Oscillations of high energy neutrinos in matter: Precise formalism and parametric resonance,''
Phys. Rev. Lett. \textbf{95} (2005) 211801
% doi:10.1103/PhysRevLett.95.211801
[arXiv:hep-ph/0506064 [hep-ph]].
%
\bibitem{MSW1}
L.~Wolfenstein, 
%\prD17,2369(1978)
Phys. Rev. D {\bf 17} (1978) 2369;
% \reference{*MSW1}
{\it Proc. of the 8th International Conference on Neutrino Physics 
and Astrophysics - ``Neutrino'78"}
(ed. E.C. Fowler, Purdue University Press, West Lafayette, 1978), p. C3.
%
\bibitem{Barger80}
% \reference{Barger80}
V. Barger {\it et al.}, 
Phys. Rev. D {\bf 22} (1980) 2718.
%
\bibitem{Langa83}
% \reference{Langa83}
P. Langacker, J.P. Leveille and J. Sheiman, 
Phys. Rev. D {\bf 27} (1983) 1228.
%

%%%%%%%%%%%% Composition of the Earth Core and Mantle %%%%%%%%%%%%%

\bibitem{Bardo:2015}
 J. Badro {\it et al.}, 
Proc. Natl. Acad. Sci. U. S. A. {\bf 112(40)} (2015) 12310-12314.

\bibitem{Kaminski:2013} E. Kaminski and M. Javoy, Earth Plan. Sci. Lett.
{\bf 365} (2013) 97-107.

\bibitem{Sakamaki:2009} T. Sakamaki {\it et al.}, 
Earth and Planetary Science 
Letters {\bf 287} (2009) 293-297.

\bibitem{EarthRef} https://earthref.org/GERMRD/datamodel/ 
(cit. on pp. 114, 115).

%%%%%%%%%%%%%%%%% Accounting for Matter effects in P_{\alpha\beta}

\bibitem{MSW2}
S.P. Mikheev and A.Y. Smirnov,
% \sjnp42,913(1985); \nc9C,17(1986)
Sov. J. Nucl. Phys. {\bf 42} (1985) 913.
% ; Nuovo Cimento {\bf 9C} (1986) 17.
%
%\cite{Fogli:2012ua}
\bibitem{Fogli:2012ua}
G.~L.~Fogli, E.~Lisi, A.~Marrone, D.~Montanino, A.~Palazzo and A.~M.~Rotunno,
%``Global analysis of neutrino masses, mixings and phases: entering the era of leptonic CP violation searches,''
Phys. Rev. D \textbf{86} (2012) 013012
%(doi:10.1103/PhysRevD.86.013012)
[arXiv:1205.5254 [hep-ph]].
%
%\cite{Lisi:1997yc}
\bibitem{Lisi:1997yc}
E.~Lisi and D.~Montanino,
%``Earth regeneration effect in solar neutrino oscillations: 
% An Analytic approach,''
Phys. Rev. D \textbf{56} (1997) 1792 % 1792-1803
% (doi:10.1103/PhysRevD.56.1792)
[arXiv:hep-ph/9702343 [hep-ph]].
%
%\cite{Capozzi:2015bxa}
\bibitem{Capozzi:2015bxa}
F.~Capozzi, E.~Lisi and A.~Marrone,
%``PINGU and the neutrino mass hierarchy: Statistical and systematic aspects,''
Phys. Rev. D \textbf{91} (2015) 073011
% (doi:10.1103/PhysRevD.91.073011)
[arXiv:1503.01999 [hep-ph]].
%
%%%%%%%%%%%%%%%%%%%%% Details of the Calculations %%%%%%%%%%%%%%%

%\cite{Capozzi:2017ipn}
\bibitem{Capozzi:2017ipn}
F.~Capozzi, E.~Di Valentino, E.~Lisi, A.~Marrone, A.~Melchiorri and A.~Palazzo,
%``Global constraints on absolute neutrino masses and their ordering,''
 Phys. Rev. D \textbf{95} (2017) 096014
% (doi:10.1103/PhysRevD.95.096014)
[arXiv:1703.04471 [hep-ph]];
Phys. Rev. D \textbf{101} (2020) 116013 (addendum)
[arXiv:2003.08511 [hep-ph]].
%
%\cite{Capozzi:2021fjo}
\bibitem{Capozzi:2021fjo}
F.~Capozzi, E.~Di Valentino, E.~Lisi, A.~Marrone, A.~Melchiorri and A.~Palazzo,
%``The unfinished fabric of the three neutrino paradigm,''
[arXiv:2107.00532 [hep-ph]].
%
%\cite{Akhmedov:1998xq}
%\bibitem{Akhmedov:1998xq}
\bibitem{ADLS1998}
E.~K.~Akhmedov, A.~Dighe, P.~Lipari and A.~Y.~Smirnov,
%``Atmospheric neutrinos at Super-Kamiokande and parametric resonance in neutrino oscillations,''
Nucl. Phys. B \textbf{542} (1999) 3 % 3-30
%(doi:10.1016/S0550-3213(98)00825-6)
[arXiv:hep-ph/9808270 [hep-ph]].

%\cite{Bernabeu:2003yp}
\bibitem{Bernabeu:2003yp}
J.~Bernabeu, S.~Palomares Ruiz and S.~T.~Petcov,
%``Atmospheric neutrino oscillations, theta(13) and neutrino mass hierarchy,''
Nucl. Phys. B \textbf{669} (2003) 255  %255-276
% doi:10.1016/j.nuclphysb.2003.07.025
[arXiv:hep-ph/0305152 [hep-ph]].

%\cite{Petcov:2005rv}
\bibitem{Petcov:2005rv}
S.~T.~Petcov and T.~Schwetz,
%``Determining the neutrino mass hierarchy with atmospheric neutrinos,''
Nucl. Phys. B \textbf{740} (2006) 1 % 1-22
% doi:10.1016/j.nuclphysb.2006.01.020
[arXiv:hep-ph/0511277 [hep-ph]].

\bibitem{Fogli:2002pt}
G.~L.~Fogli, E.~Lisi, A.~Marrone, D.~Montanino and A.~Palazzo,
%``Getting the most from the statistical analysis of solar 
% neutrino oscillations,''
Phys. Rev. D \textbf{66} (2002) 053010.

 %\cite{Esteban:2020cvm}
 \bibitem{Esteban:2020cvm}
 I.~Esteban, M.~C.~Gonzalez-Garcia, M.~Maltoni, T.~Schwetz and A.~Zhou,
 %``The fate of hints: updated global analysis of three-flavor neutrino oscilla% tions,''
 JHEP \textbf{09} (2020) 178
 % (doi:10.1007/JHEP09(2020)178)
 [arXiv:2007.14792 [hep-ph]].
%
%\cite{Hofestadt:2019whx}
\bibitem{Hofestadt:2019whx}
J.~Hofest\"adt, M.~Bruchner and T.~Eberl,
%``Super-ORCA: Measuring the leptonic CP-phase with Atmospheric Neutrinos and Beam Neutrinos,''
PoS \textbf{ICRC2019} (2020) 911
% (doi:10.22323/1.358.0911)
[arXiv:1907.12983 [hep-ex]].
%
%\cite{Kelly:2021jfs}
\bibitem{Kelly:2021jfs}
K.~J.~Kelly, P.~A.~N.~Machado, I.~Martinez-Soler and Y.~F.~Perez-Gonzalez,
%``DUNE atmospheric neutrinos: Earth Tomography,''
[arXiv:2110.00003 [hep-ph]].
%
%\cite{Denton:2021rgt}
\bibitem{Denton:2021rgt}
P.~B.~Denton and R.~Pestes,
%``Neutrino Oscillations through the Earth's Core,''
[arXiv:2110.01148 [hep-ph]].
%
%\cite{Gandhi:2007td}
\bibitem{Gandhi:2007td}
R.~Gandhi, P.~Ghoshal, S.~Goswami, P.~Mehta, S.~U.~Sankar and S.~Shalgar,
%``Mass Hierarchy Determination via future Atmospheric Neutrino Detectors,''
Phys. Rev. D \textbf{76} (2007), 073012
doi:10.1103/PhysRevD.76.073012
[arXiv:0707.1723 [hep-ph]].

%\cite{Gandhi:2008zs}
\bibitem{Gandhi:2008zs}
R.~Gandhi, P.~Ghoshal, S.~Goswami and S.~U.~Sankar,
%``Resolving the Mass Hierarchy with Atmospheric Neutrinos using a Liquid Argon Detector,''
Phys. Rev. D \textbf{78} (2008), 073001
doi:10.1103/PhysRevD.78.073001
[arXiv:0807.2759 [hep-ph]].

% \cite{Honda:2004yz}
\bibitem{Honda:2004yz}
M.~Honda, T.~Kajita, K.~Kasahara and S.~Midorikawa,
%``A New calculation of the atmospheric neutrino flux 
`` in a 3-dimensional scheme,''
Phys. Rev. D \textbf{70} (2004) 043008
% doi:10.1103/PhysRevD.70.043008
[arXiv:astro-ph/0404457 [astro-ph]].

\bibitem{Rubbia:2004tz}
  A.~Rubbia,
  %``Experiments for CP violation: A Giant liquid argon scintillation, Cerenkov
  %and charge imaging experiment?,''
  arXiv:hep-ph/0402110.

\bibitem{Bueno:2007um}
  A.~Bueno {\it et al.},
  %``Nucleon decay searches with large liquid argon TPC detectors at shallow
  %depths: Atmospheric neutrinos and cosmogenic backgrounds,''
  JHEP {\bf 0704} (2007) 041
  [arXiv:hep-ph/0701101].


\bibitem{Cline:2006st}
  D.~B.~Cline, F.~Raffaelli and F.~Sergiampietri,
  %``LANNDD: A line of liquid argon TPC detectors scalable in mass from 200 Tons
  %to 100 Ktons,''
  JINST {\bf 1} (2006) T09001
  [arXiv:astro-ph/0604548].

\bibitem{Ereditato:2005yx}
  A.~Ereditato and A.~Rubbia,
  %``Conceptual design of a scalable multi-kton superconducting magnetized
  %liquid Argon TPC,''
  Nucl.\ Phys.\ Proc.\ Suppl.\  {\bf 155} (2006) 233
  [arXiv:hep-ph/0510131].




\end{thebibliography}
\end{document}